\documentclass[prd,twocolumn,aps,preprintnumbers,letterpaper,floatfix]{revtex4}
\usepackage{amsmath}
\usepackage{tipa}
\usepackage{amssymb}
\usepackage{graphicx,subfigure,amsbsy}
\usepackage{bm}
\usepackage{enumerate}
\usepackage{color}
\newcommand{\be}{\begin{equation}}
\newcommand{\ee}{\end{equation}}
\newcommand{\bea}{\begin{eqnarray}}
\newcommand{\eea}{\end{eqnarray}}
\newcommand{\ba}{\begin{eqnarray}}
\newcommand{\ea}{\end{eqnarray}}

\newcommand{\GeV}{\,\mathrm{GeV}}
\newcommand{\MeV}{\,\mathrm{MeV}}
\newcommand{\fm}{\mathrm{fm}}

\begin{document}

\title{Heavy Ion Collisions: \\
Achievements and Challenges}

\author{Edward Shuryak}
\affiliation{Department of Physics and Astronomy, Stony Brook University,\\ Stony Brook, New York 11794-3800, USA}

\date{\today}

\begin{abstract}
A decade ago brief summary of the field could be formulated as a discovery of strongly-coupled 
Quark-Gluon-Plasma, sQGP,
making a very good liquid with surprisingly small viscosity. Since 2010 we have LHC program, which added a lot to
our understanding, and now there seems to be a need to consolidate what we learned and formulate a
list of issues to be studied next. Hydrodynamical perturbations, leading to higher
harmonics of angular correlations, are identified as long-lived sound waves. Recently studied reactions involving 
sounds include phonon decays into two (``loop viscosity"), phonon+magnetic field into photons/dileptons
(sono-magneto-luminescence), and two phonons into a gravity wave, a  
 penetrating probe of the Big Bang. The mainstream issues in the field now include a quest
to study transition between $pp,pA$ and heavy ion $AA$ collisions, with an aim to locate
"the smallest drops" of the sQGP 
displaying collective/hydrodynamics behavior. The issues related to out-of-equilibrium stage of the collisions, and mechanisms
of the equilibration, in weak and strong coupling, are also hotly debated.  \end{abstract}

\maketitle

\section{A wider picture}
Before we immense ourselves in recent discoveries and puzzles of heavy ion physics itself, one perhaps should look
in a  wider field of "strong interaction physics". Its definitions have many different meanings,
ranging from   a narrow one, (i) 
hadronic and nuclear physics phenomena based on its fundamental theory, QCD; to much wider one (ii)  aiming at understanding of 
 all gauge theories, including QCD-like and supersymmetric ones, and even 
 including dualities of those to string/gravity theories. 
We will keep mostly to the former end, with only some sections venturing to the latter definition.

QCD-like theories at finite temperature and density has a number of regimes and phases. Due  to asymptotic freedom, 
the simplest of them to understand is the high $T$ (or $\mu$ or both) limit, known as
{\em  weakly coupled} Quark-Gluon Plasma (wQGP). Well developed tools of the manybody theory -- perturbative diagrams and their re-summations --   
were used to show its main features, the electric screening and formation of finite-mass quasiparticles.  This is where the
name ``plasma" came from \cite{Shuryak:1977ut}. For early reviews on high-T QCD see 
\cite{Shuryak:1980tp,Gross:1980br}.

Prior to RHIC era, it was generally assumed that wQGP regime extends down to the phase transition region, $T>T_c$. Lattice-based  thermodynamic quantities  showed relatively small  ( 20\% or less) deviation from those
for non-interacting quarks and gluons. 
Many if not most theorist were 
skeptical, thinking that RHIC program aimed at ``production of new form of matter" would basically fail. 
Perturbative equilibration estimates indicated unrealistically long time.  Models predicted ``firework of minijets"
rather than collective effects.

Yet strong hydrodynamical flows were observed at RHIC from the first months,
supplemented by strong jet quenching.  All of that was later confirmed at the LHC, which provided 
even much more accurate
data on many details.  For example, as we will discuss below, there  are unmistakable signs that the excitation modes of the matter produces is basically the sound modes. The systematics of those lead to 
 viscosity value which is remarkably small, more than an order of magnitude smaller than that
predicted originally by the perturbative theory. So, the window in which
 matter produced at RHIC/LHC is, $T_c<T<2T_c$, was renamed into a {\em ``strongly coupled QGP"},  sQGP for short. 

So, which  physical phenomena take place there and why? What exactly happens as $T$ decreases from the wQCD to sQGP domain?
The growing coupling  induces
two different kinds of phenomena. 

(i) One preserves  the basic picture of quark and gluon quasiparticles and plasma-related phenomena,
with gradual change of parameters consistent with smooth
{\em logarithmic} running $g^2 \sim 1/log(T/\Lambda)$. 
While in  the wQCD domain the electric
screening (Debye) mass  are small compared to their energy,
$M_E/T=O(g) \ll 1$,  in sQGP domain  one learned from  lattice simulations that those are not small,
$M_E/T=2-3>1$ . So, while expansions in such ratio used in the wQGP in this domain are invalid,  perhaps 
an extrapolation using the running coupling
may still hold, at least qualitatively. Similar arguments were put forward for the  quasiparticle masses, thermodynamical and  kinetic quantities as well, with variable rate of success depending on the quantity under consideration.
 
(ii) Moreover, there are other type of phenomena, which depend  on the coupling $exponentially$, leading to
much stronger {\em power} dependences on the temperature
\be exp\left(-{const\over g^2(T)}\right) \sim ({\Lambda \over T})^{power} \ee 
Those come from gauge field configurations -- semiclassical  solitons -- which are
 qualitatively distinct from perturbative quarks and gluons. 
One such example is  the (Euclidean 4-dimensional)  topological solitons called $instantons$. Their physics is   tunneling through certain barriers
into ``valleys" of topologically distinct gauge field configurations. 
Perturbative  theory , an expansion around zero fields, knows little about 
 existence of such valleys.  (Except  in a very
subtle way, via divergences of its series.There is  recent progress in this direction, known as a {\em resurgence theory},
which however still focuses on quantum mechanical and low-dimensional scalar theories rather than gauge theories.
See for recent update in the talks at \cite{resurgence_workshop} .) 

Discussion of the instanton theory will take us too far from our main direction, so I only
mention its two major breakthroughs. In 1980-1990's it has been shown how instantons {\em break the chiral symmetries}, the $U_A(1)$ explicitly and $SU(N_f)$ spontaneously,
via collectivization of fermionic zero modes, for a review see \cite{Schafer:1996wv}. Recently  
it morphed into plasma made of instanton constituents,  Lee-Li-Kraan-van Baal (LLKvB) {\em instanton-dyons}, or instanton-monopoles.  It has been shown that those, if dense enough, can  naturally enforce $both$ confinement
and chiral symmetry breaking. For recent discussion
see talks at \cite{topology_workshop}. 

Another -- still very puzzling -- set of effects, predicted by the Operator Product Expansion (OPE), are related with
``higher twist effects" and
VEVs of various operators appearing in the QCD sum rule framework. 
The so called $renormalons$ are supposed to induce effects with various integer powers of the scale
, e.g. $T$. In the gauge theory
we still do not know which solitons, if any, are related to them, but in simpler theories there was recent progress: see again \cite{resurgence_workshop}.

An extremely important notion of the theory is that of {\em electric-magnetic duality} . Asymptotically free 
 gauge theories, QCD-like or supersymmetric ones, have weak ``electric" coupling in the UV, but in the IR
they generically flow to strong coupling. ``Magnetic" objects, such as color-magnetic monopoles,
are expected to become lighter and more numerous along this process. 
 The key observation I tirelessly  emphasize  is the good old {\em Dirac condition}.  If one wants simultaneously
 describe objects with electric and magnetic charges ($e$ and $g$ in his old notations/normalization) 
 singularities of the gauge potential of the form of the Dirac strings are unavoidable. The requirement that  should be pure gauge artifacts and thus invisible leads to
 \be  {2 e g \over \hbar c }= 1 
 \ee  
(or in general an integer in the r.h.s.). Thus $e$ and $g$ must run in unison, in the opposite directions. 
Thus one expects 
in the infrared limit of gauge theories to find the so called {\em dual description} 
 based on magnetic degrees of freedom,  which are  light and weakly coupled.   The question is
  whether one can construct a Lagrangian of this dual theory,  and also if/when one  can/needs to use it.

(One great example of the kind is provided by celebrated $\cal{N}$=2 supersymmetric Seiberg-Witten theory: near certain 
special points of the Higgs VEV the monopoles are nearly massless and weakly interacting: the magnetic theory
is then scalar QED. Its beta function, as expected, has the opposite sign to that of the electric theory. 
Even greater example is $\cal{N}$=4 theory: it is electric-magnetic selfdual, so  the beta function is equal to
minus itself and thus is identically zero.)
 
  \begin{figure}[t]
  \begin{center}
  \includegraphics[width=8cm]{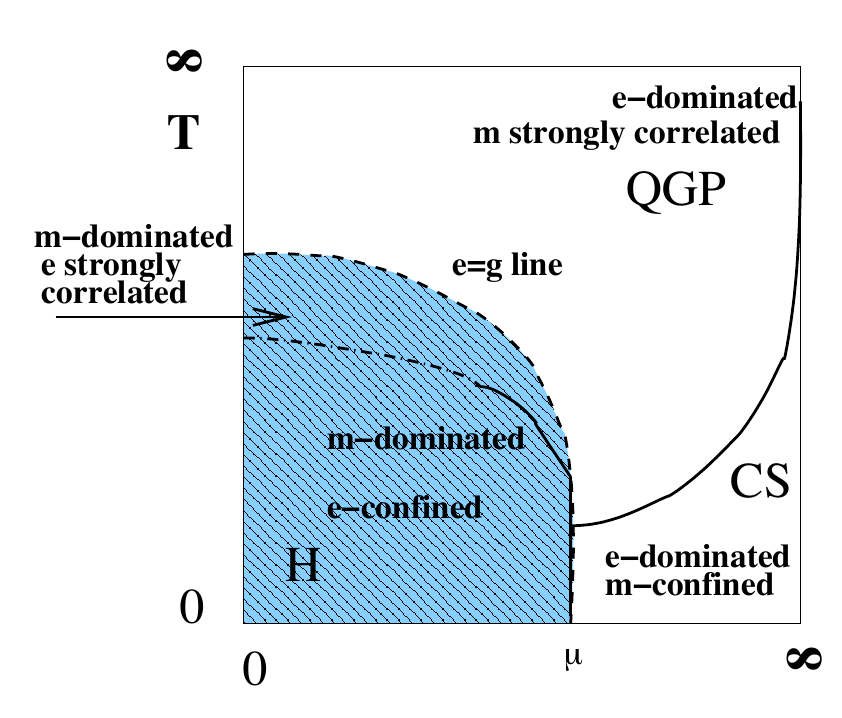}
   \caption{ A schematic phase diagram on a (ÒcompactifiedÓ) plane of temperature and baryonic chemical potential  $T - \mu$, from \cite{Liao:2006ry}. The (blue) shaded region shows Òmagneti- cally dominatedÓ region $g < e$, which includes the e-confined hadronic phase as well as ÒpostconfinedÓ part of the QGP domain. Light region includes Òelectrically dominatedÓ part of QGP and also color superconductivity (CS) region, which has e-charged diquark condensates and therefore obviously m-confined. The dashed line called Òe=g lineÓ is the line of electric-magnetic equilibrium. The solid lines indicate true phase transitions, while the dash-dotted line is a deconfinement cross-over line.
}
  \label{fig_EM_duality}
  \end{center}
\end{figure}

Returning to QCD, one can summarize the picture of the so called ``magnetic scenario" by a schematic plot shown in
Fig. \ref{fig_EM_duality}, from \cite{Liao:2006ry}.  At the top -- the high $T$ domain -- and at the right -- the high density
domain -- one finds 
weakly coupled or ``electrically dominated" regimes, or wQGP. On the contrary, near the origin
of the plot, in vacuum,   the electric fields are, 
subdominant and confined into the flux tubes. The vacuum is filled by the magnetically charged condensate,
known as ``dual superconductor". The region in between (relevant for matter produced at RHIC/LHC) is 
close to  the ``equilibrium line", marked by $e=g$ on the plot. (People for whom couplings are too abstract, can for example 
define it by an equality of the electric and magnetic screening masses.) In this region both electric and magnetic coupling
are equal and thus $\alpha_{electric}=\alpha_{magnetic}=1$: so neither the electric nor magnetic formulations of the theory are simple. 

Do we have any evidences for a presence or importance for heavy ion physics of ``magnetic" objects? Here are some arguments for that
based on lattice studies and phenomenology, more or less in historical order:

(i) In the RHIC/LHC region $T_c<T<2T_c$ the VEV of the Polyakov line $<P>$ is substantially different from 1. Hidaka and
Pisarski \cite{Hidaka:2008dr} argued that $<P>$ must be incorporated into density of thermal quarks and gluons, and thus
suppress their  contributions.  They called such matter ``semi-QGP"
emphasizing that say half of their degrees of freedom are actually operational. 

And yet, the lattice data insist that the
  thermal  energy density remains close to the $T^4$ trend nearly
  all the way to $T_c$. ``Magnetic scenario" \cite{Liao:2006ry}
 may explain this puzzle by introducing extra contributions of the magnetic quasiparticles. (They are not subject to  $<P>$ suppression because they lack the electric charge.)
  
  (ii) Lattice monopoles so far defined are gauge dependent. And yet, they
  were found to behave as physical particles. Their motion definitely shows Bose-Einstein condensation at $T<T_c$ \cite{D'Alessandro:2010xg}. Their spatial correlation
functions are very much plasma-like. Even more striking is
the observation \cite{Liao:2008jg} revealing magnetic coupling which $grows$ with $T$. Moreover, it is indeed an
inverse of the asymptotic freedom curve. 

(iii) Plasmas with electric and magnetic charges show unusual transport properties \cite{Liao:2006ry}: Lorenz force enhances
  collision rate and reduce viscosity. Quantum gluon-monopole scattering leads to large transport cross section \cite{Ratti:2008jz},
  providing small viscosity in the range close to that observed at RHIC/LHC.
  
 (iv) high density of (non-condensed) monopoles near $T_c$ affect electric flux tubed, and perhaps explain \cite{Liao:2008vj} 
 lattice observations of high tension in the potential energy (not free energy)
 of the heavy-quark potentials . We will discuss those in section \ref{sec_charm}.

(v) Last but not least, high density of monopoles near $T_c$ seem to be directly relevant for 
jet quenching. This issue- a very important part of heavy ion physics - we will discuss in section \ref{sec_quenching}.

(Completing this brief introduction to monopoles, it is impossible not to mention the remaining unresolved issues.
Theories with adjoint scalar fields -- such as e.g. celebrated $\cal{N}$=2 Seiberg-Witten theory -- naturally have
particle-like monopole solutions. Yet in QCD-like theories without scalars it has to be dynamically generated,
not yet explained.
So far lattice definitions of monopoles are gauge-dependent. A good news is that most if not all of
monopole physics can be taken care of via the instanton-dyons we mentioned above: in this case
the role of the adjoint ``Higgs"  is played by the time component of the gauge potential $A_4$. 
It is real on the Euclidean lattice, but would get $i$ in Minkowski continuation: so it cannot be directly a
``particle" in ordinary sense. )

Another famous duality is  AdS/CFT gauge-string duality \cite{Maldacena:1997re},  basically used 
at large number of colors as a gauge-gravity
duality.  It directly relates equilibrated QGP at strong coupling  to a certain black hole solutions in 5 dimensions,
with the plasma temperature being its Hawking temperature, and the QGP entropy 
being its Bekenstein entropy. In section
\ref{sec_holo_eql} we will discuss recent studies of out-of-equilibrium settings, in which black hole
is dynamically generated. Holographic models of the AdS/QCD types also lead to new 
views on the QCD strings, Reggeons and Pomerons: see section \ref{sec_Pomeron}.

This brief introduction shows that heavy ion physics did acquired a lot of outreach into many other
areas of physics.

\section{The main issues in QCD and heavy ion physics}
Like any other rapidly developing field, heavy ion physics has
 has ``growth problems". Quark Matter conferences (and their proceedings)  provide  
  regular snapshots of the field's development.  Yet those provide poor service
  for young people or outsiders who want to understand what is going on.
  The talk time is always severely limited, so
 naturally only the latest  results are reported. 
 They are kaleidoscopic set  of compressed answers
   to  questions which were never formulated, as there is no time for that. 
Discussion session are becoming extremely short (or non-existing), and 
   never recorded. Furthermore, large conferences have
multiple parallel sessions, and it is hard to follow more than one subfield. 
   
    Traditionally the ``second pillar" of any field consists of books, reviews, lecture notes.  
  Those formulate the basic questions and the
  the lessons learned. They also  provide a list of
    ``challenges", the questions which  need to be answered next.
With ongoing RHIC program and the now maturing  LHC heavy ion program, 
it is perhaps time for a review.
  Looking through the talks  
 at the latest Quark Matter conferences, I feel a
relative weakness of the theory, compared to overwhelming amount of  data from RHIC and LHC.
 and decided it is time to do it.
This review  certainly is not complete,  many important directions/achievements are not
covered. Jet quenching is nearly missing.  
 Fluctuations in the gauge topology, related to chiral magnetic effect,  are not discussed: 
it needs to be developed further before some summary emerges.  
It is a subjective overview of the field, focusing on big picture, questions and trends rather than specific results.\\

(On a personal note:  the
 first  review I wrote \cite{Shuryak:1980tp},
 on finite temperature QCD  proclaiming 
a search for QGP as a new goal for community, was written due to an advice given to me by Evgeny Feinberg.
Let me also mention here Gerry Brown: long before I met him I benefited from his vision, in founding
Physics Reports devoted to such reviews of subfields. The instanton review in RMP of 1996 \cite{Schafer:1996wv}
is in fact my most cited paper. 
In 2004, when the sQGP paradigm of ``a perfect liquid" has been  born, many people wrote summary papers.
My most extensive version was the second part of my book \cite{mybook}.)

 Let me start with few ``super-questions"  (and comments on them), which are common
to the whole strong interaction physics,  extending well beyond the boundaries of the heavy ion field.

I.{\bf Can one locate the ``soft-to-hard" boundary},  in whatever observables under consideration, where  the transition from  weak  to strong  coupling 
regimes take place?

II.  {\bf  Can one locate the `micro-to- macro" boundary }, where the transition from large mean-free-path (ballistic)
to small one (hydrodynamic) regimes happen? In particular, where in the observable discussed one finds a
transition from effects due to single-parton
distributions   to those due to collective 
explosion?

III. {\bf Can we experimentally identify signals of the QCD phase transition}, in particularly locate the {\em "softest"} and {\em the QCD 
critical point?}

Brief comments on them are:

(Ia)  Since 1970's the ``golden" test of pQCD at large momenta transfer was deep inelastic scattering (DIS), as well as
hard exclusive processes, e.g. the pion and nucleon formfactors. 
  Unlike jets in DIS, the latter have colorless final state, they can  in principle be accurately measured,  and thus promised to test the pQCD
predictions due to the specific lowest order diagrams.
 And yet, to the highest $Q^2\sim 4\, GeV$ measured so far, neither 
of them had been found to reach quantitative  agreement with  the pQCD predictions,
because even at such $Q$ the non-perturbative effects still dominate.

Closer to our field is the ``mini-jet" issue. While the identified
 jets have rather large momenta, say $p_\perp > 20\, GeV$ or so, it is widely assumed that  the
parton description is good down to much smaller momenta.  How much smaller? Following DGLAP evolution 
 toward  
small   $Q^2$ all the way to $\sim 1\, GeV^2$ one eventually reach a negative gluon density.
Many other arguments tell us that at this  scale, $1 \, GeV$, pQCD cannot be used. 
 At which scale $Q_{min}$ one has to stop is defined by the ``higher twist effects", not yet studied quantitatively.
 Thus we are  still  left with uncomfortably large  gap, between 1 and 20 $GeV$. 

(Ib) The elementary process fundamental for our field is the $pp$  scattering.  Its total cross section and elastic amplitude
is described by the so called Pomeron phenomenology.
 Elastic amplitude is function of the momentum transfer $t=-  q^2$  
and its Bessel-Fourier transform is the so called profile function $F(b)$ depending on the impact parameter $b$. 
Small $b$ is understood via perturbative BFKL Pomeron, while
large $b$ via some
 string-exchange models. In this case the experimental data actually 
 do indicate sharp transition between these regimes.  Recently an attempt to 
understand both regimes in a single model has been quite successful, in the so called AdS/QCD framework.
Furthermore, it has been suggested that the critical $b$ is related to critical temperature $T_c$
of the phase transition in the gauge theory: we will discuss this in section \ref{sec_Pomeron}. 

Ic. Proceeding from elastic to inelastic collisions, under which conditions we 
should describe the initial snapshots of hadrons and nuclei in terms of  perturbative partons (quarks and gluons) or
non-perturbative effective objects (strings,constituent quarks, etc)? 
As we will discuss, hydro sound modes survive till freezeout, and thus this make these initial snapshots visible
to the detectors. Therefore, one can now estimate at least the number of such ``objects".

(IIa) Furthermore, now we have a variety of cases --  AA, $pA$ and $pp$  collisions with widely variable multiplicities.
 All of them  show collective phenomena -- radial, elliptic and even triangular flows -at high enough multiplicity.
Pushing such observables down in multiplicity is a current frontier.  
One would like to find some regime changes there, experimentally and theoretically. 
No sharp changes are so far detected.  People apply microscopic theory at one end of the $pA$ and $pp$  collisions and macroscopic theory (thermo and hydrodynamics) at another,
 checking to what extent 
 one can justify them, but this process is far from converged so far.
  
  (IIb)
    Where exactly is the boundary between the micro and macro theories?  
    Textbook answer is that one can
    compare the micro or ``mean free path" scale $l$  to  the size of the system \be L\gg l \label{macro} \ee
  and if the $l/L$ ratio is small one can use the macroscopic theories. 
    Given a very small phenomenological viscosity $\eta/s\sim .2$ one formally get a mean free path
    smaller than interparticle distance. Does it actually make sense? Testing 
the sounds with the shortest wavelengths we can see, or the exploding systems with the smallest sizes,
 one seem to conclude that in fact it does!

\section{Sounds on top of the ``little bang"}
\subsection{Introductory comments on hydrodynamics}

Since the start of the RHIC era in 2000, it has become soon apparent
that the data on particle spectra (the radial flow) and the elliptic flow confirm nicely predictions of hydrodynamics,
supplemented by hadronic cascade at freezeout \cite{Teaney:2000cw,Teaney:2001av,Hirano:2005xf}.
All relevant dependences -- as a function of $p_\perp$, centrality, particle mass, rapidity and collision energy -- were
checked and found to be in good agreement. Since the
famous 2004 RBRC workshop in Brookhaven, with theory and experiment summaries collected in Nucl.Phys.A750 ,  the statements that QGP "is  a near-perfect liquid"
which does flow hydrodynamically has been endlessly repeated. QGP is recognized to be in the strongly coupled regime, now called sQGP, and hundreds of theoretical  papers are written, developing dynamics at strong coupling.  
Several second-order hydro+cascade models had been developed in the last decay, which do an excellent job
in describing the data. The interested readers  should look at recent reviews such as \cite{Heinz:2013th}:
we will take  this to be a ``well established domain". (Of course, the data are constantly getting more impressive with time: let me mention e.g. 
 the elliptic flow $v_2$ of the deuterons recently measured by ALICE and shown at QM2015.)

 The interest has now been shifted to understanding  the $limits$ of the hydro description. Of course, the exact boundary depends on which version of it is implied. In the simplest case, when one thinks of ideal hydro plus viscous corrections ,
  this happens when the viscosity times the velocity gradients is as large as the main terms. Now, there can be two
different  reasons why this correction gets large: (i) either the fluid is not good, viscosity-to-entropy ration $\eta/s$ is not small, or (ii)
  the gradients are too large. Studies of the former have recently been done using anisotropic hydro and exact solution
  of Boltzmann equation in Gubser setting, as well as by the so called anisotropic hydrodynamics
  : we will discuss those in section \ref{sec_anisotropy}. The latter
  situations can be approached via the so called {\em higher order hydrodynamics}. Gradient re-summation
  was in particularly attempted by Lublinsky and myself: we will discuss those
  in section \ref{sec_grad_resummation}. In all of these cases, an improved hydrodynamics is supposed to 
  shift its initiation time somewhat earlier, or promise to treat somewhat smaller systems. 
  Yet, while the ``out-of-equilibrium" initial stage get reduced,  of course it can never be eliminated. The distinction between the ``initial" and ``equilibrated" stages is a matter of 
  definition: but  physical outputs -- e.g. the total amount of entropy produced -- should ideally be independent of that.    
 Unfortunately, in practice we are still far from this ideal scenario: studies of entropy generation at an initial stage
 is still in its infancy.
 
  Few other issues  remains open till now, related with the boundary of hydrodynamical description. 
  
  One is the boundary at high $p_t$. When the first RHIC data came, it has been gratifying  to see that agreement with hydro-based 
  spectra went down more than order of magnitude and reached $p_t\sim 1\, GeV$. The it has been
  extended much further, till 
 $p_\perp\sim 3\, GeV$. This region includes more than  99.9\% of all secondaries!
 Indeed, 
   only few -- out of thousands -- particles in a heavy ion collision do come  from  hard tails of the spectra. 

 As the regime changes at  $p_\perp\sim 3\, GeV$, one needs to understand why. One important point is that
particles at high $p_t$ come from an edge of the fireball, at which the magnitude of the 
hydro flow is maximal. Using the  saddle point method for Cooper-Frye integral  \cite{Blaizot:1990zd}
one can see that the region from which such particles  come shrinks, as $p_t$ grows. (We will
return to this point in connection to high multiplicity pp collisions and HBT radii, see section\ref{sec_pp}.) 
Eventually  this region shrinks to a single  hydro cell
(defined by the mean free path), and then hydro description
 needs to be modified. 
Teaney \cite{Teaney:2003kp} introduced the viscosity effect in the cell, and noticed that due to extra derivative
these effects should be enhanced at the edge. They also should have extra power of $p_t$, he argued, and
  the deviation should be downwards, as indeed is observed in harmonic flows.  
Still, 
 physics in the window $4 < p_\perp< 20 \, GeV$,
between the hydro-dominated  and clearly jet-dominated regions is not yet understood.

  The second issue is the relative role of sQGP and hadronic stages in the expansion and flows.
  At AGS/SPS energy domain one clearly had to understand both quantitatively: otherwise
  the results got wrong. For example, the paper  \cite{Hung:1997du} was all about the sequential freezeout of different kinds of secondaries, and how it was seen in radial flow.
  It has been fairly important for radial flow at RHIC: yet elliptic 
  and other harmonics of flow were formed earlier and the hadronic stage were less important.

The last 4 years had provided  further confirmation of the {\em hydro paradigm}
at LHC. One of the benefits is that, even for radial flow, the main acceleration is done 
  in the sQGP phase,  to such an extent that one can neglect the sequential freezeout and related complications.  Let me just provide one example of how well it works, from ALICE.  
  
  Out of all individual
properties of the secondary hadron, only one -- its mass -- is important for hydro models.
 All one needs to know is that putting an object of mass $m$ into a flow with velocity $v$, the momentum will be
$mv$. Nice demonstration that it is indeed so can be made by direct comparison of the
 spectra for a pair of hadrons with the same mass, with otherwise
 completely different quantum numbers, cross sections etc. The classic example is $p$ and $\phi$: those can  hardly be more different,
the former is non-strange baryon, the latter is  a strange-antistrange meson, etc. And yet, as
e.g. ALICE measurements 
shows,  their spectra are--  with impressive precision -- identical, up to $p_T\sim 4 \, GeV/c$.
The very fact that $p/\pi,\phi/\pi$ ratios can increase by two orders of magnitude 
and reach O(1) at such $p_t $ is by itself a dramatic
hydro effect, never seen in generic (non-high-multiplicity) $pp$ collisions.



   I think it puts to rest many non-flow explanations of the particle spectra in AA collisions.
   
   Among those I count 
   a widely popularized idea of the so called ``quark scaling" of the elliptic and other flows,
that suggests that if a hadron is made of $n$ quarks then its flow harmonics should be proportional to $n$,
 $v_m^{(h)} \sim n v_m^{(q)}$, with some the ``quark" $v_m^{(q)}$.
The logic behind this idea
 \cite{Fries:2003kq} comes  
from the expression (82) of that paper which assumes that the
 quark distribution function in the phase stale is ``factorized" 
\be
w(R;p) = w_i (R;p)[1 + 2v^a_2(p_\perp )cos(2\phi)]. 
 \ee 
 Note that the bracket does not depend on the location $R$.
 So, the authors assume $v_2$ to originate from 
 some  anisotropic momentum distribution in {\em every local hydro cell} . 
This assumption is  in direct contradiction to the flow picture, which instead postulate that each cell has an equilibrium (=isotropic)
distribution of particles
\footnote{A certain anisotropy of the momentum distribution in hydro cells is created by the viscosity.
 Since QGP is a very good liquid with small viscosity 
 it is a quite small effect, and  its effect on the observed $v_2$ has the opposite sign.  Let me also remind 
 that all calculations in early 1990's 
 were originally done at zero viscosity, and yet they predicted $v_2$ very well.}
Anisotropic flows are due to different {\em number of cells} moving in different directions.
 This $n$-parton structure cannot possibly matter: a hadron can be made of 2, zero
or 22 partons: all you need to know is its mass. 

 \begin{figure}[t]
  \begin{center}
    \includegraphics[width=8cm]{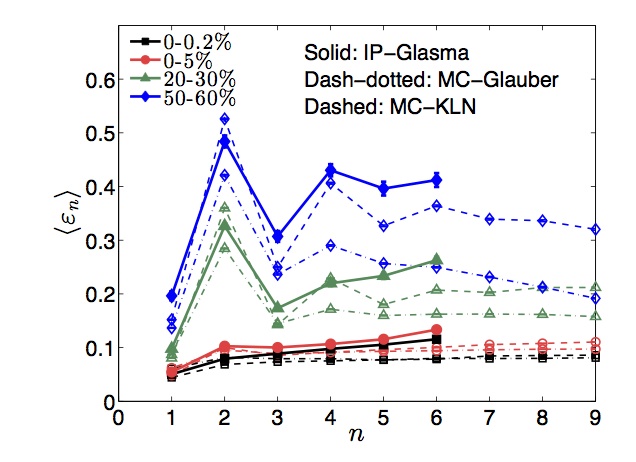}
  \includegraphics[width=8cm]{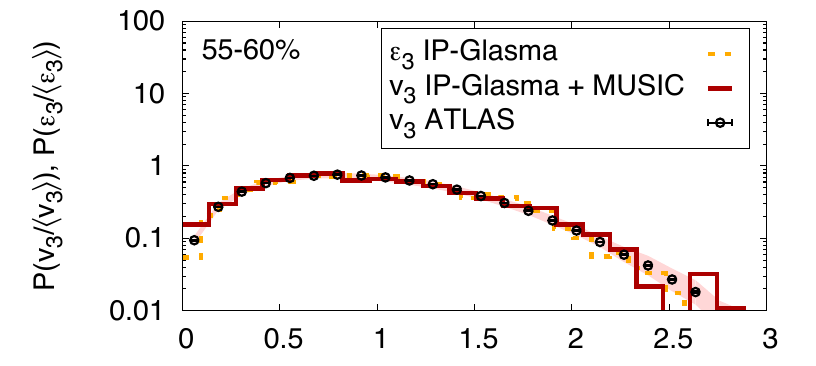}
   \caption{ (a) From \cite{Heinz:2013th}: average deformations $<\epsilon_n>=<cos(n\phi)>$  for various centralities using models indicated in the figure. 
   (b)From \cite{Schenke:2014zha} . Data points correspond to the event- by-event distribution of  v3 measured by the ATLAS collaboration, compared to the distributions of initial eccentricities in the IP-Glasma model and the distributions of $v_3$ from fluid dynamic evolution with IP-Glasma initial conditions. }
  \label{fig_P_v3}
  \end{center}
\end{figure} 

\subsection{Hydrodynamical response to perturbations}

  Now, assuming that the {\em average pattern of the fireball explosion} has been well established, 
  we are going to perturb it. The
  fluctuations and and their correlations
  is thus the next topic of our discussion.
  
 There are two schools doing this. One uses the so called
    event-by-event hydro,
  with ensemble of initial conditions defined by some model. 
  Yet I will argue that practically all one learned from such expensive studies can also be
   understood from a much simpler approach, in which one adds small and elementary  perturbations
 on top of smooth average fireball, and solve for the complete spectrum of excitations. Using an analogy,
 instead of beating the drums strongly, with both hands and all fingers, one may first touch it gently with
 a drumstick, at different locations, recording the spectra and intensities of the sounds produced. Eventually, one 
 may understand not only its spectrum, but also all the relevant modes of excitations. The one can work out linear and nonlinear superpositions of those elementary modes.
 
( The reader should understand that by no means the latter approach undermines good work which
the latter does. People 
developing stable hydro codes and ensembles of initial conditions,  averaging 
  hydro results over thousands of configurations with complicated shapes. All that  works spectacularly well:
  the $v_n$ moments of the flow perturbations in azimuthal angle $\phi$  
 as a function of transverse momentum, particle type and  centrality $v_n(p_t,m,Np)$ are all reproduced.
 Hardly any further convincing of that  is now needed:  multiple versions of the plots proving that are shown at
 all conferences and perhaps seen by a reader: so I will not discuss this and take it for granted as well.) 

    Now is the time of synthesis of all that information. In order to summarize  what we have learned
    from fluctuation/correlation studies 
     one needs to go back to the data and to the results of hydro calculation and separate their essence from
   unimportant complications. Some simple pocket formula fits to 
    analytic expressions, revealing the systematics of physical effects in question, can be very useful at this stage,
    and we will discuss them below.
       
    But before we dip into details, let me formulate the main answers, using my drum analogy. First of all,
    we did learned that perturbations on top of the little bang basically behave as sounds propagating on the drum.
    The main phenomenon is the viscous damping, and the value of sQGP viscosity is the main physics
    output of these studies. Different time evolution of different modes is the next issue: unlike the usual drum,
    the little bang is not static but exploding:  therefore oscillating behavior is superimposed with time dependence
    of the amplitudes. Different excitations are excited if the drum is struck at different places: similarly
    here, and we only start understanding the excitation modes themselves.

 Let me continue  where these calculations typically start: from angular deformations of the initial state. In Fig.\ref{fig_P_v3}(a) one finds the dependence of the mean harmonics (eccentricities) \be \epsilon_n=<cos( n \phi)> \ee where $n$ is integer and $\phi$
 is the azimuthal angle. The angular brackets mean average over events, usually for   
 certain centrality bin (indicated  
 in the upper left corner as a fraction of the total cross section, which scale as $bdb$). The one 0-0.2\%
is called the ultra-central collisions, $b\approx 0$, and $50-60$\% are some peripheral collisions.  
 The first obvious comment to this plot is that $n=2,4$ harmonics are special: they have some peaks for   peripheral bins, due to
 collision geometry.  
 However other harmonics, and  in fact, $ all$ harmonics for the central bins, are basically independent on $n$
 and centrality. What that tells us is that statistically independent ``elementary perturbations" 
  have small angular size $\delta \phi\ll  2\pi$, so one  basically deals with an angular Fourier transform of the delta function. 
 
 The next observation is that the deformations gets smaller toward the central collisions, to just few percent level. This is also natural:  larger fireballs have more particles and thus they
fluctuate less, in relative terms. In fact one expect 
 \be <\epsilon_n> \sim {1  \over \sqrt{N_{cells}} }\ee
 where $N_{cells}=A_{cell}/A_{fireball}$ is the number of statistically independent cells in the transverse plane. 
 (What exactly is a cell is not simple and we will discuss it below in the section on the initial state. Depending on the model it
 changes from a fraction to few $fm^2$.)
 
 Models of the initial state give not only the average deformation but also their distributions and correlations. 
  Going a bit ahead of myself, let me note that, remarkably, the experimentally observed  
 distributions over eccentricities of {\em particle momenta distributions} $P(v_n)$  appears to directly reflect distributions
 of  angular anisotropies $P(\epsilon_n)$ at the initial time, 
 see e.g. $\epsilon_3,v_3$ distributions  in Fig.\ref{fig_P_v3}(b).  (Thus, apparently
 no noise is generated by the hydro evolution, from the initial state $\epsilon_n$ to the final state $v_n$: 
 why  is it so we will discuss below.)

%
%
%
%
%

  \begin{figure}[t!]
  \begin{center}
  \includegraphics[width=8cm]{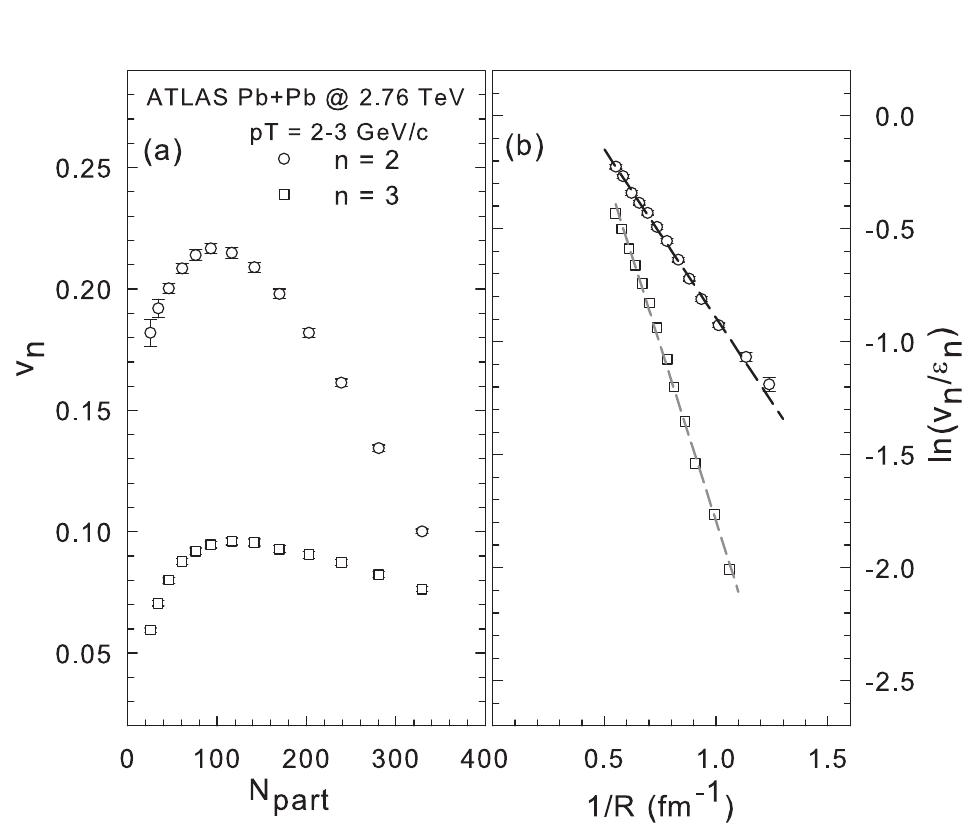}
   \caption{(a) Atlas data, from   Ref.  \cite{Lacey:2013is}, for $v_2,v_3$ vs. $N_{part}$ : (b) $ln(v_n/\epsilon_n)$ vs. $1/R$ for the same data   }
  \label{fig_L1}
  \end{center}
\end{figure}

  \begin{figure}[b!]
  \begin{center}
  \includegraphics[width=9cm]{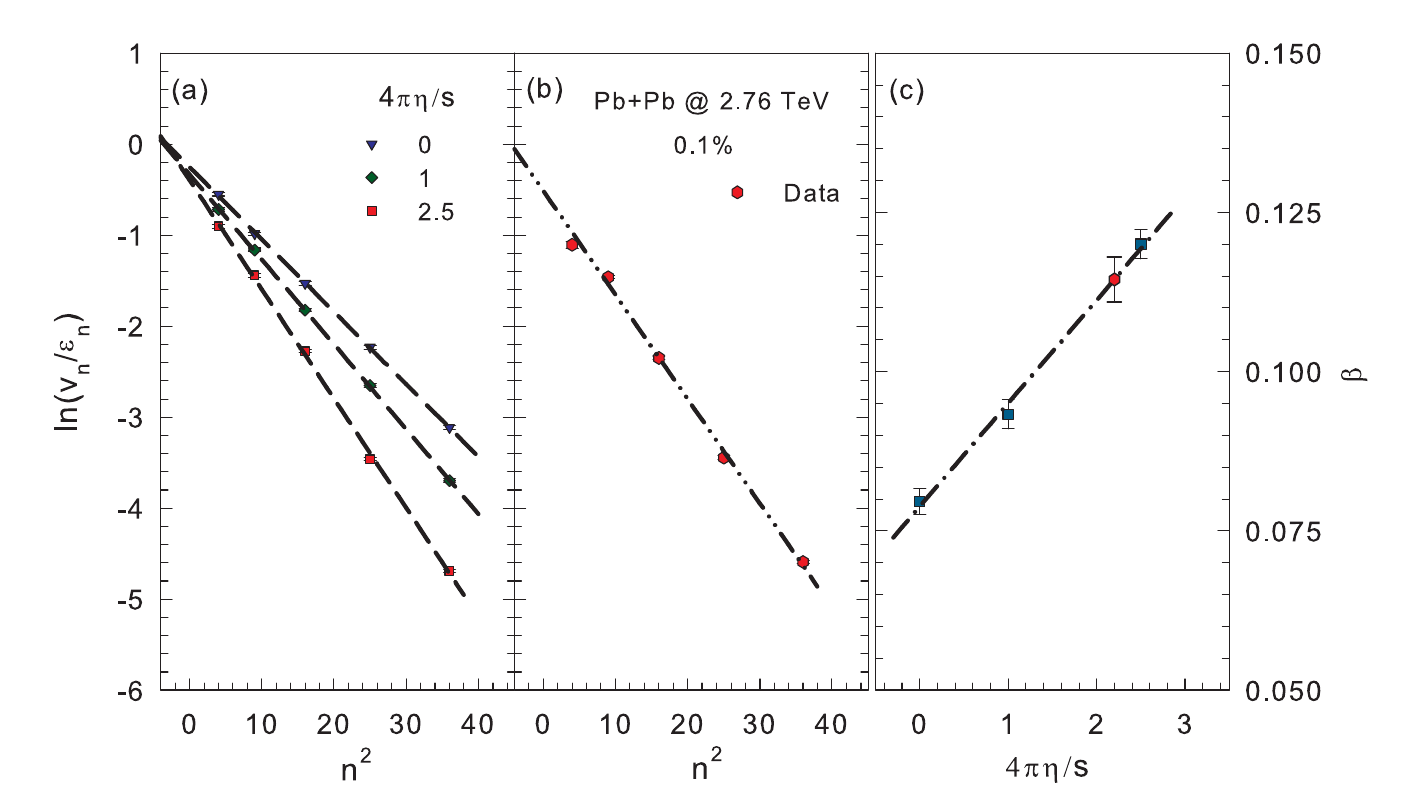}
   \caption{(a) Atlas data, from   Ref.  \cite{Lacey:2013is}, $ln(v_n/\epsilon_n)$ vs. $n^2$ from viscous hydrodynamical calculations for three values of specific shear viscosity as indicated. (b) $ln(v_n/\epsilon_n)$ vs. $n^2$ for Pb+Pb data. The $p_\perp$ -integrated $v_n$ results in (a) and (b) are from ATLAS 0.1\% central Pb+Pb collisions at
sNN = 2.76 TeV; the curves are linear fits. (c) exponent vs. viscosity-to-entropy ratio $4\pi/s$ for curves shown in (a) and (b).}
  \label{fig_L2}
  \end{center}
\end{figure}

\subsection{Acoustic systematics: the viscous damping} 
      There is a qualitative difference between the radial flow and higher angular harmonics.
   While the former  monotonously grows with time, driven by the outward pressure gradient
   with a fixed sign, the latter are
   basically sounds, or
 a (damped) oscillators. Therefore the signal observed should, on general grounds, be
  the product of the two factors: (i) the amplitude reduction factor due to viscous damping 
 and (ii)  the $phase$ factor containing the oscillation    at the freezeout.  (We will discuss the effects of the phase in the next section.)

 Let us  start with the
 ``acoustic systematics"  which includes only  the viscous damping factor. It provides good qualitative account of the data
 and hydro calculations into a simple expression, reproducing  
  dependence on the viscosity value $\eta$, the size of the system $R$  and the harmonic number $n$ in question.
Let us motivate it  as follows.  
 We had already mentioned ``naive"
macro and micro scales (\ref{macro}):
 now we   define it a bit better by inserting the famous viscosity-to-entropy ratio $ \eta/s=l T$,
 \be {l \over L}= {\eta \over s}{1 \over L T} \ee 
This ``true micro-to-macro ratio", corresponding to the mean free path in kinetic theory,
defines the minimal size of a hydro cell. 

One effect of viscosity  on sounds is the damping of their amplitude.
   The so called ``acoustic damping" formula,
  suggested by Staig
and myself \cite{Staig:2010pn} ,  is given by 
\be {v_n\over \epsilon_n}\sim {\rm exp}\left[ - C n^2  \left({\eta \over s}\right)\left({1 \over TR}\right) \right]  \label{eqn_acoustic}  \ee
where $C$ is some constant. The number $n$ appears squared because the damping includes
square of the gradient, or momentum of the wave.
So, we have the following predictions: (i) the viscous damping is exponential in $n^2$; (ii) the exponent contains  
the product of two small factors,  $\eta/s$ and $1/TR$, (iii) the exponent contains $1/R$
 which should be understood as the $largest$ gradient in the system, often modelled as $1/R=1/R_x+1/R_y$.
 


   Extensive comparison of this expression with the AA data, from central to peripheral, has been
 done in Ref.  \cite{Lacey:2013is}  from which we borrow Fig.\ref{fig_L1} and  Fig.\ref{fig_L2}.
The Fig.\ref{fig_L1} (a) shows the well known centrality dependence of the elliptic and triangular flows.
$v_2$ is small for central collisions due to smallness of $\epsilon_2$, and also small at very peripheral
bin because viscosity is large at small systems. Fig.\ref{fig_L1} (b) shows the $ln(v_n/\epsilon_n) $,
which according to the formula is the exponent. As a function of the inverse system's size $1/R$
both  elliptic and triangular flows show perfectly linear behavior.
 Further
 issues -- the $n^2$ dependence as well as linear dependences of the $log(v_m/\epsilon_m)$
 on viscosity value --
are also very well reproduced,
 see Fig.\ref{fig_L2}.  Note that this expression
  works all the way to rather peripheral AA collisions with $R\sim 1 \, fm$
and multiplicities comparable to those in the highest $pA$ binds. 
It also seem to work till the largest $n$ so far measured.




So,  the acoustic damping provides correct systematics
of the harmonic strength. This increases our confidence that -- in spite of somewhat different geometry --
the perturbations observed are actually just a form of a sound waves.  

Since we will be interested not only in large AA systems but also
in new  -- $pA$  and $pp$  -- much smaller fireballs, one may use the systematics to
compare it with the new data. Or, using it, one can estimate how many flow harmonics 
can be observed in these cases.
 For central PbPb at LHC collisions with
 \be {1 \over TR} = {\cal O}(1/10) \ee  
 its product of $\eta/s$ is $O(10^{-2})$. So one can immediately see from this expression
why  harmonics  up to  $n=O(10)$ can be observed.
Proceeding to smaller systems by keeping a similar initial temperature $T_i\sim 400 \, MeV \sim 1/(0.5\, fm)$
but a smaller size $R$,  results in a macro-to-micro parameter that is no longer small, or 
$1/TR\sim 0.5,1$, respectively. 
For a usual liquid/gas, with $\eta/s>1$, there would not be any small parameter left and one would have to conclude
that hydrodynamics is inapplicable for such a small system. However, since the quark-gluon plasma is an exceptionally
good liquid with a very  small  $\eta/s$, one can still observe harmonics up to  $m=O(\sqrt{10})\sim 3$.
And indeed, $v_2,v_3$ have been observed in the first round of measurements (for latest data see Fig.\ref{fig_pA_vn_atlas}).


     \begin{figure}[t]
\begin{center}
\includegraphics[width=7cm]{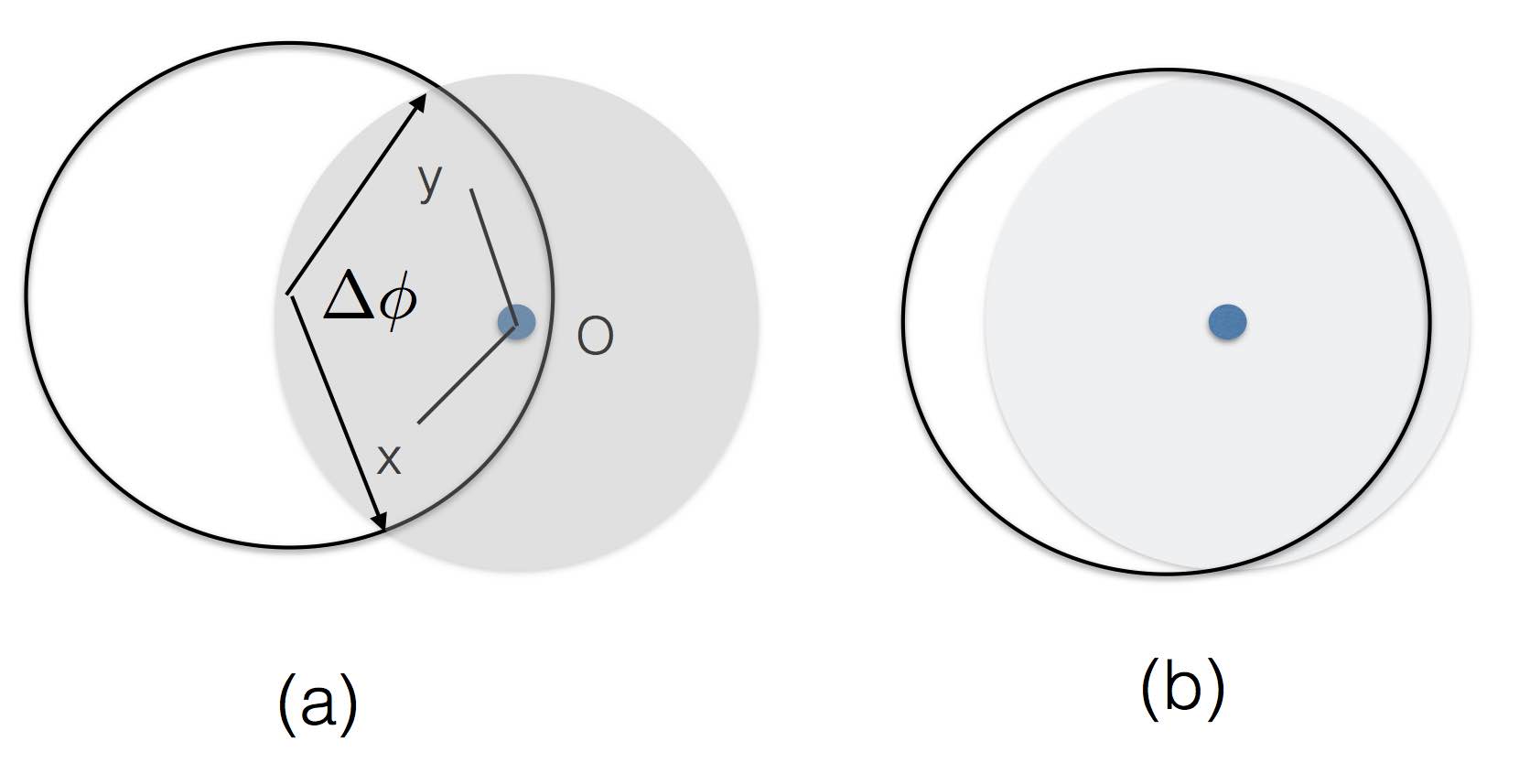}
\caption{ The perturbation is shown by small blue circle at point O: its time evolution to points $x$ and $y$ is
described by the Green function of linearized hydrodynamics shown by two lines. Perturbed region -- shown 
by grey circle -- is inside the sound horizon. The sound wave effect is maximal at the intersection points of this area with
the fireball boundary: $\Delta \phi$ angle is the value at which the peak in two-body correlation function is to be found.
Shifting the location of the perturbation, from (a) to (b), result in a rather small shift in  $\Delta \phi$.   }
\label{fig_circles}
\end{center}
\end{figure}

\subsection{The Green function: waves from a point perturbation}
The problem appears very complicated: events have multiple shapes, described by multidimensional 
probability distributions $P(\epsilon_2,\epsilon_3...)$. Except that it is not. All those shapes are however mostly a
statistical noise. 
The reason is as follows:
 rows of nucleons sitting at different locations in the transverse plane    
 cannot possibly know about each other fluctuations at the collision moment, so they must be statistically independent.   
   An ``elementary excitation"  is just one delta-function in the transverse plane (in reality, of the size of a nucleon) , on top of a smooth 
   average fireball.  In other words, the first thing to do is to calculate 
   the Green function of the linearized hydro equations.

A particular model of the initial state
expressing locality and statistical independence of ``bumps" has been formulated in \cite{Olli}: 
the correlator of fluctuations is given by the Poisson local expression 
\be <\delta n(x) \delta n(y)> =\bar n(x)\delta^2(x-y)  \label{eqn_Olli} \ee
where $\bar n(x)$ is the average matter distribution.
 The immediate consequence of this model is that,
for the central collisions on which we now for simplicity focus, $\epsilon_m$ are the same for all $m<m_{max}=O(10)$ (till
the bump size gets resolved). 

In order to calculate perturbation at later time one needs to calculate the Green functions, from the 
original location O to observation points $x$ and $y$ as shown in Fig.\ref{fig_circles}(a). That has been first done 
in \cite{Staig:2011wj}, analytically for Gubser flow. 
One finds that the main contributions come from two points  in Fig.\ref{fig_circles}, where the ``sound circle"
intersects the fireball boundary.
In a single-body angular distributions those two points correspond to two excesses of particles at the corresponding
two directions. 
 Let us call the angle between them $\Delta \phi$: at Fig.\ref{fig_circles}(a) it is about 120$^o$ or 2 rad. 
 This is because the freesout time times speed of sound  $R_h=c_s \tau_{freezeout}$  happens to be close to
 the radius of the fireball.

\begin{figure}[t]
  \begin{center}
  \includegraphics[width=5cm]{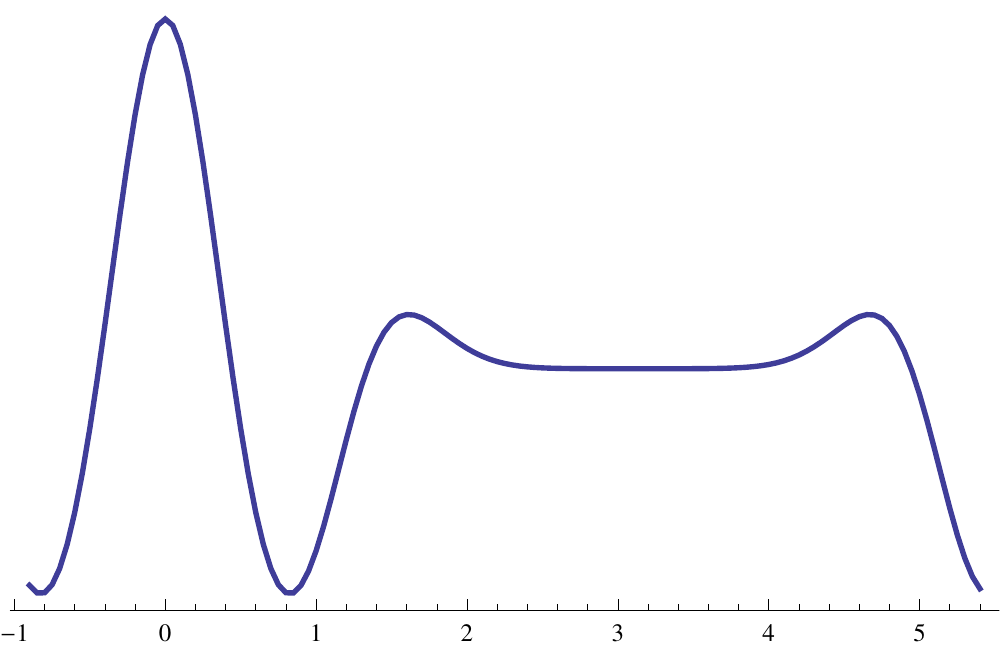}
  \includegraphics[width=5cm]{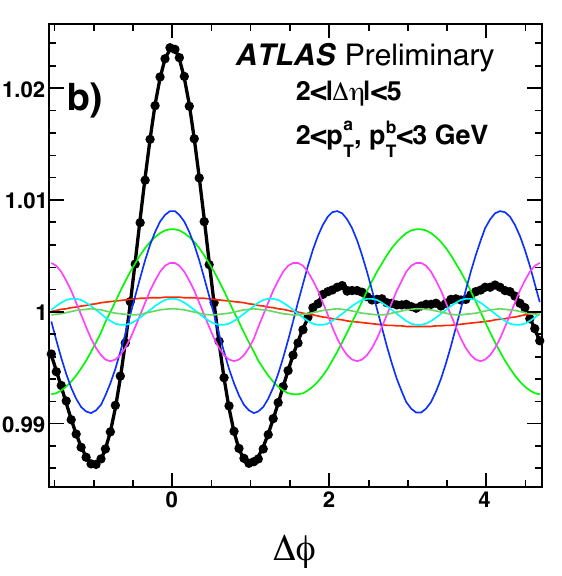} 
  \includegraphics[width=5.5cm]{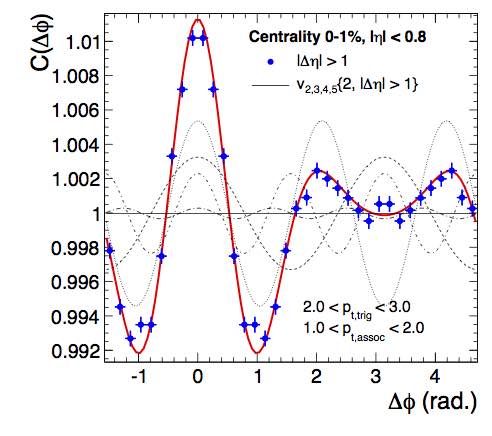}
   \caption{(a) Calculated two-pion distribution as
a function of azimuthal angle difference $\Delta\phi$, for  viscosity-to-entropy ratios $\eta/s=0.134$, from 
\cite{Staig:2011wj}.
(b) from  ATLAS \cite{ATLAS_corr}, (c) from ALICE \cite{ALICE_deltaphi}:
 All   for ultra-central collisions.}
  \label{fig_2pcorr}
  \end{center}
\end{figure}

 The correlation function calculated in  \cite{Staig:2011wj}, is
shown in Fig.    \ref{fig_2pcorr}(a). One its feature is a peak at zero $\delta\phi =0$: it is generated if both 
observed particles come from the same intersection point. If two particles come from different points, one finds
 two peaks, at $\Delta\phi = \pm 2$ rads.
( As shown in  Fig.\ref{fig_circles},  if one shifts the position of perturbation
from (a) to (b), the peak angle $\Delta\phi$ changes  toward its maximal value, $\pi$ radian, or 1/2 of the circle.) 

This calculation has been presented at the first day of Annecy Quark Matter $before$  the experimental data.
The ATLAS  correlation function  (for the ``super-central bin", with  the fraction of the total cross section 0-1\%)
 is shown in Fig.    \ref{fig_2pcorr}(b). The agreement of the shape is not perfect -- because
a model is with conformal QGP and a bit different shape -- but all predicted elements of its shape are indeed observed.

While there is no need to use Fourier harmonics -- I insist that the correlation function itself teaches us more than 
harmonics, separately studied -- one can certainly do so. Note that
  for ultra-central collisions we now discuss the largest harmonics is $v_3$ (the blue curve in Fig. \ref{fig_2pcorr}(b))
  and 
   not $v_2$ (the green one). Since starting deformations $\epsilon_n$ are basically the same for all $n$, the difference must come
   from hydro, and it does: we just explained it above, using notion of sound horizon. Alternative
   explanation can be done in terms of the freezeout phases of the harmonics $\phi^n_{freezout}$.
 
 A related arguments are quite famous  in the theory of Big Bang perturbations.
   All harmonics get excited at the same time by Big Bang -- hydro velocities at time zero are assumed to be zero for all harmonics --
   and are frozen at the same time.  The acquired phases depend on $n$ --  harmonics with a larger $n$
   oscillate more rapidly.  Binary correlator is proportional to $cos^2(\phi^n_{freezout})$
   and harmonics with the optimal phases close to $\pi/2$ or $3\pi/2$ values show maxima, with minima in between.
  Planck collaboration data on harmonics power spectrum of the 
   cosmic microwave Big Bang perturbations, shown in Fig. \ref{fig_planck}, do display a
    number of such maxima/minima.
    
%
 
\begin{figure}[t!]
  \begin{center}
  \includegraphics[width=8cm]{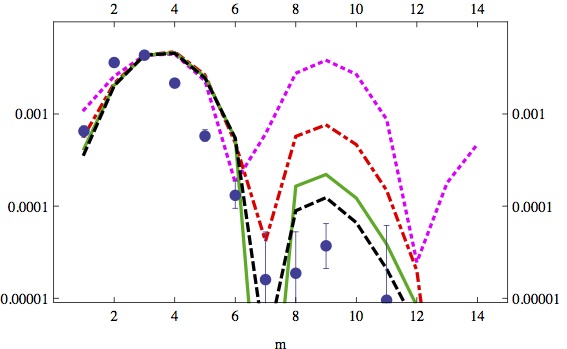}
  \includegraphics[width=9cm]{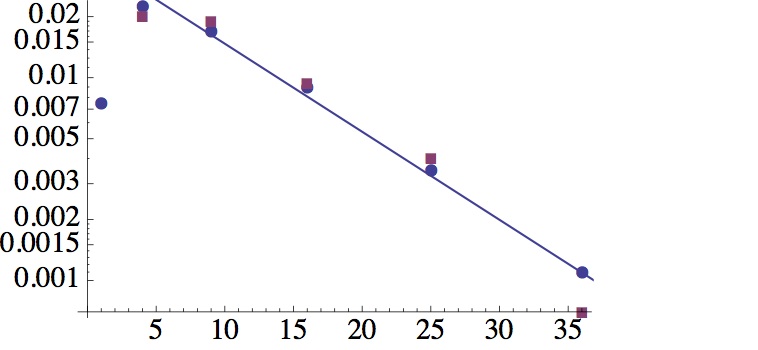} 
   \caption{(a) The lines are hydro calculations of the correlation function harmonics, $v_m^2$, based on 
   a Green function from a point source \cite{Staig:2011wj} for four values of viscosity  $4\pi\eta/s$ =0,1,1.68,2 (top to bottom at the right). The closed circles are the Atlas data
   for the ultra-central bin.
(b) $v_n\{2\}$ plotted vs $n^2$.
Blue closed circles are calculation of  via viscous even-by-event hydrodynamics \cite{Rose:2014fba},
``IP Glasma+Music",  with $\eta/s=0.14$. The straight line, shown  to guide the eye,
demonstrate that ``acoustic systematics" does in fact describe the results of this heavy calculation
quite accurately. The
CMS data for the 0-1\% centrality bin, shown by the red squares, in fact display  larger deviations,
perhaps an oscillatory ones.
   }
  \label{fig_vn_comp}
  \end{center}
\end{figure}

 Our calculation had shown such oscillations clearly, with the first peak  between $n=3,4$ (see Fig. ), the next should be at n=9, with the minimum 
 predicted to be around 7, see Fig.\ref{fig_vn_comp}(a).
 More sophisticated event-by-event   
 hydro calculation by Rose et al \cite{Rose:2014fba} does not reproduce oscillations around 
 the smooth sound damping trend, see Fig.\ref{fig_vn_comp}(b).
 One may think that averaging over many bumps in multiple configurations may indeed average out the
 freeze out phase factor. 
 
 Yet the ultra-central ATLAS data shown in  Fig.\ref{fig_vn_comp}(a), and even 
  the CMS central bin data  in 
 Fig.\ref{fig_vn_comp}(b) 
 (which include a bit larger impact parameters and thus feed more geometry-related contribution to $v_2$) 
 one can still see clear deviation from smooth damping curve $\sim exp(-n^2*const)
$. Note that the $v_3$ is higher than in all hydro output shown, while $v_6$ is lower. Let me conclude this
discussion with a note 
that while certain oscillatory deviations from ``acoustic systematics" are  there in the experimental data,
their origin and even qualitative reproduction by the hydro models remains uncertain.


  \begin{figure}[t]
  \begin{center}
  \includegraphics[width=8cm]{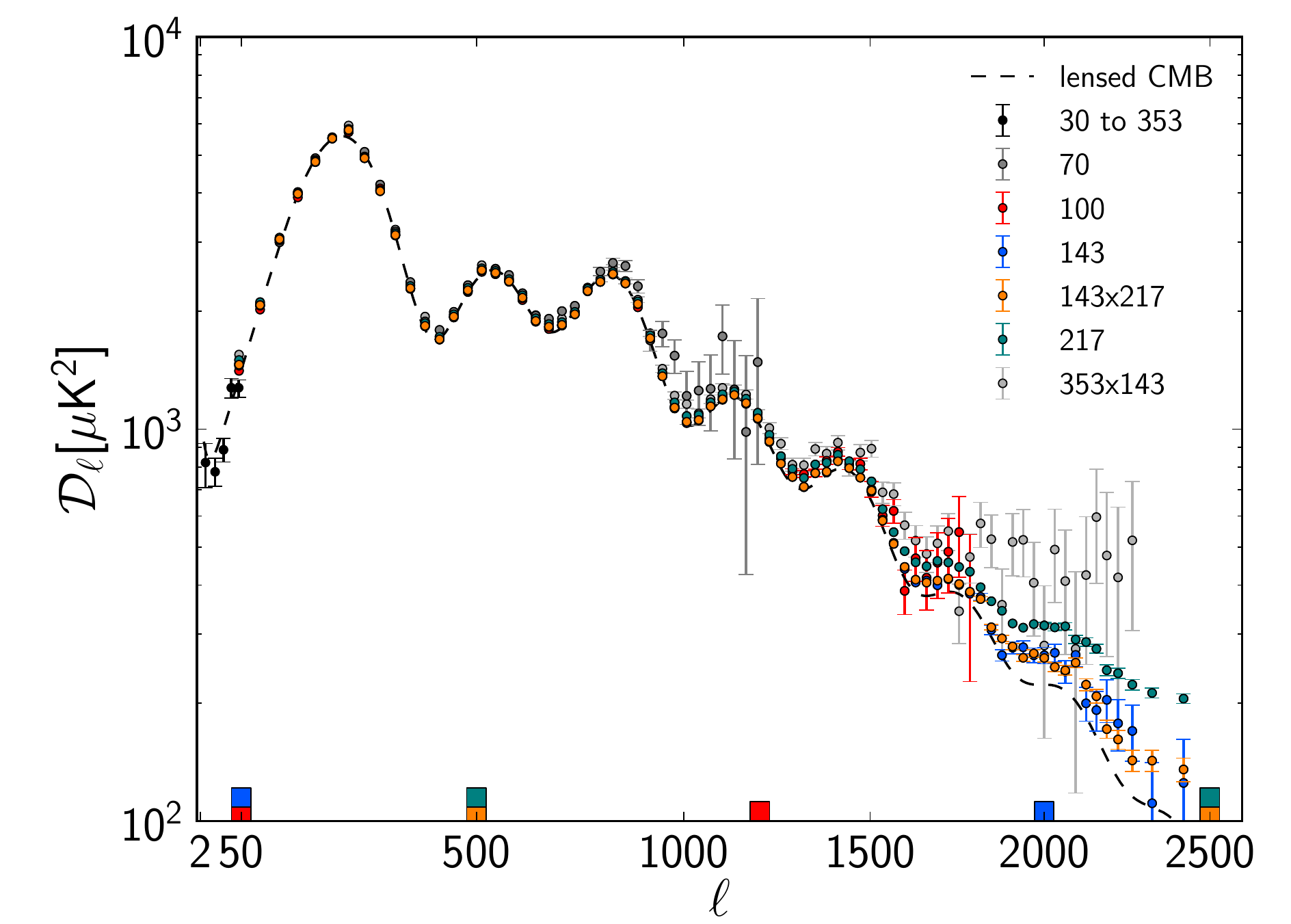}
   \caption{Power spectrum of cosmic microwave background radiation measured by Planck collaboration
\cite{Ade:2013kta}.} 
  \label{fig_planck}
  \end{center}
\end{figure}


Let me conclude this section with brief discussion of the following issue. The ``flow harmonics", solutions
of linearized equations on top of the average smooth hydro solution 
should possess a complete spectrum of perturbations, whose time evolutions are
different from each other. The perturbations of course
depend not only on the
angle $\phi$, as $e^{i m \phi}$, but of course on other coordinates  $r,\eta$ as well.

 For Gubser setting (see appendix) one can use comoving coordinates
$\rho,
\theta,\phi,\eta$ and dependence on $all$  
coordinates separate nicely, with known analytic expressions for harmonics and their time evolution. 
The flow in transverse plane is conveniently combined into a single 
angular momentum harmonics $Y_{l}(\theta,\phi)$,  combining azimuthal angle and $r$. 
Another simple eigenfunctions are plane waves with some  momenta $k$ in  the rapidity direction $\eta$.    

Can one define   similar set of independent harmonics 
 for a generic non-Gubser setting? And, even more importantly,
can those be observed?
A nice step in this direction has been recently made  by Mazeliauskas and Teaney
\cite{Mazeliauskas:2015vea}. Using experimental data (and of course hydro calculations) they define such
``subheading harmonics". Instead of describing how they do it, here is just one picture
from that work, for triangular flow,
indicating a difference between the leading and subleading flows: unlike the former, the latter gets a sign change
along the radial direction.

 \begin{figure}[t]
  \begin{center}
  \includegraphics[width=8cm]{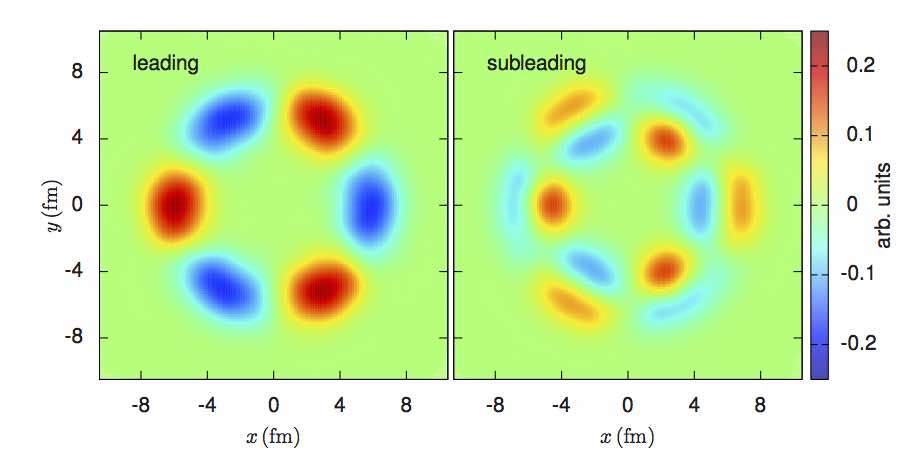}
   \caption{The leading (left) and subleading (right) harmonics of triangular flow, from \cite{Mazeliauskas:2015vea}.   } 
  \label{fig_planck}
  \end{center}
\end{figure}
\subsection{Non-linear effects for harmonics}

We argued above that one can view perturbations as superposition of uncorrelated
local (delta-function-like) sources. We then further argued that only those close to the
fireball surface are visible.
How many of them are there? For central collisions the circumference 
of the fireball $2\pi R_A\approx 40 \, fm$. The correlation length is perhaps the typical
impact parameter in NN collisions, which at LHC energies is $b\sim \sqrt{\sigma/\pi}\sim 1.6\, fm$.
Their ratio suggest existence of about
\be N_{sources} \approx {2\pi R_A \over b} \approx 25 \ee
cells on the fireball surface which fluctuate independently, up or down from the mean value.
Glauber initial conditions indeed have about a dozen sound-producing ``bumps".

The absolute scale of modulation in 
  the experimental correlators are of the order of a percent in magnitude, see e.g. Fig.\ref{fig_2pcorr}(b). 
Of course, it  comes  from incoherent sum over the single  bumps, so  one should divide by their number and get 
the individual correlation from a bump to be of $O(10^{-3})$ in magnitude. It however is quadratic in the wave amplitude: 
taking the square root of it we get sound amplitudes to be of the order of $1/30$ or so.

%
%
Summarizing: the  picture looks like that from about a dozen stones thrown into the pond. While the sound circle visibly interfere, they do not really interact with each other, since the amplitude is too small. 

Yet the conclusion that since 
 individual sounds are weak, the linear theory is completely correct, is premature. Even when
 the amplitude of the sound is small, they can produce large effects at the large $p_t$ end.
 Indeed at $p_t\sim 3\, GeV$ the elliptic flow 
 reach values which makes angular distribution 100\% asymmetric. This happens because  a 
 weak sound may sit ``on the shoulders of the giants", the explosions themselves. 

Similarly, the non-linear effects of flows at comparable large $p_t$ are non-negligible and were in fact observed.
%
%
For example $v_4$ is getting a contribution proportional to  $\epsilon_2^2$, $v_6$ from  $\epsilon_3^2$, etc.
 Detailed studies of such effects one can find in \cite{Teaney:2012ke}, including an interesting case of $v_1$ coming from $\epsilon_2*\epsilon_3$.

These nonlinear effects 
 come $not$ from nonlinear terms in the hydro equation, but from an expansion of the Cooper-Fry exponent
$exp(p^\mu u_\mu/T)$ 
in powers of the perturbation. Obviously  they become more important at high $p_t$.
Furthermore, one can see that linear terms should be linear in $p_t$, the non-linear we mentioned
should be quadratic $\sim p_t^2$, etc, see more 
    in \cite{Teaney:2012ke}.

\subsection{Event-by-event $v_n$  fluctuations/correlations} 

At the beginning of this section we had already emphasized that the main source of the $v_n$ fluctuations is that
 of the original perturbations $\epsilon_n$ themselves, see e.g. 
 Fig. \ref{fig_P_v3}(b). 
%
  Now we return to the  question:{\em Why does the ratio $v_n/\epsilon_n$, evaluated by hydro, has such a small spread?}.  While the practitioners of the event-by-event
  hydrodynamics use  huge variety of the initial configurations, it turns out that just one number -- $\epsilon_n$  --  insufficient to characterize  
 them. 
 Even adopting the simplest model advocated above -- that the perturbations come from  point-like sources -- it is nontrivial that 
 their event-by-event fluctuations of strength and locations do not create any spread in   the ratio $v_n/\epsilon_n$.   
 (Or, using my drum analogy, why does it sound the same, if hit in different locations?)
 
 Trying to understand this, let us come back to Fig.\ref{fig_circles}. The source located at the fireball edge, fig.(a), leads to correlations which are at $\Delta \phi \approx 2\, rad$,
 as we emphasized above. Projected on harmonics, it will excite $m=3$ mostly, as 2 rad is about 1/3 of $2\pi$.
 When the source moves inward,  fig.(b), the perturbations  move to the opposite points of the fireball  $\Delta \phi =\pi $, and the leading excitation becomes elliptic $m=2$.
 Estimates show that in the latter case the correlation weakens. The observed shape of the correlator,
 for ultra-central collisions  does have  a $minimum$ at $\Delta \phi =\pi $,
 supports that. As the source moves further toward the fireball center
   (not shown
 in  Fig.\ref{fig_circles}), the whole fireball becomes included inside the sound horizon. The angular correlation dilutes to all angles and its contributions to
 harmonics  gets negligible.
  In summary:  the sources located near the boundary of the fireball are mostly responsible for the harmonics we see. All of them thus generate the
 same angular shape of the correlator. What remains is  the overall strength,  a single parameter captured in the $\epsilon_n$ magnitude. 
  
   \begin{figure}[t]
  \begin{center}
  \includegraphics[width=8cm]{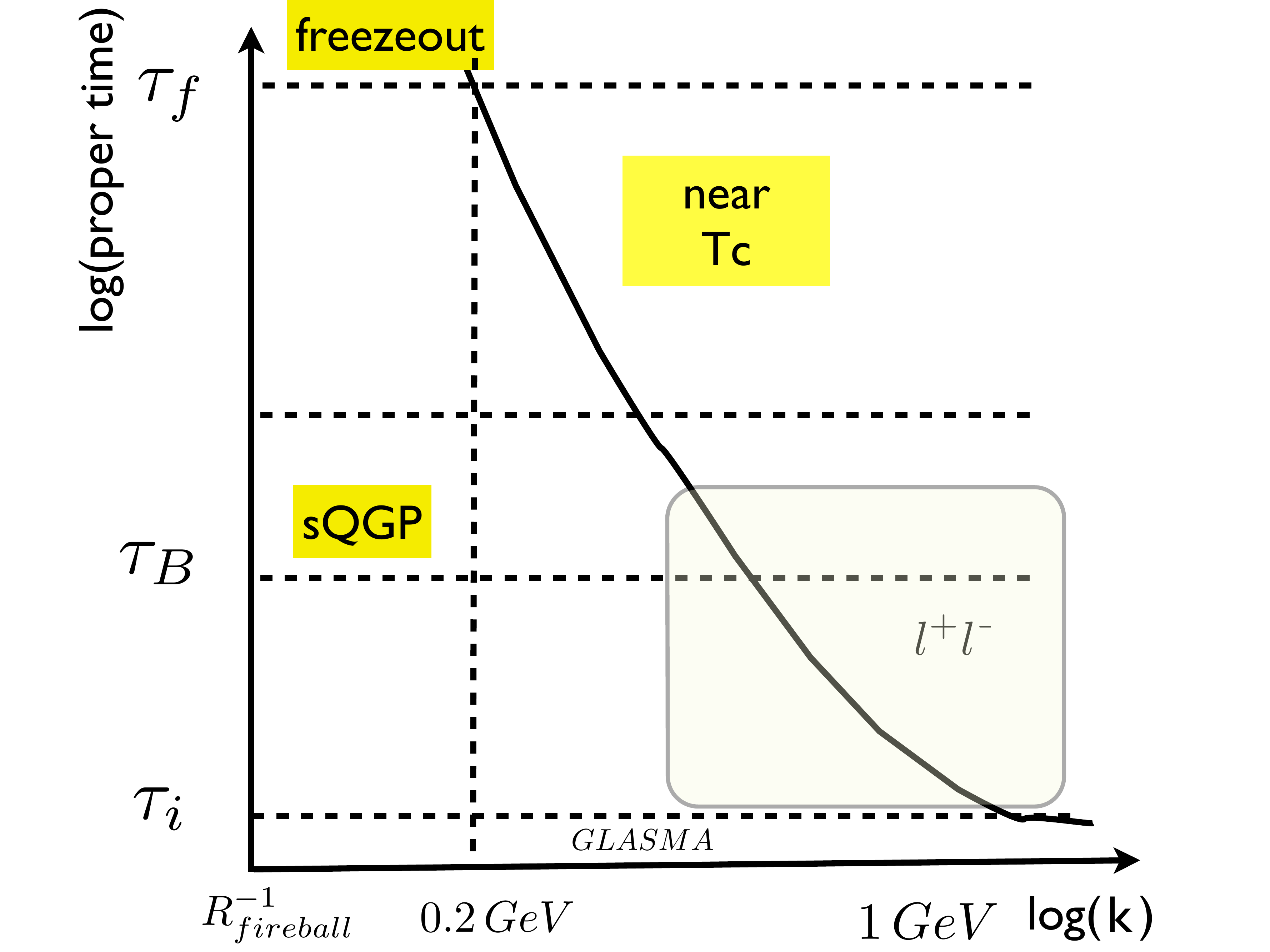}
   \caption{The log-log plane proper time $\tau$ -- sound momentum $k$. The solid curve indicates 
the amplitude   damping by a factor $e$: only small-$k$ sounds thus survive till freeze out.  The shadowed
region on the right corresponds to that in which sonomagnetoluminiscence effect may produce extra dileptons.}
  \label{fig_sound_map}
  \end{center}
\end{figure}

Harmonic correlations is a rapidly developing field. Those can be divided to correlations sensitive to relative 
phases of the harmonics and those which are not.  An example of the later is  
\be SC(m,n)=<v_m^2 v_n^2>-<v_m^2><v_n^2> \ee
on which  Alice provided good data at QM 2015 for $SC(4,2)$ and $SC(3,2)$.
In particular it is observed that
 $SC(4,2)>0$ but $SC(3,2) <0$, which is qualitatively reproduced in hydro.
 Now, in order to explain these data, one needs first to have the initial state model which is good enough to
 predict the relevant $\epsilon_n$ and their respective correlations. Suppose we do so using  the Bhalerao-Ollitrault model \cite{Olli} and the relation (\ref{eqn_Olli}): the results depend on integrals like $\int d^2r r^P \bar n(r)$
 with large powers $P=6,8$. What this tells us is that the correlation is strongly localized at the very surface
 of the system, and thus subject to significant uncertainties.  


\subsection{The map of the sounds}


We will argue below that the number of harmonics needed to describe the initial state is rather large, counted in hundreds.
 In Fig.\ref{fig_sound_map} we show a map of those, in terms of momentum (rather than angular momentum). The 
 curved line -- corresponding to ``acoustic systematics" discussed above -- show their lifetime.
 This curve crosses the freezeout time: smaller $k$ waves 
  can be observed at freezeout. Larger $k$  cannot: they are weakened due to viscous damping. (A suggestion
 to detect  those  via the MSL process will be discussed in  section \ref{sec_MSL}.)

Fluctuation-dissipation theorem tells us that 
while the initial perturbations are damped, new ones should be produced instead.
  There should be some noise produced at the hydro stage. For a paper exploring
  hydro {\em with noise} see Young et al \cite{Young:2014pka}. 
  Unfortunately, the authors have not yet 
  identified  observables which sensitive only to the $late-time$ perturbations. 
  I suggest those should be  azimuth+rapidity correlations, exploring the fact that
early time perturbations must be rapidity-independent

   This idea came from the paper \cite{Shuryak:2013uaa} in which a very specific late-time fluctuations were considered, namely
   sounds from collapsing QGP clusters inside the hadronic phase. Those collapse must happen, since this phase is
   unstable in the bulk, once $T<T_c$: its physics is similar to the celebrated Rayleigh bubble collapse. 
Another  discussed -- and as yet unobserved -- are sounds from jets depositing its energy into the ambient matter
and then propagated via hydro perturbations, known as Mach cone. Extensive recent discussion of how one can observe those can be found in 
 Ref.~ \cite{Shuryak:2013cja}.

\subsection{Sounds in the loops}
The hydrodynamical longitudinal pressure waves -- the sounds -- are the best quasiparticles 
we have. They are Goldstone modes, related with spontaneous breaking of the translation invariance by matter,
and thus  their interaction fall into certain patter familiar from physics of the pions.   
For large wavelengths they have  long lifetime,  exceeding  the freezeout time.
Therefore, in both
the Little and the Big Bangs, one can observe ``frozen" traces of the initial perturbations,
provide one looks at large enough wavelengths. 

   So, the set of sounds have their own lifetime scales, and one may wander about an ensemble of sounds, their interactions etc. 
 One way to introduce those is to
 add to hydro equations a Langevin-type noise term, with some Gaussian distribution, and then rewrite the theory with certain
 artificial fields into a QFT-like form. Progress into this direction has been recently summarized by Kovtun in a nice review 
 \cite{Kovtun:2012rj}. 
    Discussion of formal issues cannot be made in this review, however, and thus I illustrate the physics involved by one example, also due to 
    Kovtun and collaborators  \cite{Kovtun:2011np}.  Let me remind that matter viscosity can be defined via certain limit of the 
    stress tensor correlator, known as the Kubo formula. Kovtun et al calculated a
``loop corrections" to this correlator induced by the equilibrium sounds. Technically the calculation is done as follows: in the   
$<T^{\mu\nu} T^{\mu'\nu'} >$  correlator one substitute hydrodynamical expression for stress tensor containing sound perturbation velocities
and make it into a loop diagram with the ``sound propagators"
\be \Delta^{mn}= \int d^4x e^{-i p_\alpha x^\alpha } <u^m(x) u^n(0)> 
\ee
for two pairs of the velocities. (We use latin indices indicating that they are only space-like here. For shear viscosity those used are $m=x,n=y$). 
Skipping the derivation I jump to the answer obtained from this calculation, which can be put into the form of 
loop correction to the viscosity
\be \delta\eta_{loop}={17 \over 120 \pi^2} {p_{max} T (\epsilon+p) \over \eta_0 }\ee
which is UV divergent and thus includes $p_{max}$, the largest momentum for sound which still makes sense. What is 
important here is that the zeroth-order viscosity enters into the denominator.  This should not be surprising: 
a very good liquid with small $\eta_0$ support very long-lived sounds, which can transfer momentum far, which means they
produce $large$ contribution to the effective viscosity! (Like ballistically moving phonons in liquid helium or neutrinos in supernova, the most penetrating modes always dominate the  transport.)


Completing this section we would like to remind the reader about existence of other hydro modes, the rotational ones.
It is those ones which are e.g. responsible for  the atmospheric turbulence. Unlike sounds, which are always damped,
rotational modes can get excited under certain conditions. 
Very little theory work has been done in this direction so far. Floerchinger and Wiedemann \cite{Florchinger:2011qf} analyzed 
rotational modes on top of the Bjorken flow, and formulated conditions for one mode to get unstable. 
Csernai et al
\cite{Csernai:2014nva} had also found an unstable hydro mode, for non-central collisions with rotation. While this effect does not develop
into a full-scale turbulence in heavy ion collisions, due to limited time it exists, it does contribute to an overall fireball rotation
and thus it can perhaps be observed. 

  \begin{figure}[t]
  \begin{center}
  \includegraphics[width=8cm]{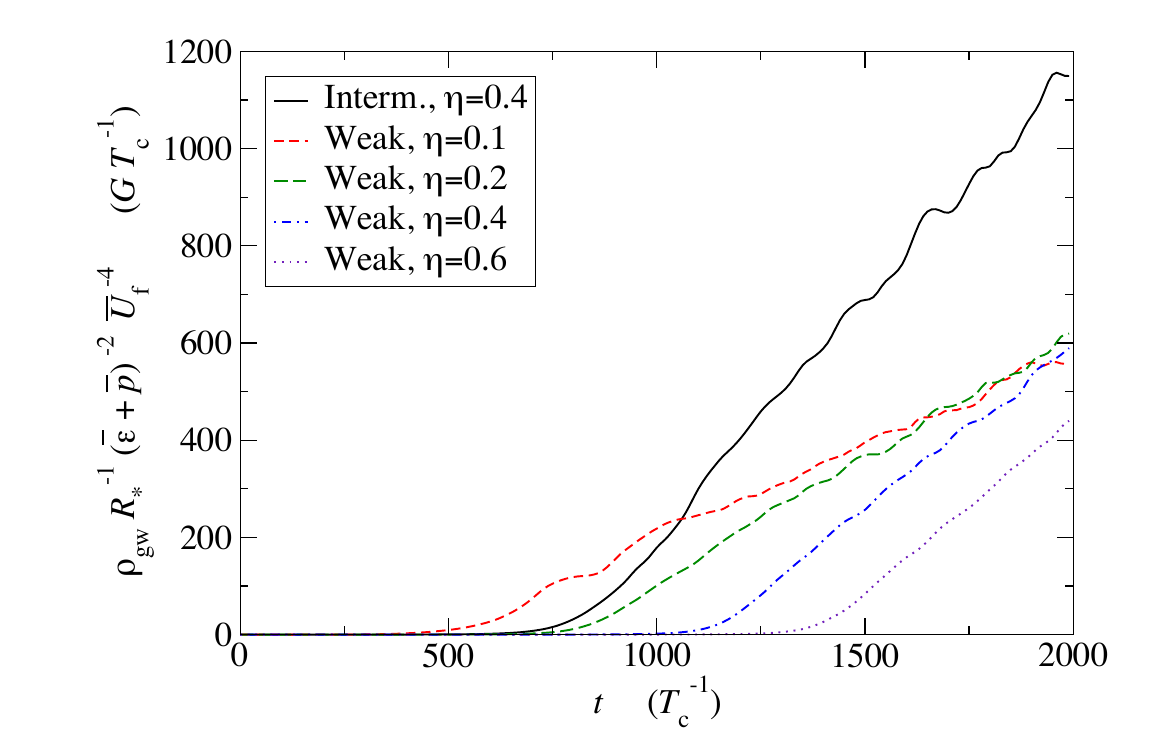} 
  \includegraphics[width=8cm]{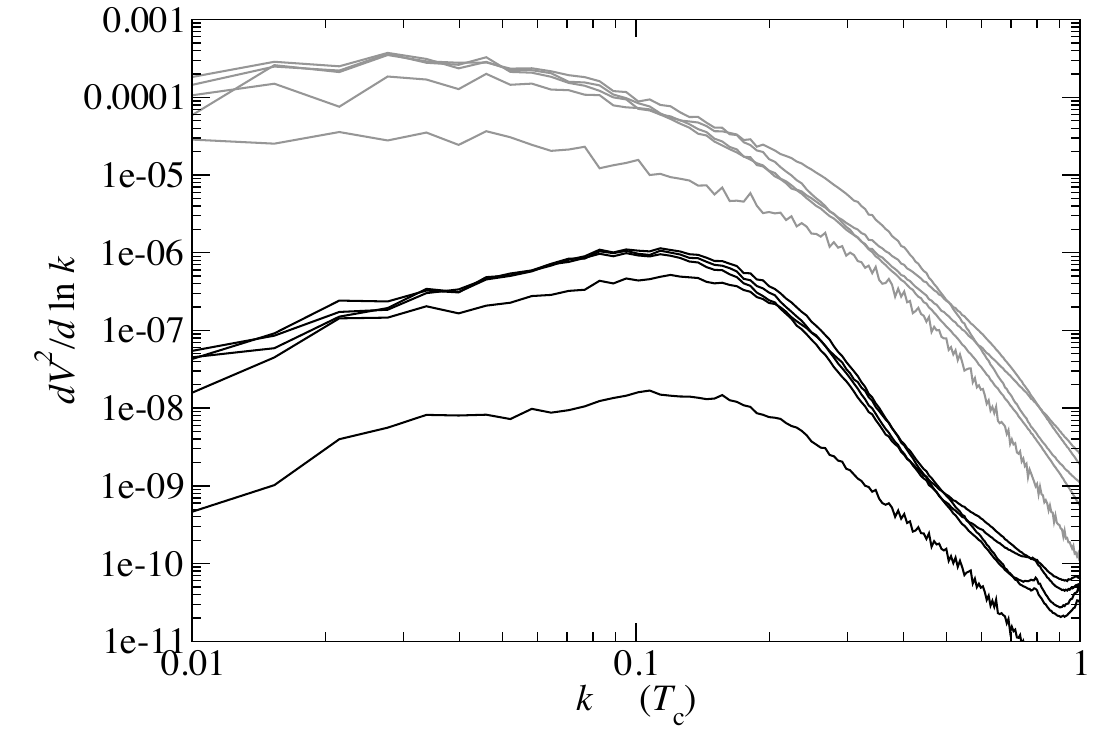}
   \caption{(From \cite{Hindmarsh:2013xza})  (a) The density of generated gravity waves $\rho_{GW}$ versus time
   $t$, in units of inverse $T_c$. The density continue to grow linearly even at the longest times, well after the phase transition itself is over. (b) The power spectrum of the velocity squared versus the wave momentum $k$. The grey upper curves are for sounds, from down up as time progresses. The black curves downwards are for rotational excitations.}
  \label{fig_hindmarsh}
  \end{center}
\end{figure}

\subsection{Cosmological gravitational waves  from sounds of the QCD phase transition } \label{sec_cosmo} 
 
 We think that our Universe has been ``boiling" at its early stages at least three times: (i) at
the initial  equilibration, when entropy was produced, at (ii) electroweak and (iii) QCD phase transitions.
On general grounds, these should have produced certain out-of-equilibrium effects.
It remains a great challenge to us, to observe their consequences experimentally, or at least evaluate their magnitude
theoretically.  The first question is then, what observable may have a chance to see through
the subsequent evolution, back to the early moments of the Big Bang.
 
  From the onset of the QGP physics in heavy ion collisions a specially important  role has been attributed to 
the ``penetrating probes", production of photons/dileptons \cite{Shuryak:1978ij}.
In this section (only) we will jump from our customary Little Bang to the discussion of the Big Bang.
So it is quite logical to start with a question whether it also possesses some kind of  a ``penetrating probe".
The answer is quite clear: while the electromagnetic and weak interactions are in this case not  weak enough,
the gravitational one is.  Thus the  only ``penetrating probe" of the Big Bang are the {\em gravity waves} (GW). 

30 years ago Witten \cite{Witten:1984rs} had discussed the cosmological QCD phase transition,
assuming it to be  of the first order: he pointed out 
 bubble production and coalescence,  producing inhomogeneities in energy distribution
and mentioned production of the gravity waves.  Jumping many years to recent time, we mention that
 Hindmarsh \textit{et al} \cite{Hindmarsh:2013xza} recently found the hydrodynamical sound waves to be the dominant  source of the GW, while
 doing numerical simulations of (variant of) the electroweak (EW) phase transition in the first order transition setting.
Since this work had triggered our interest to the subject, we start illustrating its main findings shown in
Fig.\ref{fig_hindmarsh}. The upper figure shows time evolution of the GW energy, in different simulations. In all cases
the evolution is simply linear: it means that the GW generation rate remains constant {\em long after} the phase transition itself is over. Their further studies had shown that GW originates from the long-wavelength sound waves, not from the rotational modes, which are several orders of magnitude down, see lower Fig.\ref{fig_hindmarsh}.

The sound velocity squared is 
proportional to the energy of the sound waves, which one can also view as $\sim \omega_k n_k k^3$, where $\omega_k=c_s k$ is the sound frequency
and $n_k$ is the density of the momentum distribution, to be discussed below.  So, a flat spectrum seen in  lower Fig.\ref{fig_hindmarsh} corresponds to $n_k\sim 1/k^4$. 
The spectra go 2 orders of magnitude from the UV scale $T$ down, and evolution time about 3 orders of
magnitude. The question is what happens near the IR end of the dynamical range, many orders of magnitude
away of what was simulated in that work.

Kalaydzyan and myself \cite{ourGW}  discussed sound-based GW production and argued  that generation of the cosmological GW  can be divided into four distinct
stages, each with its own physics and scales. We will list them starting from the  UV end of the spectrum
 $k\sim T$ and ending at the IR end of the spectrum $k\sim 1/t_{life}$ cutoff by the Universe lifetime at the era
: \\(i) the production of the sounds   \\(ii)
the inverse cascade" of the acoustic turbulence, moving the sound from UV to IR \\
(iii)   the final transition from sounds  to GW.

The stage (i) remains highly nontrivial, associated with the dynamical details of the
 QCD (or electroweak EW) phase transitions. 
 The  stage (ii), on the other hand, is in fact amenable to perturbative studies
 of the acoustic cascade, which is governed by Boltzmann equation. It has been already rather
 well studied in literature on turbulence, in which power attractor solutions has been identified.
 Application of this theory allows to see how small-amplitude sounds can be amplified, as one goes to smaller $k$.
 The stage (iii)  can  be treated via standard loop diagram for 
 sound+sound $\rightarrow$ GW transition, 

 Before we come to it, let us briefly remind the numbers related to the QCD and electroweak transition. Step one is to evaluate
  redshifts of the transitions, which can be done by comparing the transition temperatures $T_{QCD}=170\, \MeV$ and $T_{EW}\sim 100\, \GeV$ with the temperature of the cosmic microwave background $T_{CMB}=2.73\, \mathrm{K}$. This leads to
  \begin{align}
  z_{QCD}= 7.6 \times10^{11},\quad z_{EW} \sim 4 \times10^{14}\,.
  \end{align}
   At the radiation-dominated era -- to which both QCD and electroweak ones belong -- the solution to Friedmann equation leads to well known relation between the time and the temperature \footnote{Note that we use not gravitational but particle physics units, in which c=1 but the Newton constant
   $G_N=1/M_p^2$.}
   \begin{align}
   t=\left({ 90 \over 32\pi^3 N_{DOF}(t)}  \right)^{1/2} {M_P \over T^2}     \label{eqn_t}
   \end{align}
  where $M_P$ is the Planck mass and $N_{DOF}(t)$ is the effective number of bosonic degrees of freedom (see details in, e.g., PDG, Big Bang cosmology).
  Plugging in the corresponding $T$ one finds the
   the time of the QCD phase transition to be $t_{QCD}=4\times 10^{-5}\,s$, and electroweak $t_{EW}\sim 10^{-11}\,s$.
   Multiplying those times by the respective redshift factors, one finds that the $t_{QCD}$ scale today corresponds to
   about $3\times 10^7\,s \sim  $  year, and the electroweak to $5\times 10^4\,s \sim  $ day.

 GW from the electroweak era are expected to be searched for by future  GW observatories in space, such as eLISA.
  The observational tools for GW at the period scale of $years$ are based on the long-term
monitoring of the millisecond pulsar phases, with subsequent correlation between all of them. The basic idea is that
when GW is falling on Earth and, say, stretches distances in a certain direction, then in the orthogonal direction one expects
distances to be reduced. The binary correlation function for the pulsar time delay is an expected function of the angle $\theta$ between them on the sky. There are existing collaborations -- North American Nanohertz Observatory for Gravitational Radiation, European Pulsar Timing Array (EPTA), and Parkes Pulsar Timing Array -- which actively pursue both the search for new millisecond pulsars and collecting the timing data for some known pulsars. It is believed that about 200 known millisecond pulsars constitute only about 1 percent of the total number of them in our Galaxy.
The current bound  on the GW fraction of the energy density of the Universe
is approximately
\begin{align}
\Omega_{GW} (f\sim 10^{-8} \mathrm{Hz}) h_{100}^2<  10^{-9}\,.
\end{align}
Rapid progress in the field, including better pulsar timing and  formation of a global collaborations of observers, is expected
to improve the sensitivity of the method , perhaps making it possible in a few year time scale to  detect
GW radiation, either from the QCD
 Big Bang GW radiation
we discuss, or that from  colliding supermassive black holes.

  The temperature $T$ provides the micro (UV) scale of the problem: here the phase transition provides the sound source. 
    The cosmological horizon is the IR cutoff on the gravitational radiation wavelength:
   here two sounds generate the GW.    In between UV and IR scales 
   there is the ``dynamical range", of about 18 orders of magnitude!  
 The challenge is to understand if and how the {\em inverse acoustic cascade} can be developed there,
 and what  $n_k$ dependence is generated.  It turns out that the answer crucially depend 
on the $sign$ of the third derivative of
 the sound dispersion curve \be \omega=c_s k +A k^3+... \ee 
 which remains
  unknown, both for QGP and electroweak plasma.
  
If $A>0$ 
 sound decays 1 into 2 is kinematically possible. If so  the turbulent cascade
 based on 3-wave interaction can only develop in the $direct$ -- that is toward large $k$ or UV -- cascade,  not the one
 we are interested in.
However
 when the dispersive correction coefficient $A<0$ , is
 negative, and
 there are no binary decays of the sounds, the answer is different. 
 Turbulent cascade switches to higher order process,  of 
$2\leftrightarrow 2$ scattering  and $1\leftrightarrow 3$ processes, 
 (see e.g. ~\cite{ZLFbook})  generating an $inverse$ cascade, with
a particle flow directed to IR. In this case of
 the weak turbulence, the index of the density momentum distribution $n_k\sim k^{-s}$ is known to be
\be s_{weak}=10/3 \ee  
Furthermore, as discussed in
\cite{ourGW}, large value of the density at small $k$ leads to violation of weak turbulence applicability
condition and the regime is known as ``strong turbulence". We follow the philosophy 
of Berges et al renormalization of
similar cascades in the case of relativistic scalar with quartic interaction. Unfortunately sound waves 
have triple vertex are they are
Goldstone particles, so the problem is much more involved. The scattering is dominated by
$t$-channel diagrams with small denominators, producing small angle scattering 
with large cross sections, like it is the case for gluons. We do not yet have a complete solution
of this problem, only an estimate of the renormalized index which we think is increasing to
$s_{strong}\sim 5$. Given huge -- 18 decades for QCD --  dynamical range
of the problem at hand this will imply a significant  increase of $n_k$ at IR.   


Another key result of our paper \cite{ourGW} is the calculation of the transition rate of the sound to gravity waves. We found it to be rather
simple process, an {\em on shell} collision of the two sound waves.  The calculation of the rate \cite{ourGW} is straightforward,  following from the
evaluation of the ``sound loop" diagram (already discussed in the previous section, but in different kinematics).
Since two sounds collide, it  depends quadratically on $n_k$: so  the GW production can be hugely amplified
by the inverse acoustic cascade.


 
 \section{The pre-equilibrium state, global observables and fluctuations}
\subsection{Perturbative versus non-perturbative  paradigms}

   The requirements to any theory of the early stage can be formulated as follows: It should be able to \\(i)
   specify certain the wave function of the colliding particle, in a wide rapidity range;\\
(ii)    explain what happens at the collisions and  just after it;\\
     (iii) explain how it evolves into  the final observed hadronic state.

   It is perhaps fair to say, that approaches based on the weak coupling  (pQCD) has been able to explain (i) and (ii) but not 
    (iii).  Strong coupling ones (AdS/CFT) can do (iii) but not (i,ii). 
    
    More specifically,    
 perturbative (pQCD) regime is natural for hard processes, for which the QCD running coupling is weak. Already in 1970's 
the  pQCD  developed {\em factorization framework}, which divided production amplitudes into ``past",``during" and ``after" parts. 
The ``past" 
and ``after" parts are treated empirically, by structure (or distribution) and fragmentation functions.
 The ``during", near-instantaneous, part is described
 by explicit partonic process under consideration.
 (These three stages is the simplest example of three goals formulated above.)
 The strength of this approach is based on the separation of hard and soft scales, by some normalization scale $\mu$, 
on which the final answer should not depend. Dependence of 
 PDFs and fragmentation functions on $\mu$ is described by renormalization group  
 tool, the so called DGLAP evolution: it let us 
 tune them to the particular kinematics at hand. It  works very well for very hard processes:
  high energy physicists at LHC
looking at the electroweak scale and beyond,  $Q > 100 \, GeV$,  do not need to know
 anything else.   People who want to study mini-jets have to worry to ``higher twist" corrections to GLAP,
 not  under control yet.

Yet pQCD approach has serious weaknesses as well. The PDFs describe only the $average$  nucleon
(or nucleus). As soon as a particle is touched -- e.g. the impact parameter (multiplicity bin) is selected --
 factorization theorems are no longer applicable. The absence of good practical models describing
 partonic state with fluctuations
  remains a problem: e.g. for
 understanding  $pp$ collisions with multiplicity several times the average one.   
As we will discuss below in detail, pQCD can hardly be used for assessing the transverse plane
distributions/correlations of partons. 
   

For ``baseline" processes --
    minimally biased $pp,pA$ collisions, with rather low multiplicity -- pQCD processes can be supplemented by  string fragmentation, leading to rather successful
     ``event generators", descendent of the so called  Lund model.   In the approximation of independent string fragmentation their relative positions/correlations/interactions are unimportant. 
     However, experiments which trigger higher multiplicity bins and look for subtle  correlations
     have found phenomena clearly beyond the reach of such simple models: we will discuss those in detail.
   
      At the opposite case of  high multiplicity collisions -- central $pA,AA$  -- 
the theory of the initial state is in first approximation classical, and 
general initial conditions for  hydrodynamics is in the zeroth approximation
given just by nuclear shape and $NN$ cross section.  The main parameter
 hydro needs  is the total entropy generated: this is taken from parton density times some empirical
 coefficient, entropy/parton, not yet explained.
  In the first approximation -- including fluctuations in the positions of nucleons  in versions of the Glauber or eikonal models -- one also finds simple and reasonable predictions for $\epsilon_n$ and their correlations/fluctuations. 
 The next approximations, involving fluctuating nucleons in terms of their parton substructure,
 have been proposed, but their relevance is disputed. 

Partonic description of the initial
 state of the collision at asymptotically high parton density evolved into the so called Color-Glass-Condensate (CGC)-
 GLASMA paradigm
 originated from McLerran-Venugopalan model \cite{McLerran:1993ka}.
 Let me  briefly remind the main points. Since the number of colored objects is large, charge fluctuations will eventually also become large producing strong 
gauge fields known as 
CGC. If gluonic fields
 become so strong that the occupation numbers reach $O(1/\alpha_s)$,  the derivative and the commutator terms are comparable
 and one can use $classical$ Yang-Mills nonlinear equations for its description. 
 GLASMA is a state
 made of such random classical fields,  starting from CGC at the collision time and  then evolving as the system expands,
 till the occupation numbers reduce to $O(1)$. Self-consistency of the model is provided by the fact that
 2-d parton density define
 the characteristic
patron {\em saturation momentum} $n\sim Q_s^2$, which is treated as a scale in the evolution equations.
At early time  the charges at each ``glasma cell", of area $\sim 1/Q_s^2$,
start separate longitudinally, producing longitudinal electric and magnetic fields. 
Cells are statistically independent and fluctuate with their own Poisson-like distributions.
The explicit modelling of resulting
field, from cells in the transverse plane, is now known as an $impact-parameter$ (IP) glasma models.

     High-multiplicity initial state then evolve into sQGP: we know that because it must be complemented by hydro evolution, to describe RHIC/LHC observations. We will go over observables below in more detail, let me
     illustrate it here by one particular observable, the $elliptic$ flow $v_2$.
Suppose there is no sQGP stage: 
  patrons -- gluons and quarks --
 simply ``get real" after collision, more or less like the Weitzsacker-Williams photons do in QED, fragment into mini-jets and fly to the detector. 
   Hard
   partons at large momentum scale $Q_s$ of GLASMA belonging to individual cells 
   cannot possibly know about other cells and  
  nuclear geometry: those would be  mainly produced isotropically in the transverse plane. 
  If they re-interact later, the probability of that is higher where there is more matter -- contributing
  $negative$ correction to $v_2$. 
  Much softer 
 patrons, with  momenta $Q\sim 1/R$,  will know about the ``overlap almond" shape of the initial state:
 their distribution will be anisotropic, perhaps even with  
  $v_2$ of the order of several percents, as  observed for momentum-integrated data. Thus the prediction 
  would be of $v_2$ $decreasing$ with $p_t$, to negative values. 
    Needless to say, hydro-based theory and experiment had shown $v_2(p_t)$ to be instead growing
   up to $p_\perp\sim 3 \, GeV$, producing anisotropies as large as $O(1)$. 
   
Strong-coupling models of the initial stage and equilibrated matter fall into two categories.
One is based on classical strongly-coupled plasmas developed in \cite{Gelman:2006xw} and based on the 
notion that the ratio 
\be \Gamma ={V_{interaction} \over T} > 1 \ee
In other words, the potential energy of a particle exceeds  its kinetic energy. Simulations
and experiments with QED strongly coupled plasmas show that for relevant $\Gamma=1-10$ 
one deals with {\em strongly correlated liquids}. Screening in this regime was studied in \cite{Gelman:2006sr},
and viscosity and diffusion constant in \cite{Liao:2006ry} in a version with electric and magnetic charges.
Apart of jet quenching applications, it has not yet been used for heavy ion physics, and thus we will not discuss it here.
 


  The second -- much wider known --  strong coupling framework is based on holography and AdS/CFT 
  correspondence: we will also discuss recent progress in detail. It 
  describes rapid equilibration by black hole formation in the dual theory space: we will discuss it in significant detail below.
  
  \subsection{Centrality, $E_\perp$ and fluctuations} \label{seq_centrality}
Before diving into theory developments, let me briefly remind some basic 
facts about the global observables and their fluctuations. 
One of the first practical question for AA collisions is determination of centrality classes,
 related to observables like the number of participant nucleons $N_p$, correlated to total multiplicity $N$ or 
 transverse energy $E_\perp$.
 The $N_p$ is 
defined by forward-backward  calorimeters, while the others by the central detectors. Correlation plots between
those and precise cuts defining the centrality classes are the basic technical issues of the field.

Historically,
the ratio of the $E_\perp$ rapidity distributions for AA and pp collisions were fitted by a parameterization 
\be {dE^{AA}_\perp \over d\eta} /{dE^{pp}_\perp \over d\eta}=(1-x){N_p \over 2} +x N_{coll} 
\label{eqn_Eperp}
\ee
with a parameter $x$ interpreted as an admixture  of the 
``binary  collisions" $ N_{coll} $ to the main ``soft" term, proportional to the number of participants.
It was a nice fit: yet  we now know  that a
 ``hard"  components in particle spectra - is 3-4 orders of magnitude
down from the ``soft" exponent (thermal-hydro contribution). So, the interpretation of $x$ becomes questionable.

   One possibility  can be  that ``hard"  interpretation just mentioned remains correct at early time, yet 
with  subsequent equilibration and disappearance from spectra.   Since hydro preserves entropy, one may argue that it will still show up in the total multiplicity. It is however more difficult  to imagine preservation of $E_\perp$ at the hydro stage. 

   \begin{figure}[t!]
  \begin{center}
  \includegraphics[width=6cm]{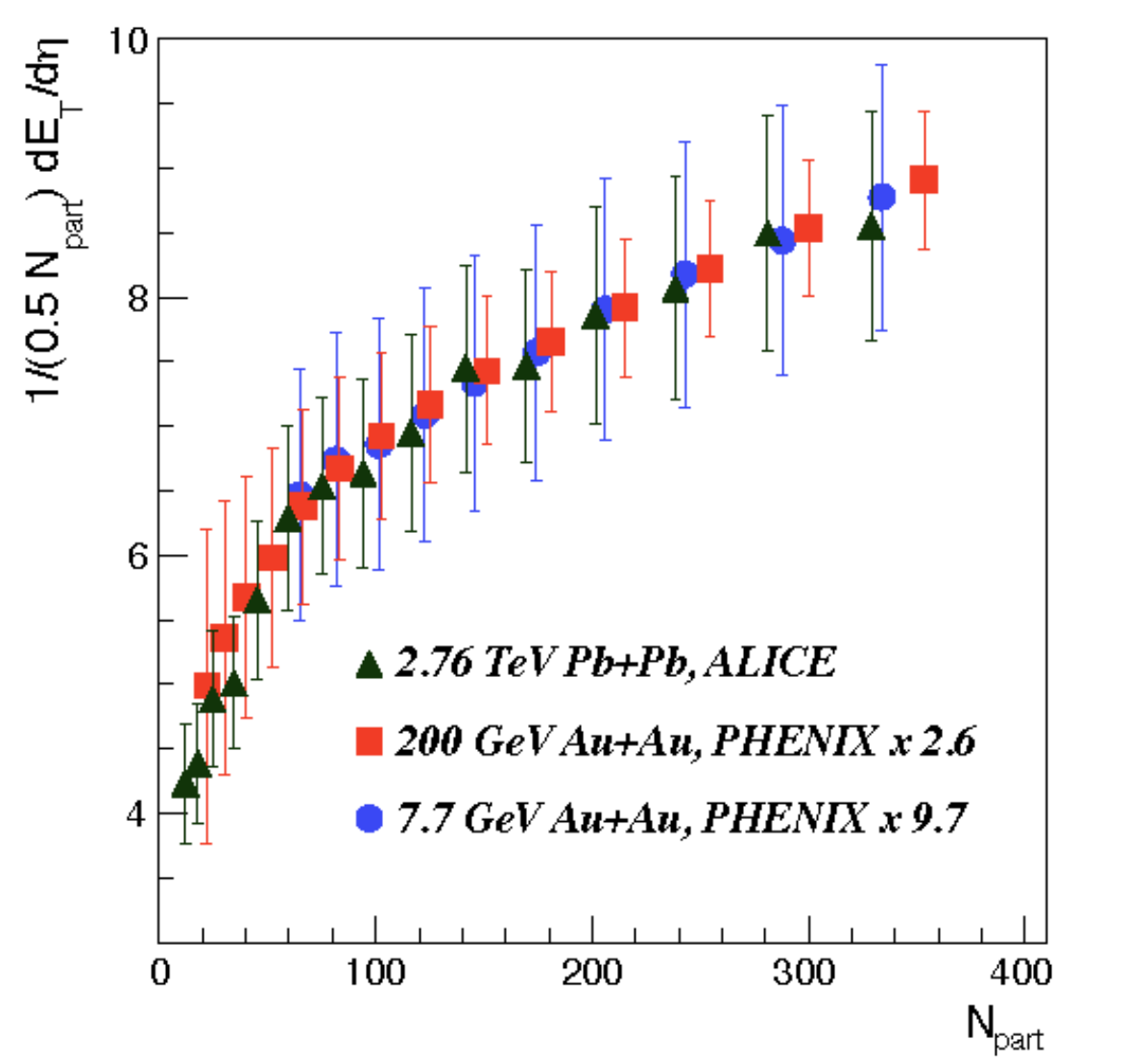}
  \includegraphics[width=8cm]{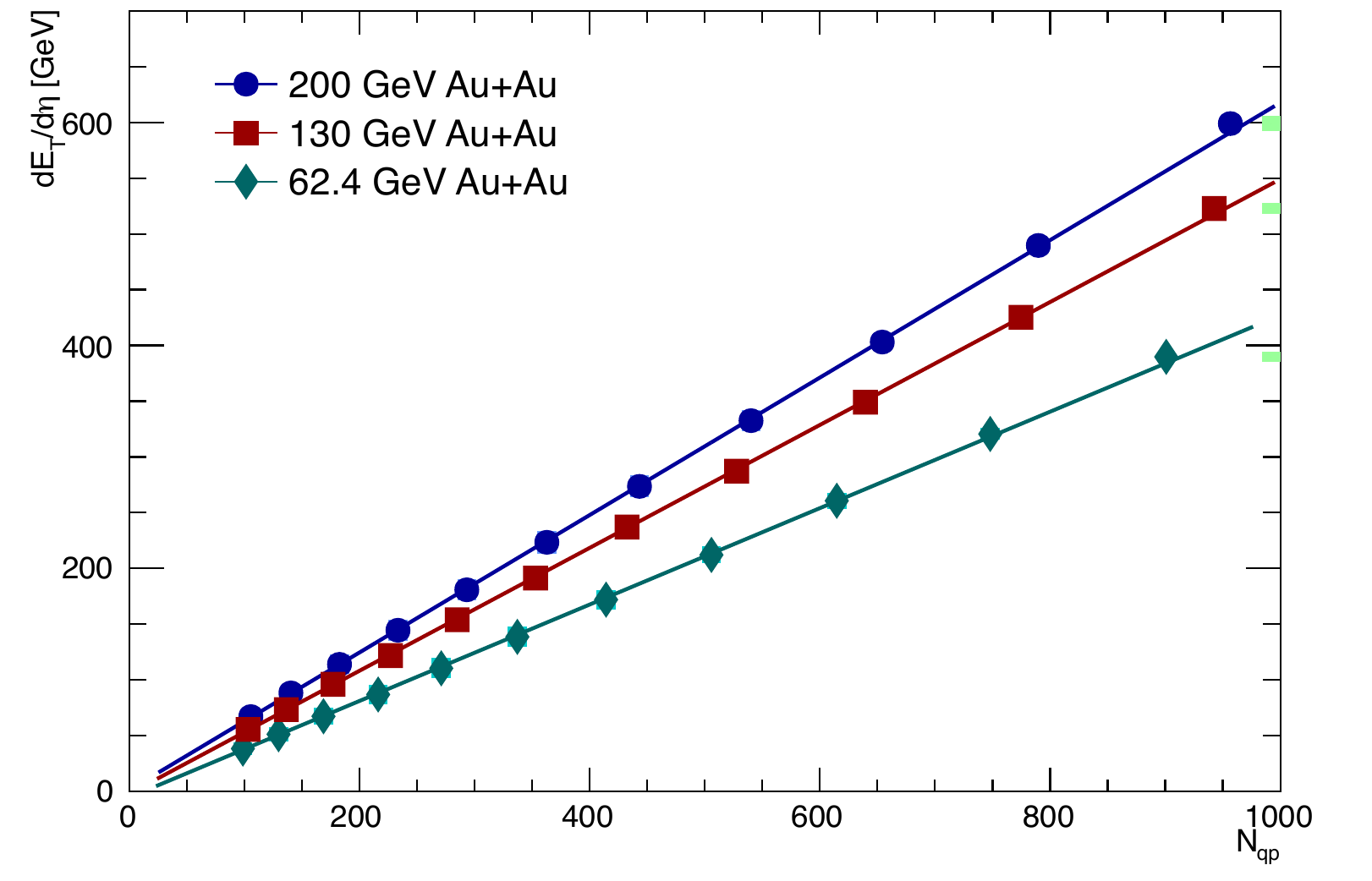} 
   \caption{ Distributions over participant nucleons (a) and participant quarks (b), from \cite{Tannenbaum}. }
  \label{fig_ETperNPvsNp}
  \end{center}
\end{figure}

But this is not the end of the story. The multiplicity and $E_\perp$ distributions were measured and connected
to the geometric distribution over 
 the r.h.s. of (\ref{eqn_Eperp}).   These studies bring along a notion of some intermediate ``clusters" (or 
ÒancestorsÓ as ALICE papers call it) which  generate the secondaries in independent  random processes
of their decay. 

Tannenbaum  \cite{Tannenbaum}  provided another, and very simple,  interpretation to these distributions. A
  nucleon is represented by 3 constituent quarks which interact separately -- the additive quark model of 1960's.
Defining the number of ``quark participants" $N_{qp}$ he showed that --
within a 1\% accuracy (!) -- it is proportional to the r.h.s. of (\ref{eqn_Eperp}). 
Thus, the $E_\perp$ is perfectly linear in $N_{qp}$, see fig \ref{fig_ETperNPvsNp}(b).
If so, each  participant quark is connected by the QCD string to the other one, and those strings are the ``clusters" or 
ÒancestorsÓ for final secondaries.  (We will return to ``wounded quarks" concept in the discussion of the
Pomeron in section \ref{sec_Pomeron}.)

On the other hand, the additive quark model does not agree with pQCD paradigm, which views
color dipoles -- not charges -- as having separate cross sections (or couplings to the Pomeron). 
   The CGC/GLASMA theory adds another 
  suggestion: it is its GLASMA cells that  are expected to be statistically independent ``clusters",
  producing entropy/secondaries.
 (The McLerran-Venugopalan "flux tubes"  differ from the 
QCD strings: unlike those  they exist  in dense  deconfined phase, are not quantized  
and they do not have universal tension). 

A view that
the CGC-glasma picture is a  high density regime, while simpler Lund-type models with
QCD strings (and perhaps constituent quarks) are in low density regime must be correct.  
The problem is we do not see transition which will help us to tell
 at which density we need to switch between those two pictures.

Let us  seek further guidance from phenomenology. Note that the number of ``clusters" $N$ 
which fluctuate  independently   define the width of observed distributions,
being $O(N^{-1/2})$. So, we take 
these three models: (i) the usual Glauber in which $N$ is the number of participant nucleons $N_p$;
(ii) its Tannenbaum's modification with  the number of participant constituent quarks $N_{pq}$; (iii) and the CGC-glasma,
and calculate the fluctuations. 

In the last case
\be N_{GLASMA}\sim (\pi R^2) Q_s^2 \ee
For central AA the area is geometrical Area=100 $fm^2$ and \be N_p\sim 400, N_{pq}\sim 1000, N_{GLASMA}\sim 10^4 \ee
For central $pA$ the area is given by the NN cross section $\sigma \sim 100 \, mb=10 \, fm^2$. So one gets very different number of ``clusters"
\be N_p\sim 16, N_{pq}\sim 40,
N_{GLASMA}\sim 10^3 \ee
 Therefore these models predict vastly different fluctuations.


There are two ways to measure fluctuations. The first is based on {\em multiplicity fluctuations}. Going to the tail of it, beyond
the most central collisions, in AA,pA and $pp$  we find some tail usually fitted by the negative binomial or similar distribution with two parameters, or convolution of two random processes with different parameters.
Its second moment should tell us how many ``progenitors" (clusters, clans, ancestors) the system goes through. 

The second one is to look at {\em angular deformations} $\epsilon_n$. 
If created by statistically independent small-size objects, they are all very similar and again parametrically $O(N^{-1/2})$ \footnote{
Even $\epsilon_2$ is the same as others, as we now  for simplicity focus on central collisions.
Note further that of one select the so called super-central collisions, like below 0.1 or 0.01\%, larger multiplicity
relates to larger $N$ and all  $v_n$ do decrease as  $O(N^{-1/2})$.}.
 We return to them in section \ref{sec_inital_angular}.




For $E_t$ distributions,  one finds that participant quark model describes AuAu and dAu 
data extremely well, while in $pp$  it clearly  underpredicts the tail of the distribution. Even 6 participant quarks -- the maximal of the model --
is not enough. Recalling our density estimates above, one may think that the highest multiplicity $pp$ 
is the first case when ``soft" models become insufficient.   
The models which have pQCD gluons in the wave function are doing better on the ``tails".
Popular  quantitative model with pQCD component is
Pythia (pQCD+strings): the version used by CMS  describes the
multiplicity tail of  $pp$  reasonably well, while it underpredicts a bit the tail in $pA$.
 (see e.g. CMS pages with public info).


\subsection{Anisotropy of the excited matter and the boundaries of hydrodynamics \label{sec_anisotropy}}
Partonic initial state has small $p_t$ and large longitudinal momenta: so
initial out-of-equilibrium stage of the collision is highly anisotropic. However,
after collisions, patrons are  naturally separated in time according to different rapidities,
and create ``floating matter" in which cells have a spread of longitudinal momenta $smaller$ than
the transverse one, reversing the sign of momentum anisotropy.  At later hydrodynamical
stages the viscosity effects produce local anisotropy of the particle distributions, which is however small
due to small viscosity of the fluid.

What happens in between is still a matter of debates. Weak coupling approaching --partonic cascades --
predict anisotropy to be rising to quite large values, while the strong coupling (holographic) approaches
lead to rapid convergence to small values, consistent with hydrodynamics. (For more detailed discussion see e.g. \cite{Martinez:2009ry}).

\begin{figure}[t]
  \begin{center}
  \includegraphics[width=8cm]{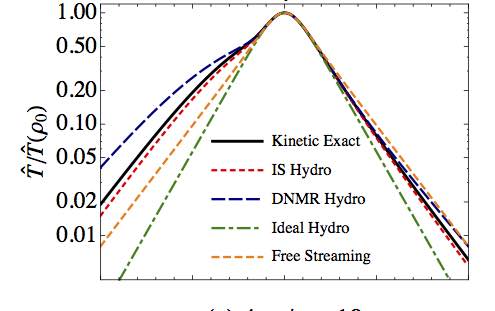}
    \includegraphics[width=7cm]{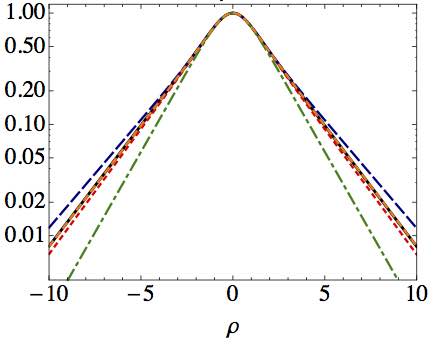}
   \caption{ From \cite{Denicol:2014tha}: the normalized temperature for $4\pi \eta/s=1, 100$,  upper and lower plots, respectively.
   The meaning of different curves is explained in the upper plot. }
  \label{fig_G_Boltz_T}
  \end{center}
\end{figure}

\begin{figure}[b]
  \begin{center}
  \includegraphics[width=6cm]{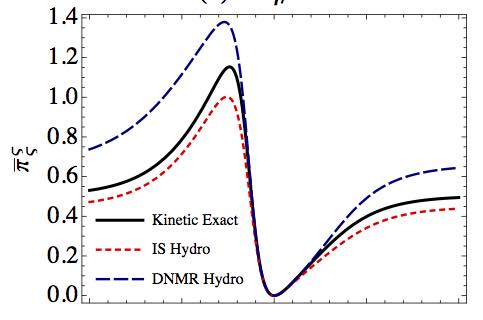}
    \includegraphics[width=6cm]{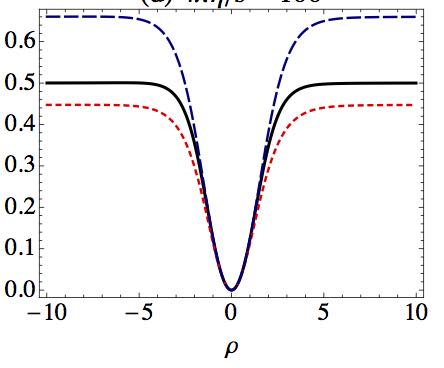}
   \caption{From \cite{Denicol:2014tha}: shear stress for $4\pi \eta/s=1, 100$,  upper and lower plots, respectively }
  \label{fig_G_Boltz_Pi}
  \end{center}
\end{figure}

The issue of anisotropy has two practical aspects. The experimental one -- to which we return in section \ref{sec_anisotropy} -- is a question how
one can experimentally monitor the anisotropy of matter, at various stages of the evolution.
The theoretical question is whether one can extend the hydrodynamical description for strongly anisotropic
matter. Recently there were significant development along the line of  the so called {\em anisotropic hydrodynamics},
or aHydro.
The idea  \cite{Martinez:2009ry,anysotropic} is to introduce the asymmetry parameter into the 
particle distribution, and then derive separate equation of motion for it from Boltzmann equation. 
More recently, solutions of various versions of hydrodynamics were compared to the exact solution of the Boltzmann equation
itself, derived for Gubser geometrical setting in \cite{Denicol:2014tha}. This paper contains many instructive plots,
from which I selected the normalized temperature shown in Fig.\ref {fig_G_Boltz_T} and the sheer stress $\Pi^\xi_\xi$
shown in Fig.\ref {fig_G_Boltz_Pi} . In both cases the pairs of points correspond to small and very high viscosity
values, separated by two orders of magnitude and roughly representatives of the  strongly and weakly coupled regimes.  
 
 Gubser's variable $\rho$ is the ``time" coordinate. At all four plots one can see that all curves coincide in the interval $-2<\rho< 2$,
 but deviate from each other both at large negative values, corresponding to the very early stage, and
for  large positive ones, corresponding to very late times. In fact all practical applications of hydrodynamics
were indeed made inside this interval of $\rho$, with other regions being ``before formation" and ``after freezeout". 

Solutions for two -- hugely different - viscosities show a very similar trends. Israel-Stuart hydro 
seems to follow Boltzmann in the most accurate way. Even the free streaming regime is not very far
from all hydros and exact Boltzmann: this would be surprising for the reader if we would not 
already know that the radial flow -- unlike higher harmonics -- can indeed be ``faked". 
If these authors would calculate the elliptic and higher flows, the results would be quite different.
It is not done yet, but one expects that for $4\pi \eta/s=100$ those would be completely obliterated.

The plots for the shear stress show different behaviors for small and large viscosity, but again all
curves coincide inside  the ``hydro window" of $-2<\rho< 2$. Even going well outside that domain, we never see discrepancies between them by more than say 20\%.

The  overall conclusion one can draw from all of those impressive works is quite simple: all versions
of hydro used in practice are very accurate for realistic viscosities $ 4\pi \eta/s\approx 2$ and the times hydro is actually 
applied in practice.

\section{The smallest drops of QGP}
  We have described above some
 successes of hydrodynamics for description of the flow angular harmonics, showing that those are basically
  sound waves generated by the initial state perturbations. We also emphasized
a significant 
 gap which still exists between approaches based on weak and strong couplings,
 in respect to equilibration time and matter viscosity. 
  Needless to say, the key to all those issues should be found in experimentations
  with systems smaller than central AA collisions. They should eventually clarify {\em the limits of hydrodynamics}
  and reveal what exactly happen in this hotly disputed ``the first 1 fm/c" of the collisions.
  
Let us start this  with another look at the flow harmonics. What is the spatial scale corresponding to
the highest $n$ of the $v_n$ observed? 
%
A successful description of the $n$-th
 harmonics along the fireball $surface$ implies that hydro still works at a wavelength scale $2\pi R/n$: taken the nuclear radius $R\sim 6\, fm$ and the largest harmonic studied in hydro $n=6$ one concludes that
this scale is  still few fm. So, it is still large enough, and it is impossible
to tell the difference between the initial states of the Glauber model (operating with nucleons) from 
those generated by parton or glasma-based models  (operating on quark-gluon level) .
And indeed, 
 we don't see harmonics with larger $n$ simply because 
 of current statistical limitations of the data sample. Higher harmonics suffer stronger
 viscous damping, during the long time to  freezeout. In short, non-observation of $v_n,n>6$
  {\em are unrelated} to the limits of hydrodynamics.
 
 (Except in the region of high $p_\perp\sim 4 GeV$ in which indeed the relevant region becomes
 too thin to be treated macroscopically. Unfortunately, there is still little
  progress in  understanding $v_n$ at and above such momenta.)


In principle,  one can study $AA$ collisions for  smaller and smaller systems,
looking for the  lighter nuclei which still see  flow harmonics. 
Note that in such case the time to freeze out scales with the radius, so angular correlations stay
about at the same angles and $n$. In our pocket formula for viscous damping,
it changes $T R$ downward, allowing
to understand the sound damping phenomena from another angle.

However, it is not the way of the actual development, as
unexpected discoveries of flows in very small systems --  $pp$ and $pA$ collisions, with high multiplicity trigger.
It is those which we discuss  in this chapter. As we will see, there are similarities but also important differences of the two cases.
  
   \begin{figure}[t!]
  \begin{center}
  \includegraphics[width=8cm]{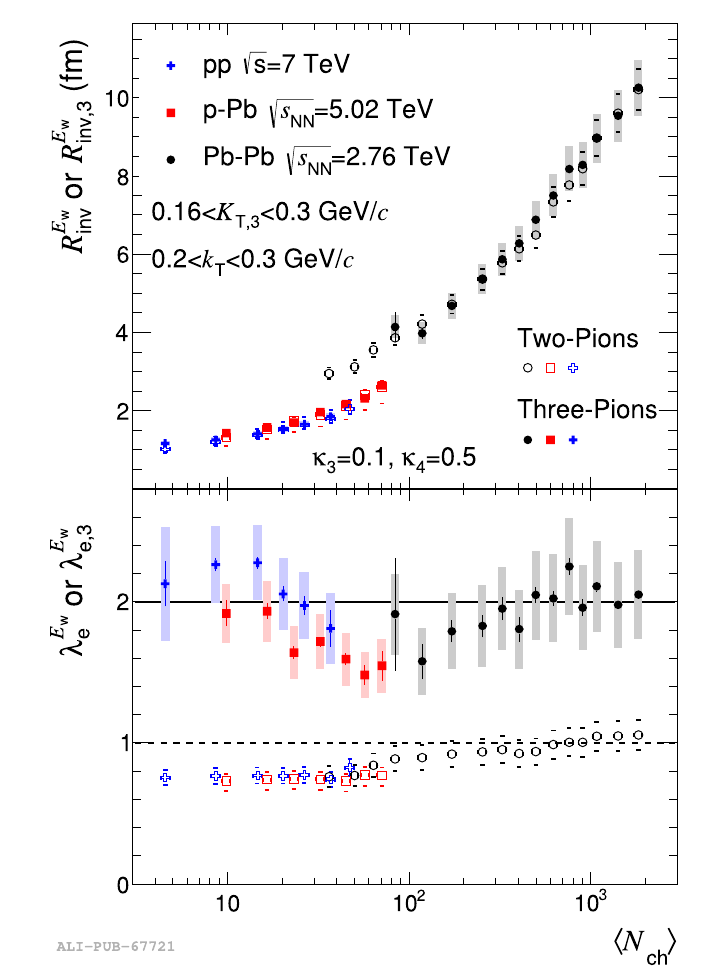}
   \caption{Alice data on the femtoscopy radii (From \cite{ALICE_qm20141})  (upper part) and ``coherence parameter" (lower part)
   as a function of multiplicity, for $pp,pPb,PbPb$ collisions. }
  \label{fig_A_HBT}
  \end{center}
\end{figure}

Before we go into details, let us try to see how large those systems really are.
At freezeout the size can be directly measured, using femtoscopy  method.
(Brief history: so called Hanbury-Brown-Twiss (HBT) radii. This interferometry method  came from radio astronomy.
The influence of Bose symmetrization of the wave function of the observed mesons in particle physics 
was first emphasized by Goldhaber et al \cite{Goldhaber:1960sf} and applied to proton-antiproton annihilation.
Its use for the determination of the size/duration of the particle production processes had been proposed 
by Kopylov and Podgoretsky \cite{Kopylov:1973qq} and myself \cite{Shuryak:1974am}. 
With the advent of heavy ion
collisions this ``femtoscopy'' technique had grew into a large industry. Early applications for RHIC 
heavy ion collisions were in certain tension with the hydrodynamical models, although this issue was later
resolved \cite{Pratt:2008qv}.)

The corresponding data are shown in Fig.\ref{fig_A_HBT}, which combines the
traditional 2-pion with more novel 3-pion correlation functions of identical pions. 
An overall growth of the freezeout size with multiplicity, roughly as $<N_{ch}>^{1/3}$,
is expected already from the simplest picture, in which the freezeout density is some universal constant.
For AA collisions this simple idea roughly works: 3 orders of magnitude of the growth in multiplicity correspond
to one order of magnitude growth of the size. 
 
 Yet the $pp, pA$ data apparently fall on a different line, with significantly  smaller radii, even 
 if compared to the peripheral AA collisions at the same multiplicity. Why do those systems 
 get frozen at higher density, than those produced in AA? 
  To understand that
 one should recall the {\em freezeout condition}:
 ``the collision rate becomes comparable to the expansion rate"
 \be <n\sigma v> =\tau_{coll}^{-1}(n)\sim  \tau_{expansion}^{-1}= { dn(\tau)/d\tau \over n  (\tau)} \ee
Higher density means larger l.h.s., and thus we need a larger r.h.s.. So, 
more ``explosive" systems, with larger expansion rate, freezeout earlier.
We will indeed argue below that $pp,pA$ high-multiplicity systems are in fact more ``explosive":
it is seen from radial flow effects on spectra as well as HBT radii.  

But $how$ those systems become ``more explosive" in the first place? Where is the room for that,
people usually ask, given that even the final size of these objects is  smaller than in peripheral AA,
which has a rather weak radial flow. 
Well, the only space left is at the beginning:  those systems must {\em start accelerating earlier},
 from {\em even smaller} size. Only then they would be able 
to get enough acceleration, and eventually strong collective flow, by their  freezeout.  

\subsection{Collectivity in small systems}

  Let us briefly recall the time sequence of the main events. 
The first discovery --  in the very first LHC run -- was due to CMS collaboration \cite{Khachatryan:2010gv} which found a ``ridge" correlation 
in high multiplicity $pp$ events, enhanced by a special trigger. That was required because, unfortunately, 
the effect was first seen only in events with a  probability $P\sim 10^{-6}$
\footnote{ Dividing the cost of LHC, $\sim 10^{10}\$$ by the number of recorded $pp$ events  $\sim 10^{10}$
one finds that they cost about a dollar each. These  high multiplicity $pp$ events we speak about thus cost  about a million dollar each, and one needs thousands of them to report a correlation function!}.

Switching to most central $pA$ CMS \cite{CMS:2012qk} and other collaborators had observed a similar ridge there, 
now with much higher --  few percent -- probability. By subtracting high multiplicity and low multiplicity correlators
CMS and ALICE groups soon had concluded, that ``ridge" is accompanied by the ``anti-ridge" in other hemisphere, and thus it is basically a familiar elliptic flow $v_2$.

PHENIX collaboration at RHIC also found a ridge-like correction in central $dAu$ collisions, Furthermore, their
 $v_2$ is $larger$ than in $pPb$ at LHC, by about factor 2. This was soon attributed to
 different initial conditions, for $d$ and $p$ beams, 
 since the former have ``double explosion" of two nucleons. Quantitative prediction came from
 pioneering 
hydrodynamical studies of such collisions by Bozek \cite{Bozek}, and then many others.
Hydro predicts the effect correctly: and that
 was the first indication for the collectivity of the phenomenon.
Another contribution of  PHENIX was the observation that $dA$ HBT radii 
display the famous decreasing trend with $p_t$ well known
for $AA$ collisions, which is another -- and very direct -- evidence for presence of the collective flow. 

\begin{figure}[t]
\begin{center}
\includegraphics[width=8cm]{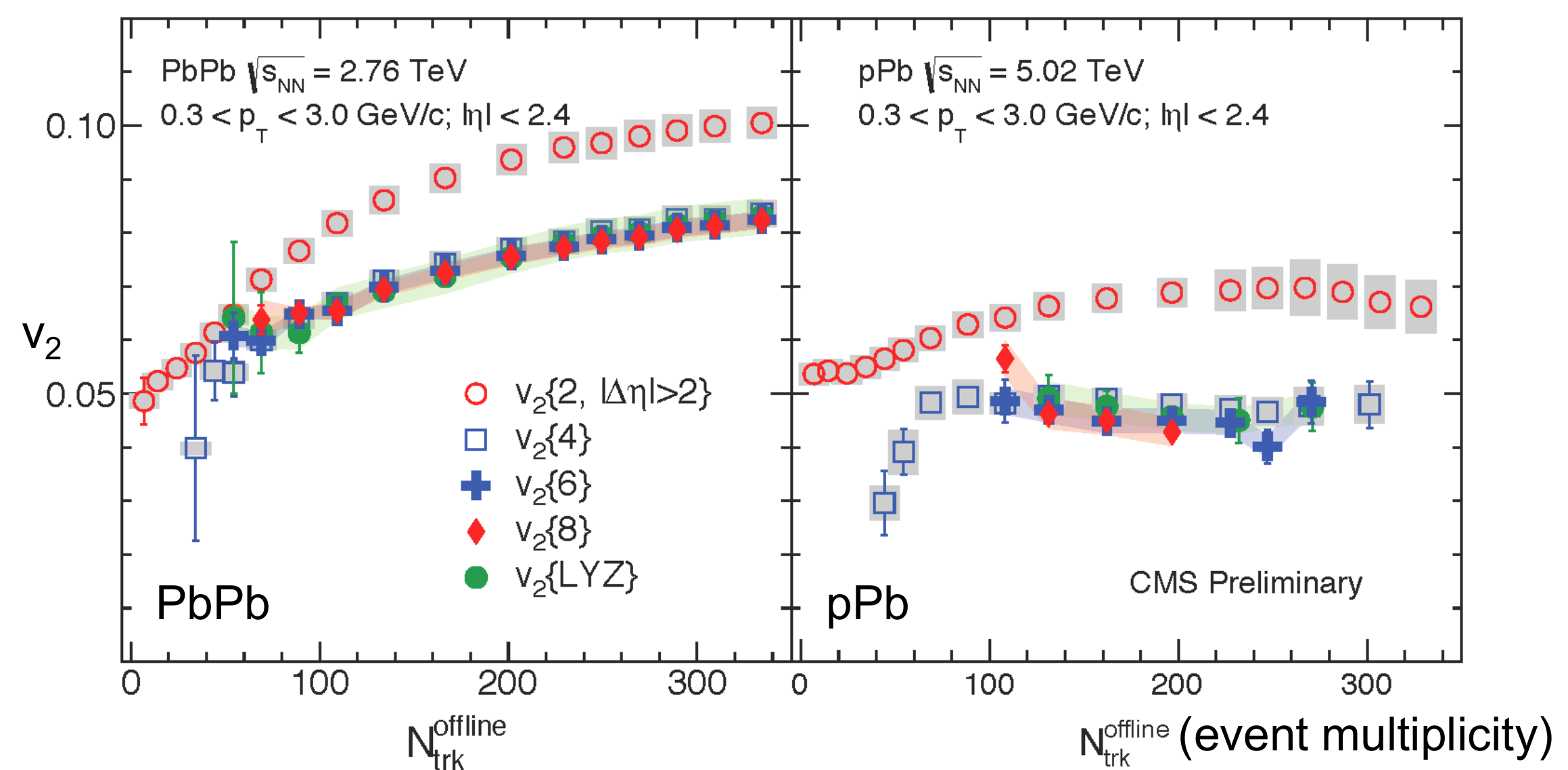}
\caption{  CMS data \cite{CMS_v2} for $v_2$ calculated using 2,4,6, 8 particle correlations, as well as Lee-Yang zeroes (basically all particles). Good agreement between those manifest collectivity of the phenomenon.   }
\label{fig_CMS_v2}
\end{center}
\end{figure}

Truly amazing set of data  came from  CMS \cite{CMS_v2}. Their $v_2$ measurements from 4,6
and even 8 particles are shown in Fig.\ref{fig_CMS_v2}. Previous data for AA collisions had shown perfect agreement
between those, and new data for $pA$ are in this respect the same. This establishes $collectivity$ of the flow
in pA, ``beyond the reasonable doubt".  

Taken collectivity for granted, one can further ask if the $v_2$ observed is caused 
by the pre-collision correlations  or by  the after-collision collective flows. 
A very nice control experiment testing this is provided by $d A$ and
 $He^3 A$ collisions. Two nucleons in $d$ are in average far from each other and 2 MeV binding
 is so small that one surely can ignore their initial state correlations. So, whatever is the ``initial shape"
 effect in $pp$, in $dA$ it should be $reduced$ by $1/\sqrt{2}$
 because two shapes cannot be correlated. It should be
  reduced further in  $He^3 A$ by  $1/\sqrt{3}$, if the same logic holds.
  
 Hydrodynamical  predictions are opposite: double (or triple) initial explosions lead to a common fireball, with the anisotropies  $larger$ than in $pA$.  
  Data from RHIC by PHENIX and STAR on $ dAu, He^3 Au$  do indeed show such an $increase$ of
the $v_2, v_3$, relative to $p Au$, again in quantitative agreement with
 hydro  \cite{BozekHe3,Nagle}. It looks like this issue is now settled.
 
\begin{figure}[t]
  \begin{center}
  \includegraphics[width=6cm]{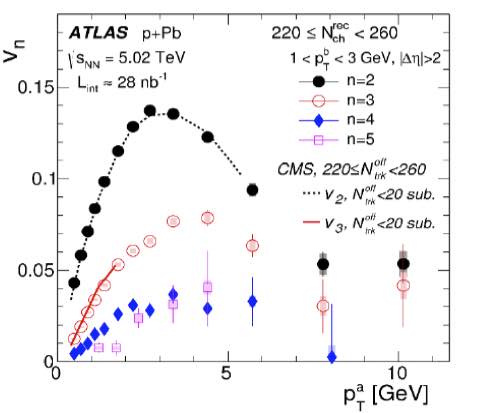}
   \caption{$v_n$ for $n-2,3,4,5$ versus $p_t$ in GeV, for high multiplicity bin indicated on the figure.
The   points are from Atlas, lines from CMS (presentation at QM2015).}
  \label{fig_pA_vn_atlas}
  \end{center}
\end{figure}

 Recently Atlas was able to perform the first measurements of higher harmonics $v_n,n=4,5$ in central $pPb$,  see Fig. \ref{fig_pA_vn_atlas}. Except at very high $p_t$, those two harmonics seem to be comparable in magnitude:
 it is the first contradiction to ``viscous damping" systematics seen so far. (No idea so far why does it happen.)  


     \begin{figure}[t]
\begin{center}
\includegraphics[width=7cm]{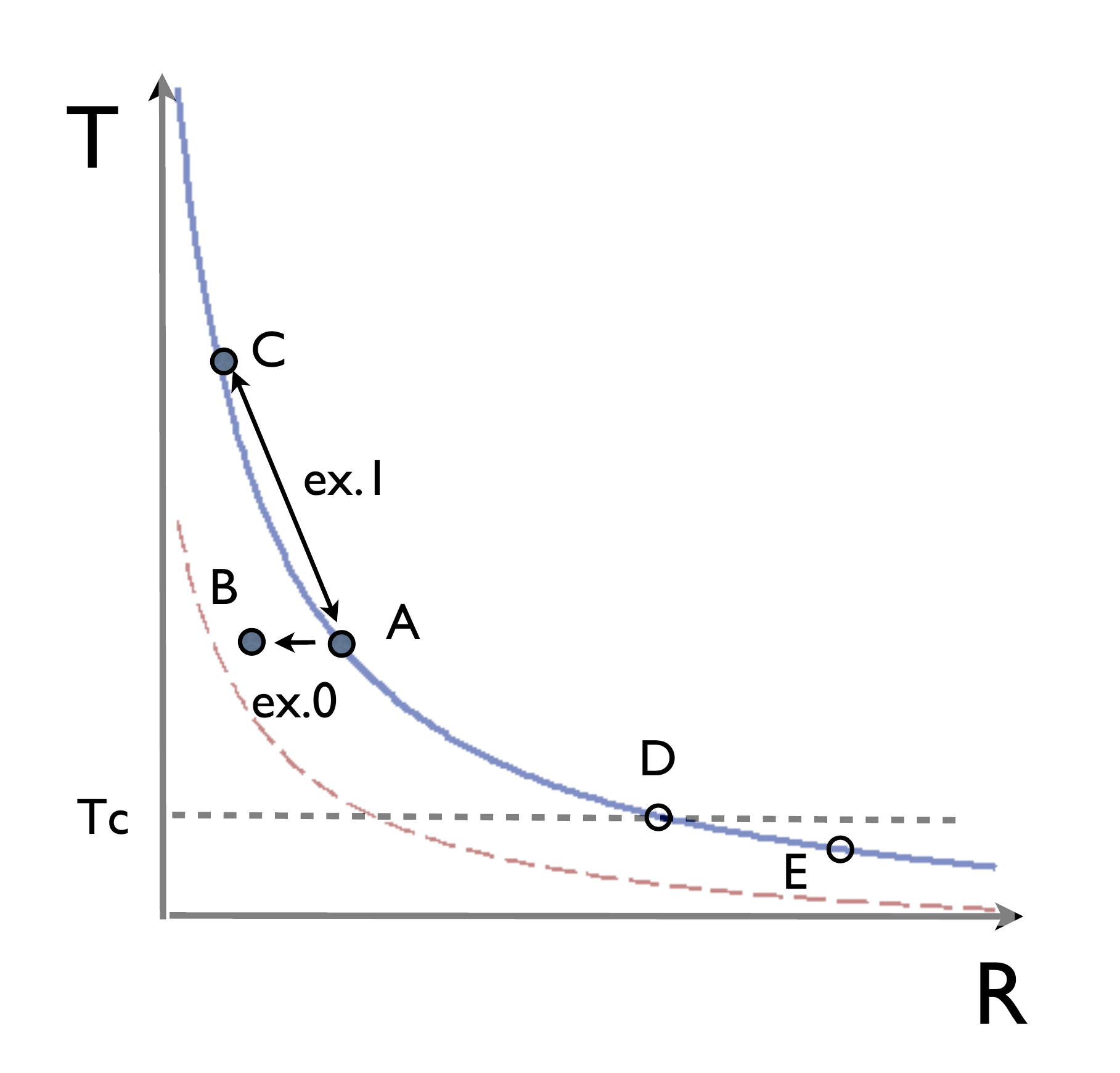}
\caption{(color online)  Temperature $T$ versus the fireball size $R$ plane. Solid blue line is the adiabatic $S=const$,
approximately $TR=const$ for sQGP. Example 0 in the text corresponds to reducing $R$, moving left $A \rightarrow B$.
Example 1 is moving up the adiabatic $A \rightarrow C$.  Example 2 corresponds to adiabatic expansion, such as
$A \rightarrow E$,$C \rightarrow E$. If in reality $C$ corresponds to $pA$, the freezeout occurs at the earlier point $D$.      }
\label{fig_TR}
\end{center}
\end{figure}

\subsection{Pedagogical digression: scale invariance of sQGP and small systems} \label{sec_conf}

Acceptance of hydrodynamical treatment of ``small system explosions"  need to pass a psychological barrier: people repeatedly ask 
if it is indeed possible to treat a  system of less than 1 fm in size as macroscopic.
(And indeed, just 15 years ago the same question was asked about systems of 6 fm size.)
So, let us take a step back from the data and consider the issue of scales.

If one takes smaller and smaller cells of ordinary fluid -- such as water or air --
eventually one reaches the atomic scale, beyond which water or air, as such, do not exist:
just individual molecules.
QGP is $not$ like that: it is made of essentially massless quarks and gluons which have no scale of their
own. The relevant scale is given by only one parameter $T$ -- thus QGP is approximately 
scale invariant. (The second scale $\Lambda_{QCD}$ only enters via logarithmic running of the coupling,
 which is relatively slow and can in some approximation be ignored.)  

As lattice simulation show, above  the phase transition $T>T_c$ QGP thermodynamics soon
becomes scale invariant $\epsilon/T^4,p/T^4\approx const(T)$. The comparison of LHC to RHIC data 
 further suggests that it is similarly true for viscosity as well $\eta/s \approx const(T)$
 (although with less accuracy so far). Thus, QGP does not
have a scale of its own, and thus would show {\em exactly the same behavior} if conditions related by the
scale transformation
 \be R_A/R_C=\xi,T_A/T_C=  \xi^{-1} \ee
are compared.

   Consider  a {\em thought experiment 1}, in which we compare two systems 
     on the same adiabatic $A$ and $C$.  For scale invariant  sQGP the points  $A,C$ are
  related by this scale transformation mentioned above, and have the same entropy. 
    Since the scale transformation is approximate symmetry, we expect the same output.
   A smaller-but-hotter plasma
   ball $C$ will explode in exactly  the same way as its larger-but-cooler version $A$.

Let us now proceed to the {\em thought experiment 2}, which is the same as above but in QCD, with a running coupling.
In the sQGP regime it leads to (very small, as lattice tells us ) running of $s/T^3$, some (unknown) running of $\eta/T^3$, etc. 
The most dramatic effect is however not the running coupling {\em per se}, but the lack of supersymmetry,
which allows for the  chiral/deconfinement phase transition, out of the sQGP
phase at $T=T_c$ to hadronic phase.  The end of the sQGP explosion $D$ thus has an {\em absolute scale}, not subject to scale transformation!

So let us consider two systems $A$,$C$ of the same total entropy/multiplicity, initiated in sQGP with conditions related by scale transformation
and left  them explode.
The sQGP evolution would be related by nearly the same set of intermediate states (modulo running coupling)  till $T\approx T_c$,
after which they go into the ``mixed" and hadronic stages, which are $not$ even close to be scale invariant!  Thus the 
result of the explosions are not the same. In fact the smaller/hotter system will have an advantage over the larger/cooler one,
since it has $larger$  ratio between the initial and final scales $T_i/T_f$. 

(In the language of holographic models the scale is interpreted as the 5-th coordinate $x^5$, and evolution is depicted as gravitational falling of particles,strings, fireballs etc toward the AdS center. The ratio of the scales is the distance  along the
5-th coordinate: thus in this language two systems fall similarly in the same gravity, but smaller system starts ``higher" and thus got
larger velocity at the same ``ground level" given by $T_c$.) 

The hydro expansion does not need to stop at the phase boundary $D$. In fact large systems, as obtained in central AA
collisions are known to freezeout at $T_f<T_c$, down to 100 \, MeV range (and indicated in the sketch by the point $E$. 
However small systems, obtained in peripheral AA or central $pA$ seem to freezeout at $D$, as we will show at the end of the paper.

Short summary of these thought experiments: not only one expects hydro in  the smaller/hotter system to be there, it should be very similar
to that in larger/cooler system, due to approximate scale invariance of sQGP.

\begin{figure}[t!]
\begin{center}
\includegraphics[width=7cm]{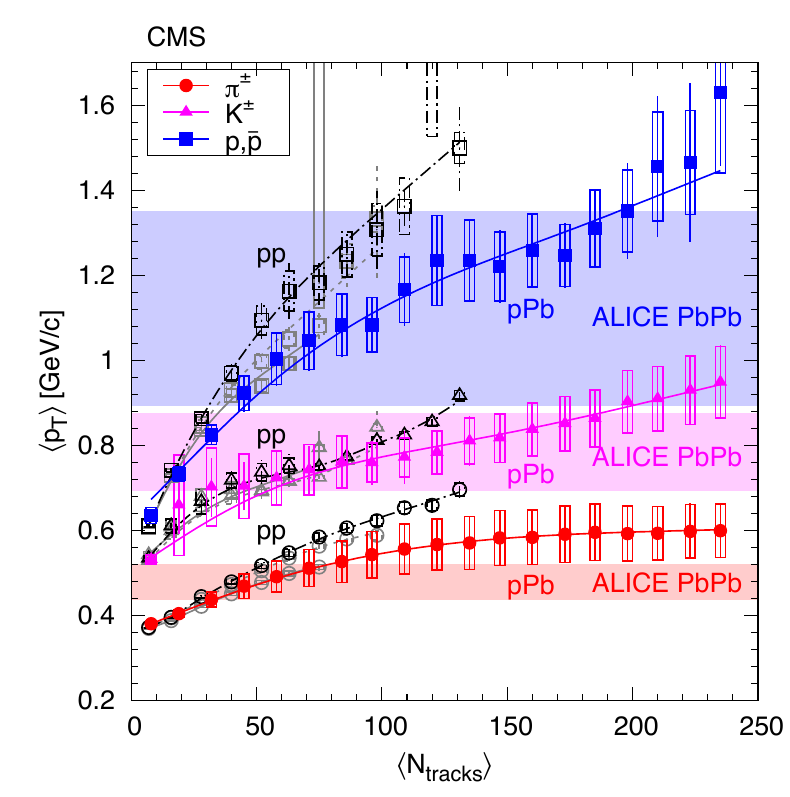}
\includegraphics[width=7cm]{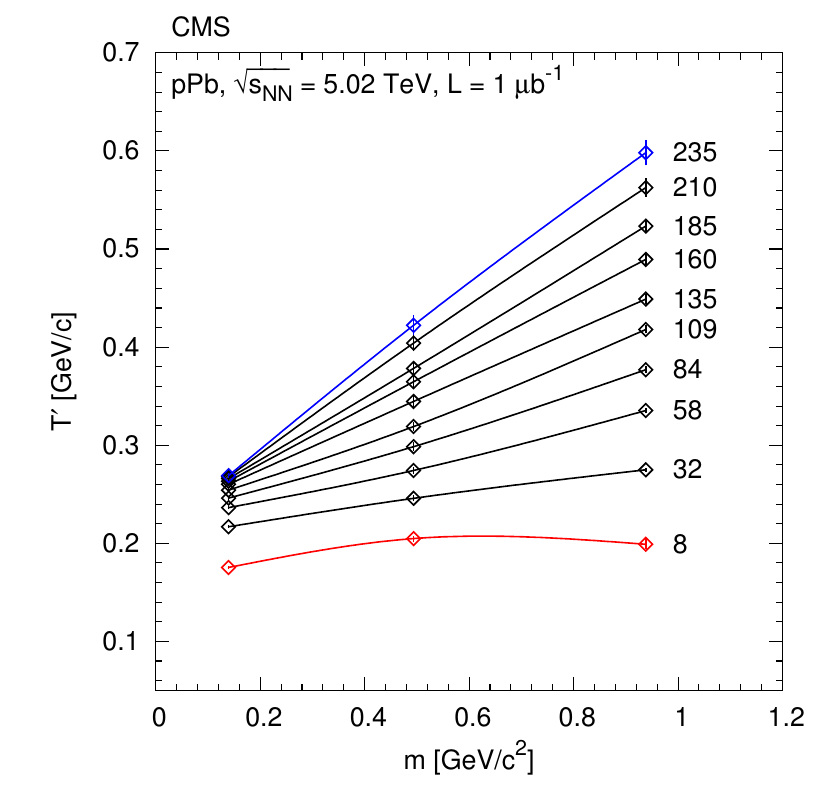}
\caption{(color online)  (From   
\cite{Chatrchyan:2013eya}.) (a) Average transverse momentum of identified charged hadrons (pions, kaons, protons; left panel) and ratios of particle yields (right panel) in the range $|y| < 1$ as a function of the corrected track multiplicity for 
$|\eta| < 2.4$, for $pp$  collisions (open symbols) at several energies, and for pPb collisions (filled symbols) at $\sqrt{s_{NN}} = 5.02\, TeV.$ (b)
The slopes of the $m_\perp$ distribution $T'$ (in GeV) as a function of the particle mass. The numbers on the right 
of the lines give
the track multiplicity.}
\label{fig_slopesa}
\end{center}
\end{figure}

\subsection{Preliminary comparison of the peripheral AA, \\ central  $pA$ and high multiplicity $pp$ }

Now is the time to go  from thought experiment with some ideal systems  to the real ones. We will
do it in two steps, first starting in this section with  ``naive" estimates for three cases at hand,
 based on standard assumptions about the collision dynamics, and then returning to more realistic
studies of the last two cases in the next subsections. 

We aim at {\em initial} transverse radii and density, thus 
we use the initial size of the nuclei rather than that of the fireball at freezeout, after hydro expansion.
The multiplicity is the final one, but due to (approximate) entropy conservation during the hydro stage we think of it as a proxy for
the entropy at early time as well. (Entropy generated by viscosity $during$ expansion is relatively small and can be corrected for, if needed.)

(i) Our most studied case, the central AuAu or PbPb, is the obvious benchmark. With the total multiplicity about $N_{AA}\approx 10^4$ and transverse area of 
nuclei  $ \pi R_A^2\approx 100\, fm^2 $ one gets the density per area \be n_{AA}={N \over \pi R_A^2} \sim 100  \, fm^{-2}\ee
which can be transformed into entropy if needed, in a standard  way. 

(ii) Central $pA$ (up to few percent of the total cross section) has CMS track multiplicity of about 100. Accounting for unobserved range of $p_t,y$ and neutrals increases it by about factor 3,
so $N_{pA}^{central} \sim 300$. The area now corresponds to the typical impact parameter $b$  in $pp$  collisions, or $\pi< b^2>=\sigma_{pp} \approx 10\, fm^2$. 
The density per area is then \be   n_{pA}^{central}={ N_{pA}^{central}  \over \sigma_{pp}} \sim 30 \, fm^{-2} \ee 
or 1/3 of that in central AA. Using the power of LHC luminosity CMS can reach -- as a fluctuation with the probability $10^{-6}$ --- another increase of the multiplicity, by about factor 2.5 or so,
reaching the density $N_{pA}^{max} /\sigma_{pp}$ in AA. 
Another approach used is a comparison of central $pA$ with peripheral AA of the same multiplicity, or more or less same number of participants.  
Similar matter density is obtained. 

(iii) Now we move to the last (and most controversial) case, of the high multiplicity $pp$ collisions. (Needless to say the density is very low for min.bias events.) ``High multiplicity" at which CMS famously discovered the ``ridge"
starts from about $N_{pp}^{max} > 100*3$ (again, 100 is the number of CMS recorded tracks and 3 is extrapolation outside the detector covered). 

The big question now is: {\em what is the area?}  Unlike in the case of central pA, we don't utilize standard Glauber and full cross section (maximal impact
parameters):
 we address now
a fluctuation  which has small probability. 
In fact, nobody knows the answer to that. Based on the profile of $pp$  elastic scattering (to be discussed in section \ref{sec_Pomeron})  I think it should correspond
to impact parameter $b$ in the black disc regime. If so $\pi b^2_{b.d.}\sim 1/2 \, fm^2$, which leads to density per area
 \be n_{pp}^{max}\approx  {N_{pp}^{max}  \over \pi b^2_{b.d.} }\sim 600 \, fm^{-2} \ee
Other evidences about glue distribution in a proton comes from HERA diffractive production, especially of $\gamma\rightarrow J/\psi$: they also suggest a r.m.s. radius of only $0.3 \, fm$,
less than a half of electromagnetic radius. 

Let us summarize those (naive) estimates: in terms of the initial entropy density
one expects the following order of the densities per area involved
\be {dN^{pA}_{maximal} \over dA_\perp} \sim {dN^{AA}_{peripheral} \over dA_\perp}  \ll  {dN^{AA}_{central} \over dA_\perp}  \ll  {dN^{pp}_{maximal} \over dA_\perp}  \ee   
One may  expect that the radial flow follows the same pattern: yet the data  show it is $not$ the case.

 \subsection{The ``radial flow puzzle"  for central  pA}
      The simplest consequence of the radial flow is increasing mean transverse momentum. CMS data on those, as a function of multiplicity, are shown in 
 Fig.\ref{fig_slopesa}(a). While $pp$ and $pA$ data are shown by points, the AA ones (from ALICE) are shown by shaded areas:
 the central ones correspond to its upper edge.  
 While one invent many mechanisms of the mean $p_t$ growth -- e.g. rescattering, larger saturation momentum $Q_s$  at higher multiplicity, etc --  
    those explanations generally do to explain why heavier particles -- protons -- get this effect stronger than the pions.  
 
      True experimental signatures of the radial flow are based on the observation that collective flow affect  spectra of secondaries of different mass differently.  While (near) massless pions retain exponential spectra, with blue-shifted slope,  massive particles have spectra of a modified shape. 
      Eventually, for very heavy particles (e.g. $d$ or other nuclei) their thermal motion gets negligible and their 
      momenta become just $mv$ where $v$ is the velocity of the flow. Its distribution  has a characteristic peak at the fireball's edge.  
      
      More specifically, a measure proposed in  \cite{Shuryak:1979ds} is to look at the so called ``violation of the $m_\perp$ scaling". 
The  $m_\perp$ slopes $T'$ are defined by the exponential  form (above certain $p_t$)
  \be {dN \over dy dp_\perp^2} = {dN \over dy dm_\perp^2} \sim exp( -{m_\perp \over T'} ) \ee
 and they are
 the best indicators of the radial flow.      
 A sample of such slopes for $pA$ collisions, from CMS, is shown in  
Fig.\ref{fig_slopesa} (similar data from ALICE but for smaller multiplicities are also available, see Fig.\ref{fig_alice_blast}).  
   The multiplicity bins (marked by $8$ and $32$ at the bottom-right)   show the same $T'$ for all secondaries: this is the $m_\perp$ scaling indicating that the flow
   is absent.   This  behavior is natural  for independent string fragmentation, rescattering or glasma models.

   Flow manifests itself differently. For pions $T'$ is 
simply the freezeout temperature, blue-shifted by the exponent of the transverse flow rapidity
\be  T'=T_f e^\kappa \ee 
For more
   massive particles --   kaons, protons, lambdas, deuterons etc -- the slopes are mass-dependent .  As seen from  Fig.\ref{fig_slopesa}(b),  
 they are growing approximately linearly with the mass, and the effect gets more pronounced with multiplicity.
This is the signature  of the collective flow.
 Furthermore, the central $pA$ bin has slopes
 $exceeding$  those in central PbPb collisions at LHC,  the previous record-holding! 
 (Predicted to happen  before experiment: see
  version v1 of  \cite{Shuryak:2013ke}.)
 
 This gives rise to what we call {\em the radial flow puzzle}. Indeed,  naive estimates of densities
 in the previous subsection may suggest that explosion in highest multiplicity 
  $pA$ case should still be  $weaker$ than in AA. Indeed, both the system
 is smaller and  the initial entropy density seem to be smaller as well. Yet the data show the opposite:
 the observed radial  flow strength 
follows a different pattern
\be y_\perp^{AA,central} <  y_\perp^{pA,central} <  y_\perp^{pp,highest} \ee


     \begin{figure}[t]
\begin{center}
\includegraphics[width=7cm]{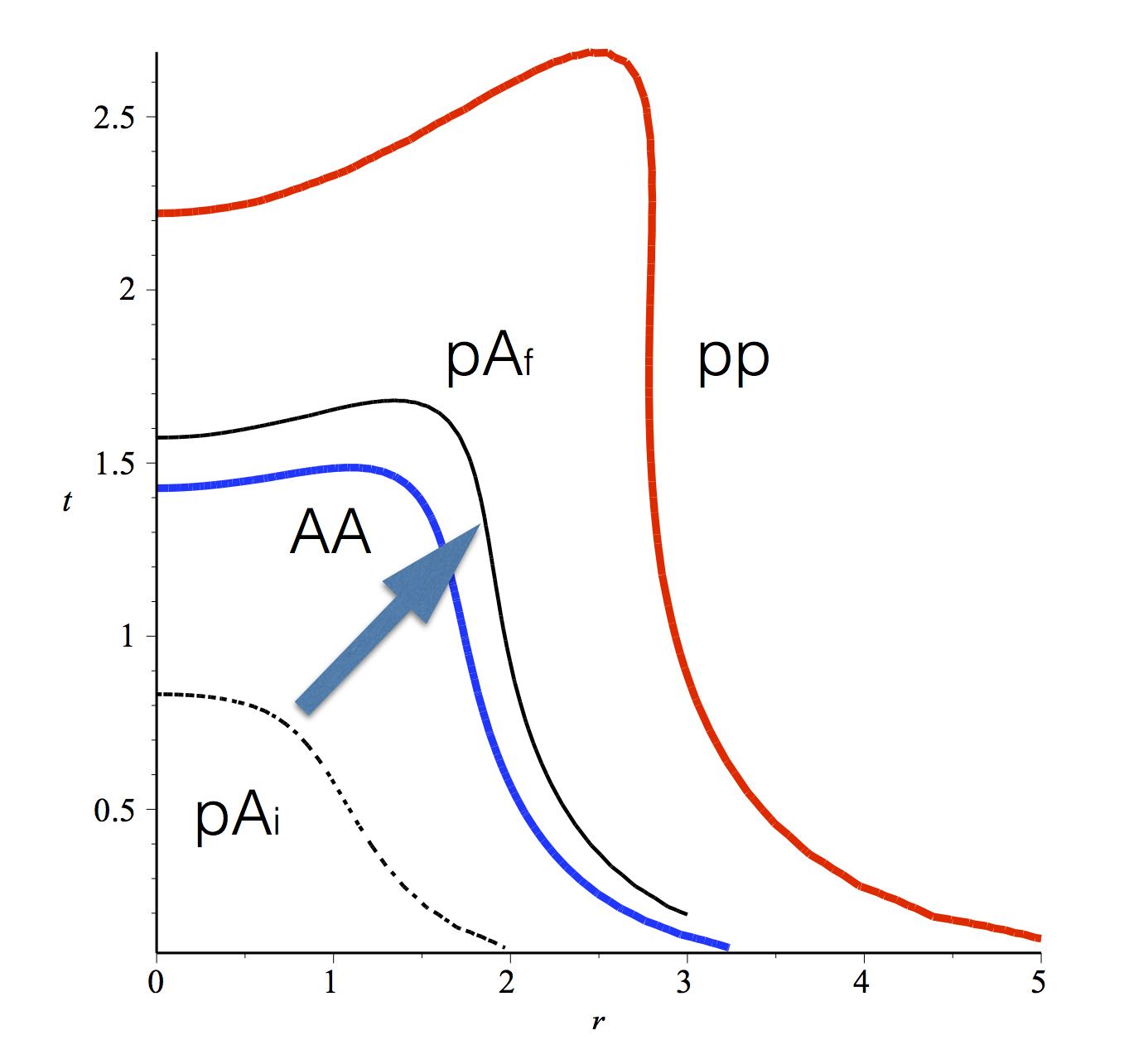}
\caption{(color online) The freezeout surface in universal dimensionless time $t$ and radial distance $r$ coordinates. (Blue) thick solid line in the middle corresponds to central AA (PbPb) collisions, (red)  thick solid line  on the top to the highest multiplicity $pp$ . Two (black) thin ones correspond to central p Pb case,
before and after collapse compression, marked $pA_i, pA_f$ respectively. The arrow connecting
them   indicates the effect of multi string collapse.    }
\label{fig_freezeouts}
\end{center}
\end{figure}

Hydrodynamics is basically a bridge, between the initial and the final properties of the system. For the radial flow
  dependence on the size of the system  it is convenient to follow Ref. \cite{Shuryak:2013ke} based on Gubser's flow , see section \ref{seq_Gubser} .
One single analytic solution describes all cases considered: we will proceed from the
 dimensional variables $\bar{\tau},\bar{r}$ with the bar
to dimensionless variables
\be  t=q \bar{\tau}, \,\,\,r=q \bar{r} \ee
using rescaling by a single  parameter $q$ with dimension of the inverse length.
In such variables there is a single solution
 of ideal relativistic hydrodynamics, which for the transverse velocity 
 (\ref{G_v} )
 and the energy density (\ref{G_energy}) dependence on time and space is known.
The  second equation fixes the shape of the freezeout surface, usually an isotherm $T=T_f$.

Before turning to actual plot of such surfaces, let us 
recall our thought experiment 1 in the
subsection \ref{sec_conf}: two collisions which are  conformal copies of each other
would look the same in dimensionless variables. And indeed, 
the blue line marked AA in Fig.\ref{fig_freezeouts} 
corresponds not only to central PbPb
collisions at LHC, but actually to any $AA$ collisions. 
(Its parameters, for the record, are
 $q=1/4.3 \, fm, \hat{\epsilon}_0 = 2531, T_f=120\, MeV$, a benchmark with ``realistic"
radial flow.)
Two  black lines are for the pPb case: they both have $T_f=170\, MeV$ and the same multiplicity 
but different scale parameters: $q=1/  1.6 \,fm$ for the lower dotted line but twice  smaller spacial scale 
$q=1/0.8 \, fm$ for the upper thin solid line. As an arrow  indicates, 
in order to explain the data one has to
start
hydro not from the ``naive" initial  size, the former line, but  from the ``compressed" size,
according to ``spaghetti collapse" scenario we will discuss in section \ref{sec_conf}. 
If this is done, the freezeout surface ``jumps over" our
 AA benchmark blue line, and its radial flow gets stronger.  The  maximal transverse velocities on these curves (located near the turn of the
freezeout surface downward) are
\be  v_\perp^{pAu} = 0.56 \, < \,v_\perp^{AA} = 0.81 \, < \, v_\perp^{pAu,f}  = 0.84 \ee
The upper red line is our guess for the maximal multiplicity $pp$  collisions, assuming its $q=1/ 0.5\, fm$:
it has even stronger radial flow, with maximal $v_\perp^{pp}  \approx 0.93$.
So, paradoxically, small systems are in fact larger than AA in appropriate dimensionless variables, 
and that is why their radial flow is better developed.

 In summary: the observed pattern of radial flow magnitude can only be explained if
 the initial size of the $pA$ system is significantly reduced, compared to the naive estimates 
 in the preceding section.
 

 \subsection{Radial flow in high multiplicity $pp$ } \label{sec_pp}
According to our preliminary discussion of the densities per area, in this case it is much higher than in AA 
collisions. 
 The initial state must be in a GLASMA state, if there is one. Yet we do not have theoretical guidance about the size.
( It is not at all surprising: those are 
    fluctuations with small probability $\sim 10^{-6}$, and understand their precise dynamics is difficult.)  Lacking good theory guidance,
 one may invert our logical path, and proceed as follows: (i) extract the magnitude of the flows -- radial, $v_2,v_3$ -- from the data, at freezeout. Then (ii) ``solved hydro backwards",
 deriving the initial conditions  needed to generate such a flow.
 
 One phenomenological input can be the mean $p_t$ and spectra of the identified particles in high multiplicity $pp$:
 some of those we have already shown in
Fig.\ref{fig_slopesa}. 
%

\begin{figure}[t]
\begin{center}
\includegraphics[width=60mm]{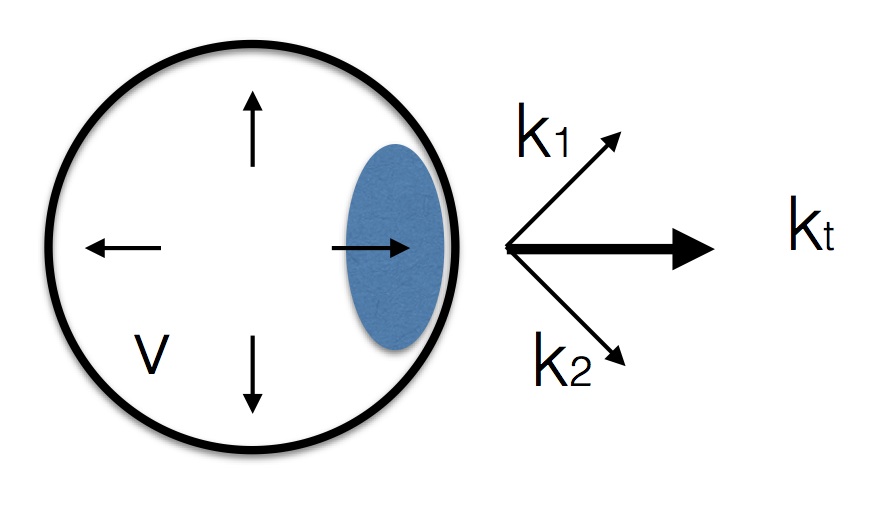}
\end{center}
\caption{Sketch of the radial flow (arrows directed radially from the fireball center) explaining how it influences the
HBT radii. At small $k_t$ the whole fireball (the large circle) contributes, but at larger $k_t$ one sees only the
part of the fireball which is co-moving in the same direction as the observed pair. This region -- shown by shaded ellipse
-- has a smaller radii and anisotropic shape, even for central collisions.
}
\label{fig:sketch}
\end{figure}

Another approach is to use is
the femtoscopy method. It allows 
 to detect the magnitude and even deformations of the flow.
 Makhlin and Sinyukov \cite{Makhlin:1987gm} made the important observation that HBT radii
decrease with the increase of the (total) transverse momentum $\vec{k_1}+\vec{k_2}=\vec{k_t}$ of the pair.
Modification of their argument for our purposes is explained in a
 sketch shown in Fig.\ref{fig:sketch}. At small  $k_t$ the detector sees hadrons emitted from the whole fireball,
but the larger is $k_t$, the brighter becomes its 
small
  (shaded) part in which the radial flow is (i) maximal and (ii) has the same direction
 as $\vec{k_t}$. This follows from maximization of the Doppler-shifted thermal
 spectrum $\sim exp\left(p^\mu u_\mu/T_{freezeout}\right)$. One way to put is is to note that
 effective $T$ in it is increased by the gamma factor of the flow.

\begin{figure}[t!]
\begin{center}
\includegraphics[width=7.5cm]{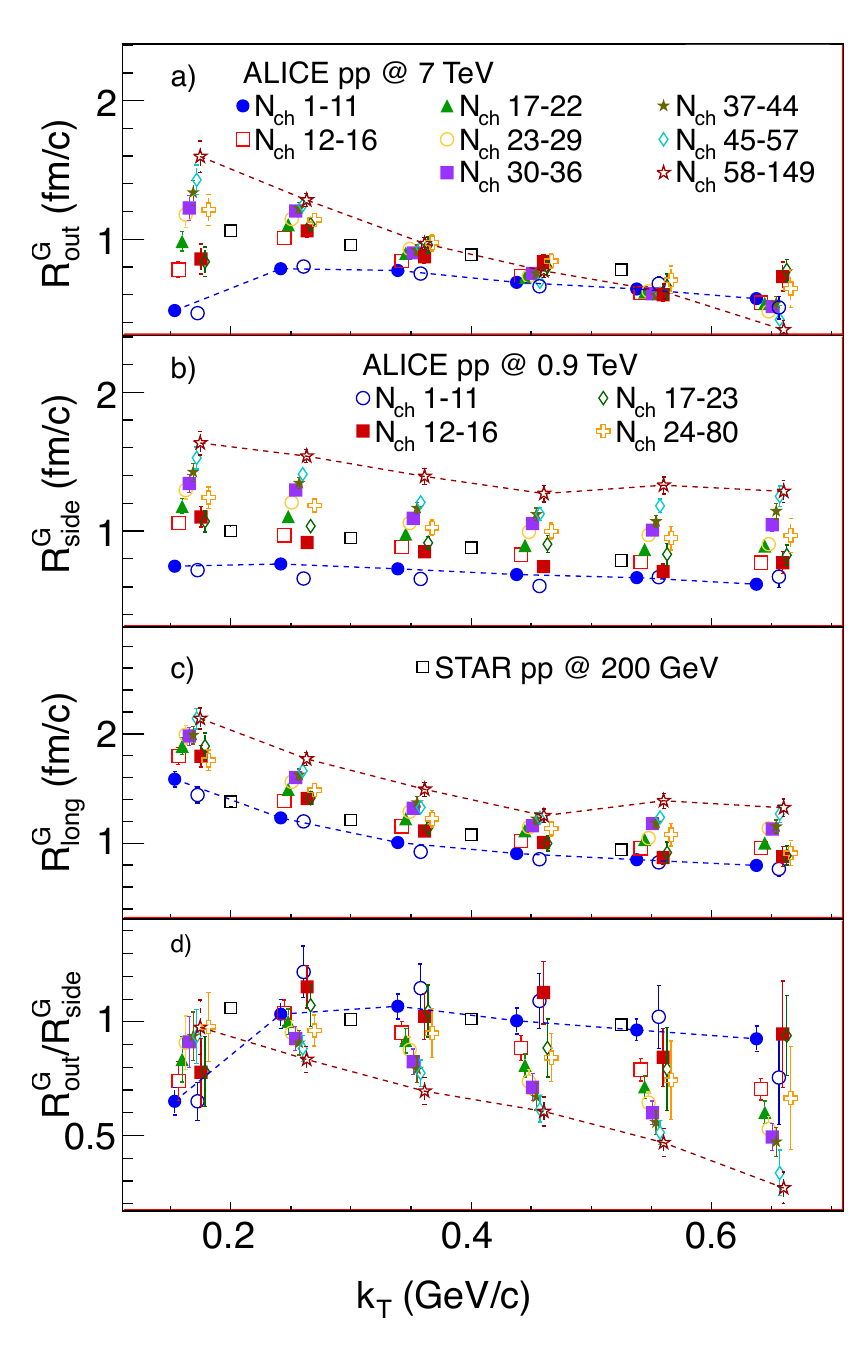}
\end{center}
\caption{HBT radii versus the pair transverse momentum $k_T$, for various multiplicities of the  $pp$ collisions, from ALICE \cite{Aamodt:2011kd}.}
\label{fig_alice_hbt}
\end{figure}

\begin{figure}[b!]
\begin{center}
\includegraphics[width=80mm]{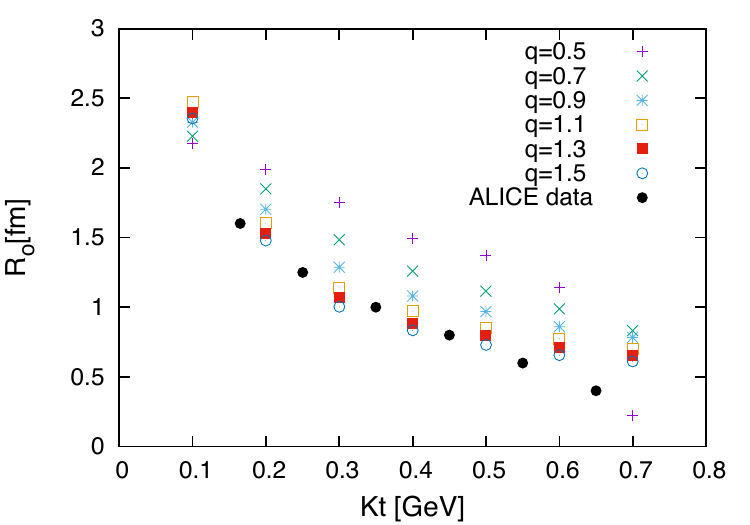}
\end{center}
\caption{From \cite{Hirono_HBT}: HBT radii compared to ALICE data (closed circles), for solutions starting
with different initial size of the fireball, indicated by Gubser scale parameter $q$ (which is inversely proportional to the size).
} \label{ro-kt}
\end{figure}

 Hirono and myself \cite{Hirono_HBT}
  had calculated such effect and compare the results with the  ALICE HBT data \cite{Aamodt:2011kd}  shown in Fig.\ref{fig_alice_hbt}. We  deduced the magnitude of the flow
in high multiplicity $pp$ collisions, directly visible in those data. 
The effect is best seen in the so called ``out"-directed radius $R_{out}$ (the top plot). While low multiplicity
data (connected by the blue dashed line) are basically independent on the pair momentum, at high multiplicity 
(stars and red dashed line) they are decreasing, by a rather large factor. Another consequence of the flow
is anisotropy of radii. In the bottom plot the ratio of two radii are shown: at small multiplicity it is always 1 -- that
is the source is isotropic -- but at high multiplicity the source becomes anisotropic, the radii
in two directions are quite different, with their ratio 
dropping to about $1/3$, at the largest $k_t$. It means, only 1/3 of the fireball emits pairs of such momenta,
a direct consequence of the flow.

In Fig.\ref{ro-kt} we show a series of calculations in which the initial QGP stage of the collision is 
modelled by numerical hydro solution close to Gubser analytic solution with variable parameter $q$.
(The late stages need to deviate from Gubser since near $T_c$ the EOS is very different from 
conformal $\epsilon=3p$ assumed in Gubser's derivation).

Let us summarize  what we learned in this subsection so far.
Unlike  central $pA$,  the highest multiplicity $pp$ events are significantly denser/hotter than central AA. 
Very strong radial flow, seen in spectra of identified particles and HBT radii, require very small, sub femtometer,
initial size of the system. In spite of high cost associated with those events, their studies are
of utmost importance because here we produce the most extreme state of matter ever created in the lab.

     \begin{figure}[t]
\begin{center}
\includegraphics[width=8cm]{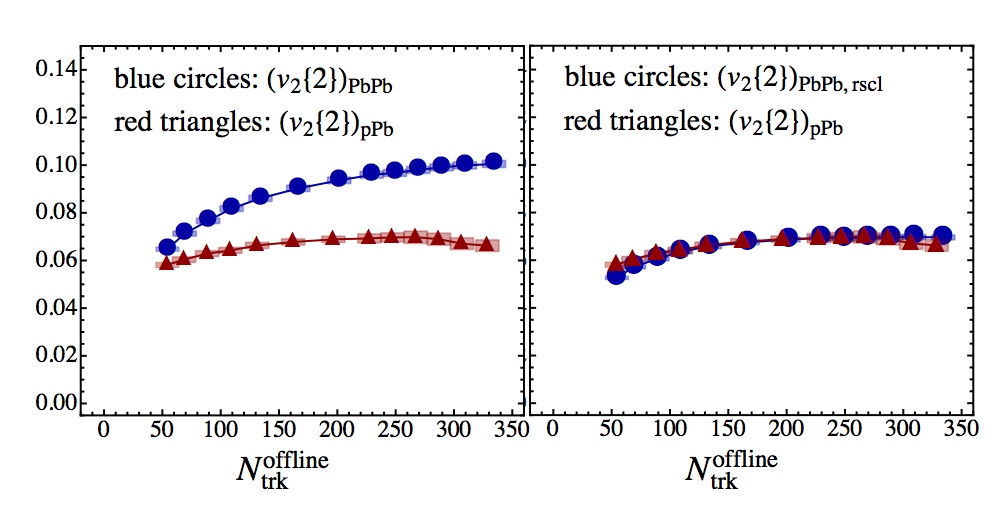}
\caption{  The integrated $v2\{2\}$ for PbPb and pPb vs. multiplicity from [23]. Left: Original values. Right: The fluctuation dependent elliptic flow, with the geometrical part subtracted.  This geometrical part was calculated using the Phobos Glauber Model  and is not a fit.    }
\label{fig_BT}
\end{center}
\end{figure}

\subsection{Can flows in small systems be ``fake"? }
The question and subsequent development is due to Romatschke \cite{Romatschke:2015dha},
who proposed a particular model of what I call  ``the fake flow". In this scenario quarks and gluons
at the QGP phase have no interactions and free stream from the point of the initial scattering.
Only at some hadronization surface the system switches to hadronic cascade.

     \begin{figure}[t]
\begin{center}
\includegraphics[width=8cm]{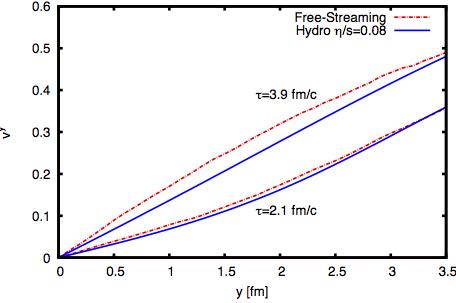}
\caption{  Comparison of the flow profile, for hydrodynamics and free streaming , from   \cite{Romatschke:2015dha}. }
\label{fig_Rom_vy}
\end{center}
\end{figure}

In Fig.\ref{fig_Rom_vy} one see a comparison of the radial flow profiles of the two cases, with and without
interaction at the QGP phase. One can see that crudely the profiles are in fact very similar,
becoming linear Hubble-like as time goes on. (In fact
  free streaming generates even a bit stronger flow, because 
  free streaming uncouples from longitudinal direction, and equilibrated hydro medium
  does not.) Comparing particle spectra and HBT radii Romatschke shows that this ``fake" radial flow 
  is indeed $indistinguishable$ from the hydro one.

\begin{figure}[b]
\begin{center}
\includegraphics[width=8cm]{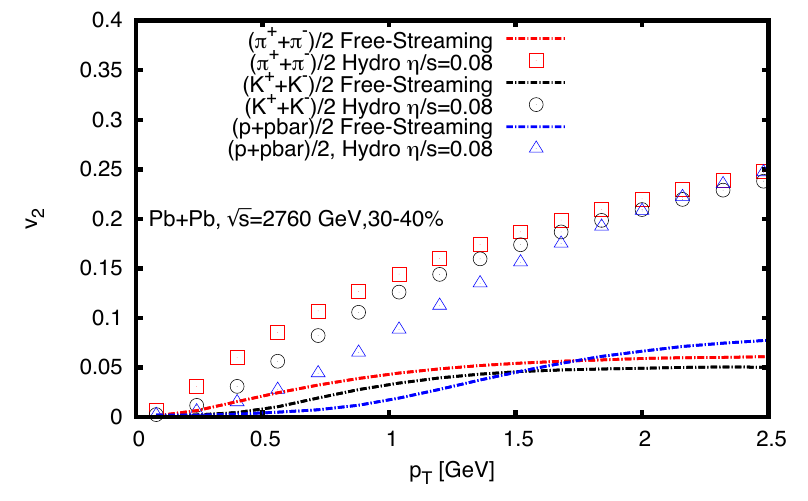}
\includegraphics[width=8cm]{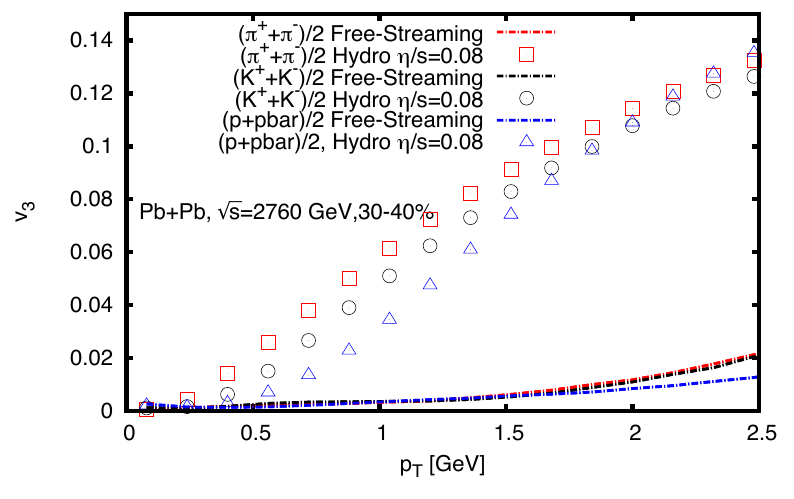}
\includegraphics[width=8cm]{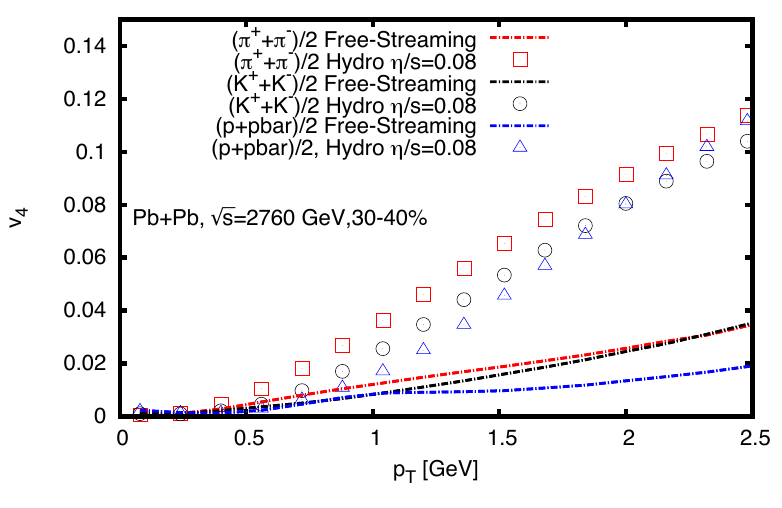}
\caption{  Comparison of the flow harmonic, for hydrodynamics and free streaming for PbPb
collisions, from   \cite{Romatschke:2015dha}. }
\label{fig_Rom_v234}
\end{center}
\end{figure}

What about flow harmonics? The results for PbPb collisions are shown in Fig.\ref{fig_Rom_v234}.
As one can see, without hydro those (crudely speaking) disappear. It is not 
surprising since if 
there is no hydrodynamics then there should be no sound waves, and initial bumps 
are simply dissolved without trace.

(In fact it is an interesting question how  $any$ $v_n$ can be generated in the free streaming.
The initial momentum distribution of partons is isotropic, and so it must be related to the interaction 
after hadronization.   
Romatschke found that indeed before hadronization they are absent. However this happens because
 two component of $T_{\mu\nu}$, the
flow $\sim u_\mu u_\nu$ part and the dissipative $\Pi_{\mu\nu}$  part, have nonzero values
but cancel each other in sum. After hadronization the hadronic interaction  kills the second component
$\Pi_{\mu\nu} \rightarrow 0$  
and reveals the effect of the first one.)

Not only the ``fake" flow harmonics are small, they do not show two important
features of the ``true" hydro ones: (i) they do not show strong increase with $p_\perp$; and (ii)
they do not show strong decrease with the number $\sim exp(-n^2)$ induced by the viscosity during
the time before hadronization

\begin{figure}[b]
\begin{center}
\includegraphics[width=8cm]{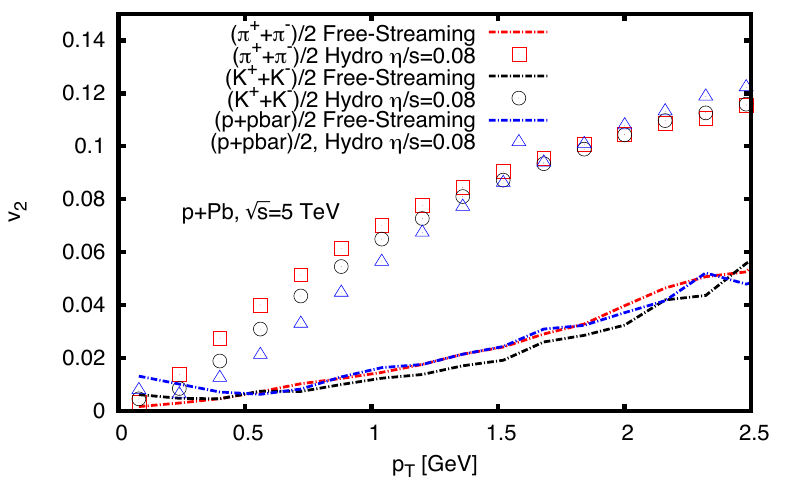}
\includegraphics[width=8cm]{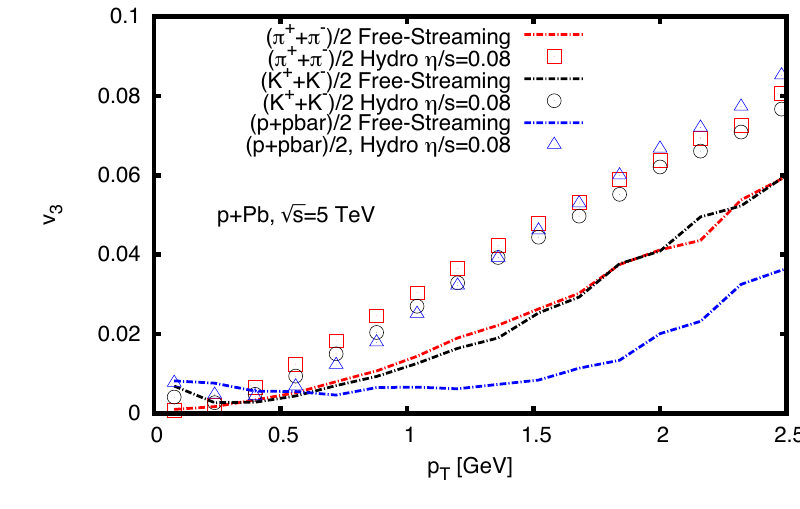}
\includegraphics[width=8cm]{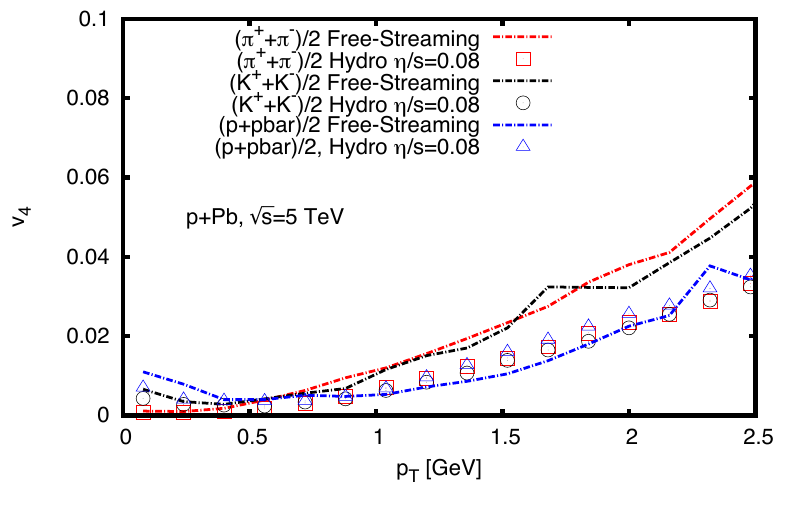}
\caption{  Comparison of the flow harmonic, for hydrodynamics and free streaming  in the pPb
central bin , from \cite{Romatschke:2015dha}. }
\label{fig_Rom_pPb_v234}
\end{center}
\end{figure}

So, we now see that, unlike the radial flow,  higher harmonics in large (PbPb) systems cannot be faked.
What about smaller systems? Romatschke gives the results for $pPb$ at LHC  and $dAu$ and $He^3Au$ for RHIC
energies. We show the first case in Fig.\ref{fig_Rom_pPb_v234} , the rest can be looked at the original paper. 
Again the free streaming model seems to be failing for $v_2$, is somewhat marginally possible for $v_3$, except for protons. The $v_4$
are comparable but both too weak to be observed. 

So, flow harmonics seem not to be ``faked" even for smaller systems. Yet,
taking into account remaining 
uncertainties of the initial stage models and thus $\epsilon_n$ values, this conclusion is  
not as robust as for the AA. Perhaps some scenarios intermediate between equilibrated hydro and free streaming,
with larger viscosity, may still be possible in this cases.

\subsection{Shape fluctuations:  central $pA$ vs peripheral AA} 
    Scaling relation between central $pA$ and peripheral AA has been proposed and tested by Basar and Teaney
\cite{Basar:2013hea}. Step one of their paper has been prompted by the fact (noticed in the CMS paper already):   at the same multiplicity,
$v_3$ in  central $pA$ and peripheral AA are basically the same. Some people called for new paradigms based on
this fact: but in fact
it is hardly surprising:  equal multiplicity means equal number of  participant nucleons, and thus equal fluctuations of the shape. 
After  Basar and Teaney   removed the geometrical contribution to  $v_2$ in peripheral AA, they found that the remaining -- fluctuation-driven -- part of elliptic flow
is also  the same in both cases, see Fig.\ref{fig_BT}.

  Their second proposal is that the $p_t$ dependence of (the fluctuating part) of the $v_n$ has an universal shape, and AA and $pA$ data are only different by a scale of mean $p_t$
  \be v_n^{pA} (p_t) = v_n^{pA} ({p_t \over \kappa} ) \ee 
  where the scaling factor is defined as
  \be \kappa= {<p_T >_{pPb}  \over <p_T >_{PbPb} } \approx  1.25 \ee and is due to difference in the radial flow. This relation also works well.

 two possible effects, as the multiplicity grows:\\
 (i) an increases the initial temperature $T_i$. Since the final one  is fixed by hadronization near the
 phase transition   $T_f\approx T_c$, the contrast between them gets larger and hydro flow increases;\\
 (ii) the initial size of the fireball $decreases$, increasing the initial temperature $T_i$ even further.

\section{Equilibration in QCD-based models}

\subsection{ CGC and turbulent GLASMA} \label{sec_weak_equilibr} 
The idea of continuity, from a  state before collision to early time after it, is most directly realized in the so called CGC-GLASMA approach. Technically it is based on McLerran-Venugopalan argument \cite{McLerran:1993ka}
that high density of partons leads to large color charge fluctuations, which should create
strong color fields. 
If fields are strong
enough, then  classical Yang-Mills equations are sufficient, and those can be solved numerically.
It is important that at this stage the fields get  strong, the occupancy of gluons $n_g\sim 1/\alpha_s \gg 1 $ and rescaling them one can get the coupling out of YM
equations. It means that GLASMA is non-perturbative, in spite of weak coupling.
It remains so till gluon occupation numbers drop to their thermal magnitude  $n_g\sim 1$.

    \begin{figure}[b]
\begin{center}
\includegraphics[width=7cm]{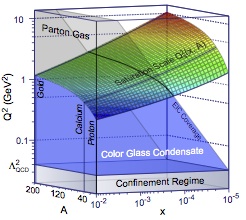}
\caption{  The CGC phase diagram: the
saturation momentum $Q_s$ as a function of fraction of momentum $x$ and the atomic number $A$, from  \cite{Lappi_QM14}.   }
\label{fig_Qs}
\end{center}
\end{figure}

When density of gluons gets large enough and nonlinear effects  become important, glasma is in its ``dense regime".
Its boundary in a ``glasma phase diagram"  is shown in Fig.\ref{fig_Qs}. It is defined by
 the saturation momentum $Q_s(x) $, separating it from the dilute or partonic phase.  
 It is expected to grow with collision energy (smaller $x$) and higher atomic number $A$.  

A good news from the plot is that at the highest LHC energies and A $Q_s^2\sim 10\, GeV^2 $,
perhaps in the perturbative domain. I however would not accept that
the ``confining regime" at the bottom of the figure 
is delegated to very small $Q^2< 0.03 \, GeV^2$. In fact
 the boundary of pQCD is
 a factor 30 or more higher: there are no
 gluons with virtuality below 1 GeV! Nambu-Jona-Lasinio noted (already in 1961!)
 that some strong non-perturbative forces turns on at  $Q<1\, GeV$, creating chiral symmetry breaking.
 By now we have plenty of evidences from the lattice simulations that all correlation
 functions of the gauge fields do not behave perturbatively at distances larger than 0.2 fm or so.
 Heavy quark potential is also nor Coulombic but string-like, and the string width is only .2 fm or so. 

   Theoretical study of parton equilibration in weak coupling domain has a long history.
 The so called ``bottom up" approach by Baier, Mueller, Schiff and Son \cite{Baier:2000sb} was based on
 soft gluons radiated by scattered hard partons.  The name means that
thermal occupation starts from the IR end. (Note that it is opposite to the ``top-down" equilibration in
holographic models we will discuss below.) 
The main prediction of that model was 
the equilibration time and the initial temperature scaling with the coupling
\be \tau_{eq}\sim 1/(\alpha_s^{13/5} Q_s), \, \, \,  T_i\sim \alpha_s^{2/5} Q_s \ee
Some details were changed as later one tried to incorporate Weibel, Nielsen-Olesen
and any other instabilities which may occur into the model. 
 The validity domain of this theory is restricted by its core assumption of small angle scattering
of the gluons, justified by large impact parameters of the order of inverse (perturbative) Debye mass.
 Perturbative means  $M_D\approx gT \ll T$ or small $g\ll 1$. If so,  $g=1,\alpha_s=1/4\pi$ 
 and one finds very long equilibration time
$ \tau_{eq}Q_s\sim 700$ exceeding duration of the collisions. (Not surprisingly, the pre-RHIC 
perturbative predictions have been that no QGP will be produced, just some ``fireworks of mini-jets",
 with small rescattering corrections.) One may take ``more realistic" $\alpha_s =1/3,1/2$ and get shorter time,
 but then the question is whether pQCD series are at such coupling meaningful, and not divergent starting
 already from the first term.

In the last few years several groups made  significant efforts to do numerical  studies of parton equilibration 
using both Boltzmann equation or the field equations, for
 many field harmonics.
   Typically, in such
studies the coupling constant is taken to be extremely small. In fact, 
so small that one can treat not only powers of $\alpha_s$ but even its log as a large parameter 
$log(1/\alpha_s)\gg 1$, allowing  the total GLASMA evolution scale
 \be \tau_{GLASMA}\sim {log^2(1/\alpha_s) \over Q_s}\ee
to be considered  large.

Significant progress in this directions has been  incorporation of certain ideas from general theory of 
 turbulent cascades. Not going into its long history, let me just mention 
 Kolmogorov-Zakharov stationary power-like solutions for
a vast number of systems with nonlinear interaction of waves   for Boltzmann equation,
and time-dependent self-similar solutions.
(For sound cascades in early universe some of that we already discussed in section \ref{sec_cosmo}.) 
 J.Berges and collaborators developed it for scalar and gauge fields, pointing out different regimes for UV and IR-directed
cascades, and identifying such regimes in impressive  numerical simulations. 

New paradigm, resulting from this body of work,   is that the pre-equilibrated plasma may 
 be captured by a {\em nontrivial turbulent attractor} -- certain self-similar power solution,
before progressing toward thermal equilibrium.
 let me comment on one aspect of the result, on the momentum anisotropy
 $p_l/p_\perp$. 
 Epelbaum and Gelis \cite{Epelbaum_QM14}
performed a next-order GLASMA 
simulation, with $g=0.5$, in which $p_l/p_\perp$ keep approximately constant value
during the whole time of the simulation $\tau Q_s =10-40$. 
 Berges,  Schenke, Schlichting, Venugopalan
\cite{Schlichting_QM14} found that, at $g=0.3$, the longitudinal 
pressure $p_l/\epsilon$ remains close to zero at similar times. 
These results are shown in Fig.\ref{fig_ETperNPvsNp}.

    \begin{figure}[t]
  \begin{center}
    \includegraphics[width=6cm]{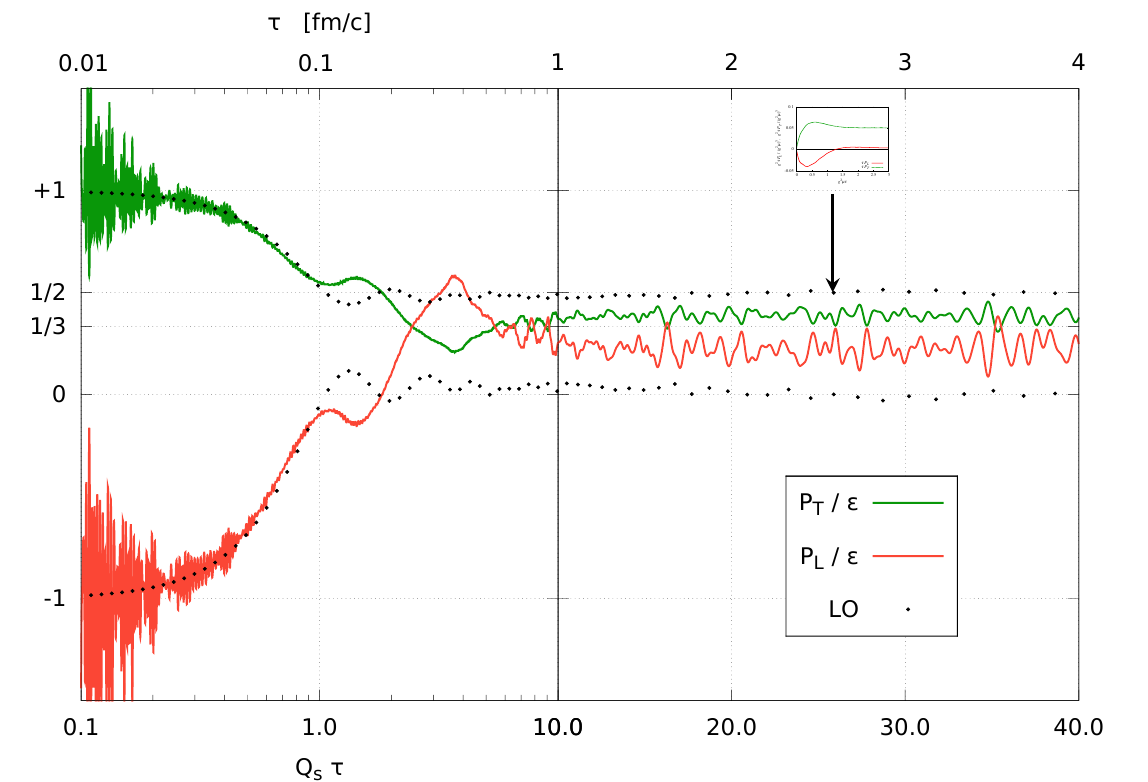}
  \includegraphics[width=6cm]{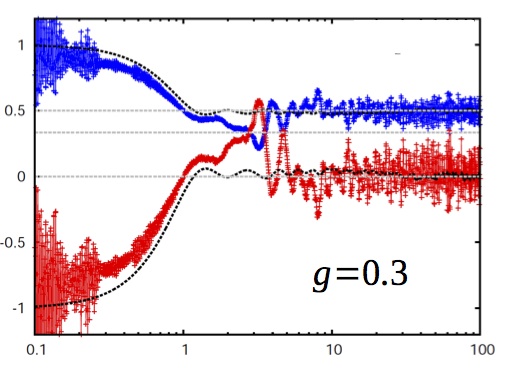}
   \caption{ (a) Upper green curve is $p_t/\epsilon$, lower red is $p_l/\epsilon$, as a function of time in
   units of saturation scale, $\tau Q_s$,  at $g=0.5$, from   \cite{Epelbaum_QM14}. \\
   (b) Upper blue curve is $p_t/\epsilon$, lower red is $p_l/\epsilon$, at $g=0.3$, from   \cite{Schlichting_QM14}. }
  \label{fig_ETperNPvsNp}
  \end{center}
\end{figure}

\begin{figure}[t]
  \begin{center}
  \includegraphics[width=8cm]{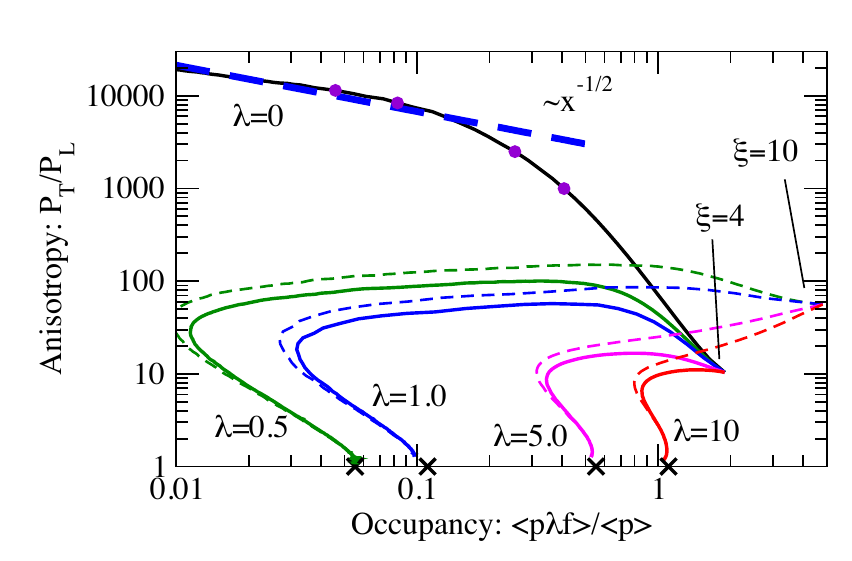}
   \caption{From  \cite{Kurkela:2015qoa}: Trajectories of the systems on the  occupancy-anisotropy plane for various settings. The parameter $\lambda$ near curves show the corresponding 't Hooft coupling constant $\lambda=g^2 N_c$.  All solid line originate
   from one initial distribution characterized by the anisotropy parameter $\xi=4$, the dashed lines originate from a different point with $\xi=10$. .
   }
  \label{fig_Kurkela}
  \end{center}
\end{figure}

Recently Kurkela and Zhu \cite{Kurkela:2015qoa} performed
cascade simulations at larger couplings. In Fig.\ref{fig_Kurkela} from this paper one can see two sets of trajectories,
shown by solid and dashed lines, starting from two different initial distributions. At zero coupling 
(upper left curve) the longitudinal momenta of particles gets very small compared to transverse one,
and anisotropy steadily increases. This is a scaling-like classical regime with a nontrivial fixed point.
However all other paths stay more or less at the same initial anisotropy, and then rapidly turn downward,
to locally isotropic distributions (marked by diagonal crosses at the bottom). 

It is a pity there are no simulations done with $\lambda$ between 0 and 0.5: perhaps at some critical coupling
a bifurcation
of the trajectories happens, separating those who proceed toward the new and the equilibrium fixed points.
Yet the issue 
 is rather academic, since the realistic
 relevant coupling value $\alpha_s=g^2/4\pi\approx 0.3$ corresponds to 't Hooft coupling constant 
$\lambda=\alpha_s N_c 4 \pi\approx 10$, which is  the $largest$ value shown in this figure (bottom right). The corresponding
 curve   rapidly approaches  the equilibrium
 point, with coordinates (1,1) at this figure.

\subsection{From Glasma to hydro} 

   Summarizing the previous subsection in few words, one may say that weak coupling cascades predict
highly anisotropic pressure $  p_l\ll p_t$. The question is whether it persists for the time of few fm/c,
or is rapidly reduced to the values prescribed by viscous hydrodynamics. The latter scenario is predicted by strong coupling approaches, to be discussed later. We do not know when exactly a transition between these regimes happen.
This question is  central for our field, and one should not jump to conclusions without serious
evidences,one way or the other. 

(Example of a premature statement:``No need for strong coupling to get hydrodynamization" is conclusion
of  Epelbaum's QM14 talk. Hydrodynamics is not just some instantaneous value of the stress tensor,
but the equations to follow. The value of viscosity (mean free path) is central to their validity:
and simulations done at  $g=0.5,\alpha_s= 0.02$  correspond to viscosity hundreds 
 times larger than the actual values of sQGP extracted from data. Anyone who doubt that should 
just try to derive harmonic flows from such cascades, without hydro.)

 Surprisingly, there are little discussion of how to test for persistence of such
anisotropy in experiment, so let me start with some comments on that.
First, one can indeed measure an early-time anisotropy, via dilepton polarization, see section \ref{dilepton_polarization}. 
Second,  one can study the effect of the longitudinal pressure  on the rapidity distribution.
Historically, the original papers of Landau  focused on  the longitudinal expansion, driven by
gradient of $p_l$. However Landau's initial condition -- the  instantaneous stopping -- looks rather unrealistic today.
Since  all hydro modeling starts after certain time -- when partons already passed  each other -- it needs certain
initial distribution. Its width in rapidity, appended by the calculated longitudinal hydro effect, was compared with the data. 
Uncertainties, in particular related with removal of the fragmentation regions near beam rarities, 
makes an accuracy of   $p_l$ extraction poor. 

  Since 1990's hydro effects focused on the transverse motion, the radial and elliptic flows.
  Since sQGP is believed to be near-conformal, the trace of stress tensor  remains zero
  even out of equilibrium: so even  for vanishing $p_l=0$ one finds $p_t=\epsilon/2$ instead of $\epsilon/3$  in equilibrium. If strong anisotropy be there for  $\tau\sim 2 \,fm$ or so,
this difference would already be seen in the magnitude of the elliptic flow. This issue deserves however dedicated studies.

 \begin{figure}[h]
  \begin{center}
   \includegraphics[width=6cm]{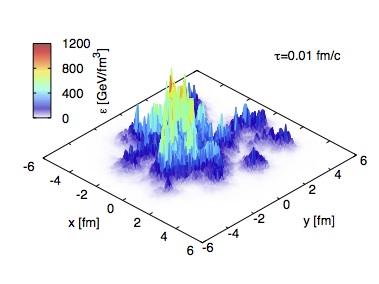}
   \includegraphics[width=6cm]{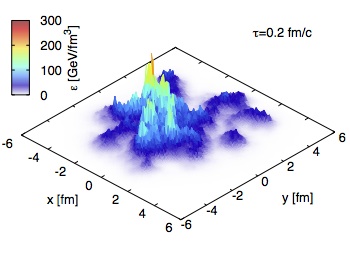}
  \includegraphics[width=6cm]{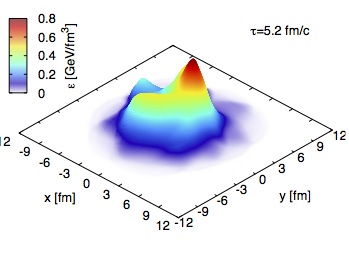}
   \caption{ From \cite{Schenke:2014zha}:
   Transverse energy profile from the IP glasma model for a semiperipheral (b = 8 fm) Au+Au collision at s = 200 A GeV, at times $\tau$ = 0.01, 0.2, and 5.2 fm/c. From  $\tau$= 0.01 fm/c to 0.2 fm/c the fireball evolves out of equilibrium according to the Glasma model.}
  \label{fig_in123}
  \end{center}
\end{figure}
 
  Let us look what practitioners actually do while combining glasma and hydrodynamics.
Fig.\ref{fig_in123} illustrate what is done  in the so called impact parameter (IP) glasma approach.  
An important feature of glasma is independent fluctuations of color in different cells,   
which seeds the harmonic flows. 
At certain proper time -- 0.2 fm/c in this example  -- glasma evolution is stopped  and then the energy momentum tensor  is matched to ideal fluid. For technical reasons the value of the viscous tensor is put to zero.

How important is the time 0.2 fm/c selected?
Note first that the second picture is hardly  different from the first, except the overall scale of the energy density is reduced.
Indeed, even moving with a speed of light one can only shift by 0.2 fm, a very small distance relative to the nuclear size. 
Glasma is diluted by purely longitudinal stretching: thus transition time can be shifted up. Many other
practitioner start hydro at 0.6 fm/c or so.
 
 By  starting hydro  right from the second picture Schenke and collaborators implicitly assume that 
 hydro cells can indeed  be as small as 0.2 fm, and that their code can cope with 
 huge gradients between the cells. 

(Note that
 the bottom figure, at time 5.2 fm/c, looks very different. The original bumps  disappeared and
instead a new one is formed. The reason is  sound refuses to stand still and is moving with a speed of sound.  At intersections of ``sound circles" from 
the primary bumps random enhancements of the density are observed. Yet since the bumps are statistically unclorrelated, those should get averaged out,  at least in 2-particle correlations, and only correlations from the same circle will stay.)

How many harmonics are needed to describe pictures like that shown in Fig.\ref{fig_in123} ?
Taking  0.2 fm as a resolution and 4 fm as a size, one finds that the upper picture requires about 20*20=400 pixels  to be represented by certain stress tensor components. 
At the  freezeout there are  only several angular harmonics observed, 
so 99\% of the information inclosed in those pictures does not survive till the freezeout. 
In the hydro simulation just described those disappear predictably, via viscous damping.
It is possible that this systematics will fail at shorter wavelength: so it is worth trying to measure 
higher harmonics which perhaps deviate from it. Other ways to observe density waves can
perhaps be invented: one  of such is potential observation of those  in the dilepton  via the so called MSL process.


%



%

 \begin{figure}[t]
  \begin{center}
  \includegraphics[width=8cm]{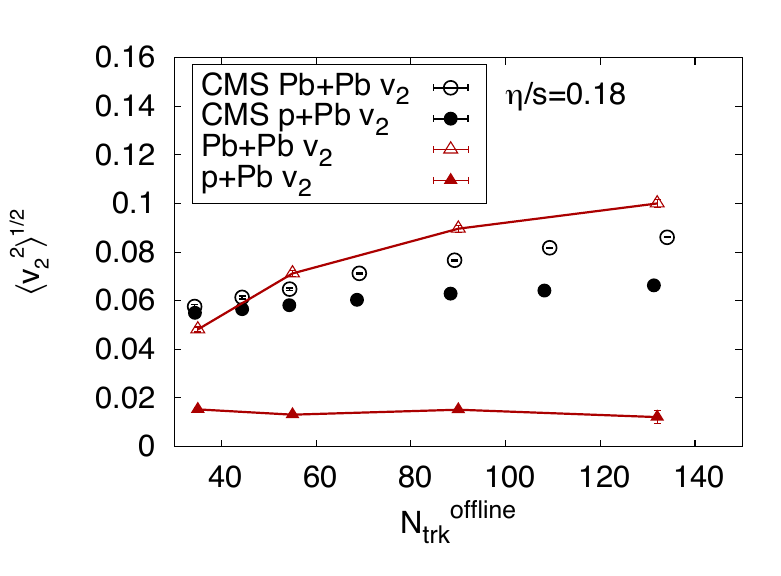}
   \caption{(Color online) Multiplicity dependence of the root-mean-square elliptic flow coefficient v2 in Pb+Pb (open symbols) and p+Pb collisions (filled symbols) from the IP- Glasma+music model (connected triangles) \cite{Schenke:2014zha} compared to experimental data by the CMS collaboration.}
  \label{fig_SchVenu}
  \end{center}
\end{figure}

 \begin{figure}[b]
\begin{center}
\includegraphics [height=5.cm]{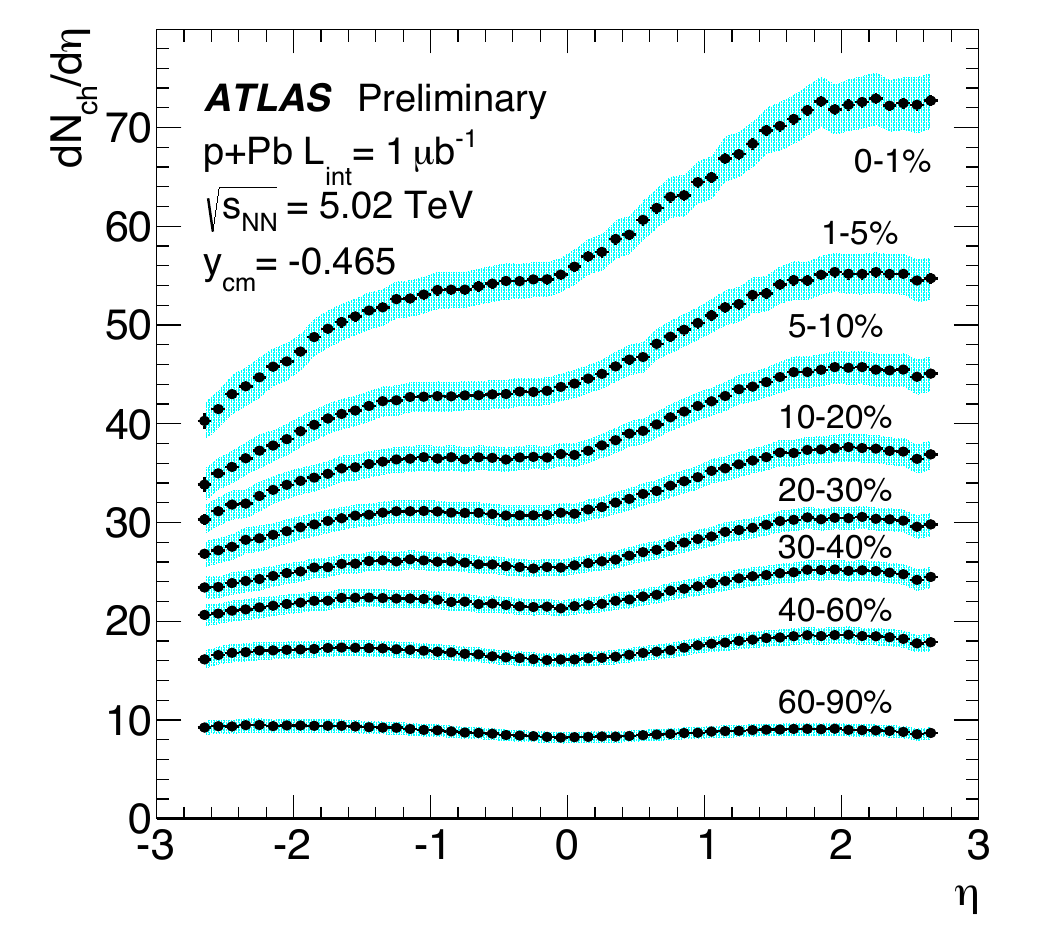}
\caption{Rapidity distribution in pPb collisions for different centrality classes, from \cite{ATLAScentrality}
 .}
 \label{fig_ATLAS_rapidity}
\end{center}
\end{figure}

\subsection{The initial state and angular correlations} \label{sec_inital_angular}

Importance of the initial state information can be demonstrated further using ``small systems", such as
central  $pA$ collisions. When
a nucleon is going along the diameter of heavy nucleus  the mean number of participant nucleons is
\be <N_p>= n_0 \sigma_{NN} 2R_A \ee
so for pPb at LHC one gets $ <N_p>\approx 16$. 
The question however remains, where exactly the deposited energy is?

\begin{figure}[t]
  \begin{center}
  \includegraphics[width=6cm]{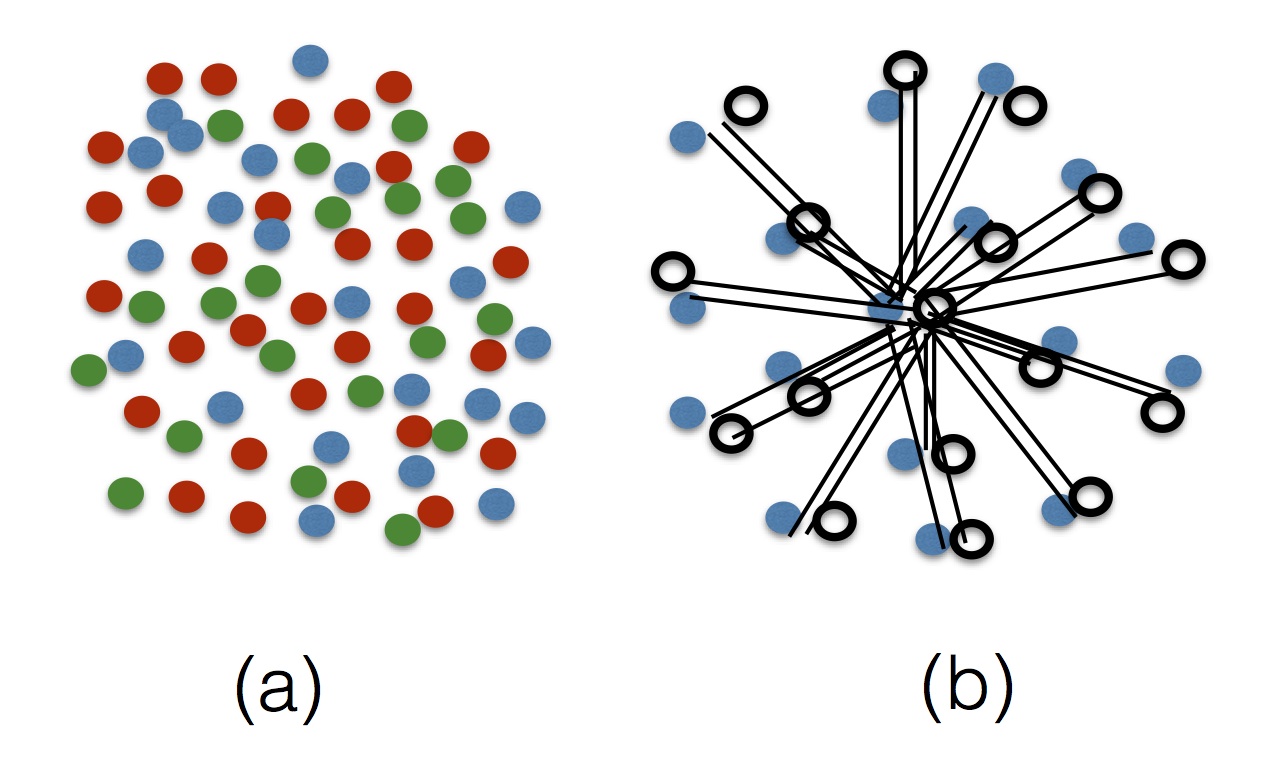}
   \caption{Sketch of the initial state in central $pA$ collisions. The plot (a) corresponds to IP-glasma model, with colored circles representing multiple gluons.
   Fig.(b) is for $N_p=16$ Pomerons, each represented by a pair of cold strings. The open circles are quarks and filled blue circles are diquarks.}
  \label{fig_sketch_pA}
  \end{center}
\end{figure}

In Fig.\ref{fig_sketch_pA} we sketched two opposite models of the initial state.
 In (a) we show each of the $N_p$ participants 
represented by $N_g$ gluons  (ignoring sea quarks and antiquarks) from their PDFs each, so the total number of partons  $N_p N_g$. 
We assume all gluons uncorrelated in the spot of the size of pp cross section.
In (b) we
had shown an alternative picture, the stringy Pomeron,
in which there are no gluons but $2N_p$ QCD strings instead. Since those are
``cold" (unexcited) we show those by straight lines. Note that the picture a bit exaggerates the ratio
of two parameters involved. Those are (i) the mean impact parameter between the nucleons $b\sim \sqrt{\sigma_{NN}/\pi}\approx 1.6\, fm$ and (ii) the size of the quark-diquark dipole $d\sim .4\, fm$, so $b\gg d$.  

Let us estimate the deformation of the initial state in each case. Since it is central collision, there is no mean geometrical
effects and all deformations comes from fluctuations. As discussed above, for all $n$ one expects the same
magnitude
\be \epsilon_n \sim {1 \over \sqrt{N} } \ee 
where $N=N_p N_g$ for (a) and only $N=N_p $ for (b).  Evaluating $N_g$ from PDF's at LHC energy
includes integration from $x_{min}\sim 10^{-3}$ to 1: one gets roughly
the ratio 
\be { \epsilon_n^{(b)} \over  \epsilon_n^{(a)}} \sim {1 \over \sqrt{N_g}}\sim 4 \label{eq_eps2_ratio} \ee

Let us now switch to practical calculations and comparison to data. Schenke and Venugopalan
\cite{Schenke:2014zha} had recently studied $v_2,v_3$ flows in (very peripheral) AA and 
central  pA. They found that
 the IP-glasma model they developed does a very good job for the former and 
 underpredicts them in the latter case, see Fig. \ref{fig_SchVenu}.
 
 As we already discussed above, in the  peripheral AA $\epsilon_2$ is large, O(1), in any model, and 
in order to get the right $v_2$ one has to have correct viscosity -- which apparently these authors have.
The central  $pA$ is indeed the test case: we argued above that the density is not yet large enough to apply 
  the IP-glasma model, and that stringy Pomerons should be more applicable one.  If so, using (\ref{eq_eps2_ratio}) we should increase the $v_2$ by a factor of 4, which brings it to
 an agreement with the CMS measurements. 
   We thus conclude that the stringy model Fig.\ref{fig_sketch_pA}(b) is preferable over the picture (a), the uncorrelated gluons.

(The actual IP glasma model is not the uncorrelated gluons: it should be uncorrelated glasma cells of $1/Q_s$
size: so our model (a) is just a straw man holding place before more consistent calculation is done.)

%

%
   
   Above we simplistically assumed a Glauber picture in which  each wounded nucleon (or a participant)
   interact with the projectile proton by the usual single Pomeron. This means 
    one color exchange with each participant, who therefore is connected to the projectile by (at least) $two$ strings. 
   
   If this picture be correct and the strings were simply independently fragmenting,
    the rapidity distributions of secondaries would be flat  (rapidity independent) for all centrality classes. This is not the case in reality, as is
    seen in ATLAS data shown in Fig.\ref{fig_ATLAS_rapidity}. As one can see, the peripheral bins have flat rapidity distribution: this corresponds to
   few strings extending from one fragmentation region to the other one without a change. Yet central bins have
   rather asymmetric distributions, with larger multiplicity at the Pb side.

   In the Pomeron language it is explained by
   the so called ``fan diagrams" in which one Pomeron can split into two. 
   The ``triple Pomeron  vertex" is however small and we do not have 
   developed ``Pomeron cascades". The multiplicity 
 difference between the r.h.s. an the l.h.s. of the plot is not too dramatic, certainly not factor $N_p\sim 20$
 as the old wounded nucleon model would suggest. 
   For example, for the most commonly used 
   centrality bin 1-5\% the rapidity density $dn_{ch}/d\eta$ changes from about 35 to 55, across the rapidity interval shown in this figure.
   If on the $Pb$ end there are say $N_s>2N_p\approx 40$ strings, then on the $p$ end there are not one or few, but
   still  20 strings or so. Since the area on the l.h.s. is reduced by an order of magnitude or so, and the number only by
   factor 2, it is by far more dense system than the r.h.s.!  
   
   If one thinks of that, one may conclude that flows and 
   development of collectivity should strongly depend on rapidity.   
Yet, at least in some crude sense, this is not the case: for example, the famous $v_2$ ``ridge"
is rather flat in rapidity. So, the issue needs more attention.

%
%
%
%
   
   Finally, let me mention high multiplicity $pp$ . We do not yet know $\epsilon_n$ in this case. Experiment
should do 4,6 particle correlators and separate non-collective 2-particle correlations from collective  ones. Hydro
practitioners still have to do $v_n/\epsilon_n$ and establish its viscosity dependence and accuracy. Theoretical predictions for $pp$  cover the
whole range: from elongated transverse string 
\cite{Bjorken:2013boa} predicting large $\epsilon_2\sim 1$ to a IP-glasma or ``string ball" picture 
\cite{Kalaydzhyan:2014tfa} which predicts very small  $\epsilon_2$ instead.


 \begin{figure}[b]
  \begin{center}
  \includegraphics[width=6cm]{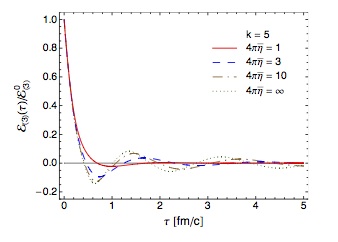}
   \caption{Oscillation of the energy density in simulations starting from ``glasma"-like initial conditions. $k=5$ is the number of fluxes through
   the flux tubes, from \cite{Florkowski_QM14} .}
  \label{fig_Florkowski}
  \end{center}
\end{figure}

\subsection{Multi-string state: spaghetti} \label{sec_spaghetti}
A version of the initial state theory, alternative to glasma picture,
 is old Lund model, used in event generators like PYTHIA. It is suppose to be applicable for
lower matter density, remaining in the confined phase. Multiple color charges, moving
relativistically from each other after collision, are in such case connected by multiple QCD strings.
As those are rapidly stretched longitudinally, strings become nearly parallel to each other,
and we will call this state ``spaghetti" for short.

Transition from such a spaghetti state to GLASMA can be relatively smooth, since in both pictures the color fields have similar longitudinal structure.
 Florkowski at a talk at QM14 \cite{Florkowski_QM14} discussed dynamics of 
   spaghetti-like GLASMA state. Interesting oscillations
of such system is predicted, see Fig.\ref{fig_Florkowski}.

Transition between two picture is naturally expected when the diluteness of the QCD strings become of the order 1,
so they can no longer be separated 
\be  {N_{string} \over Area} \sim {1 \over \pi r_{string}^2} \sim 10 \, fm^{-2} \ee
where in the numerical value we use the field radius in the string $r_s\approx 0.17 \, fm\sim 1/GeV$. 


   \begin{figure}[t!]
  \begin{center}
  \includegraphics[width=6cm]  {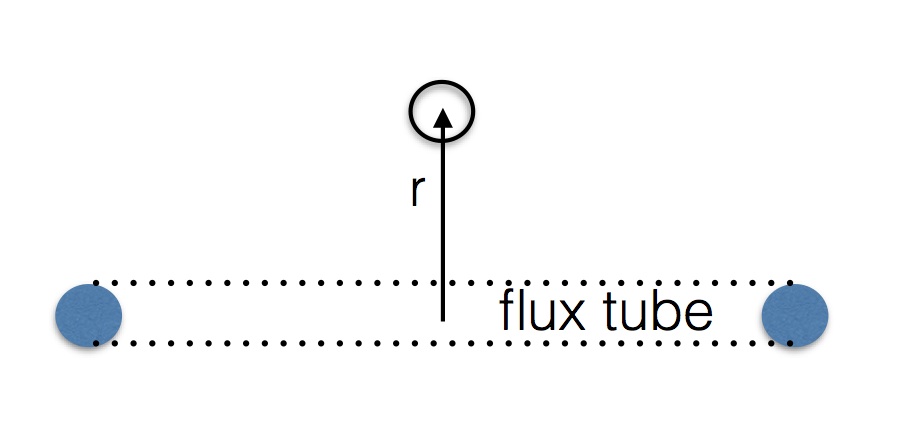}
  \includegraphics[width=6cm]{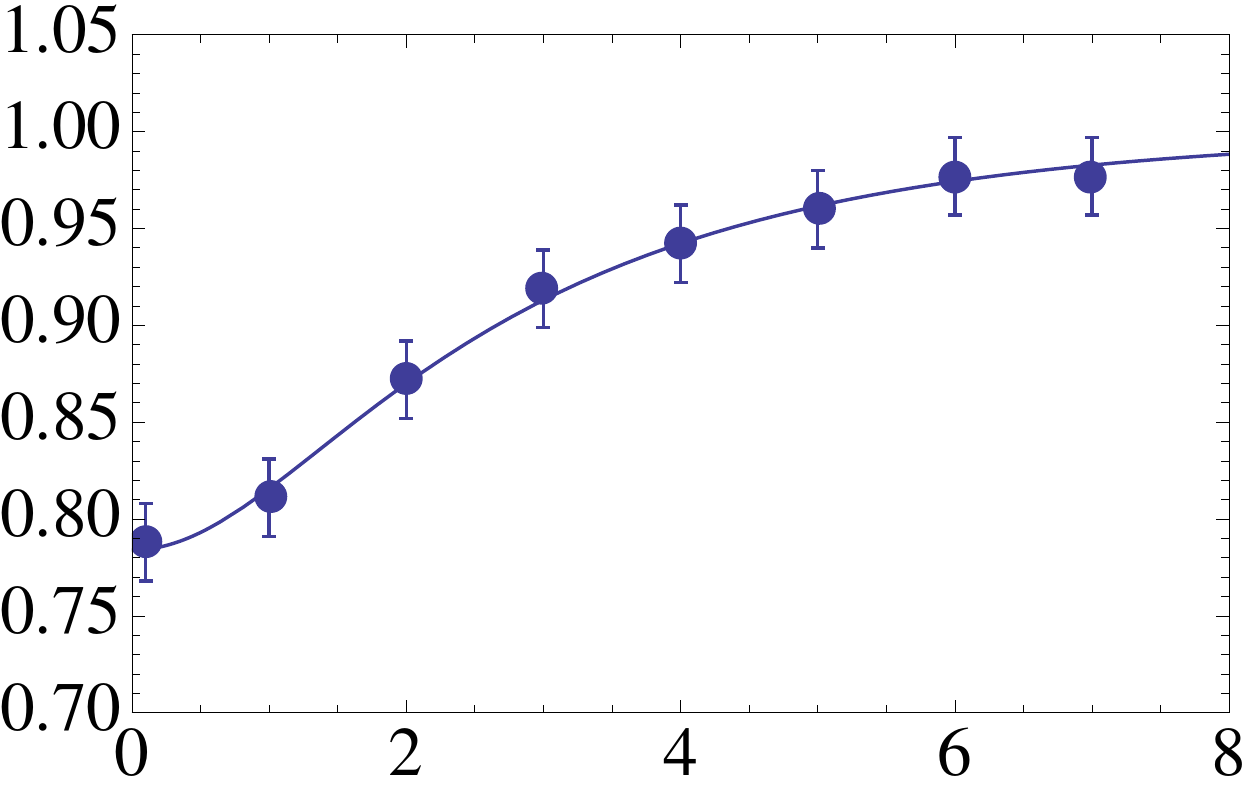}
  \caption{(Color online).
  (a) Static quark-antiquark pair are indicated by shaded circles: those are connected by the 
  flux tube (QCD string). At distance $r$ from the tube one place the operator  $\bar{q} q$, whose average
  is measured, with and without the static quark-antiquark pair.
  (b) Normalized chiral condensate as a function of the  coordinate $r$ transverse to the QCD string (in lattice units). Points are from the lattice calculation \cite{Iritani:2013rla}. The curve is the expression (\ref{eqn_pot}) with $C=0.26$,
  $s_{string}=0.176\, \fm$.}
  \label{fig_lat}
  \end{center}
\end{figure}

\begin{figure}[b]
  \begin{center}
  \includegraphics[width=6cm]{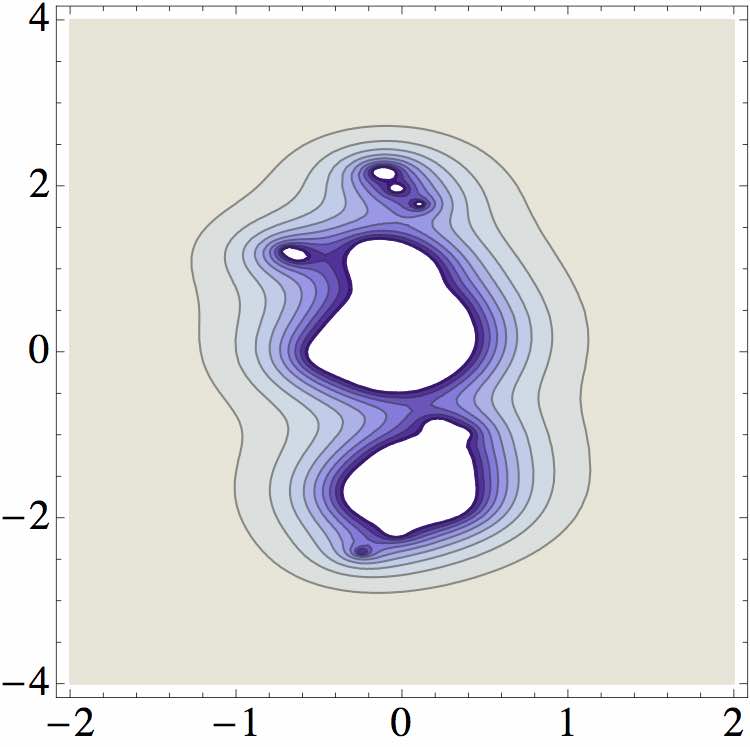}\hspace{0.5cm}\includegraphics[width=1.05cm]{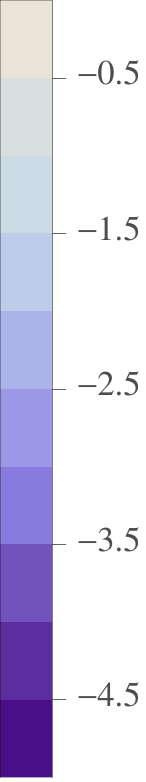}
  \caption{ (Color online)  Instantaneous collective potential in units $2 g_N \sigma_T$ for an $AA$ configuration with $b=11\,\fm$, $g_N \sigma_T = 0.2$, $N_s=50$ at the moment of time $\tau = 1\,\fm/c$. White regions correspond to the chirally restored phase.}
  \label{allAA_pot}
  \end{center}
\end{figure}

   The interaction between the QCD strings in a spaghetti state has been studied in recent paper by Kalaydzhyan and
   myself \cite{Kalaydzhyan:2014tfa}. First of all, 
   we observed that
 the string interaction is, as expected, mediated by the lightest scalar $\sigma$. 
 Indeed, the theoretical expression
\begin{align} {\langle \bar q q(r_\perp) W \rangle \over \langle W \rangle \langle \bar q q \rangle} = 1-  C K_0( m_\sigma \tilde{r}_\perp)\,, \label{eqn_pot}
\end{align}
 (where $K_0$ is the modified Bessel function and the ``regularized" transverse distance $\tilde{r}_\perp$ is
 \begin{align}
 \tilde{r}_\perp= \sqrt{ r_\perp^2+s_{string}^2}\,,  \label{eqn_tilde_r}
 \end{align}
 which smoothens the 2D Coulomb singularity $\sim \ln(r_\perp)$ at small $r_\perp$)
 described lattice data well, see
 Fig.~\ref{fig_lat}. The sigma mass here was taken to be  $m_\sigma=600\, \MeV$ as an input, and not fitted/modified). According to our fit, the ``intrinsic'' width is $s_{string}\simeq 0.176\,\fm$, confirmed by other
 lattice results.
 
 Since the strings in a spaghetti are almost parallel to each other, the problem is reduced to the set of
point particles on a plane with the 2D Yukawa interaction. From the fit (\ref{eqn_pot}) one can see \cite{Kalaydzhyan:2014tfa}, that the main parameter of the string-string interaction (in string tension units) is numerically small,
 \begin{equation}
 g_N \sigma_T = \frac{\langle \sigma \rangle^2 C^2}{4 \sigma_T} \ll 1\,,
 \end{equation}
 typically in the range $10^{-1}- 10^{-2}$.
 So, it is correctly neglected in the situations, for which the Lund model has been originally invented --
 when only $\mathcal{O}(1)$  strings are created.

 The interaction starts playing a role  when   this smallness
 can be  compensated by a large number of strings. As seen from Fig.~\ref{fig_lat},
a magnitude of the quark condensate  $\sigma= |\langle \bar q q \rangle|$ at the string
 position is suppressed by about  $20\%$ of its vacuum value.
So, in a ``spaghetti" state one should think of the quark condensate suppression of about 0.2 times the diluteness, which is still less than 1.

 On the other hand,  about 5 overlapping strings would be
 enough to eliminate  the condensate and  restore the chiral symmetry.
 If $N_s > 30$ strings implode into an area several times smaller than $\sigma_{in}$ (which is the case as we will argue below), then  the chiral condensate will be eliminated inside a larger
 region of  1~$\fm$ in radius, or about 3~$\fm^2$ in area.
 This is nothing but a hot QGP fireball.


As discussed above, the strings can be viewed as a 2D gas of particles (in transverse plane) with unit masses at positions $\vec{r}_{i}$. The
forces between them are given by the derivative of the energy (\ref{eqn_pot}), and so
\be \ddot{\vec{r}}_i= \vec{f}_{ij}= {\vec{r}_{ij}\over \tilde{r}_{ij}} (g_N \sigma_T) m_\sigma  2 K_1(m_\sigma \tilde{r}_{ij}) \ee
with $\vec{r}_{ij}=\vec{r}_{j}-\vec{r}_{i}$ and ``regularized" $\tilde{r}$ (\ref{eqn_tilde_r}).

In our simulations we used a classical molecular dynamics code.
%
The evolution consists of two qualitatively distinct parts: (i) early implosion, which converts potential energy into the kinetic one, which has its peak when
fraction of the particles ``gravitationally collapse" into a tight cluster; and (ii) subsequent approach to
a ``mini-galaxy" in virtual quasi-equilibrium. Only the first one is physical, as the imploded spaghetti
has density sufficient for production of QGP fireball, and after that explodes hydrodynamically.
The whole scenario thus resembles the supernovae: implosion leads to more violent explosion later.

In Fig.~\ref{allAA_pot} we show an example of the instantaneous collective potential produced by the strings in the transverse plane. The white regions correspond to the values of potential smaller than $-5\cdot 2 g_N \sigma_T (\fm^{-1}) \approx -400\,\MeV$, i.e. the chiral symmetry can be completely restored in those regions.
Large gradient of this potential at its edge can cause quark pair production, similar to Schwinger process in electric field: one particle may flow outward and
one falls into the well. Such phenomenon is a QCD analog to Hawking radiation at the black hole horizon.

%
%
%
%
%

\section{Holographic equilibration} \label{sec_holo_eql}
\subsection{ Near equilibrium}
Since this development is with heavy ion community for over a decade,
no general introduction is perhaps needed. And relating it to string theory background
from which AdS/CFT correspondence came will take us too far from the topics.
I will only comment on a couple of issues to be relevant to discussion later in this chapter.

 The equilibrium setting includes the so called ``AdS-black hole" metric, with its horizon located at the 5-th coordinate 
  $z_h=1/\pi T$,  so the gauge theory -- located at the $z=0$ boundary feels  the  Hawking radiation temperature $T$.
(Hawking radiation in asymptotically flat background leads to evaporation of
  BH. Indeed, one BH cannot heat up an infinite Universe. 
   But  AdS asymptotics means basically a finite box, and in this case
 BH can get into an equilibrium state with the Universe.)

    \begin{figure}[t]
  \begin{center}
  \includegraphics[width=6cm]{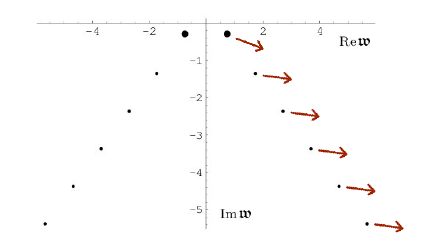}
   \caption{A set of frequency modes, on $\omega$ complex plane, from \cite{Kovtun:2005ev}. Dots are for a particular wave vector $k$,
   arrows indicate the direction of motion as $k$ increases.}
  \label{fig_modes}
  \end{center}
\end{figure}

 Various gravity waves propagating in this background  have dispersion relations $\omega(\vec k)$ with calculable real and imaginary parts,
 see an example in Fig.\ref{fig_modes}.  (Such quasinormal\footnote{Like wave functions of the
$\alpha$-decaying nuclei, when energy is complex the wave function grows in space and is not
normalizable-- thus the name. Physicists called them quasistationary states. 
 } modes are known for various examples of BHs for a long time, these particular ones were calculated  by  Kovtun and Starinets \cite{Kovtun:2005ev} ). In this  channel the lowest eigenvalue, shown by larger dots,  is close to the origin and describes the sound mode. For reference let me mention several known  terms
  at small $k$ (from \cite{Lublinsky:2009kv})
   \begin{align}
 &{\omega\over 2\pi T} =\pm{ \tilde k \over  \sqrt{3}} \left[ 1+ \left({1\over 2}-{\ln 2 \over 3}\right) \tilde k^2-0.088\, \tilde k^4\right]\nonumber\\
&- {i \tilde k^2\over 3} \left[1-{4-8\ln 2+\ln^2 2 \over 12}\, \tilde k^2 - 0.15\,\tilde k^4\right]\,,
\label{eqn_sounddisp} 
\end{align}
where $\tilde k \equiv\left(k/ 2\pi T\right)$. 
 First, note that at small $k$ the imaginary viscous term is $Im\omega \sim k^2$: we already
used that fact in discussion of the ``acoustic damping" phenomenology. Second, the dispersive correction
to the speed of sound has positive coefficient: thus sound wave can decay into two: we already discussed that
in discussion of the cosmological acoustic turbulence. Third: note that higher order corrections to viscosity
are both negative. This is in contrast to some popular second order ad hoc schemes such as Israel-Stuart.  

All ``non-hydrodynamical modes" have $Im(\omega)/(2\pi T)=O(1) $, indicating that during the  time of the order of $z_h\sim 1/(2\pi T)$ 
they all ``fall into the black hole". (It is amusing to note that a
mystifying process of QGP equilibration is, in AdS/CFT setting, reduced to
the problem number one in physics history, the Galilean falling stone in gravity field. 
)
The
 essence of the AdS explanation of the rapid equilibration is thus simple: any initial objects
 gets irrelevant,  all of those are absorbed by a black hole. 
 The only memory which remains is their total mass,  which the BH transforms into
 corresponding amount of Bekenstein entropy. (Plus conserved charges, if they are there.)

  \begin{figure}[t]
  \begin{center}
  \includegraphics[width=8cm]{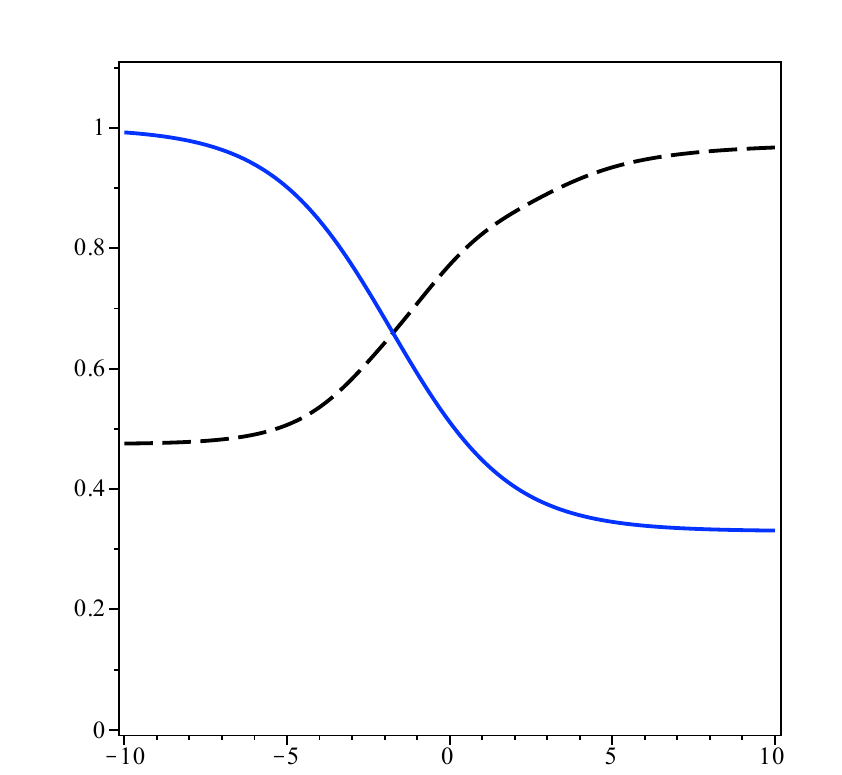}
   \caption{An example of a strong shock in QGP, according  to the Navier-Stokes hydro. 
 Pressure shown by (blue) solid line is in units of its initial value: thus
the curve starts near 1 on the left side. The (black) dashed line is
    the rapidity.  From \cite{Shuryak:2012sf}.}
  \label{fig_shock}
  \end{center}
\end{figure}

 \subsection{Out of equilibrium 1: the shocks}
 
Shocks are the classic objects used in studies of out-of-equilibrium phenomena. They traditionally
are divided into   weak and strong shocks. In the former case
there is small difference between matter before and after the shock: so those can be treated hydrodynamically,
e.g. using the Navier- Stokes (NS) approximation. Strong shocks have finite  jumps in matter properties.
Their profiles have large gradients: so one needs some more powerful means to solve 
the problem, not relying on hydrodynamics, which is just an expansion in gradients.
 
 The reason we put this example as number one is because it is the only one which 
 can be considered in {\em stationary approximation}. Indeed, in the frame which moves with the
 velocity of the shock, its profile is {\em time-independent}. 
 
Strong shocks in AdS/CFT setting were discussed in my paper
 \cite{Shuryak:2012sf}. A particular example worked out in detail 
starts with NS profile shown in Fig.\ref{fig_shock}. From left to right the pressure decreases, and rapidity increases:
the energy and momentum fluxes are tuned to be the same. One may think of it as a process
in which higher-density QGP on the left, moving more slowly, ``burns" into a lower density QGP  floating out
with higher rapidity. 
 
In the AdS/CFT setting
 one however can solve the problem from first principles, by solving the Einstein equations. 
 Of course the setting has an extra holographic dimension, and so all the functions depend
 not on one but on two variables: the longitudinal coordinate $x$ and $z$. 
 So I plugged in the   NS solution, and worked out corrections to it 
 (by a somewhat novel  variational approach). Not going to the details of that,
 let me  jump directly to the surprising conclusion: those corrections happen to be
  small, at a scale of few percent, even without any apparent small parameter in the problem.
 
 Another tool used to correct the NS solution was the so called ``re-summed hydrodynamics" by Lublinsky and myself:
 it also lead to corrections at the percent level. (Unfortunately, the accuracy I had on AdS/CFT solution was insufficient
 to tell whether both agree or not). 
 
 The lesson:  all higher order gradient corrections to the NS solution
 have strong tendency to cancel each other. 
 
\subsection{Out of equilibrium 2: the falling shell }


This setting has been proposed by Lin and myself \cite{Lin:2008rw} and it
is in a way orthogonal to the previous one: there is  time dependence but no dependence on space point.
The motion occurs only along the holographic 5-th direction.

The physical meaning of this motion is as follows. First of all, recall that the 5-th  coordinate $z=1/r$
corresponds to a ``scale", with small values near the boundary (large $r$) corresponding to the UV end of 
the scales, and large $z$, small $r$  corresponding to the IR or small momenta end. 
Since everything happens much quicker in UV as compared to IR, the equilibration process
should naturally proceed from UV to IR, also known as ``top-down" equilibration.

\begin{figure}[t]
  \begin{center}
  \includegraphics[width=8cm]{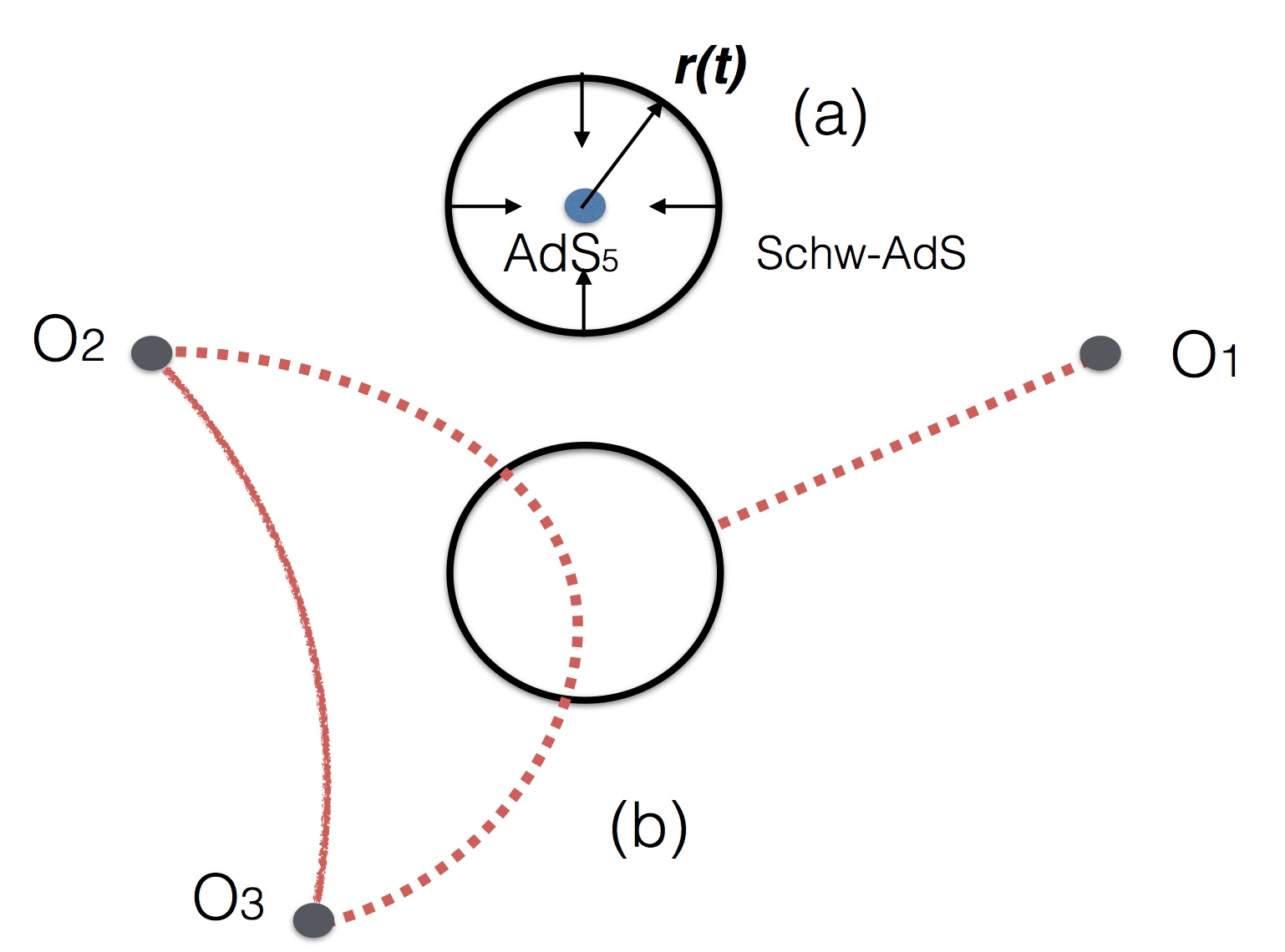}
   \caption{(a) A sketch of a falling shell geometry. Its radius $r(t)=1/z(t)$ which is used in the text. 
   (b) Single-point observer $O_1$ and the two-point observers $O_2,O_3 $ }
  \label{fig_shell}
  \end{center}
\end{figure}

One can imagine that this process can in some sense be reduced to a thin ``equilibration shock wave",
propagating in the $z$ direction. The key idea of the paper   \cite{Lin:2008rw}
 was that this shock can be thought of as a certain material objects -- a shell or an elastic membrane -- falling 
 {\em under its own weight}. (See the sketch in  Fig.\ref{fig_shell}(a)). 
 If this is the case, the total energy of the membrane is $conserved$ (potential energy goes into kinetic). 
 The consequences of that are very important: while the shell is falling toward the AdS center,  the 
 metric -- both above and below the membrane -- is actually time independent, as it can only depend on 
 its total energy. So there is no need to solve the Einstein equations. 
 (Recall the Newton's proof that a massive sphere has the outside  field  the same as a point mass, and that
 there is no gravity inside the sphere. This is also true if the sphere is falling. It also remains true in GR.)
In the case of an extreme black hole at the AdS center (the blue dot    in  Fig.\ref{fig_shell}(a))
 the solution consists of   (i)
  thermal   Schwarzschield-AdS metric above the shell and (ii) ``empty vacuum" or the $AdS_5$ solution below it.

The only equation of motion which needs to be derived and solved are those of the shell itself, $r(t)=1/z(t)$. 
It is not so trivial to derive it, since the coordinates below and and above the shell are discontinuous. 
Fortunately, a thin shell collapse has already bend solved in GR literature: the key to it 
is  the so called ``Israel junction condition".
The shell velocity (in time of the  distant observer $t$) is given by the following expression
\be\label{trajectory}
\frac{dz}{dt}=\frac{\dot{z}}{\dot{t}}
=\frac{f\sqrt{(\frac{\kappa_5^2 p}{6})^2+(\frac{3}{2\kappa_5^2 p})^2(1-f)^2-\frac{1+f}{2}}}{\frac{\kappa_5^2 p}{6}+\frac{3}{2\kappa_5^2 p}(1-f)}
\ee
where $\kappa_5^2 p$ is the 5-dim gravity constant and shell elastic constant and $f=1-z^4/z_h^4$ is the ``horizon function"
 in thermal AdS. 
The qualitative behavior of the solution is as follows:
The shell start falling with zero velocity from certain hight get accelerated, to near the speed of light, and then 
 is ``braking" toward the horizon position $z_h$,
 which the shell asymptotically approaches with velocity zero.

 After solution is found, one can calculate what
 different  observers  at the boundary -- that is, in the gauge theory -- will see. 
In particular, one may ask if/how
 such an observer can tell a static black hole (the thermal state with stationary horizon)  
 from that with a falling shell?
 
A {\em ``one-point observer"} $O_1$ Fig.\ref{fig_shell}(b)
  would simply see stress tensor perturbation induced a gravitational propagator 
  indicated by the red dashed line. Since the metric above the shell is thermal-AdS, he will
find  {\em time-independent} temperature,
 pressure and energy density
 corresponding to static final equilibrium.
Yet more sophisticated 
{\em ``two-point observers" }  $O_2,O_3$ can measure certain correlation functions of the stress tensors.
They will see effects from gravitons flying along the line shown by the solid line above the shell,
that is in the thermal metrics, as well as from gravitons flying over the path shown by the dashed line which
 penetrate $below$ the shell: those would notice deviations from equilibrium.
  Solving for various two-point functions
in the background with falling shell/membrane we
found such deviations.  They
happen to be oscillating in frequency around thermal ones. This observation is explained  \cite{Lin:2008rw} by certain ``echo" times due to  a signal reflected from the shell.
So, here is the lesson: one can in principle experimentally observe an 
echo, from the 5-th dimension!

For further discussion of the scenarios of top-down equilibration, with  infalling scalar fields etc -- the reader is referred to
 \cite{Balasubramanian:2011ur} and subsequent literature.

  \begin{figure}[t]
  \begin{center}
  \includegraphics[width=8cm]{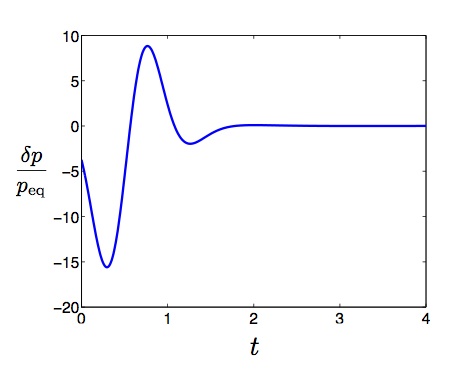}
   \caption{Pressures anisotropy as a function of time, from \cite{Chesler:2013lia}.}
  \label{fig_chesler}
  \end{center}
\end{figure}

\subsection{Out of equilibrium 3:  anisotropic plasma}
  Our next example, due to 
  Chesler and Yaffe, 
   is a setting in which one starts with some $anisotropic$ but homogeneous metric, and then follow its relaxation to
   equilibrium. It is nicely summarized by Chesler \cite{Chesler:2013lia}, so I will be brief. 
The metric is of the form of diagonal, time dependent but space-independent components,
and Einstein equation is solved numerically. 
Rapid relaxation to AdS-Schwarzschield-like solution is observed.
The typical behavior found is displayed in Fig.\ref{fig_chesler}. 
The characteristic relaxation time is shorter than $1/T$, even when the system is initially quite far from equilibrium.

A number of initial states can be compared,
with the condition that all of the evolve to the same equilibrium energy density (or horizon, or T). 
While at early time the momentum asymmetry
can be very large -- say, an order of magnitude --  it  becomes very small exponentially in time.
Any deviations from equilibrium get strongly red-shifted as they approach the horizon. 

If a more detailed analysis is made, one can locate few specific quasi normal modes. There are no hydro modes since the setting is homogeneous, and the  lowest is $\lambda=(2.74+i 3.12)\pi T$. So, the basic lesson is that in the strong coupling regime
the isotropization time is as  short as \be \tau_{isotropization}\sim {0.1 \over T} \label{eqn_isotropization} \ee

\begin{figure}[t]
\centerline{ \includegraphics[width=10cm]{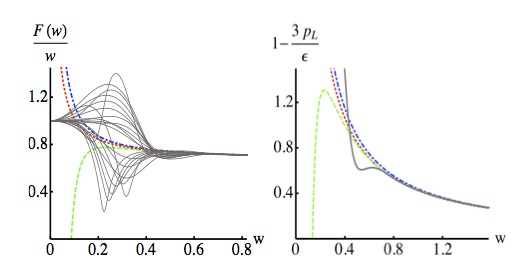}}
\caption{(left) From Heller et al \cite{Heller:2012je}: The temperature evolution combination $dlog(w)/dlog\tau$ for different initial conditions (black thin curves) converging into a universal function of $w=T\tau$, compared to hydro. (right) The pressure anisotropy
for one of the evolutions compared to 1-st (NS), 2-nd and 3-ed order hydrodynamics.
} 
\label{fig_equilibration}
\end{figure}

\subsection{Out of equilibrium 4: rapidity independent collisions  \label{sec_grad_resummation}}

 Various ``bulk" objects  may fall into the AdS center. 
In an  early paper
 Sin, Zahed and myself~\cite{Shuryak:2005ia}
 argued that all ``debris" created in high energy collisions form bulk black hole which then 
falls toward the AdS center. Its hologram is the
 exploding/cooling
fireball seen by observer on the boundary. The specific solution 
 discussed in that paper was spherically symmetric (and thus  more 
 appropriate for cosmology than for heavy ion
applications). 

 Subsequent series of papers by Janik
and collaborators  (see e.g. \cite{Janik:2005zt} and references therein)
had developed  rapidity independent  ``falling" horizons, corresponding to Bjorken 
hydro solution. In this case one has homogeneous x-independent horizon falling as a function of time $z_h(\tau)$. At late time it converges into hydro and \be z_h(\tau)=1/\pi T(\tau) \sim \tau^{1/3} \ee
The variable $w=T\tau$ has the meaning of the macro-to-micro scale ratio: at late time it grows
indicating that the system becomes more macroscopic and hydro more accurate. 
The question is $when$ hydro description becomes valid and {\em with what accuracy}.

 Fig.\ref{fig_equilibration} from Heller et al \cite{Heller:2012je} displays time evolutions of 
many initial states, all approaching  the same hydrodynamical solution.
Fig.\ref{fig_equilibration}(left) shows that this happens via convergence
 to certain {\em universal function} of the variable $w=\tau\,T$ defined by
\be  {d w\over d\ln \tau}\,=\,F(w)\,,\ee
Existence of such universality is the essence of the ``re-summed hydro" proposed by Lublinsky and myself \cite{Lublinsky:2009kv}. Depending on accuracy required, one may
assign specific value of $w$ at which ``hydrodynamics starts".  Its  value is in the range $w_i=0.4..-0.6$. The plot on the right
demonstrates that at such time the anisotropy is still large and viscosity is important. 

The lesson from this work can be better explained by comparing its result to
"naive expectations" , that hydro starts when macro and micro times are the same, $w=\tau T> 1$, and that the accuracy of hydro should be bad O(1),
say 100\%. Calculations show instead that  at {\em twice smaller} time  $w\sim 1/2$ the accuracy of
(the lowest-order  Navier-Stokes) hydro
suddenly becomes  good, at few percent level! 

Why is it so? Note, that while  gradients are not yet small at that time,
but the combined effect of all of them is. Lublinsky-Shuryak re-summation provides an answer:
higher gradient series have alternating signs and can be Pade re-summed a la geometric
series to a decreasing function.

The issue has its practical aspect, related to one of the first observations made at the first LHC  PbPb run. It was found that  the (charged hadron) multiplicity in PbPb
collisions grow with energy a bit more rapidly than  in $pp$ :
\be
{dN^{PbPb}\over dy}(y=0,s)\sim s^{0.15}  \hspace{1cm}  {dN^{pp}\over dy}(y=0,s)\sim s^{0.11} 
\label{eqn_multiplicity}
\ee
From the RHIC energy ($E=0.2\, TeV$) to the LHC, the double ratio 
  \be \label{dr}
  {{dN\over d\eta}|_{PbPb,LHC}\,\,/\,\, {dN\over d\eta}|_{pp,LHC} \over  {dN\over d\eta}|_{AuAu,RHIC}\,\,/\,\, {dN\over d\eta}|_{pp,RHIC}} = 1.23\,. \ee
shows a  noticeable change with the energy, which  calls for an explanation.

Lublinsky and myself \cite{Lublinsky:2011cw}  proposed a simple form for the function $F(w)$ and calculate the entropy produced, from the time $w_i$ on. It turns out to be about 30\%.
Furthermore, we get the following expression for the contribution
to this double ratio  $\approx 1+{3[\bar\eta(LHC)-\bar\eta(RHIC)]\over  2w_i+3\bar\eta(RHIC)} $ and show, that the observed growth can be naturally explained by the viscosity growth, from RHIC to LHC, predicted by a number of phenomenological models.

\section{Collisions in strong coupling} 
\subsection{Trapped surfaces and the entropy production}
The simplest geometry to consider is the wall-on-wall collisions, in which there is no dependence on two
transverse coordinates, and only the remaining three -- time, longitudinal (rapidity), and the holographic direction
-- remain at play. Needless to say, it is a very formidable problem,   solved
by Chasler and Yaffe via clever ``nesting" of Einstein equations.  The reader
can find explanations in a summary  \cite{Chesler:2013lia}.

\begin{figure}[t]
 \includegraphics[width=8cm]{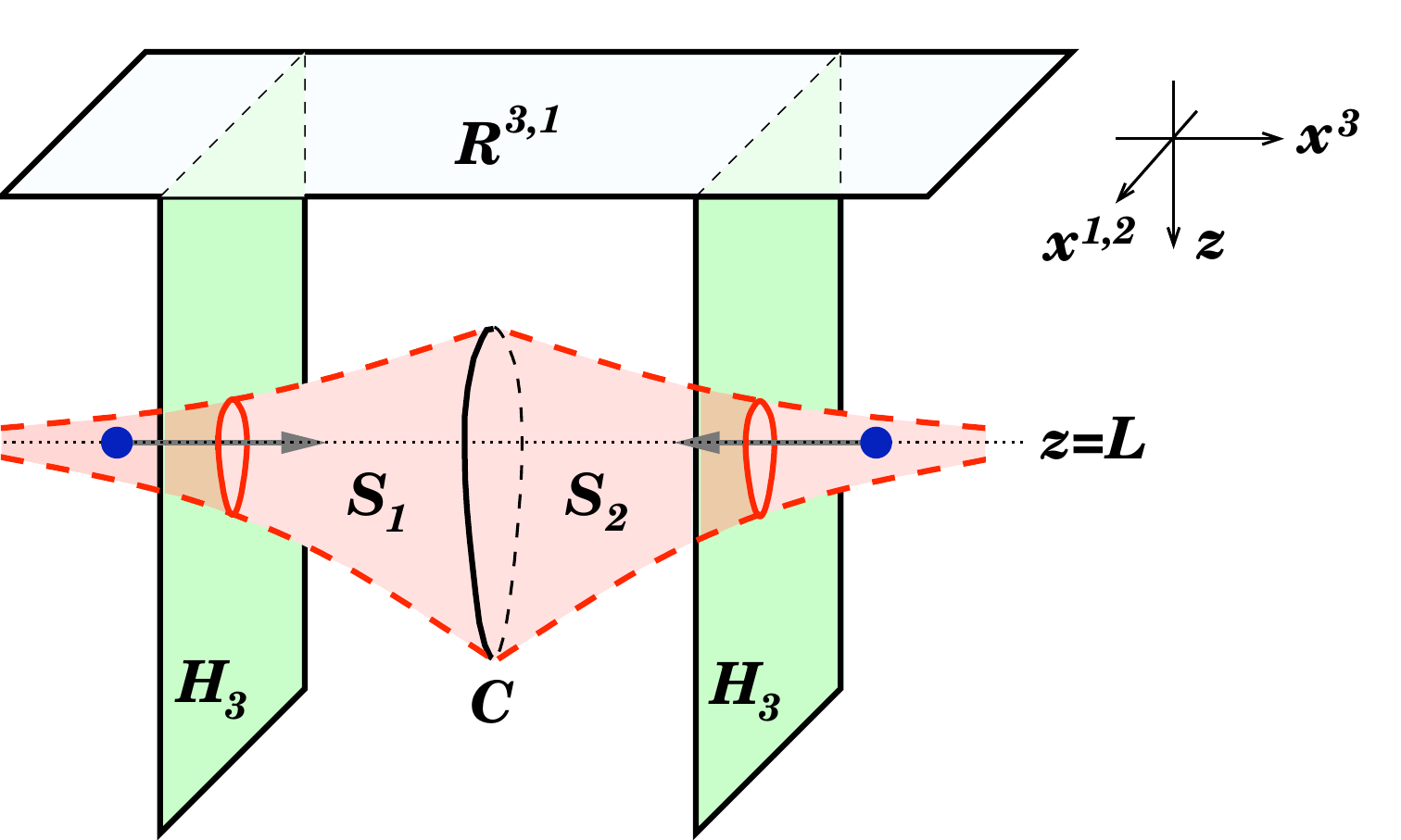}
 \includegraphics[width=8cm]{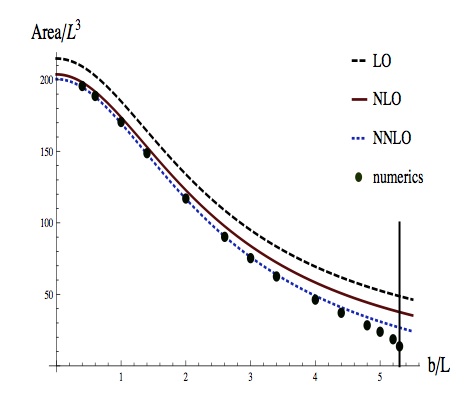}
\caption{(a) From \protect\cite{Gubser:2008pc}:A projection of the marginally trapped surface onto a fixed time slice of the AdS geometry.  
(b) The area of the trapped surface versus impact parameter, with the comparison of the numerical studies  \protect\cite{Lin:2009pn} shown by points and
analytic curves from \protect\cite{Gubser:2009sx}. The vertical
line shows location of the critical impact parameter $b_c$ beyond which there is no trapped surface.} 
\label{fig_trapped}
\end{figure}

Collisions of finite size objects are even more difficult to solve, 
but  those  historically brought into discussion very important issues of {\em trapped surface} formation and  the {\em entropy production}. It was pioneered by
 Gubser, Pufu and Yarom           
\cite{Gubser:2008pc} who considered
head-on (zero impact parameter) collisions of point black holes. 
The setting is shown in Fig.\ref{fig_trapped}(a). 

Trapped surface is a technical substitute to the horizon -- for this review it is not too technical to discuss the 
difference -- and its appearance in the collision basically means that there exists a black hole
inside it. Classically, all information trapped inside it cannot be observed from outside, and lost information is $entropy$.
(Needless to say, for known static black hole solutions this area does give the black hole Bekenstein entropy.)
 So, locating this surface
allows one to limit the produced  entropy  {\em from below}, by simply calculating {\em its area}.
The reason why this entropy estimate is from below is because the trapped surface area 
is calculated at the $early$ time $t=0$ of the collision, not at its end. No particle
 can get out from trapped surface, but some can get into it $during$ the system's evolution, increasing
the black hole mass and thus its entropy.  


 The setting of Gubser et al have the parameter $L$, the distance separating colliding b.h. from the
boundary: we will discuss its physical meaning below. Naively, central collisions have only axial O(2)
symmetry in the transverse plane $x_\perp$, but
  using global AdS coordinates
these authors found higher O(3) symmetry of the problem, which becomes  
black hole at the collision moment at its center,
thus  in certain new coordinate \be q={\vec x_\perp^2+(z-L)^2\over 4 z L } \ee
 the 3-d trapped surface C at the collision
moment becomes a 3-sphere, with some radius $q_c$. 
One can find $q_c$ 
and determine its relation to CM collision energy and
Bekenstein entropy. For large $q_c$ these expressions are just 
\be E\approx {4 L^2 q_c^3 \over G_5}, \hspace{.2cm} S\approx {4 \pi L^3 q_c^2 \over G_5},\ee
from which, eliminating $q_c$, the main conclusion follows:
the entropy grows with the collision energy as
\be S\sim E^{2/3} L^{5/3}  \label{eqn_entr}\ee
Note that this power is in general $(d-3)/(d-2)$  directly relates to the $d$=5-dimensional gravity, and is different
from the 1950's prediction of Fermi/Landau who predicted 
$E^{1/2}$ as well as from the data, which indicate the power of about $0.30$ (\ref{eqn_multiplicity}).

   Let me now return to the meaning of the parameter $L$. Gubser et al relate the ``depth" of the colliding objects with the nuclear size, $L\sim 1/R_A$  which cannot depend on the energy. 
Lin and myself  \cite{Lin:2009pn} argued that $L$ should rather be identified 
with the inverse ``saturation scale", the typical
 parton's momenta in the wave function of the colliding objects
 \be L\approx {1 \over Q_s(E)} \sim E^{-\alpha}\ee
 It becomes especially clear if one would like to turn back to wall-on-wall collisions,
 in which the nuclear size $R_A$ goes to infinity while $Q_s$ remains fixed. 
 If so,  $L$ has a completely different scale, it is not of the $O(10\, fm)$ scale but rather $O(0.1\, fm)$. 
  Furthermore, it is expected to $decrease$ with the energy $L\sim 1/Q_s(E)$ with certain empirical index $\alpha\approx 1/4$.  Including this dependence into (\ref{eqn_entr})
   one  gets
   \be S\sim E^{(2/3)-(5/3)\alpha}\sim s^{0.125} \ee 
   now in
   reasonable agreement with the observed multiplicity.

The generalization of this theory to non-central collisions, by Lin and  myself
 \cite{Lin:2009pn}, have  technical details I would omit here
 and proceed directly to the results shown in Fig.\ref{fig_trapped}(b).
The figure shows
dependence of the trapped surface area on the impact parameter is actually
from the paper  by Gubser, Pufu and Yarom \cite{Gubser:2009sx}, comparing our 
results (points) with analytic series of curves obtained by these authors.

From  gravity point  of view the qualitative trend shown is clear: two colliding objects may merge
into a common black hole only provided that the impact parameter is less than some critical value $b_c(E)$, depending on the collision energy. Indeed, with $b$ rising, the trapped energy decreases while the total
angular momentum increases: so at some point Kerr parameter exceeds 1 and thus no black hole can be formed. 
Interestingly, the calculation had shown that it happens as a jump --  the first order transition in impact parameter. 
Just a bit below this impact parameter reasonable trapped surface and black hole exist and nothing obvious indicates that at large $b$ none is formed.

 (Note that  conclusion about  the first order transition is  consequence
of the large $N_c$ approximation and classical gravity: those perhaps may be smoothened at finite $N_c$. Furthermore,
the problem of trapped surface in quantum gravity  is way too complicated,  not studied so far.)

This transition should be interpreted as follows: at $b<b_c$
a
 creation of QGP fireball takes place, while the system is in hadronic phase for peripheral $b>b_c$ collisions. 
 Still, from a physics point of view  a jump in entropy as a function
of impact parameter is surprising. Interesting, that the  experimental multiplicity-per-participant plots 
do show change between non-QGP small systems and QGP-based  not-too-peripheral $AA$ collisions. 
It would be interesting to compare it with recent information on small systems, undergoing
transition to explosive regime.

In the same paper  \cite{Lin:2009pn} we have pointed out that the simplest
geometry of the the trapped surface would be that for a wall-wall collision, in which
there is no dependence on transverse coordinates $x^2,x^3$. Thus a sphere becomes just two points in $z$, above and below the colliding bulk objects.
We elaborated on this in  \cite{Lin:2010cb}, considering
collision of two infinite walls made of material with different ``saturation scales"
(e.g. made of lead and cotton) and studied conditions for trapped surface formation.

 \begin{figure}[t]
  \begin{center}
  \includegraphics[width=6cm]{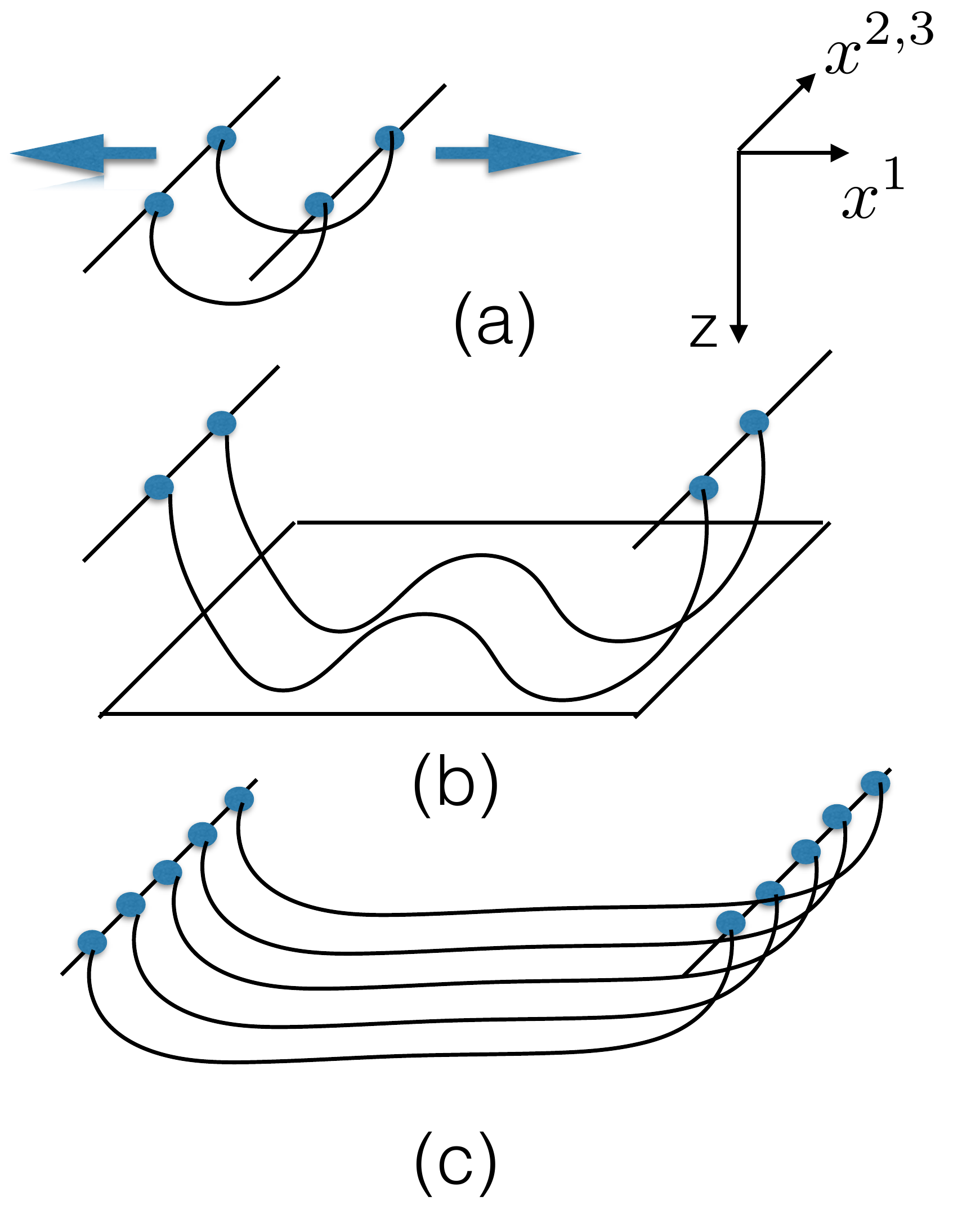}
   \caption{ (a) An early time snapshot of a pair of strings created after one color exchange. The coordinates are
   explained on the right: the colliding objects move with a speed of light away from each other and strings are
   stretched. (b) Later time snapshot, in QdS/QCD background. Strings reached the ``levitation surface"
   shown by a rectangular shape and start oscillate around it; (c) In case of high density of many strings
   they can be approximated by a continuous membrane.
   }
  \label{fig_strings_in_AdS}
  \end{center}
\end{figure}

\subsection{From holographic to QCD strings}
AdS/CFT is a duality with a string theory, so fundamental strings are naturally present  in the bulk.
In fact already the first calculation, made in Maldacena's original paper,  the ``modified Coulomb law", was based on
 the evaluation of the shape and total energy of a ``pending string", sourced by ``quarks" on the boundary.
(This example is so classic, that must be known to most readers.)

 Lin and myself in two papers \cite{Lin:2006rf,Lin:2007fa} extended it to non-static strings.
We first solved for a falling string with ends moving away from each other with velocities $\pm v$, and the second
calculating its hologram (stress tensor distribution) at the boundary. This study can be thought of as 
a "strongly coupled version of the
$e+e-$ annihilation into two quarks". The hologram showed 
 a near-spherical explosion: the early indication  that at strong coupling  setting {\em there are no jets}.  

These works used the setting associated with conformal gauge theory: in    $AdS_5$ string  falling  continues forever.  
   This is of course $not$ what we observe in the real world, in which there is confinement and there are jets. Modern  strong coupling models moved into what is collectively known  as  AdS/QCD
 , for review see e.g. \cite{Gursoy:2007cb,Gursoy:2007er}.  In contrast with the original AdS/CFT, 
 the background metric is not conformally invariant and incorporates both
 $confinement$ in the IR and the {\em asymptotic freedom} in UV. These models use additional
  scalar (``dilaton") field, with certain potential, complementing a gravity background. 
  
  In such settings bulk
  strings (and other objects)
 can ``levitate", at some position $z_*$, at which the downward gravity force is compensated by the
 uplifting dilation gradient. The hologram of such levitating string at the boundary is {\em the QCD string}:
 its tension, width and stress tensor distribution inside are calculable. 
  The  potential between point charges is given by a pending string: in the AdS/QCD background
  it changes from Coulombic at small $r$ to linear at large $r$. Furthermore, 
 allowing fundamental fermions in the bulk -- via certain brane construction --and including their back reaction
 in a  consistent manner one
 can get quite nice description of the so called Veneziano-QCD ($N_c,N_f\rightarrow \infty, x=N_f/N_c=fixed$)
 \cite{Arean:2013tja}. 

 Since in the UV these models also possess a weak coupling regime, one can also model perturbative glasma, by 
putting certain density of color ``sources" at the two planes, departing from each other. In such setting there would be smooth transition between
 two alternatives description of the initial state we discussed above -- from the perturbative glasma made 
of longitudinal field cells to a ``spaghetti" made of the QCD strings. When time $\tau$ is small,
Fig.\ref{fig_strings_in_AdS}(a),
strings are in the UV domain (at small distance $z\sim 1/Q_s$ from the boundary) their hologram is
Coulombic or glasma-like. 
 When strings fall further and reach the ``levitation point" $z^*$
 they  start oscillating around it \cite{strings_adsqcd}, Fig.\ref{fig_strings_in_AdS}(b). This is very
similar to oscillations discussed by  Florkowski  (see sect.\ref{sec_spaghetti})
without AdS/QCD.  

At this point one may wander about AdS/QCD predictions for string-string interaction. This subject is now under investigation in \cite{strings_adsqcd}. We already discussed this issue in the QCD context above, and concluded
that its long-range attraction is dominated by the $\sigma$ meson exchanges (just like between nucleons, in nuclear forces): so we need to see if it is indeed the case in the  AdS/QCD, and whether one can learn something
new using it. This is indeed the case, and in order to explain why we need to briefly remind the reader 
some basics on how hadronic spectroscopy comes out in this setting.

 AdS/QCD has very few fields in the bulk -- gravity, dilation and (quark-related) ``tachion".
 Each of them (or their mixtures) produce towers of 4-dimensional hadronic states via the quantization  
 of their dynamics in the 5-th coordinate. One may say that hadrons are standing waves in it, and their masses are just quantized 5-th momentum. So, a new perspective on hadrons one gets from this approach is 
 that one can calculate not only the masses but also the {\em wave functions} in a scale space.
 
 So the issue studied in detail in \cite{strings_adsqcd} is the mechanism of hadronic flavorless scalars, which includes
the $\sigma$ and others. Without any changes in the setting of AdS/QCD we found very good description
of (quite involved) mixing pattern of the scalars, which puzzled spectroscopists for decades. 
 Since strings are gluonic objects and $\sigma$ interact strongly with quarks, understanding of such mixing
  is crucial for obtaining realistic string-string forces.

  Note that there is so far no temperature or entropy in the problem:
  the dynamics is given by classical mechanics of strings moving in certain backgrounds.
  
   That is because we so far considered a single (or few) strings. If there are very many charge exchanges and the number (or density) of strings is high
  enough, Fig.\ref{fig_strings_in_AdS}(c), one should perhaps include the back reaction of their gravity and dilaton field, or even include
   mutual attraction of strings. 
   Such AdS/QCD version of multi-string dynamics from \cite{Iatrakis_etal} 
   of what we discussed in section \ref{sec_spaghetti} following
   \cite{Kalaydzhyan:2014zqa}.  
   
   Not going into detail, let me comment on just one issue. Unlike GR, in which there is
   only gravity and thus universal horizon definition, in holographic models physics is more
   complex and interesting. In particularly, gluons are strings vibrations.
   String have their 2D worldvolumes which can be curved and have black hole
   of their own. What it means is the wave on the string can have modes propagating left and right,
   but also can have regions where both waves move right. So gluons can fall into those,
   but cannot escape. Such feature can be induced, in particularly, by collective gravity/dilaton
   of the strings themselves. Its meaning is a QGP fireball, and its experimental signature is hydro explosion.

\subsection{Holographic Pomeron} \label{sec_Pomeron}
Pomeron description of hadronic cross sections and elastic amplitudes
originates from phenomenology, starting from 1960's. 
Veneziano amplitudes 
derived in a ``resonance bootstrap" ideology were the original motivation for 
existence of QCD strings.

\begin{figure}[t!]
\includegraphics[width=7cm]{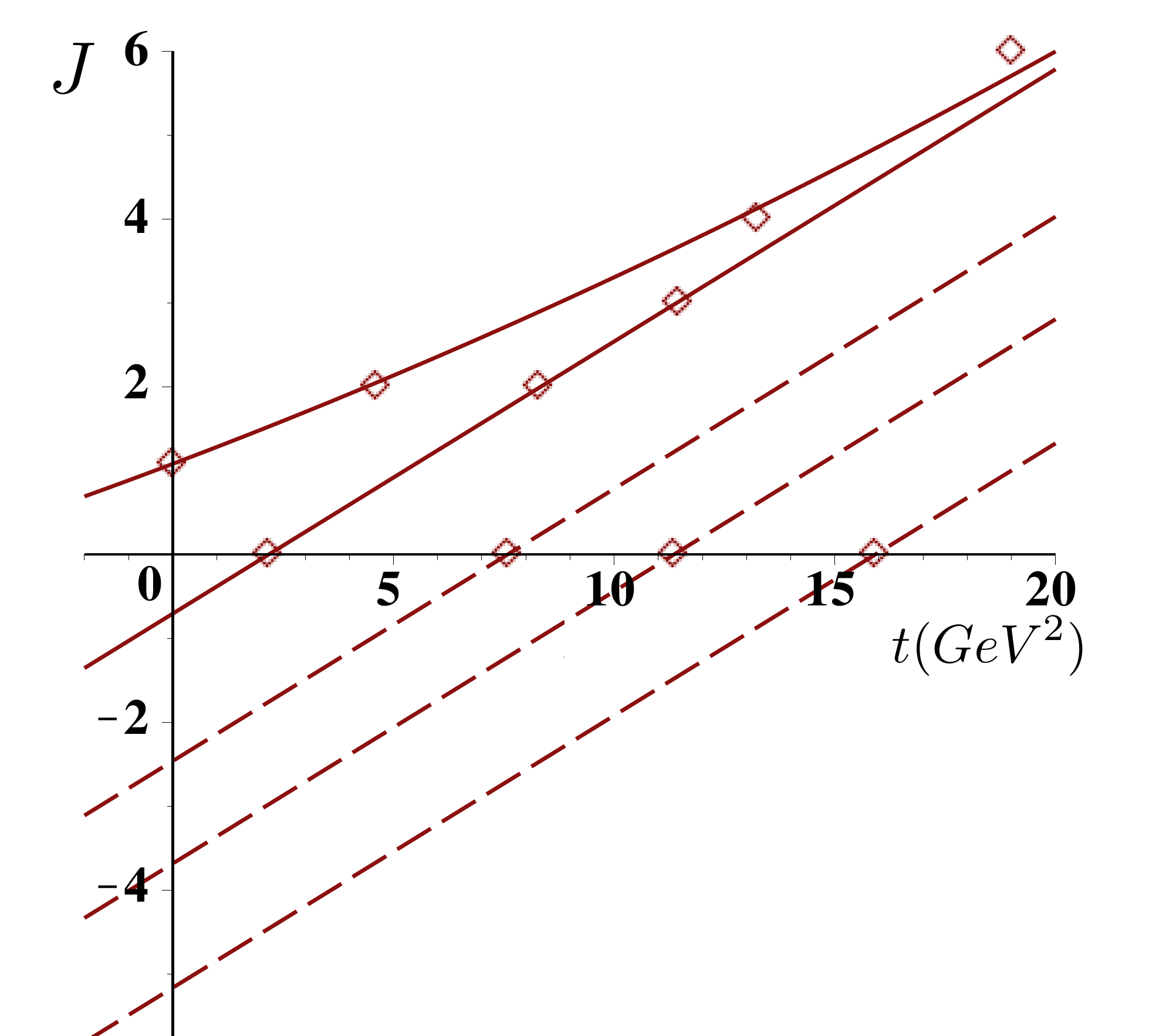}
  \caption{ Glueball masses calculated on the lattice (diamonds) organized in Regge trajectories (lines). (From  \cite{Shuryak:2013sra}).}
  \label{glueballs}
\end{figure}

The Pomeron is an effective object corresponding to the  ``leading"   Regge trajectory $\alpha(t)$
which dominates the high energy asymptotics of the hadron-hadron cross sections.
 Fig.\ref{glueballs} is a recent version of the Regge plot (angular momentum $J$ versus the mass squared $m^2$
 for glueballs. The Pomeron corresponds to scattering and thus has small non-physical mass $t<0$
  and a non-integer $J$ slightly above 1: the trajectory $\alpha(t)$ has of course physical states as well.
  It enters the elastic cross section in a form 
\be 
{d\sigma \over dt} \approx \left( \frac{s}{s_0}\right)^{\alpha(t)-1 } \approx e^{{\rm ln}(s)[(\alpha(0)-1)+\alpha ' (0)t]} 
\ee

  \begin{figure}[h!]
  \begin{center}
  \includegraphics[width=6cm]{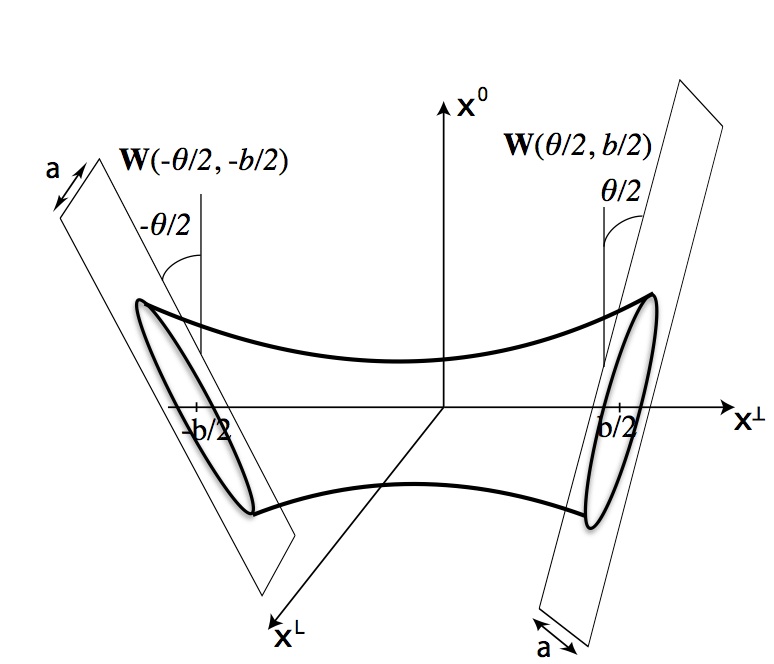}
    \includegraphics[width=5cm]{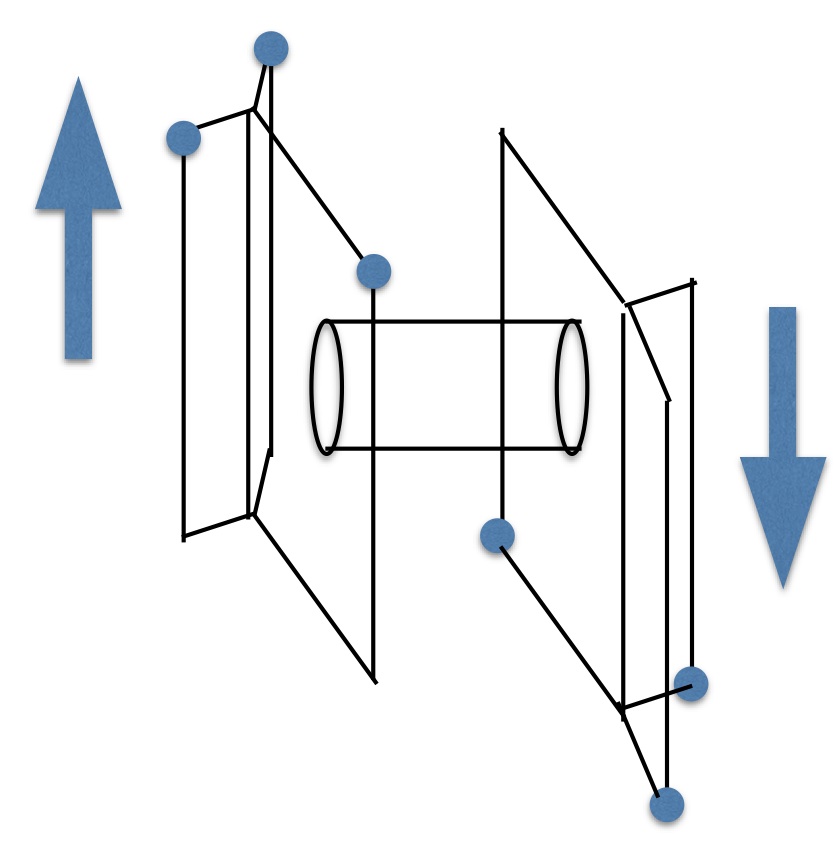}
  \caption{(Color online)
(a)  Dipole-dipole scattering configuration in Euclidean space. The dipoles have size $a$ and are $b$ apart. The dipoles are tilted by $\pm \theta/2$ (Euclidean rapidity) in the longitudinal $x_0 x_L$ plane.
 (b) A sketch illustrating Pomeron exchange for baryon-baryon scattering: only one pair of quarks become ``wounded quarks".  }
  \label{fig_tube2}
  \end{center}
\end{figure}

Dualism of the Pomeron is seen from the very different roles which two main parameter play, the intercept $ \alpha(0)$
and the slope $\alpha'(t=0)$. The former is a dimensionless index, describing the power with which the
total cross section rises.
Perturbative description of the Pomeron is  well developed: the so called BFKL \cite{BFKL}
 re-summation of the gluon ladders provides perturbative $O(\alpha_s)$ value for it. 
  The Pomeron slope $\alpha'(t=0)$ has dimension $[mass^{-2}]$ and  is nonperturbative
  for sure,  related to the string tension.
  (By the way, the  ``string scale" in the fundamental
    string theory  still is traditionally called $\alpha'$, the only piece of QCD phenomenology left in it, for historic reasons.)  
  The meson/baryon Reggion slope is related to a single open string and their
  $\alpha'=1/(2\pi \sigma)$, glueballs such as Pomeron are closed strings should naively have
  the double tension or half the slope: phenomenologically it is closer to 1/3. It
  perhaps indicates changes due to {\em two string interaction}.
   
 Another way to see dual nature of the   Pomeron phenomenology is when one considers
 the so called collision profile as a function of the impact parameter $b$.  At small $b$
 people think of color dipoles and gluon exchanges. 
    At large impact parameter $b=1-2\, \fm$ is naturally given in terms of a (double) string exchange  (see below).
  
  Holographic models of the Pomeron aim at describing both regimes inside the same model, in the spirit of
 AdS/QCD.  As the main representative of the related work let me mention two works of Zahed and collaborators 
  \cite{Stoffers:2012zw} and   \cite{Basar:2012jb}. Strings live in the 5-dimensional space, which is weakly coupled in UV and strongly coupled in IR. Furthermore,  a semiclassical (instanton-like) derivation of the Pomeron amplitude
  is given  in terms of closed string production -- similar to Schwinger pair production in an electric field. 
  The string worldvolume has the shape depicted in Fig.\ref{fig_tube2}(a): it is a ``tube" connecting two flat strips,
the world volume of propagating color dipoles.   In Fig.\ref{fig_tube2}(b) we sketch a Pomeron in a collision
  of two nucleons, now consisting to 3 quarks and 3 string, joined by a string junction: in this case 
  a Pomeron tube ``punctures" one of the three surfaces. This produces one ``wounded quark",
  as we discussed in connection to Tannenbaum's description of fluctuations. 
    
   Direct semiclassical derivation of the scattering amplitude was obtained in \cite{Basar:2012jb}
  using  minimization of the Nambu-Goto action
  (the tube's area). Fig.(a) indicate Euclidean setting in which difference in rapidity is represented by 
  twisted angle $\theta$ between the direction of two dipoles. Fig.(b) illustrates a baryon-baryon scattering,
  in which the Pomeron tube can be connected to any of the available dipoles, explaining the concept
  of the ``wounded quarks" we mentioned in section \ref{seq_centrality}. More than one Pomeron means more tubes,
  connecting perhaps other quarks.
   
  The classical action of this configuration provides the $\alpha'$ term, while the
  intercept deviation from 1 is due to
  the next order (one-loop) corrections due to string vibrations. 
  It gives the amplitude squared: if cut in half by the unitarity cut the ``tube" provides two strings of certain shape,
  which should be taken as the shape which ``jumps from under the barrier" and should be used as the initial conditions to real-time evolution.
  
  Not going into details, we just note that the amplitude has alternative derivation from
  string diffusion equation, and that 5-d space is actually important numerically.
  Scattering pp data as well as deep-inelastic $ep$ (DESY data)
  are well reproduced by this model, see  \cite{Basar:2012jb}.

  It was further argued by Zahed and myself \cite{Shuryak:2013sra} that 
 because the ``tube" has a periodic variable, resembling the Matsubara time, its fluctuations take the thermal form.
 Appearance of an effective temperature and entropy was new to the Pomeron problem:
 but once recognized the analogy to thermal strings can be exploited. We in particularly argued that   
  between two known regimes of the Pomeron -- (cold) string exchange at large impact parameter $b$ and perturbative 
  gluon exchange at small $b$ --
  there should be a third
   distinct regime in which a string is highly excited due to the Hagedorn phenomenon. This is what we know to happen
    at finite temperature gluodynamics,  which possesses  a``mixed" phase between the confined (hadronic) and the deconfined (partonic) phases. As it is known from decades of theoretical and numerical (lattice) research,
   this is  described most naturally by the {\em near-critical strings}, namely the strings in the Hagedorn regime.

\begin{figure}[t]
  \begin{center}
 \includegraphics[width=6.5cm]{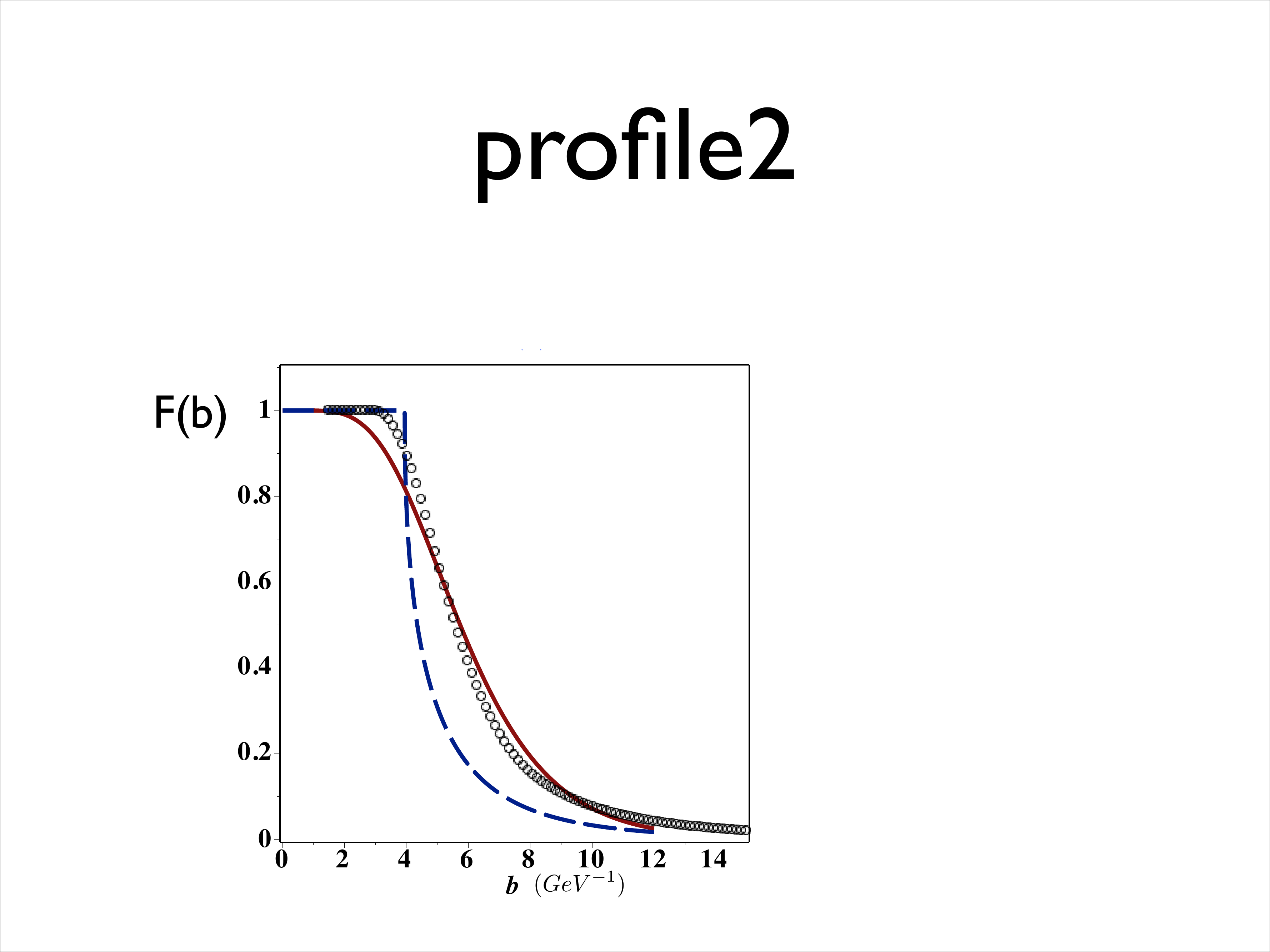} \\
    \includegraphics[width=6cm]{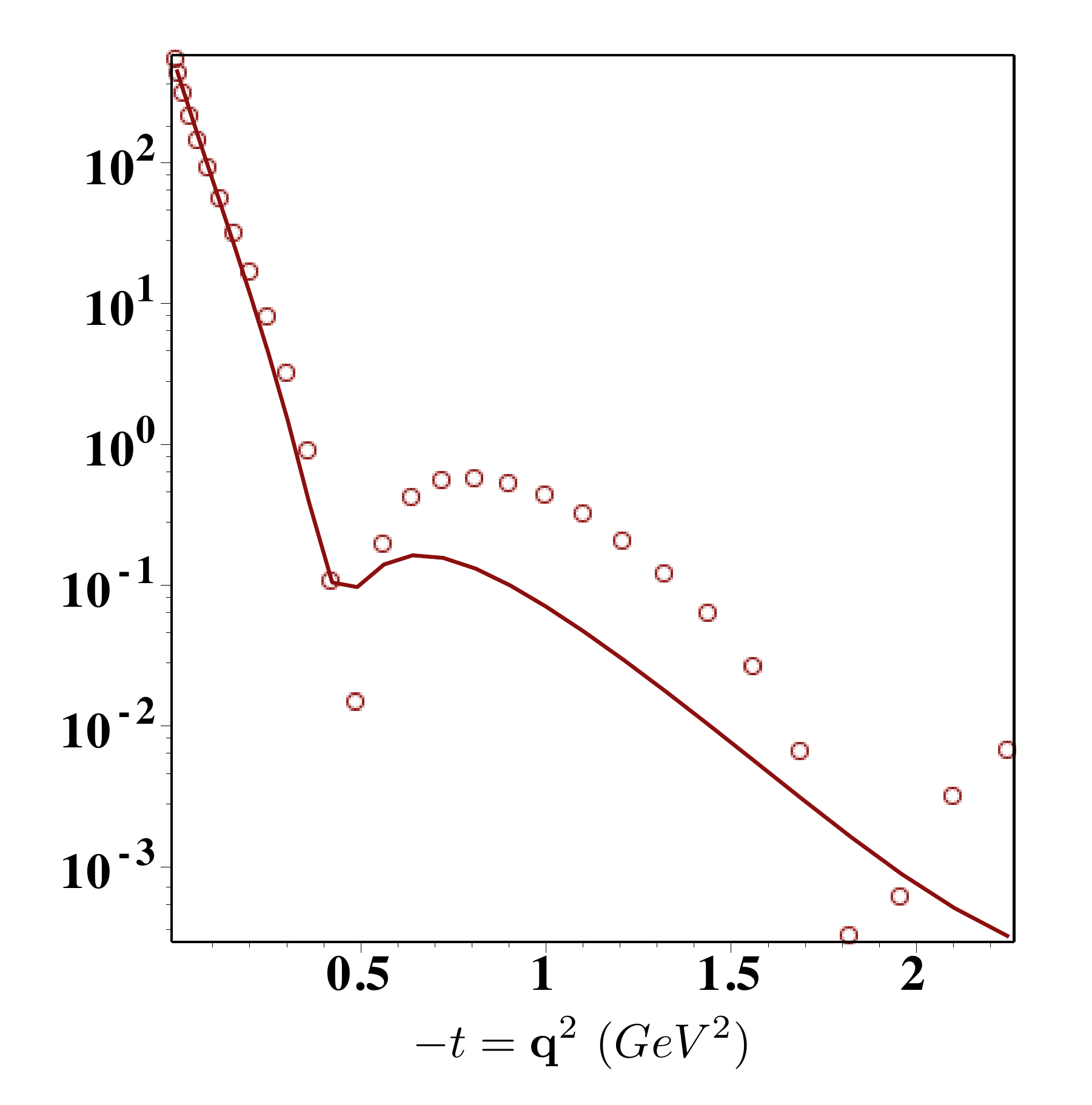}
  \caption{
  (a)
  The solid line is the empirical  LHC data parametrization.
  The dashed line is the shape corresponding to the ``excited string" approximation  for fixed
  sizes of the dipoles, while the circles correspond to the profile averaged over the
  fluctuating dipoles. (b) The corresponding elastic amplitude (the absolute value squared of the profile Bessel transform)
  as a function of the momentum transfer. Model prediction agrees with parameterization well at small $t$,
  up to the dip.
  }
  \label{fig_profile2}
  \end{center}
\end{figure}

In Fig.\ref{fig_profile2}(a) one finds the so called profile of the elastic scattering, the Bessel transform of the imaginary part of the elastic amplitude. Basic prediction of the model is shown by the dashed line, which has 
the ``first order transition". By circles we show more realistic prediction of the model coming from averaging
of that expression over the dipole sizes: it should be compared to empirical fit (solid line) to the LHC elastic scattering data. The difference is relatively small, although it becomes noticeable in scattering with  momentum transfer  
 $Q$ beyond the minimum, see Fig.\ref{fig_profile2}(b). Apparently the predicted ``transition edge" in $b$ is
 a bit  sharper than in the experimental data. 
 
Concluding this section: decades-old Pomeron amplitude can be derived in a ``stringy" manner, in the AdS/QCD setting.
Surprisingly, this approach relates the Pomeron to the  thermal theory of excited strings, with an  effective temperature
 proportional to the inverse impact parameter.  Rapid ``phase transition" in $b$, from black to gray, 
corresponds to the deconfinement transition, with an intermediate regime dominated by  highly excited strings (``string-balls").  
 So far  only for the tunneling (Euclidean) stage of the system is considered, predicting
 elastic amplitude.
The remaining challenge is to describe the subsequent evolution of the system and to extend
 this theory to $inelastic$ collisions.

\subsection{Collisions at ultrahigh  energies}

Discussions about Pomeron regimes ultimately drive us to the old question: what happens at the
ultrahigh energies, well above the LHC reach?

The highest observed energies, by 
Pierre Auger Observatory and similar cosmic ray detectors,  reach 
\begin{align}
E_{\mathrm{lab}} \lesssim  E_{\mathrm{max}} \sim 10^{20} \mathrm{eV} \,.
\end{align}
limited by the so-called Greisen-Zatsepin-Kuzmin (GZK) bound
at which interaction with cosmic microwave photons become important.  

 For future comparison with the LHC observation it is convenient to
 convert the laboratory energy into the energy in the center of mass frame and use a standard Mandelstam invariant, assuming it is a $pp$ collision,
 \begin{align} \sqrt{s_{\mathrm{max}}}=(2E_{\mathrm{max}}m_p)^{1/2}\approx 450 \, \mathrm{TeV}\,. \end{align}
While significantly higher than current LHC $pp$ energy $\sqrt{s_{LHC}}=8$ TeV,   the jump to it from LHC is
 comparable to that from Tevatron $\sqrt{s}=1$ TeV or  RHIC $\sqrt{s_{RHIC}}=0.5$ TeV.  In view of smooth small-power
 $s$-dependence of many observables, the extrapolation to LHC worked relatively well, and further extrapolation
  may seem to be a rather straightforward task.
 And yet, smooth extrapolations using standard event generators plus, of course, the cascade codes do not reproduce
 correctly the experimental data of the Pierre Auger collaboration (e.g., the so called muon size of the showers).

This makes the issue subject to speculations. As an example of an exotic model let me mention the paper 
\cite{Farrar:2013sfa} which suggest that somehow freezeout is entirely different and 
  the pion production becomes suppressed. According to their simulations, a model in which mostly nucleons are produced explains the Pierre Auger data better.

  \begin{figure}[t!]
\begin{center}
\subfigure[\label{curves2:a}]{\includegraphics[width=6.3cm]{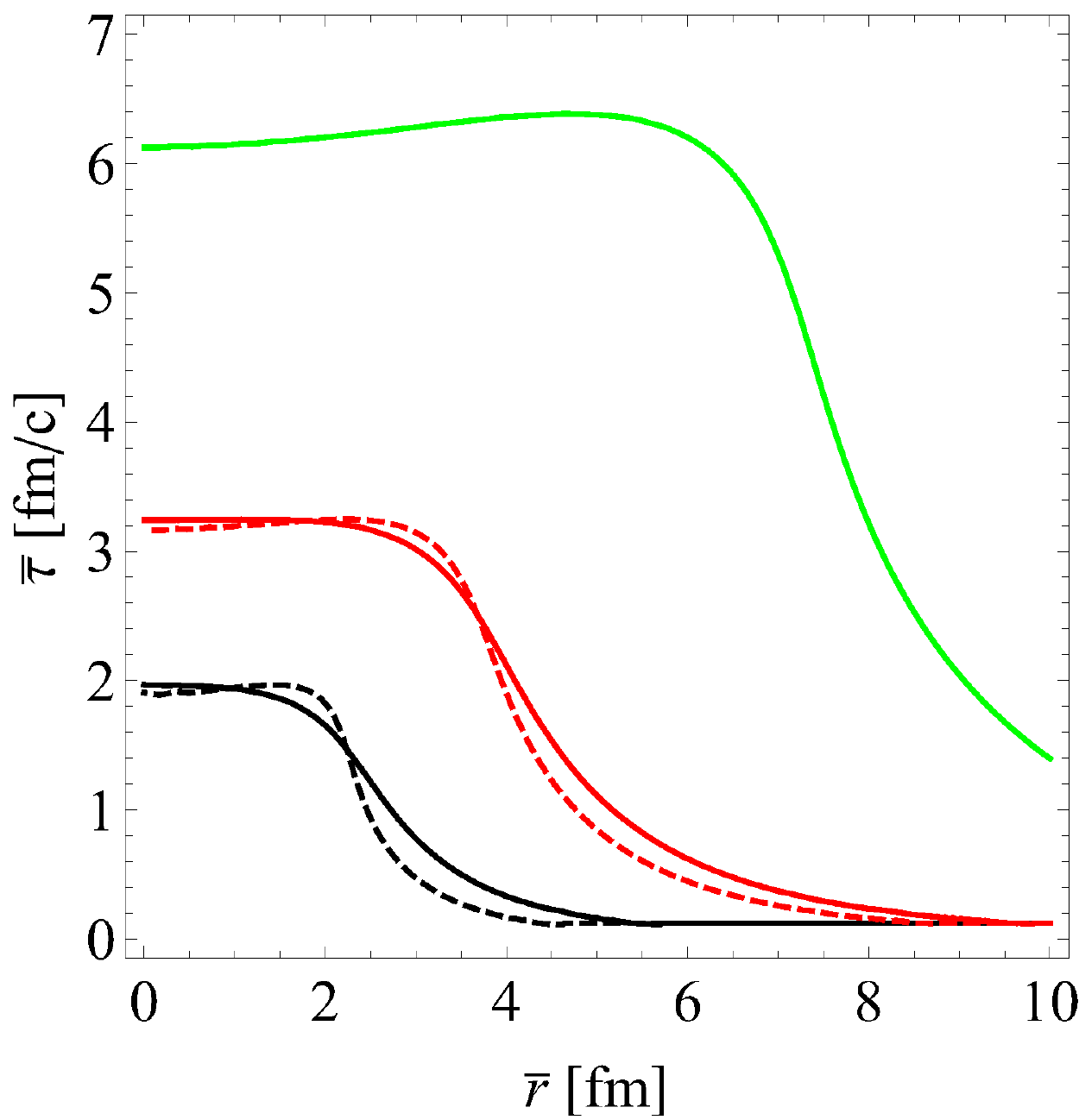}}\\
\subfigure[\label{curves2:b}]{\includegraphics[width=6.5cm]{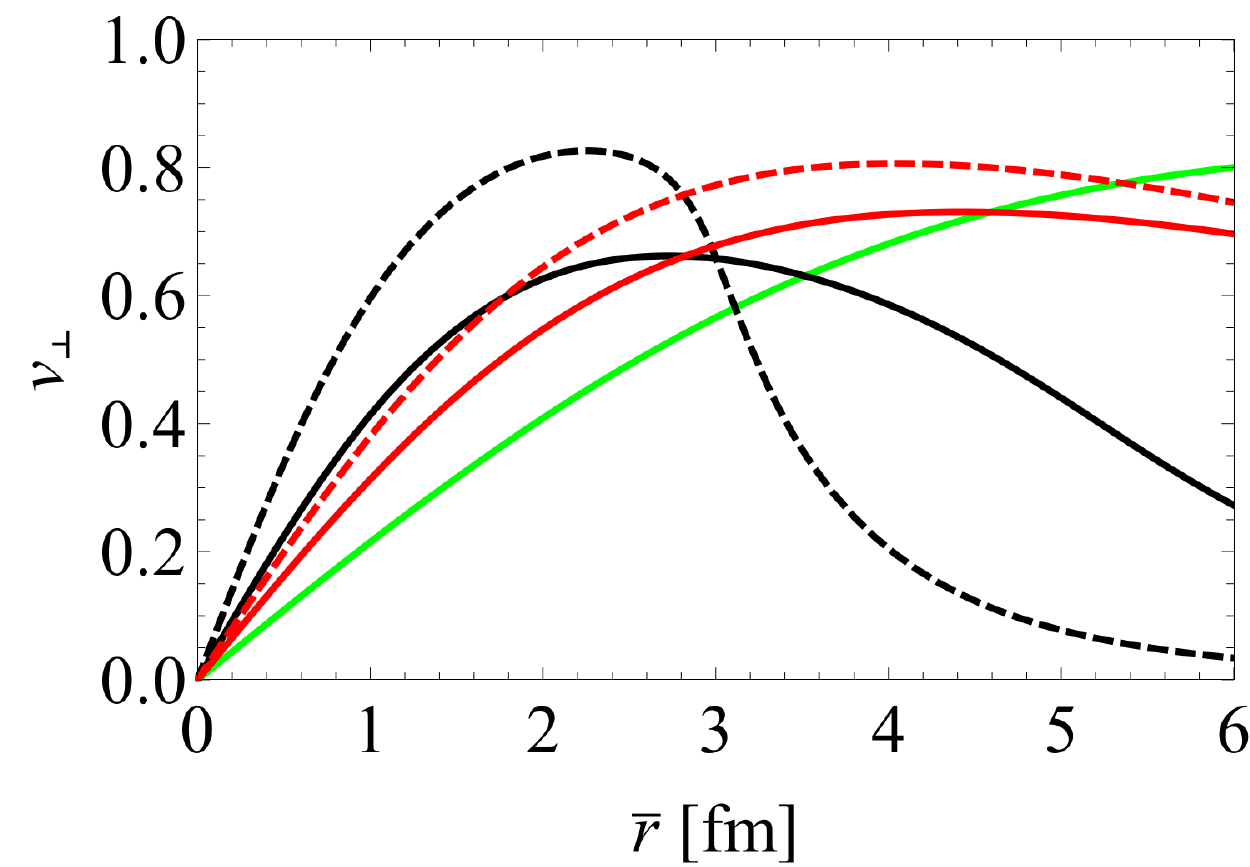}}
\caption{(color online) (a) The freezeout surfaces in the $(\bar\tau, \bar r)$ plane
(no rescaling) and (d) the distribution of the transverse flow velocity on those surfaces.
 In both plots the green solid curve at the top is our ``benchmark", the central $PbPb$ collisions at LHC. Black solid line is for light-light nuclei
collisions, black dashed (coincident with green by chance) are light-light collisions with the size compression.  Similarly, red solid and red dashed are heavy-light collisions
without and with the size compression, respectively. }
\label{fig_freezeouts_KS}
\end{center}
\end{figure}

More modest (but still significant) change
 between LHC and ultrahigh energies, has been proposed by Kalaydzhyan and myself
 \cite{Kalaydzhyan:2014uha}. Our main statement is that at the ultra-high energies  $\sqrt{s_{\mathrm{max}}}$ observed in cosmic rays, the
   ``explosive regime"  even in $pA$ collision is expected to change from a very improbable $P\sim 10^{-6}$ fluctuation to
   the mean behavior, with $P=\mathcal{O}(1)$.  The reason for it is
   simply an increase in mean particle (entropy) density with energy $\sqrt{s}$. Extrapolations suggest an increase of 
   $dN/dy$ by about factor 3. In Fig.\ref{fig_freezeouts_KS} we show the free out surfaces and hydro velocities
at them, corresponding to our calculations. 
By ``compression" we mean the phenomenon of ``spaghetti collapse" discussed above.
The mean transverse momenta of secondaries is shown in the Table.

Another generic reason for a change is that both primary collisions and subsequent cascade of ultra high energy cosmic rays all happen in the Earth atmosphere,
 so the targets are not protons but light ($N$ or $O$) nuclei. Furthermore, the projectiles themselves are also most likely
 to be not protons but some nuclei.    It is
  either also some light nuclei or some mixture including heavier ones, believed to be up to $Fe$.

 Taking into account large $pp$ cross section at ultra high energies, $\sim 150\, \mathrm{mb}$, one finds that its typical impact parameters
 $b\approx 2 \, \fm$. Thus the range of the interaction in the transverse plane is comparable to the radius of the light nuclei (oxygen $R_O\approx 3\, \fm$) and therefore even in the $pO$ collisions
most of its 16 nucleons would become collision ``participants". For light-light AA collisions like $OO$ the number of participants changes from 32
 (central) to zero. Accidentally, the average number of participants is comparable to the
average  number of participant nucleons $\langle N_p \rangle\approx 16$
  in central $pPb$ collisions at the LHC.
  
  \begin{table}[t!]
\centering
\begin{tabular}{| c || c | c | c | c | c |}
\hline
particles &  \,\,\,FeO\,\,\, & \,FeO comp.\, & \,\,\,pO\,\,\, & \,pO comp.\, & \,\,PbPb\,\,\\ \hline \hline
$\pi^\pm$   	&  0.56    &  0.69    &  0.53    &  0.76    &  0.73        \\\hline
$K^\pm$ 	    &  0.71    &  0.88    &  0.66    &  0.96    &  0.92        \\\hline
$p,\, \bar p$ 	&  0.90    &  1.09    &  0.83    &  1.17    &  1.13         \\ \hline
\end{tabular}
\caption{Mean $p_T$ [GeV/c] for pions, kaons and protons obtained from the particle spectra.
By ``comp" we mean compressed initial state, as explained in the text.
}\label{pttable}
\end{table}

  The question of principle still remains: and what should happen when the collision energy goes to infinity?
Qian and Zahed \cite{Qian:2014rda} had proposed an argument which would stop 
multiplicity growth. The argument is that when the produced string becomes so long and heavy
that it basically collapses due to its self-interaction into a black hole, the entropy
produced would be limited by the Bekenstein bound. In their opinion, not only the parton density growth,
but also growth of the cross section, should stop at certain collision energy.

\section{Electromagnetic probes} \label{sec_em}

\subsection{Brief summary }

  Let us on the onset remind standard  terminology to be used below. The sources of the dileptons
 are split  into the following categories: \\(i)  instantaneous $\bar q q$ annihilation, known as the Drell-Yan partonic process;
 \\   (ii) the pre-equilibrium stage, after the nuclei pass each other; \\
 (iii) the sQGP  stage, in which matter is usually assumed to be equilibrated.\\
 (iv) 
 hadronic stage in which rates are calculated via certain hadronic models;\\
 (v)  ``cocktail" contribution of leptonic decays of secondaries occurring $after$  freezeout.
   
   The corresponding windows in the dilepton mass  are: $M>4 GeV$ for (i), $1< M< 3 \, GeV$ for (iii) and $M<1\, GeV$ for 
(iv). So the early stage dileptons mostly fall into the 3-4 GeV window. While it contains also $J/\psi,\psi',\psi" $ states,
one should rely on their subtraction, which depends on mass resolution of the detector. This coincidence
adds difficulty to the measurements.

Status of the
dilepton/photon measurements, in brief, is as follows.

At SPS: NA60 has been very successful, in particularly:
(a) Large enhancement at small masses $M<m_\rho$ and deformation of the spectral density
 of the electromagnetic current is documented; \\
 (b) using $p_t$ slope as a function of
$M$ one sees that   $M<1\, GeV$ are produced when flow is developed, \\
(c) while intermediate mass dileptons
(IMD) $1\, GeV < M < 3\, GeV$ are produced early when it is still absent; \\
(d) the IMD come from 
QGP thermal radiation rather than charm decays. 

At RHIC: (a) Low mass dileptons are enhanced, the magnitude still somewhat disputed between STAR and PHENIX in the most central bin. 
\\(b) IMD are well measured but the contribution from charm/bottom decays remains unsubtracted. PHENIX is solving it with new
vertex detector, STAR with $e-\mu$ correlations due to new muon system now in place; \\
(c) 
direct photons have spectra consistent with standard rates and hydrodynamics in shape, but not in absolute magnitude.\\
(d) 
unexpectedly large elliptic flow $v_2$ of direct photons persists. 

LHC only starts to address these issues. Let me only mention one thing: ALICE confirmed  large value of the direct photon $v_2$ found originally by PHENIX, and puzzling theorists since. 

\begin{figure}[t]
  \begin{center}
  \includegraphics[width=9.cm]{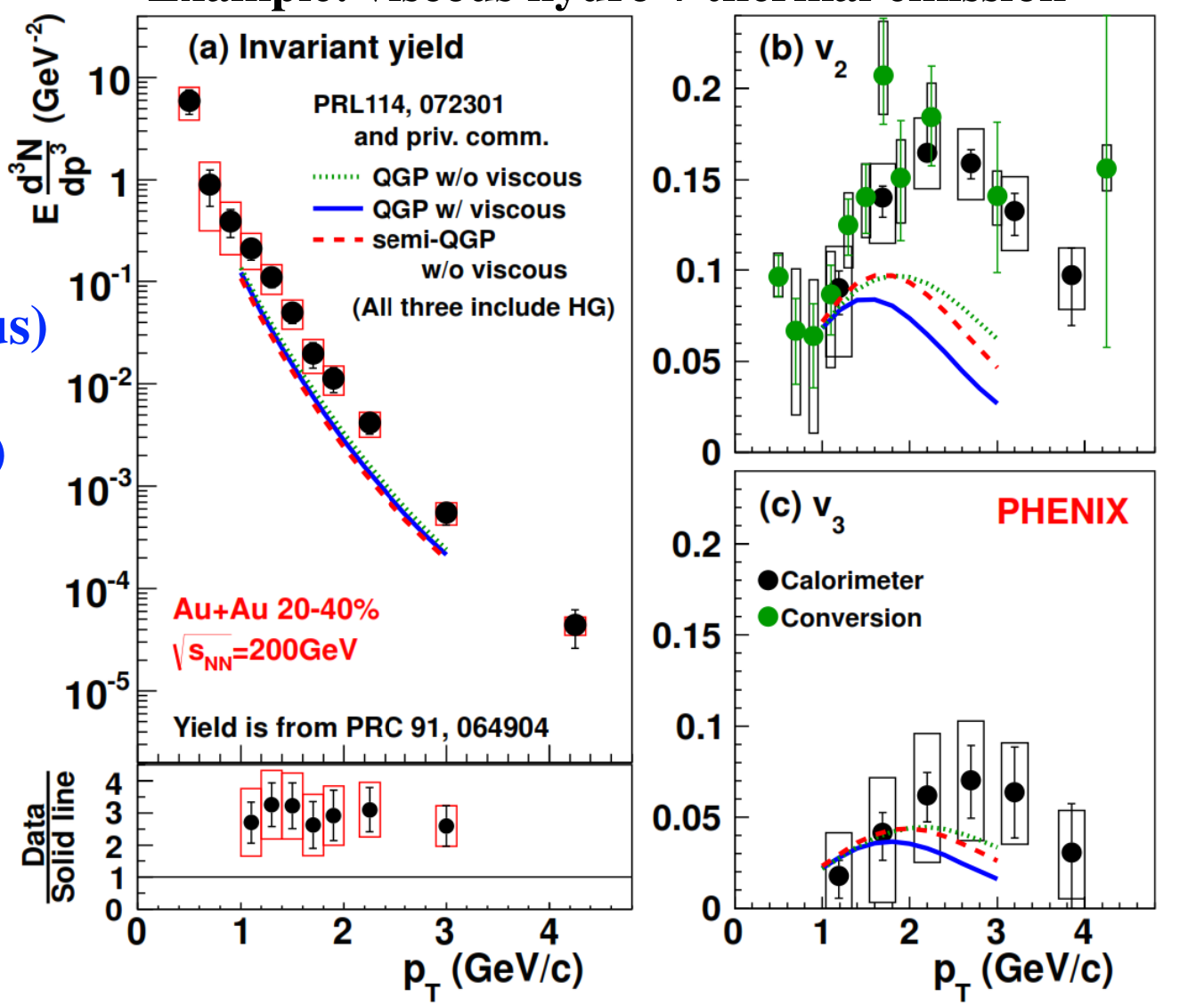}
   \caption{Illustrations to ``direct photon puzzle" (from A.Drees, PHENIX presentation at QM2015).}
   The yield (a), $v_2$ (b) and $v_3$ of the direct photons. Points are data and curves are from 
   theory based on hydrodynamical model, the reference indicated on the figure.
  \label{fig_direct_photon}
  \end{center}
\end{figure}

Both (c,d) together are known as a  ``direct photon puzzle",
illustrated by the latest data in Fig. \ref{fig_direct_photon}. 
My comment is that this field still uses a paradigm set up from its very beginning \cite{Shuryak:1980tp},
using basically perturbative rates or those evaluated from hadronic models. Now we know that
such approach fails elsewhere, in viscosity and jet quenching, and it is perhaps time to
approach the issue in more flexible 
phenomenological way. If  the photon production rates
at the late (near $T_c$) stage of the collisions is increased, both (c) and (d) problems will disappear.
Like for jet quenching parameter $\hat q$ I would argue that may be due to 
additional scattering on monopoles, not present yet in any models.

\subsection{New sources of photons/dileptons: multi-gluon or phonon+magnetic field} \label{sec_MSL}

Theoretically,  the production of photons and dileptons is tied to the presence of quarks, and is thus sensitive to the
issue of {\em quark chemical equilibration} and its timing.
 The initial stages of the  high energy collisions are believed to be dominated by gluons. 
 Old perturbative arguments \cite{Shuryak:1992bt} show that chemical equilibration via quark-antiquark pair production
is relatively slow and should be delayed relative to thermal equilibration of the glue. 
This idea led to a ``hot glue" scenario in which the quark/antiquark density at early stage  is suppressed by powers of quark fugacity $\xi_q<1$. 
  The basic process of the dilepton production
 \be q+\bar{q} \rightarrow \gamma^* \rightarrow l_++l_- \ee
  is expected to be suppressed quadratically, $\sim \xi_q^2$.

This scenario has been challenged: higher order processes with {\em virtual quark loops} can produce electromagnetic effects as well, even without
on-shell quarks. First, 
contrary to general expectations, the quark loop effect in GLASMA
has been suggested to be significant \cite{Chiu:2012ij}, enhanced  due to multigluon -to virtual quark loop -to dilepton
processes like
\be ggg \rightarrow ({\rm quark \, \, loop}) \rightarrow \gamma^* \rightarrow l_++l_-. \ee
To see how much correction to production rates these processes 
contribute one  needs more explicit calculation of the absolute rates
of such processes, which are still lacking.

An explicit calculation \cite{Basar:2014swa} of the rate of such type has been made, the
two gluon to two photon rate $gg\rightarrow \gamma\gamma$,
in which one of the photons is the ambient magnetic field and  the gluons are  combined
into a colorless   matter stress tensor, and one of the photons is rotated into the initial state 

\be T^\mu_\mu + \vec{B}  \rightarrow  ({\rm quark \, \, loop})   \rightarrow \gamma^* (=dileptons) \ee

 The terminology introduced in this paper is as follows:
The process in which glue appears as 
``average'' matter stress tensor $<T_{\mu\nu}>$, producing photons ( real or virtual) due to time-dependent
magnetic field, is called   {\em Magneto-Thermo-Luminescence}, MTL for short. 
By ``average" we mean that the value of the stress tensor is averaged over the fireball and is nearly constant, with negligible
momentum harmonics $p\sim 1/R$.

Individual events, however, are known to also possess fluctuations  of the  matter stress tensor $\delta T_{\mu\nu}$,
with complicated spatial distribution and thus non-negligible momenta. Although these fluctuations include both longitudinal and transverse 
modes, in a somewhat a loose way we will refer to all of them as ``sounds".   
We will thus call the interaction of the ambient electromagnetic field and the fluctuations  of the  matter stress tensor that produces photons and dileptons 
{\em Magneto-Sono-Luminescence}, MSL. 

The paper is rather technical, and it is probably not very useful to present here specific lengthy formulae,
so let us only make some general comments on its potential importance. If observed, 
the 
MSL process tests both the amplitudes of the short-wavelength sounds and also harmonics of the magnetic field.
We already discuss many uses of sounds above: let us here only comment on the magnetic field. 
It is easy to evaluate its early values, resulting from Maxwell equations and the currents due to the spectators
in peripheral collisions. Since sQGP is believed to be a good conductor, these fields are expected to
create currents capturing a fraction of the field inside the plasma \cite{Tuchin:2013apa}, perhaps lasting for many fm/c. 
Magnetic field evolution is important to know for other applications as well, e.g. chiral magnetic effects and the like.
  
So far, luminosity and acceptance limitations had lead experiments to focus on most 
luminous central collisions. Now it is perhaps time to look at dileptons in semi-peripheral 
collisions as well. 
 
\subsection{Dilepton polarization and the (early time) pressure anisotropy}
\label{dilepton_polarization}

It is well known that when spin-1/2 particles (such as  quarks) annihilate and produce
lepton pairs, the cross section is not isotropic but has the following form
\be  {d\sigma \over d\Omega_k}\sim (1+a \cdot cos^2\theta_k) \ee
where the subscript correspond to a momentum $k$ of,say, the positively charged lepton
and $\theta_k$ is its direction relative to the beam. 
The anisotropy parameter $a$ in the  Drell-Yan region --  stage (i) in the terminology introduced at the beginning of this section-- is produced by 
 annihilation of the quark and antiquark partons, are collinear to the beams, and therefore $a=1$. 

In my  note \cite{Shuryak:2012nf} it is suggested that 
parameter $a$ can be used to control anisotropy of the
 early stage of the collision. In particular, 
if it is anisotropic so that longitudinal pressure is small relative to transverse, $p_l<p_t$, the annihilating quarks should mostly move $transversely$
to the beam, which leads to $negative$ $a<0$. 
(Such regime is expected because such anisotropic  parton distribution with small differences
in longitudinal momenta is produced by a ``self-sorting" process, in which partons with different rapidities get
spatially separated after the collision.)

For illustration I used  a  simple one-parameter angular distribution of quarks over their momenta $p$ in a form
\be W\sim exp[-\alpha cos^2\theta_p]    \label{W} \ee
 and calculate $a(\alpha)$ resulting from it. 
It does show that $a$ may reach negative values as low as -0.2 at stage (ii), before it vanishes,  when equilibration is over, at stages (ii-iv). 

\section{Charm and charmonia} \label{sec_charm}
\subsection{Charmonium theory: an overview}

Quarkonia  have always been among the first classic QGP signature, although
rather confusing one. The
views on what QGP actually does for charmonia yield had changed in a complicated path, which proceeded via a couple of turns of the famous logical spiral. Thus reviews which  follow the historical development 
usually are rather confusing. 

Instead, ignoring the historical order, we will  proceed pedagogically, from simpler to more complicated 
settings. We will proceed though three fundamentally different settings:\\ (i) time independent
 equilibrium state of charmonia; \\
(ii) equilibration processes and rates; \\
(iii) heavy ion collisions.

\begin{figure}[t!]
  \begin{center}
  \includegraphics[width=6cm]{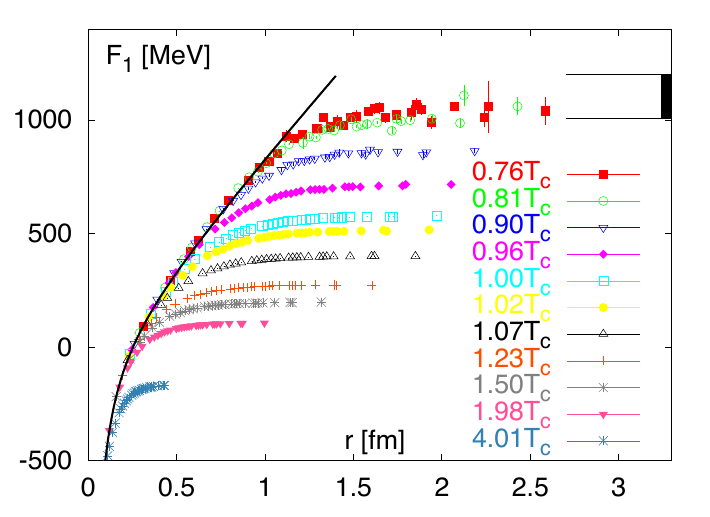}\\
   \includegraphics[width=6cm]{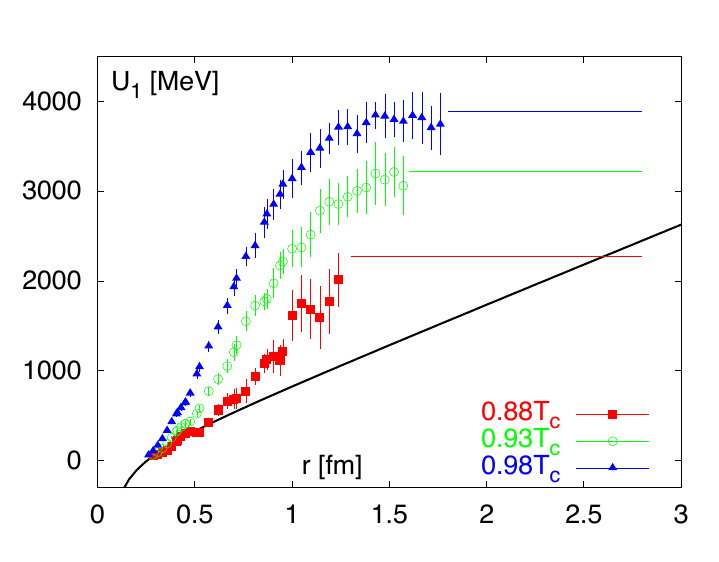}\\
    \includegraphics[width=6cm]{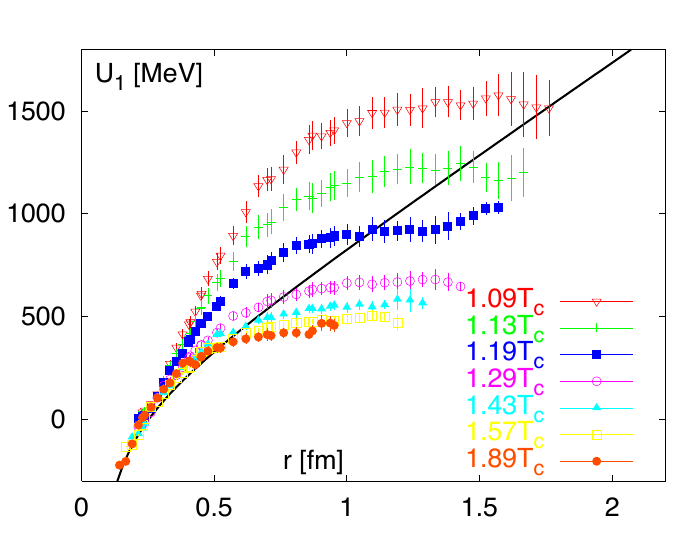}
   \caption{ from \cite{Kaczmarek:2005zp}. Free energy singlet potentials  $F_1(T,r)$ (top plot)
   and the potential energy $U_1(T,r)$ below and above $T_c$.
 }
  \label{fig_pots}
  \end{center}
\end{figure}

But before we discuss those, let us start with even simpler forth setting, (0) the case of static heavy quark. 
 The so called {\em static  potentials}  which has been measured
 on the lattice, at zero and nonzero temperatures
 . A sample of 2-flavor QCD results from Bielefeld group  \cite{Kaczmarek:2005zp}
 are shown in Fig.\ref{fig_pots}. The linear potential at zero $T$ is indicated at all plots, for comparison. 
 Static potential gives the free energy, related to entropy and internal energy $V(T,r)$ by
 \be V(T,r)=F(T,r)-T{\partial F \over \partial T}=  F(T,r)+T S(T,r) \ee
 So, determining the entropy term one can subtract $TS(T,r)$ term and
 get the {\em internal energy}. As one can see from Fig.\ref{fig_pots}. they are rather different. The force -- potential gradient -
 has a maximum just below $T_c$, and 
 $U$ itself reaches 4 GeV in magnitude. What are their physical meaning and which one, if any, should one use
 as the potential in the charmonium problem? 
 
 Liao and myself \cite{Liao:2008vj}
had proposed a model  describing these lattice findings, based on monopole-generated flux tubes. 
 Its detailed discussion is out of context here: let me just  mention how
 an entropy associated with static quarks can be understood.  The key idea is the Òlevel crossingÓ phenomena, occurring while the separation between charges is changed. Suppose a pair of static charges (held by external ÒhandsÓ) are 
 slowly moved apart in thermal medium at certain speed $v = \dot{L}$. For each fixed $L$, there are multiple configurations of the medium populated thermally. When L is changed, the energies of these configurations are 
 crossing each other, and at each level crossing there is certain probability to change population of the states, depending on the speed of separation $v = dL/dt$. If the motion is adiabatically slow, then all the level crossing processes happen with probability 1: in thermodynamical context this leads to maintained equilibrium and maximal entropy/heat generation. If however the pair is separated  fast, then the level crossing happens probabilistically, and certain entropy is developed.
The medium is no longer in equilibrium with the pair. The amount of entropy generated is less than in the adiabatic case. In the extreme case one may expect that the pair, if moving on a time scale much much shorter than the medium relaxation time scale, decouples from the media and produce negligible entropy. It is plausible, therefore, to identify the adiabatic limit as probing the free energy $F (T , L)$ measured on the lattice with the presence of static $Q\bar{Q}$ pair. The Òinternal energyÓ $V (T, L)$, on the other hand, is different from $F (T , L)$ by subtracting the entropy term and thus can be probed in the extremely fast limit in which possible transitions among multiple states via level crossing do not occur and no entropy is generated.
We emphasize that such phenomenon in thermal medium is a direct analogue of what exists in pure quantum mechanical context. Perhaps the oldest example is the so called Landau-Zener phenomenon of electron dynamics during the vibrational motion of two nuclei in a diatomic molecule. Specific electron quantum states are defined at fixed $L$ (the separation between the two nuclei) with energies $E_n(L)$, and certain levels cross each other as the value of $L$ changes. The issue is the probability of the transition during such crossing of two levels. Consider two levels with their energies given approximately by $E_1(L) = \sigma_1L+C_1$ and $E_2(L) =\sigma_2 L+C_2$ near the crossing point. When the two nuclei approach the crossing point adiabatically slowly $v = dL/dt\rightarrow 0$, the electrons always change from the lowest state to the lowest other. But if $dL/dt$ is finite, then the transition to $both$ levels at crossing point may happen, with certain probability. This is how a pure state becomes a mixture, described by the density matrix,
and entropy is produced.  

  (More quantitatively, Landau and Zener solved the problem and showed that the probability to remain in the original state (i.e. no transition) is exponentially small at small velocity $v$
\be P= exp[-{2\pi H_{12} \over  v|\sigma_1-\sigma_2|}]
\ee
where $H_{12}$ is the off-diagonal matrix element of a two- level model Hamiltonian describing the transition be- tween the two levels. In the opposite limit of rapid crossing, the system remains in the original state, and no entropy is produced again. So it has maximum at some speed. )

For more discussion of the role of the ``entropic force" in charm motion, see paper by Kharzeev \cite{Kharzeev:2014pha}.

Now we return from static quarks to physical charmonium, and discuss cases (i-iii) subsequently.

(i) Suppose one put a $J/\psi$ in matter at some temperature $T$. However small is its value, transitions from
  $J/\psi$  to its excited states will happen. Eventually those will come to $\bar{D}D$ threshold and propagate further.
  Since any subsystem tends to equilibrium, at given $T$, and $\bar{D}D$ can eventually occupy an infinite volume, they will win over the bound states. So,  given enough time, a given initial $J/\psi$ will  dissolve always,
  at $any$ $T$.  
   
  On the other hand,  random thermal transitions should also proceed in the opposite direction
  as well:  $\bar{D}D$  can produce charmonia and those can get de-excited back to $J/\psi$. Thus
  $recombination$ of charmonia states should also happen. Given infinite time, and equilibrium
  density of $J/\psi$ will be reached.
  
  Pure thermal occupation rate for $J/\psi$  $\sim exp(-M_{J/\psi}/T)$ is tiny at all temperatures of interest. But
   heavy quarks $c,b$ are produced in hard processes, not in thermal reactions. 
   Since heavy ion collision time is small compared to weak decays of those,
   their number is conserved. This leads  to the concept of charm chemical potential $\mu_c$.
It is not coupled to charm quantum number, is  the same for $c$ and $\bar c$, and thus the overall charm neutrality is still intact. Now one may ask what is the density of certain quarkonia states $i$ $n_i(T,\mu_c)$
in equilibrium with a particular $\mu_c$.
  
  While at low $T$ this is perfectly good question, but it becomes difficult as $T$ increases:
the notion of a state gets ambiguous. Proper field theory object to study is a correlation
function of local gauge invariant operators, such as charmed scalar/vector currents
\be K(x,y)=<\bar{c} \Gamma c(x) \bar{c} \Gamma c(y)> \ee
where the average is over the heat bath and $x$ or time $x^0=t$ can be Minkowskian or Euclidean.
In both cases it is related with the same {\em spectral density} $\tilde K(\omega,k)$ characterizing
all physical excited states with given energy and momentum. At low $T$ 
  one can see
individual states as certain peaks at the lines $\omega=\omega_i(k)$ in the spectral density,
but with increasing $T$ they merge into a smooth continuum.

Euclidean correlation functions
had been numerically calculated on the lattice: for a recent review 
see \cite{Mocsy:2013syh}. Unfortunately the problem of 
{\em spectral density reconstruction} from those is mathematically
badly defined. In practice, highly accurate (and very expensive)
Euclidean 
correlation functions are
 converted into a relatively poorly defined spectral density.
 Even when the individual states are seen, as some peaks in spectral density, there is hardly
 any accuracy to define their widths. Above certain $T$  
  all peaks corresponding to charmonium states
merge into one ``near-threshold bump",
the imprint of the Sommerfeld-Gamow enhancement due to an attractive potential $\sim e^{-V/T}>1$. .

 (ii) A more detailed -- time dependent -- set of questions can be asked about {\em transition rates}
 in equilibrium matter. Those has been addressed (at least) at three levels,  (a) real time QFT; (b) quantum mechanical; (c) classical diffusion.
 
 Real time QFT, also known as Schwinger-Keldysh formalism, can follow a system from
 some initial to some final state using the full Hamiltonian
 \be < i | Pexp(-\int_i^f dt H) | f > \ee
which is viewed as a sum of that for the subsystem in question $H_0$ and matter perturbation $V$. 
Diagrammatic expansion, including two-time contours as well as Matsubara portion of an Euclidean time 
for thermal media are widely used in conduced matter problems, but they are not much used so far
in the problem we discuss.

If $H_0$ corresponds to non relativistic quantum mechanical description of quarkonium,
we will call it quantum mechanical approach. One can evaluate matrix elements of $V$
over various quarkonia states. 
Already my first 1978 paper on QGP signals \cite{Shuryak:1978ij} had charmonia (called psions in its title).
The first considered in it was  $J/\psi$ {\em excitation} to unbound states of $\bar{c}c$ 
due to photoeffect-like reactions of one gluon absorption $J/\psi+g\rightarrow \bar{c} c$.
For heavy quarkonia the {\em diagonal part} of the real and  imaginary part of the perturbation $V$
can be considered as a modified potential:  
 for recent review see \cite{Brambilla:2013dpa}. 
 More generally, one can define transition rates between states and write a {\em rate equation}. 
 The fundamental  question here is of course whether the ``matter perturbation" $V$
 is small or not. (We will argue below that at very low and very high $T$ perturbative approach may work, but
  at least for charmonium in the near-$T_c$ matter the answer
 to this question is negative).

 Suppose  the ``perturbation" $V$ is not small comparable with the interparticle interaction: 
 then quantum quarkonium states are no longer special and one can as well use for $H_0$ just
 free particles. Using mass as large parameter, one can argue \cite{Svetitsky,Moore:2004tg} that even in strong coupling
 setting the heavy quark motion should be described by classical stochastic equations,
 the Langevin or Fokker-Plank type.   Let me only mention two crucial consequences of the argument.
 First, motion is diffusive, with $x\sim \sqrt{t}$ as it happens in random uncorrelated directions. Second,
 each step in space is very small.
 Suppose a perturbation delivers a kick of the order $T$ to a heavy quark of mass $M\gg T$. Its velocity
 is changed little, by $\Delta v \sim T/M$ and by the time next kick comes $\Delta t\sim 1/T$ the shift in coordinate
 is small $\Delta x \sim 1/M$.

 Suppose a quark needs to diffuse a distance large enough so that the gradient of the potential no longer
 pulls it toward the antiquark. From internal potentials displayed above one can see that the distance it needs
 to go is about 1.5 fm, or $\sim 10 \Delta x$ jumps it can make. However, since it is moving diffusively, 
 to get that far the quark would need $\sim 10^2$ jumps, which can well be larger than time available.
  Quantitative study  of classical diffusion to charmonium made by Young and myself \cite{Young:2008he}
  confirmed that  to climb out of the attractive potential in multiple small steps is
 hard. Contrary to common prejudice, using the realistic charm diffusion constant
 we found that the survival probability of $J/\psi$ is not small but is of the
 order of 1/2 or so. 
  
 (iii) Finally, in order to model the fate of heavy quarks/ quarkonia in heavy ion collisions,
 one need to follow them all the way, from initial hard collisions to the freezeout. 
 In classical diffusion approach one starts with pair distribution in the phase space,
 as defined by the parton model, and at the end project the resulting distribution
 to the charmonia states using their Wigner functions.

%
%
%
%
%

If the reader is insufficiently puzzled by all that, let me finish this section presenting two opposite
answers to the question: What is the effect of the  QGP production on the charmonium survival?
In 1986 Matsui and Satz \cite{Matsui:1986dk} famously argued that QGP, via the Debye screening
of the color potential, kills charmonium states  $sequentially$, from excited states down to the ground one. 
In 2008 Young and myself \cite{Young:2008he} argued that strongly coupled QGP 
helps to preserve charmonium states, since small diffusion constant prevents 
$Q\bar{Q}$ to move away from each other. There are perhaps no identifiable charmonium states
during the process: but so what? Close pairs at the end get projected back to the bound states. 

\subsection{Charmonium composition}

Let me start with the most radial model of all, that of statistical hadronization of charmonium at chemical
freeze out \cite{Andronic:2007ff}. It assumes that
charmonium states, like all light hadrons,  are produced in thermal equilibrium state at this time.  
If true, all questions discussed in the previous subsection are completely irrelevant, since
 all charmonium history during the intermediate stages is simply forgotten. 

The data at RHIC and LHC show that this is only partially true, and so we witness two component
of the charmonium population, the ``survived" one and the ``recombination" contribution. 
Observations of the recombination component 
 is one of the most important recent results from heavy ion experiments. Instead
 of showing the actual data, let me ask the reader to look at the original sources, such as 
  Andronic's summary talk at QM2014, which does it very well.

 Let me proceed to next order questions, related in particular with relative population of charmonium states.
 If all of them come 
 the statistical hadronization at chemical freezeout, the consequence should be
 \be {N_{\psi'} \over N_{J/\psi} }=exp\left(-{M_{\psi{2s}} - M_{J/\psi}\over T_{ch}} \right)\ee  
and similar formulae for other states. 
However if we have two out-of-equilibrium components, with different 
history, the answer should be different. The ``survival" component, with its flow of probability from small to large $r$,
should be richer in lower states. The ``recombination" component flows the opposite way,
and it should have more higher states instead. In general, two components have different
centrality and $p_t$ dependences, and in principle can be separated.

Experimental data are rather incomplete: there is however puzzling double ratios found by CMS
\cite{Khachatryan:2014bva}
which hint that the situation can be rather interesting.

\subsection  
 { 
New types of ``quarkonia states" in a strongly coupled  medium }

 Like in other parts of heavy ion theory, there is a weakly coupled and strongly coupled points of view
 on this question. Which one to chose is now parametric: if the quark mass is very large $M \gg T$,
 quarkonia are nonrelativistic and perturbative. In zeroth approximation one thus starts from the unmodified vacuum states,
 with effects such as excitations and Debye screening included subsequently. Those can be described by corrections to real part of the
 potential $\delta Re V$ and appearance of its imaginary part $Im V$, see e.g. \cite{Brambilla:2013dpa} for nice summary of this theory.

  
 The opposite picture of very strong coupling has been discussed in  \cite{Young:2008he}. 
  If $Im V$ is  large, using the  initial vacuum states make no sense. 
 Indeed, there are no 2-particle bound states in a liquid, just certain correlations between the charges. 
 Classical theory has two input parameters. The first is 
    the diffusion constant $D$ of the charm quark, and the second
is
 the ``interaction strength", usually characterized by
 the ``plasma parameter"
 \be \Gamma={ mean \,potential\,  energy \over T} \ee 
 in which the ambient temperature $T$ stands for typical kinetic energy. Classical strongly coupled systems have $\Gamma>1$.

 
 Let us start with a
   simplified limit
  at infinite coupling and  zero diffusion. Thus $\bar{c}c$ pair, locally produced,  remains stuck at the same place
 during the sQGP period of time.  At the end matter returns to hadronic phase and
     the ground state $J/\psi$ is recreated, simply
because it has the largest wave function at zero. 
 If so, there should be {\em no} $J/\psi$ suppression at all!  sQGP helps preserve the charmonia.

More realistic simulations with finite diffusion constant can be set using Langevin or Fokker-Planck (FP) equations.  
One setting starts with close $\bar{c}c$ pair, which makes $positive$ flow toward large relative distance $r\rightarrow \infty$. 
The other -- recombination -- setting starts with  originally unrelated $\bar{c}c$ at large $r$, with
negative  diffusion current toward small $r$. 
Both are followed for some time of sQGP era, and  
 at the hadronization the obtained distributions in the phase space 
 are projected back to the vacuum quarkonia states, using their Wigner functions.

Which picture is closer to the truth, weak or strong coupling, depends on parameters such as the quark mass: perhaps bottonium
is closer to the former regime while charm perhaps closer to the second one.

  Not going into details here, let me just say that
both the diffusion coefficient for a charm quark in QGP 
and the potential are reasonably well defined\footnote{Up to a timescale issue, which potential to use, $F$ or $V$ or their combination? } and the simulations were done in  \cite{Young:2008he} and in subsequent paper for the recombination: the difference is the direction of the diffusion, equations are the same.

%

\begin{figure}[t]
  \begin{center}
  \includegraphics[width=6cm]{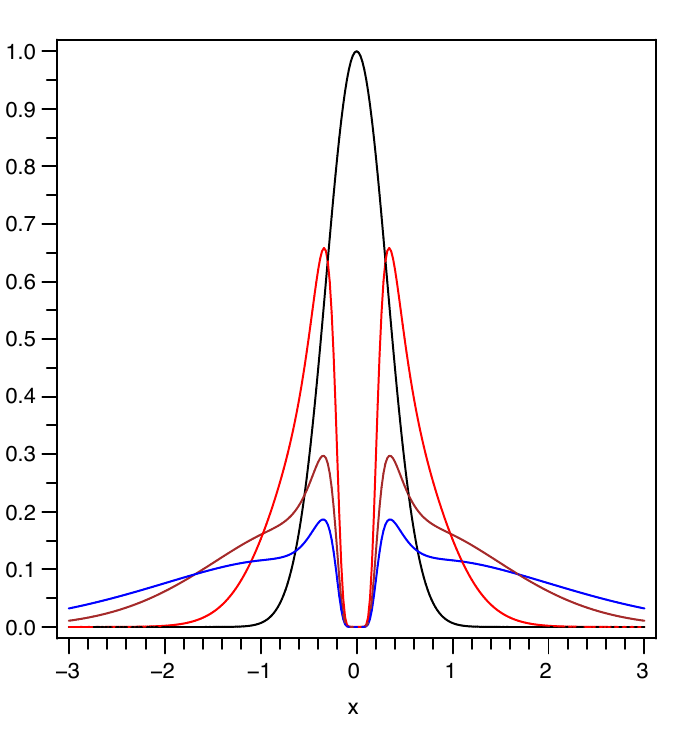}
   \caption{ from \cite{Young:2008he}: 
one-dimensional Fokker-Planck equation for an interacting c?c pair. The relaxation of the initial narrow Gaussian distri- bution is shown by curves (black, red,brown,green,blue, or top to bottom at r=0) corresponding to times t = 0, 1, 5, 10 fm, respectively. }
  \label{fig_FP}
  \end{center}
\end{figure}

    The point I try to make does not depend on the details of those papers but concerns their qualitative observations:
 both the inward/outward diffusions are very slow, and the reason for that is quite interesting. The spatial distribution rapidly reaches certain shape which persists 
   with only slow growth of its tail. The example is shown in Fig. \ref{fig_FP}. 
   In this case the attractive Coulomb-like potential
   has been complemented by a repulsive quantum effective potential $\sim \hbar^2/m r^2$ which generates the hole 
   in the distribution at small $r$ and prevents classical falling of the charge on the center.

   We called   solutions  with a nearly-permanent shape and small flux  ``quasi-equilibrium" solutions. 
   This concept is -- to my knowledge -- not yet been noticed in this particular field, but it deserved to be.
   Let me show how quasi-equilibrium solutions works for the quark diffusion problem, using the FP equation
\be {\partial P \over \partial t}= {\partial \over \partial \vec r} D  ({\partial P \over \partial \vec r }
+\beta P { \partial V \over \partial \vec r}) \label{eqn_FP}
\ee
  where $P(t,\vec r)$ is the distribution over the $\bar{c} c$ separation  $\vec{r}$ at time $t$, D is the diffusion constant, $\beta=1/T$ and $V(r)$
  is the interquark potential. First of all, when \be P\sim exp(-\beta V) \ee 
  two terms in the r.h.s. bracket cancels: this is stationarity of the true equilibrium. The bracket in the r.h.s. of (\ref{eqn_FP})  -- the particle flux -- is then zero.
  
  Note however what happens when this bracket is nonzero but 
  {\em constant(r)}: the divergence of the flux  still makes the r.h.s. to vanish. Then the l.h.s.
  is also zero: these solutions are stationary. Such solutions, while stationary, still possess  a constant flow of particles. The direction depends on the sign of the constant,
  it can be from small to large $r$ as in charmonium suppression problem, or from large to small $r$ for recombination.

  \begin{figure}[t]
  \begin{center}
  \includegraphics[width=6cm]{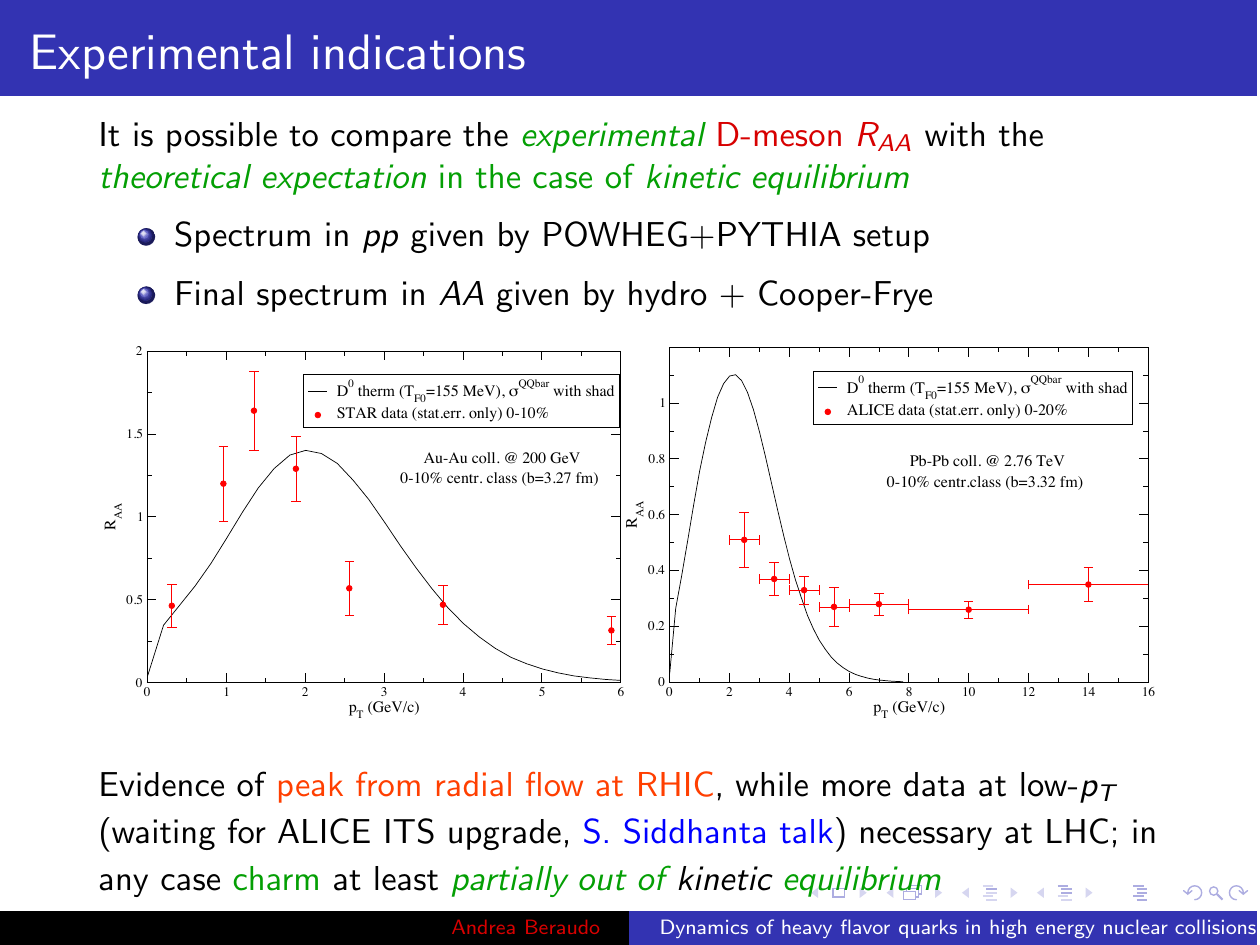}
  \includegraphics[width=6cm]{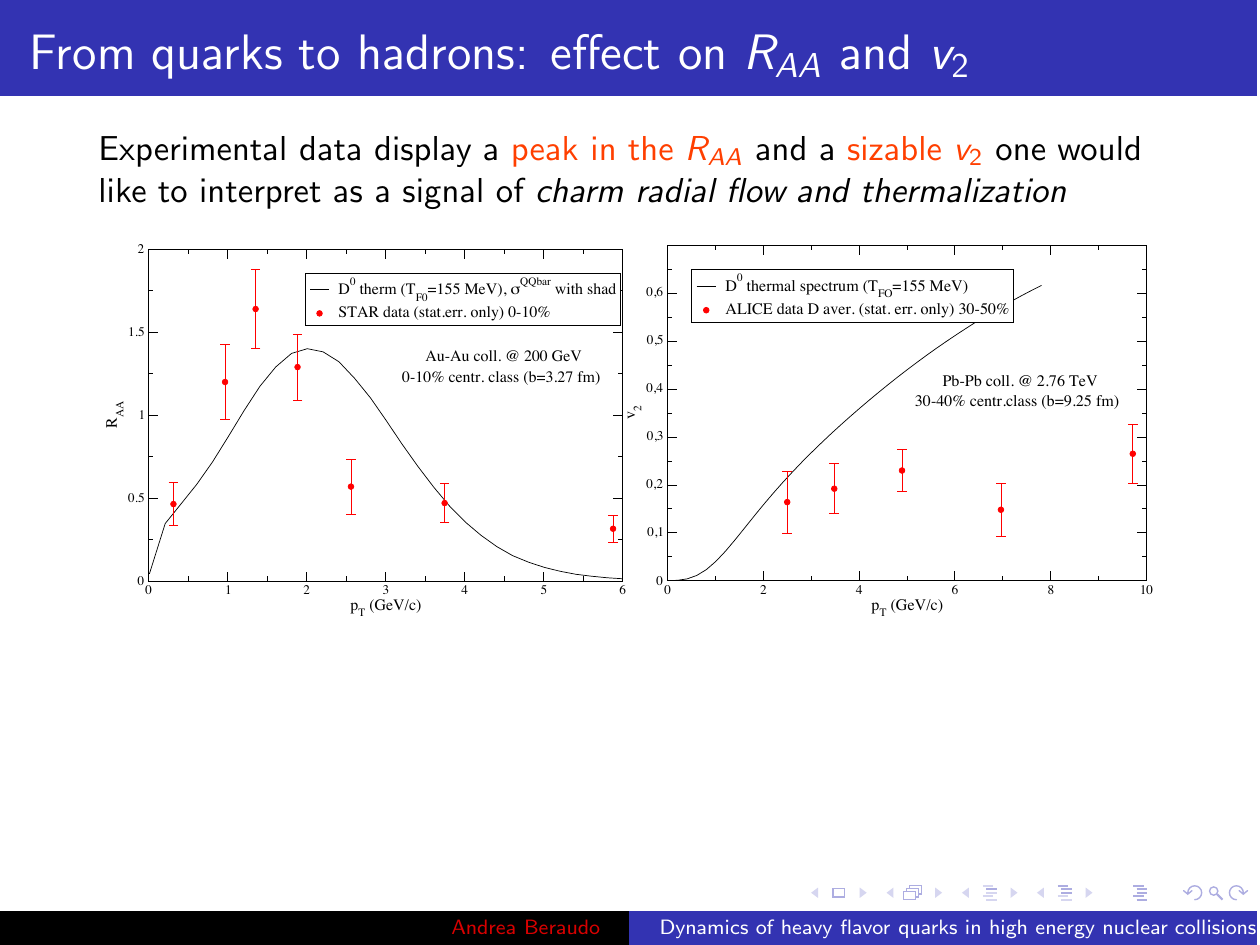}
   \caption{STAR data on RAA of D mesons, and ALICE v2 of D, both from \cite{Beraudo_QM14} }
  \label{fig_D_RAA_STAR}
  \end{center}
\end{figure}

\subsection{Do charmed mesons and charmonia flow?}
Let me start with an observation that D mesons have a maximum in $R_{AA}$ shown in Fig. \ref{fig_D_RAA_STAR}(a) resembling that for protons:
and we know the latter is of hydrodynamical origin. 
However 
the peak is
at $p_t=1.5\, GeV$, lower than for the nucleon, while   the mass of D is higher. 
Direct comparison to hydro at Cooper Fry (the line) does not do such a good job
as it does for light flavor hadrons. 
The observed elliptic flow  shown in Fig.\ref{fig_D_RAA_STAR}(b)
is nonzero but  smaller that hydro predicts (curve).
So, (i) either charm is heavy enough to be out of equilibrium with matter,  
or (ii) it is in kinetic equilibrium with it but spectra were not correctly calculated. 
(The meaning of the second option will be soon clarified.)
 \begin{figure}[b]
  \begin{center}
  \includegraphics[width=6cm]{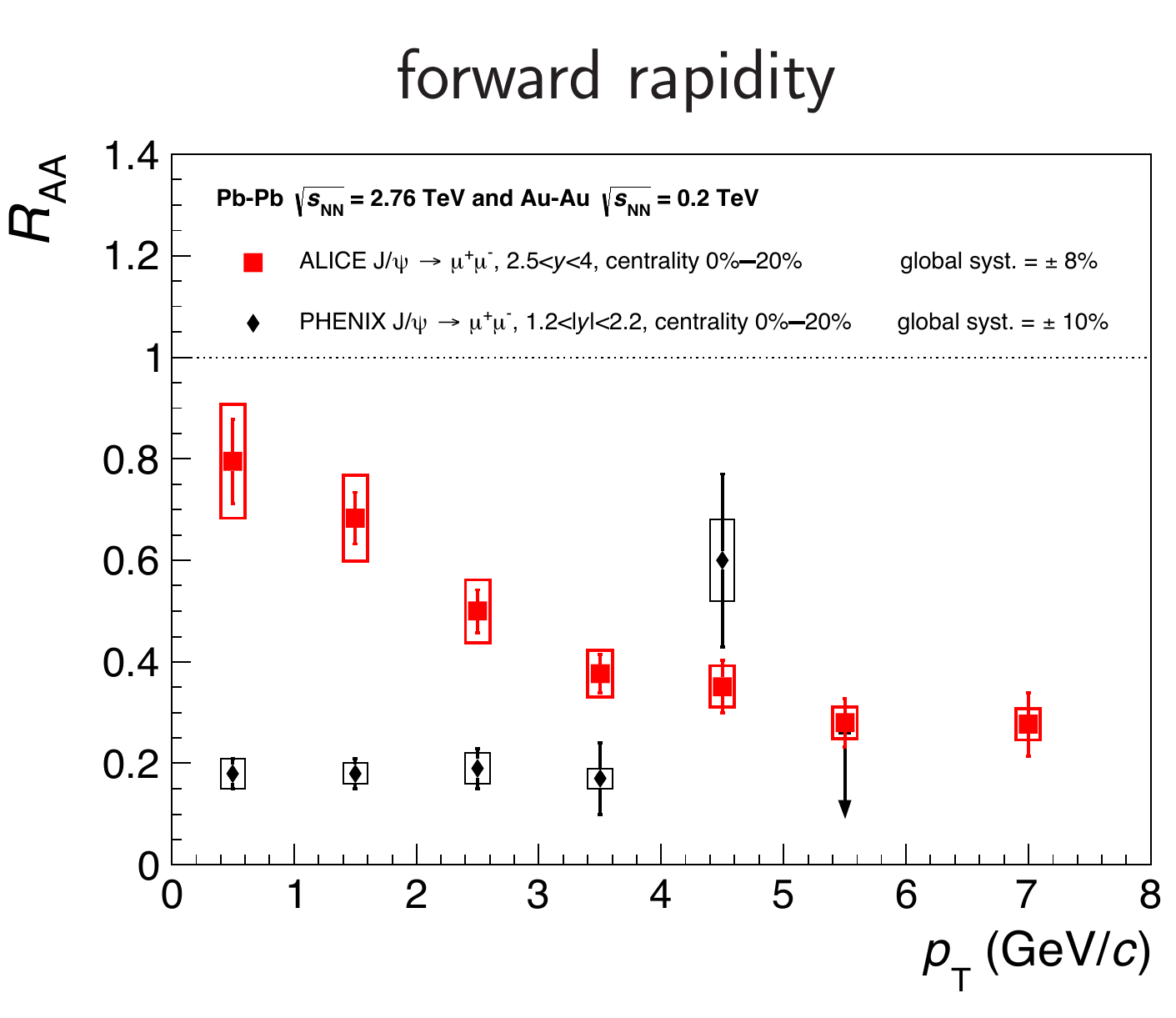}
  \includegraphics[width=6cm]{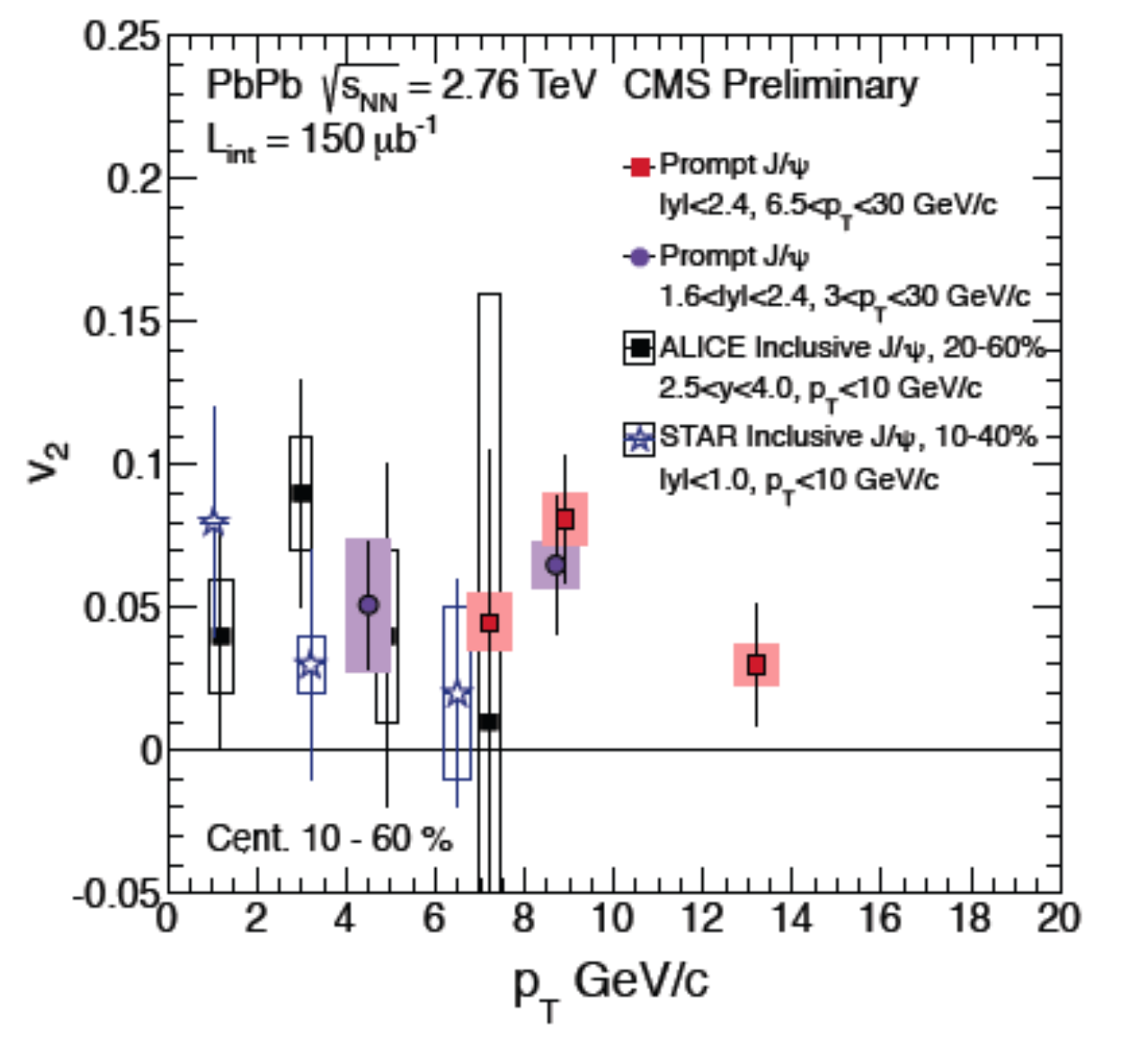} 
   \caption{ Comparison between ALICE and Phenix  charmonium $R_{AA}$ (a) and charmonium elliptic flow (b), both from \cite{Beraudo_QM14}.}
  \label{fig_Jpsi}
  \end{center}
\end{figure}

LHC data on charmonium dependence on $p_\perp$ is transitionally done in the form 
of $R_{AA}^{J/\psi} (p_\perp)$,
normalizing it a la parton model.
 A comparison of PHENIX,RHIC and ALICE,LHC data,  shown in Fig.\ref{fig_Jpsi}(a)
(from \cite{Andronic_QM14}) display drastic change. There are much more charmonia at LHC than at RHIC
at small $p_t$:  those presumed to be due to
 the ``recombined" ones.
 The observed $v_2$ of $J/\psi$ Fig.\ref{fig_Jpsi}(b) is non-zero but smaller than hydro predicted.

 The recombined component is produced from matter and thus it is maximal at small $p_\perp$, while
 the primordial are mostly fast 
 (relativistic) $P/M>1$. It is presumed that recombined component flows with matter. 
 The question is whether the primordial component also do so, or ``slip" relative to the medium:
 the data seem to favor the latter option.

 \subsection{Inhomogeneous distributed charm} 
 
 He, Fries and Rapp \cite{He:2011qa} had demonstrated that if one artificially increases the drag
 coefficient, one reproduces the $p_t$ distribution obtained from hydrodynamical flow of charm.
 In other words, they nicely checked that at high enough scattering cross section   charm is 
  kinetically equilibrated with matter.  In this calculation the original
 particles were  homogeneously distributed over the fireball at some early time, which is
 indeed needed to get consistency between 
 the two calculations.
 
 Yet what would happen if the drag/cross section be still large, but the initial charm be distributed over the fireball
 in a $non-uniform$ way. Introducing charm fugacity $f=exp(\mu_c/T)$ in a standard manner, one should think of it
  as some function of time and space point, depending on the production mechanism.
Charm recombination into charmonium states is calculated
at the chemical freezeout surface, and it  should  be proportional to an average of over it  $<f^2(t,\vec{r})>$, not
square of the average $<f>^2$ as it is done now. 
In short, I wander what one should do for thermal recombination when the charm fugacity is not homogeneously
distributed over the system, as is usually assumed.

This lead to nice hydro exercise I will briefly tell about. Let me start with the longitudinal hydro equation written as 
entropy conservation
\be \left[ s(t,x) u^\mu \right]_{;\mu}=0 \ee
with the semicolon meaning the covariant derivative. The homogeneous distribution means that 
 $n/s=const(t,x)$, and thus any solution for the entropy generates solution for $n(\tau,r)$ as well.
If one would like to put an arbitrary distribution of charm (or any other conserved quantity) density
at some initial time, in order to find its distribution later one has to solve its conservation 
equation $\left[ n(\tau,r) u^\mu\right]_{;\mu}=0$
at all times. I started with Gubser flow, in coordinates proper time transverse distance $\tau,r$, for which
$s(\tau,r)$ is known and put  
  $n(\tau,r)=f(\tau,r)*s(\tau,r)$.  The resulting ``fugacity equation" reads
\be 
 2 r \tau \partial_r f+ (1+r^2+\tau^2)  \partial_\tau f=0
\ee
It has a solution with an arbitrary function $\phi$ of the following argument 
\be f=\phi({2 r \over 1 - r^2 + \tau^2 }) \ee
The function $\phi$ can be fixed from the initial conditions at $\tau=0$. Using it, one finds the
corresponding distribution of our quantum number (charm) at the freezeout surface as well.

The problem was solved, but the meaning of this answer looked mysterious. It was further  clarified
when I used the  co-moving coordinates (see appendix)
 $\rho, \theta$.  All one needs to know
is that $\rho$ has meaning of time and $\theta$ of space, and  the particle conservation in a comoving frame
reads simply as
\be {\partial n \over \partial \rho}=0 \ee
meaning that any function of $\theta$ independent on coordinate $\rho$  solves the problem. And constant theta implies constancy
of the combination I obtained in the original coordinates.   
The lesson I got from this is that even in numerical solutions of the hydro equations one perhaps
should try to find the co-moving coordinates. Those would lead to very natural distributions of 
external objects like charm.

I did checked that for recombinant $J/\psi$ the effect of  $<f^2>\neq <f>^2$ is actually small and 
can perhaps be ignored. I however further found that
 the inhomogeneity effect suppresses $v_2$ of charm, especially at higher $p_t$,  because the fireball edge is a bit 
 less populated by charm than its middle. My  
 calculation gives \be {v_2^{inhomogeneous} \over  v_2^{homogeneous}}|_{p_\perp=3\, GeV}=0.65 
\ee
which brings hydro prediction closer to experimental ones on $v_2^{J/\psi}$. Perhaps recombined charmonia do flow with the matter,
after all.

\section{Jet quenching} \label{sec_quenching}

Let me mention on the onset that hard QCD is expected to be perturbative, and this is not my field.
I thus would not go into gluon radiation, in vacuum and in matter, in any detail. 
This small section contains only some remarks on recent theory developments.
\subsection{``Quasi-equilibrium" in jets}

I think it has been an important development  relating  jet-in-matter physics with a more general
 turbulence theory.
In it I will follow ref.\cite{Blaizot:2014ula} by Blaizot et al. 
For a large enough medium, successive gluon emissions can be considered as independent: multiple emissions can be treated as probabilistic branching processes, with the BDMPSZ spectrum playing the role of the elementary branching rate. 
The inclusive gluon distribution function
 \be {dN \over dlog(x)d^2\vec{k} } = {D(x,\vec{k},t)  \over (2\pi)^2} \ee
satisfy certain diffusion-brunching equation. Integrating over transverse momentum one gets the so called zeroth moment
$D(x, t) = \int_k D(x, k, t)$ on which we focus for simplicity. This satisfies the equation

\be t_* {\partial D(x,t) \over \partial t}=\int dz K(z) \left[\sqrt{z\over x}D({x\over z},t)-{z \over \sqrt{x}}D(x,t)\right] \ee
with the gain and the loss terms. The details such as the shape of the kernel K and time parameter $ t_*$ can be found in  
ref.\cite{Blaizot:2014ula}. The central feature I want to focus on is the analytic solution
 \be D(x,t)={ (t/t_*) \over  \sqrt{x}(1-x)^{3/2}} exp(- {\pi t^2 \over t_*^2 (1-x)}) \ee
Essential singularity at $x=1$ is expected: it is known as the Sudakov suppression factor. The remarkable
news is that apart of the exponent, the shape of the x-dependence remains the same at all times, only the 
normalization changes.

 Let us see how it works in the most important small $x<<1$ region. If one simply plugs in  the leading $x^{-1/2}$ dependence,
 the gain and loss terms simply cancel. This is the quasi-equilibrium solution under consideration.  
As a result,  jets in matter are expected to approach some universal shape,  not determined by the particular
initial conditions, but by the quasiequilibrium solution to which it gets attracted as the process proceeds.   

It is essentially the same phenomenon as we have seen in Fig.\ref{fig_FP} for diffusing charmonium: the shape itself is dictated by
the balance of the gain and loss. Both are ``quasi-equilibrium" attractor solutions: their main feature  is  {\em constant flux} of certain quantity, from
one end of the spectrum to the other. 
(The flux in decaying charmonium comes from small to large distance between quarks, in the jet case it comes from large to small $x$.)
Once again, the constancy of the flux in such solutions is the key idea going back to Kolmogorov's theory of hydrodynamical turbulence,
which I find quite remarkable. As Einstein once observed (approximate quote from memory): ``... the number of good ideas in physics is so
small, that they keep being repeated again and again in various context".

\subsection{Is jet quenching dominated by the  near-$T_c$ matter?}
 
\begin{figure}[t!]
\begin{center}
\includegraphics[width=7cm]{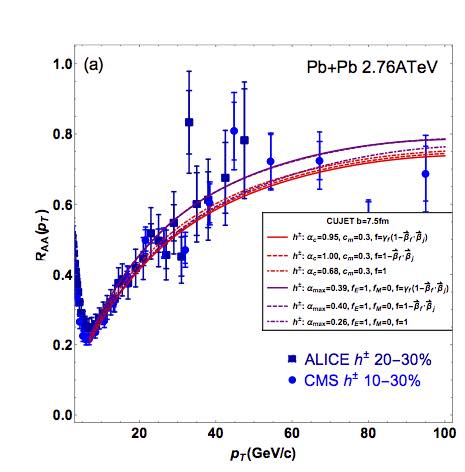}
\includegraphics[width=7cm]{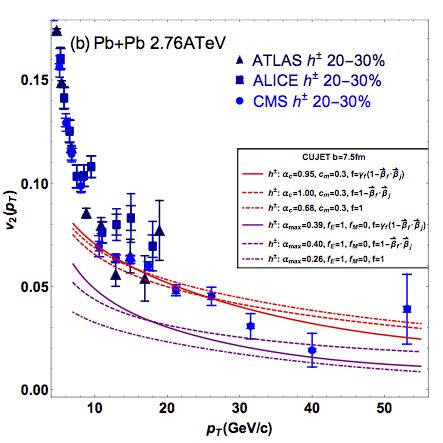}
\caption{(color online) From \cite{Xu:2015bbz}: jet suppression and elliptic  parameter $v_2$, data versus
models.  }
\label{fig_BG}
\end{center}
\end{figure}

 Let me start with Fig. \ref{fig_BG}, emphasizing some features of the LHC (ALICE) data. The first one shows the suppression factor  $R_{AA}$ compared with
 theory based on pQCD expressions. The first obvious  feature I want to point out is the pronounced dip in the data, clearly separating
 two distinct regimes. Although the magnitude is centrality dependent, its location at $p_t^{dip}\sim 6\, GeV$ changes a little.
 (It also about the same as at RHIC, where the dip is minor.). The region above the dip $p_t>p_t^{dip}$ is 
 well described by 
 jet quenching models, while below the dip it is perhaps some tail of hydro-related phenomena, yet to be understood.
 
 Now we proceed to angular distribution of jet quenching, fig.(b) on which $v_2(large\, p_t)$ is plotted. 
Already   at the very beginning of RHIC era I noticed \cite{Shuryak:2001me}
 a problem: 
  direction-dependent in or out-of reaction plane by
 as \be R_{AA}^{in/out}= R_{AA}(1\pm 2 v_2) \ee
 apparently is
 incompatibly large to most theories of jet quenching. 
 As one can see from the second  Fig. \ref{fig_BG} , it is still
  very much true at LHC. Theory curves which describe so well $R_{AA}$, 
 all the way down to the dip $p_t^{dip}$, lead to $v_2$ (at the dip) much smaller than needed!
  What is new  here are LHC measurements at much higher $p_t >  30\, GeV$, demonstrating 
 $v_2$ values which are consistent with the pQCD models. 
  Yet it is still true that in the range $6<p_t<  30\, GeV$
 something significant is missing. (Steeply falling another contribution Fig. \ref{fig_BG} (a) below the dip makes it unlikely 
 to be important.)
 
  A solution has been proposed by Liao and myself \cite{Liao:2008dk}: we found that one can get
  near the observed $v_2$ value if the jet quenching is strongly enhanced at the near-$T_c$ matter.  
  We motivated it by presence of magnetic monopole in the near-$T_c$ region.

\begin{figure}[t!]
\begin{center}
\includegraphics[width=7cm]{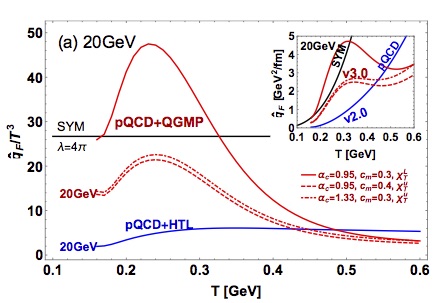}
\caption{(color online) the NTcE model with enhanced near-$T_c$ quenching}
\label{fig_BG2}
\end{center}
\end{figure}

  A model based on this idea can be reconciled with RHIC and LHC data. Let me jump to the
  recent paper \cite{Xu:2015bbz} in which the corresponding model is well developed. 
  In Fig.\ref{fig_BG} one finds fits of the $R_AA$ and $v_2$ of the jets for models with and without 
  magnetic component.
In   Fig.\ref{fig_BG2} we show the $\hat q(T)$ of the model for a jet with 20 GeV.

 \begin{figure}[t]
\begin{center}
\includegraphics [height=8.cm]{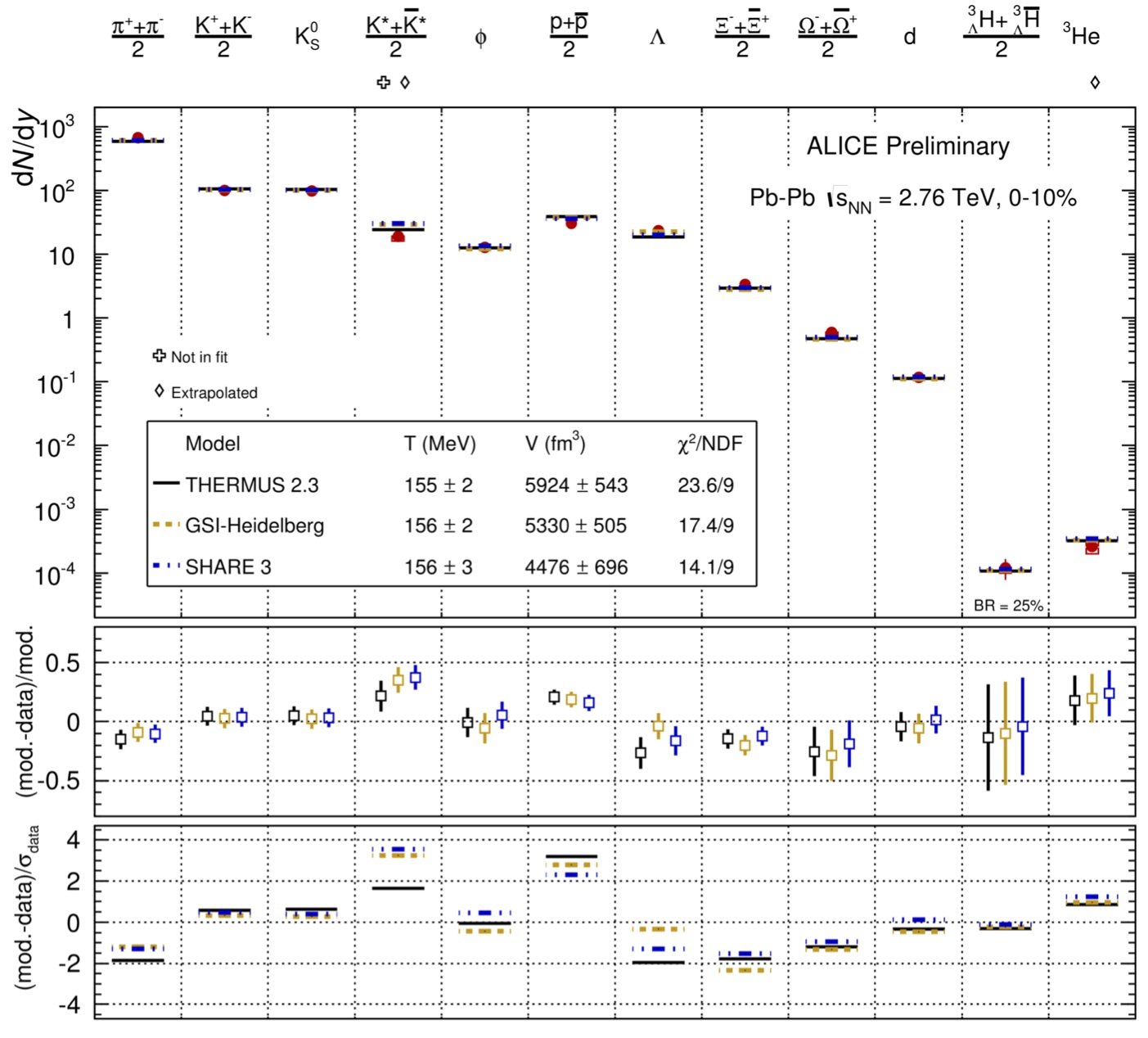}
\caption{
 Alice data on particle yields compared to thermal model \cite{Andronic:2007ff}. The main fit parameters are indicated
 in the upper plot.}
 \label{fig_alice_thermal}
\end{center}
\end{figure}

\section{Near the phase boundary: fluctuations and  the freezeouts}

\begin{figure}[b]
\begin{center}
\includegraphics [height=6.cm]{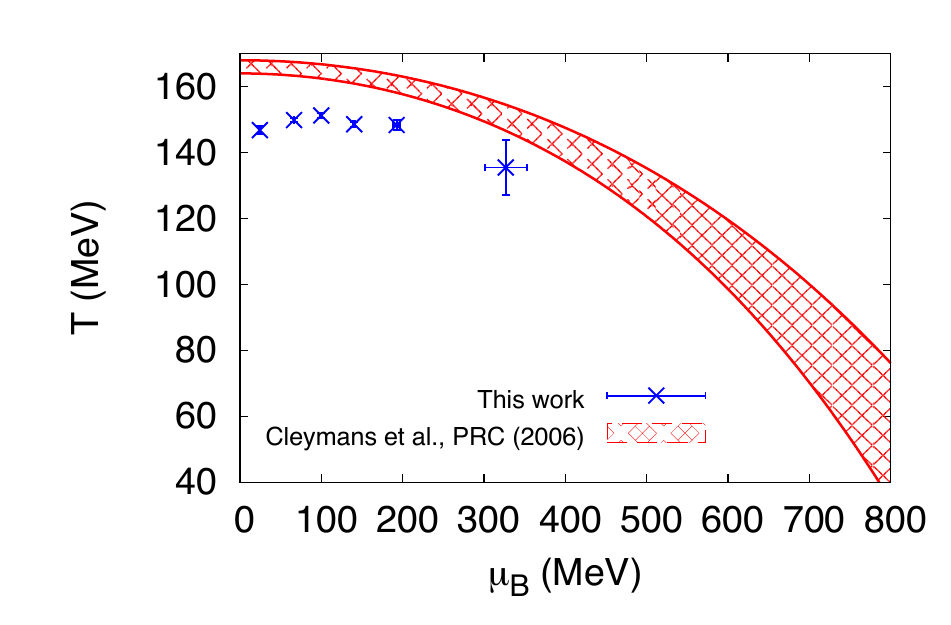}
\includegraphics [height=6.cm]{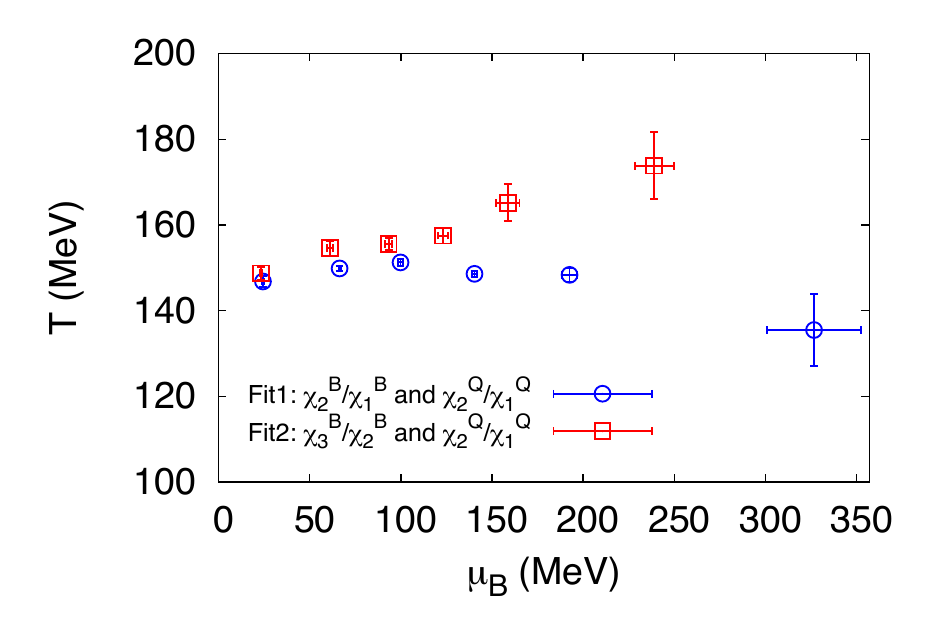}
\caption{ (Color online) (a) Freeze-out parameters in the ($T - \mu_B$ ) plane: comparison between the curve obtained in Cleymans et al (red band) and the values obtained in the present analysis from a combined fit  for net-electric charge and net protons (blue symbols). (b) The freezeout parameters fitted from two different set
of rations, as shown in the plot}
\label{fig_Ratti}
\end{center}
\end{figure}

\subsection{Chemical freezeouts}

Statistical equilibration is a famous success story, and it hardly needs to be emphasized here.
Let me just show ALICE data with thermal fits in Fig.\ref{fig_alice_thermal}, presented at CPOD 2014 by A. Kalweit.

Note that even light nuclei -- $d,t, He^3$ and their antiparticles are also included. One may wander how $d$, with its 2 MeV binding energy, can be found in an environment with ambient temperature $T\sim 160 \, MeV$. It is literally finding
a snowflake jumping out of a hot oven. Shouldn't $d$ be instantly destroyed in it?
The answer to this and similar questions is well known: 
 thermodynamics does not care about $d$ lifetime. As it is
destroyed with some rate, perhaps large,   in equilibrium it is recreated by the inverse process with the same rate, so
that its  average population is conserved. What thermodynamics gives us is that average. 

Note that one deviation is $K^*$: the model predicts more than observed. This is expected: it is short
lived resonance which decays when the density is still non-negligible, the products can be re-scattered
and their invariant mass moved out of the peak. Corrections to that can be made using 
 any cascade codes.
 
 Another deviation is for $p+\bar{p}$. Some argued that one should take into account possible 
 annihilation on the way out, and again use 
  available practical solution like UrQMD or similar code to calculate it. Here however comes an objection:
  all of them include annihilation $p+\bar{p}\rightarrow n \pi, n\sim 5$ but $not$ the inverse reaction. 
  This is an old story \cite{Rapp:2002uf}: contrary to popular beliefs the inverse reactions are not suppressed,
  in fact at equilibrium their rate are exactly the same as the direct one. Using a cascade code which
  does not respect the detailed balance -- and thus the very concept of thermal equilibrium -- to
  calculate corrections to thermal model is just logical nonsense.

I would not show the freezeout points on the phase diagram, which has been done many times. 
Let me just remind that those seem to be remarkably close to the phase boundary,
defined on the lattice. Why should this be the case? 
An answer suggested by
 Braun-Munzinger, Stachel and Wetterich
\cite{BraunMunzinger:2003zz} is also related with the multiparticle
reaction rates. Since those depend on very high power of the temperature, they all 
should decoupled very close to critical line
 \be {|T_{ch}-T_c | \over T_c}  \ll 1 \ee 

The concept of two separate freeze outs is based on separation of elastic and inelastic 
rates at low $T$, in magnitude.

New development is usage of event-by-event fluctuations. As it has been first pointed out in
my paper \cite{Shuryak:1997yj}, those in general provide information about
multiple susceptibilities, or higher derivatives of the free energy over $T$ or various chemical potentials. 

This idea further was applied toward location of the critical point in  \cite{Stephanov:1999zu},
suggesting the low energy scan program. The results of the RHIC scan are however still too fluent
to be reviewed, even after QM14.

Let me instead comment on another development, on the interface of  the lattice and heavy ion communities.
Calculation of susceptibilities on the lattice, including  the most difficult near-$T_c$ region,  reached significant maturity. 
  Karsch stressed at many meetings, that higher susceptibilities now can be compared to the data 
  and the freezeout curve be reconstructed, even 
  without using the particle ratios. I will follow here \cite{Alba:2014eba} from which Fig.\ref{fig_Ratti}
is taken: Fig(a) compares new set of freezeout parameters (points) compared to earlier one from the particle ratios.
Fig.(b) shows that as $\mu$ grows a consistency between different ratios is getting worse.
It is however generally believed that at chemical freezeout the Hadron Resonance Gas model is generally 
describing the QCD thermodynamics, expect perhaps in the vicinity of the QCD critical point.

\begin{figure}[t!]
\begin{center}
\includegraphics[width=7cm]{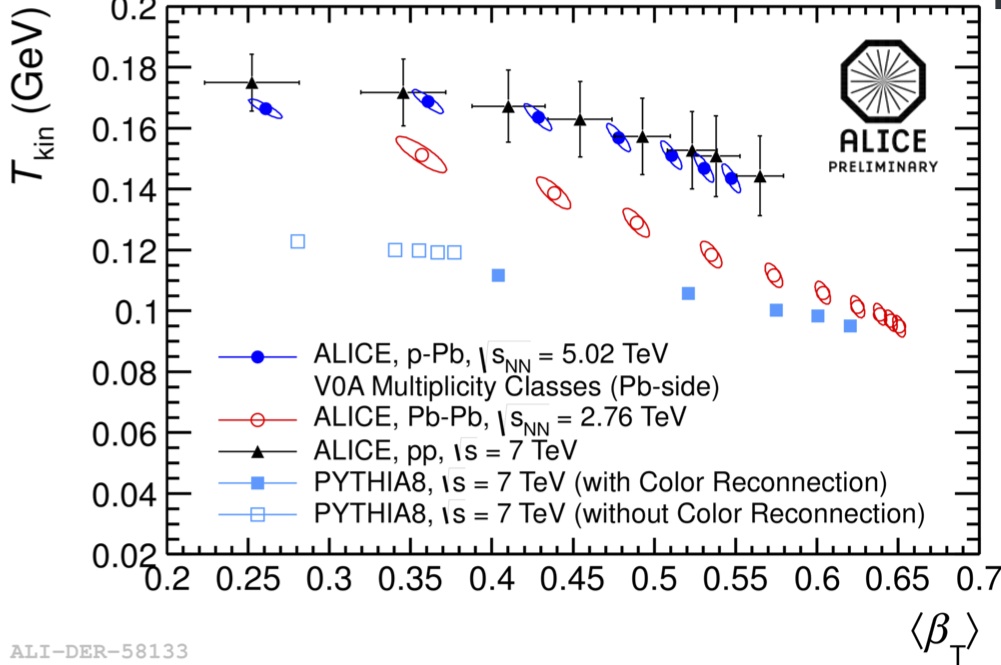}
\caption{(color online) The temperature of the kinetic freezeout versus mean velocity
of the radial flow, fitted to ALICE spectra of the identified secondaries ($\pi,K,p,\Lambda,\Xi,\Omega$). }
\label{fig_alice_blast}
\end{center}
\end{figure}

\subsection{From chemical to kinetic freezeouts}

Separation in magnitude of the elastic and inelastic (low energy) hadronic reactions
is the basis of the ``two freezeouts" paradigm, with separate chemical $T_{ch}$ and kinetic    $T_{kin}$ 
temperatures.

Its effectiveness became even more clear at LHC, which we
illustrate by Fig.\ref{fig_alice_blast} containing the ``blast wave" fitted parameters
to the ALICE spectra of .  
Unlike   $T_{ch}$, the kinetic  one  $T_{kin}$ strongly depends on centrality of PbPb
collisions, decreasing  to values below 0.1 $GeV$ for the most central bins.  
Cooling from 0.16 to 0.1 $GeV$ may not look so dramatic: but thermodynamical
quantities in this region are proportional to high power of $T$ and they do change strongly.

First, does on understand the dependences displayed in Fig.\ref{fig_alice_blast} at all?
The most central AA collisions produce the largest systems, which have the highest $T$
at the early stages, and also the lowest $T$ at the end. One popular way to explain it is to mention, that the
energy conservation negatively correlates the flow velocity $<\beta_T>$ and  $T_{kin}$, so  more ``explosive" systems
cool more.

 While correct, this simple idea seems to be in contradiction to the $pp, pA$ data also shown in 
Fig.\ref{fig_alice_blast}, and for which  $T_{ch}$ and    $T_{kin}$ are close. An  explanation to that,
discussed in the middle of this review, is that high multiplicity bins, displaying strong radial flow, 
achieve it via more extreme conditions at their early stage, by increasing the initial $T_i$. 
In other words, higher multiplicity in this case means large energy/particle, which reveals itself in the flow
without cooling. 
(In AA
collisions $T_i$ also depends on centrality, but much weaker.)

The kinetics of the freezeout 
-- that the rescattering probability on the way out should be about 1/2 -- 
is of course central to defining $T_{kin}$. Cold fireballs created in central PbPb 
have thousands of particles and 
freezeout late, at times reaching 15 fm.  The highest multiplicities observed
 in  $pp, pA$ have 10-20 times less particles, and  freezeout time/size of 3 fm or so. 

In summary, contrary to belief of many, in PbPb LHC collisions one does have an extended  phase
 of the collision, in which matter cools deep into the {\em hadronic phase}. This opens some interesting
 questions related to hadronic phase, which can now be addressed experimentally.

Let me mention only one issue here. Since between $T_{ch}$ and    $T_{kin}$ the particle numbers are conserved,
one should introduce new nonzero chemical potentials, not associated with
conserved quantum numbers like charge and baryon number. In particular, there should be
  nonzero chemical potentials for pions. 

\begin{figure}[b]
\begin{center}
\includegraphics [height=5.cm]{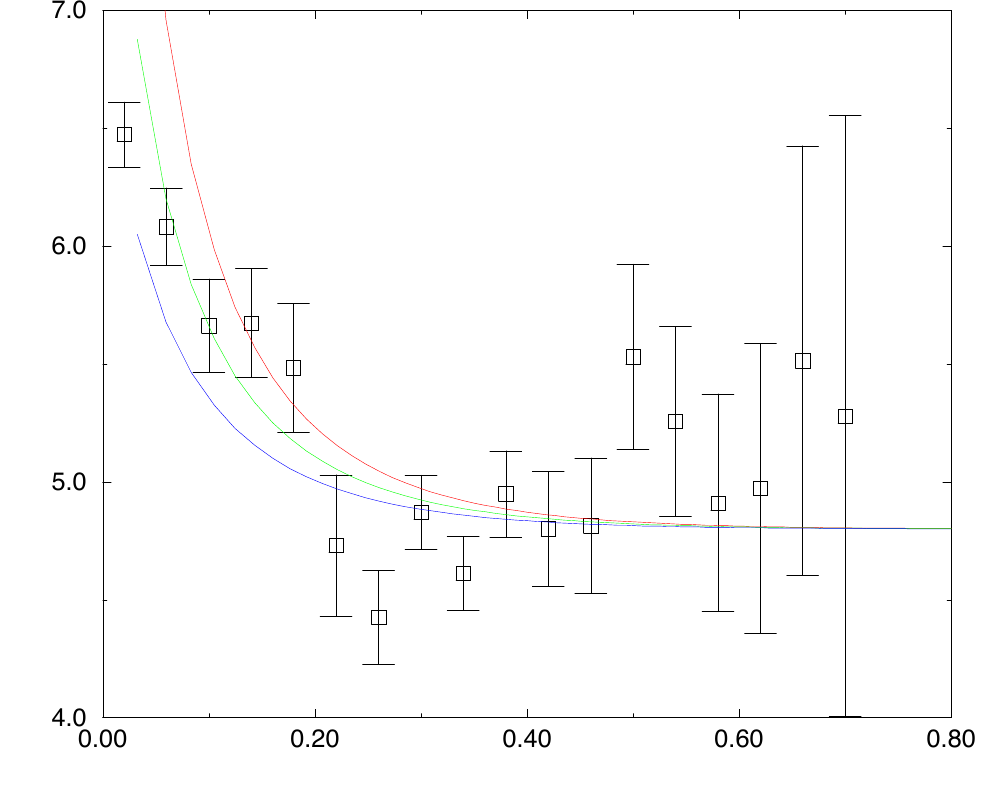}
\includegraphics [height=8.cm]{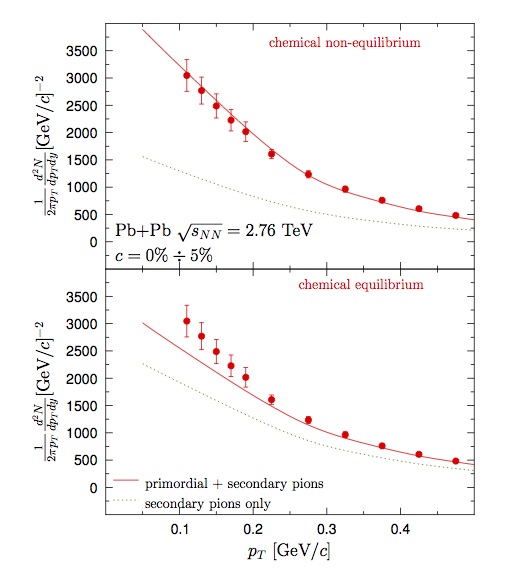}
\caption{
 (Color online) (a) Points show the ratio of PbPb to SS spectra, from NA44. Three curves are for
 pion chemical potential $\mu_\pi=60,80,100\, MeV$, from \cite{Hung:1997du}.
 (b) (From \cite{Begun:2013nga}) Dots are ALICE transverse momentum spectra of pions in the low-$p_t$ region, compared to the
  model with (upper plot) and without (lower plot) pion chemical potential.}
 \label{fig_begun}
\end{center}
\end{figure}

Whether there are nontrivial fugacity factors at the kinetic freezeout can be directly observed in the
pion spectra, because in this case Bose enhancement can be measured.  This idea
is at least 20 years old \cite{bebie}: its experimental manifestation was displayed in the paper by me and Hung
\cite{Hung:1997du} is shown in Fig.\ref{fig_begun}(a). (The
reference  SS collisions is much smaller system than PbPb, and thus
its chemical and kinetic freeze outs should be close.)
Fig.\ref{fig_begun}(b)  shows that the same effect shows up, now  at LHC.
The fit $without$ chemical equilibrium, with nonzero pion $\mu$ on top, provides a better description 
to the spectra at small  $p_t<200 \, MeV$.

Interesting that the parameter of the fit in this last work gives $\mu_\pi\approx m_\pi$, so the authors
speculated if the conditions for
 pion Bose-Einstein Condensation (BEC)   were actually  reached. 
  If this indeed becomes true, it has been many times suggested previously that the femtoscopy parameter $\lambda$
  should show it, as it is sensitive to ``degree of coherence" of the pion source.   
 Recent femtoscopy data on 2 and 3 identical pions from ALICE, discussed in \cite{alice_napa},
 can be indeed fitted with a coherent source. The fraction of coherent pions coming from this fit
  is as large as 23\% $\pm$ 8\%. 
  
  Do we actually witness BEC formation at LHC? In order to answer this question it is useful to recall
  BEC discovery in experiments with ultra cold atoms a decade ago. As the atomic system does evaporating cooling
  and its temperature decreases, the measurements of the momentum distribution (by switching off the trap)
  revealed appearance of new and much more narrow component. Unlike the usual thermal component,
  its width was independent of $T$, and related to the inverse spatial size of the BEC cloud.
  
  This indeed sounds like what is observed in heavy ion collisions:
  as one goes to most central collisions and the kinetic freezeout $T_{kin}$ get below 100 MeV,
   the $p_t$ spectra do become enhanced at small momenta.  The difference however is in the shape:  
   we don't see a new Gaussian, as in the atomic experiments, but thermal spectrum modified by $\mu$.
  
   The condensate should be a separate component, with $\mu$ being exactly $m_\pi$
   and independent on $T$.  
  If BEC cloud contains about 1/4 of all pions, 
  its diameter should be large, at least of the order of 2-3 fm.
  The corresponding width of momentum distribution, from uncertainty relation, should  be as small 
  as say $<p_t> < 0.1\, GeV$.   Looking back to Fig.\ref{fig_begun}(b) one however finds,
  that such soft secondaries seem to be $outside$ of  the acceptance. So, even  if BEC
  component is there, we so far cannot see it, neither with ALICE nor with any other LHC detectors!
 How then can we get their influence in the femtoscopy?
 
 This issue can perhaps  be clarified by a
  short dedicated run,  in which the ALICE detector
 switches to smaller (say 1/2 of the current value)  magnetic field, to improve the low $p_t$ acceptance.
 (Yes, recalculating all the efficiencies is a lot of extra work, but perhaps it is worth clarifying this interesting issue.)

\subsection{The search for the critical point and RHIC low energy scan }

\begin{figure}[t]
\begin{center}
\includegraphics [height=3.cm]{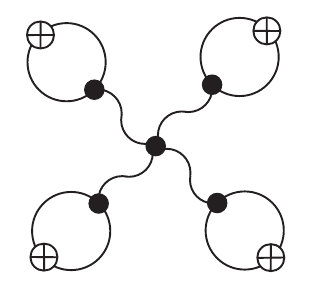}
\caption{
 The enhanced contribution to 4 particle correlator, from \cite{Stephanov:2008qz}.}
 \label{fig_Misha4}
\end{center}
\end{figure}

The main idea \cite{Stephanov:1999zu} is 15 years old and  well known: critical point -- if exists --
should lead to large correlation lengths and enhanced e-by-e fluctuations, similar to critical opalescence  
known in many cases.
Mathematically speaking one may go
for effects given by diagrams which have as many critical propagators as possible. 
 Stephanov \cite{Stephanov:2008qz} pointed out that since  the $n$-particle correlators
may contain up to $n$ such propagators, they are more sensitive to large correlation length:  3 particle correlators are $\sim \xi^6$, 4-particle  ones $\sim \xi^8$, see Fig.\ref{fig_Misha4}, 
and so on. The wavy line at zero 4-momentum is $\sim 1/m_\sigma^2\sim \xi^2$: but the prediction
is not just the power of the propagators because the coupling of critical modes  by itself vanishes as certain power of $\xi$ given by the critical indices. The  quartic one in the diagram considered  is $\sim 1/\xi$ so the total power is 7, not 8. 

Let me first retell the same story in a simpler language. The critical field we here call $\sigma$ 
should be viewed as some stochastic/fluctuating background field coupled to fluctuations in particle number
(circles with crosses in the diagram above). One can view it being proportional to some stochastic potential $\Delta V(x)$
which enter the probability in the usual way $P\sim exp(-\Delta V(x)/T)$, so that in its minima
the probability is larger and more particles -- e.g. 4 mentioned above -- all gather there. 
The critical point is special in that the scale of the correlation length $\xi$ of this potential increases,
and thus more particles have a chance to get into the same fluctuations.

Which particles we speak about? In principle sigma is scalar-isoscalar, so any one of them. In our paper 
  \cite{Stephanov:1999zu} and in \cite{Stephanov:2008qz} the simplest coupling was considered as an example, the
  $\sigma \pi\pi$ one, and so the particles were pions. 

Let me now argue that using the nucleons should  work even better. First, the powers of the baryon density $n_B=N_N-N_{\bar{N}}$ correlated together are the susceptibilities calculated on the lattice as derivatives over $\mu_B$.
Second, we know from the nucleon forces -- e.g. the simplest version of the Walecka model -- that
$\sigma$ is the main component of the attractive nuclear potential which binds
the nuclei. 
\be \Delta V= {g_{\sigma NN}^2 \over 4\pi r} exp(-m_\sigma r) \ee
In vacuum the typical mass $m_\sigma\sim 600 \, MeV$ and the inter-nucleon distance $r\sim 1.5 fm$
are combined into small suppression factor $\sim exp(-5) \ll 1$
explaining why the nuclear potential scale $\Delta V\sim -50 \, MeV$  
is much smaller than the nucleon mass, in spite of strong coupling. (At smaller distance $r$ 
 the repulsive omega contribution dominates  the attractive sigma one.).   

Can it be so, that at the QCD critical point $m_\sigma\rightarrow 0$ and this small exponent disappears?
If so, one should expect much deeper  $\Delta V$, perhaps even larger than the freezeout temperature $T$.
Furthermore, if say $\xi=2 \, fm$, the volume $4\pi\xi^3/3\sim 40 fm^3$ is large enough to collect many nucleons,
not just 3 or 4, as Stephanov suggested. 
So, such clustering of the nucleons should produce large nuclear fragments,  a new cute signal of the critical point! 

The argument, unfortunately, is rather naive. The critical mode which gets long-range is not just the
$\sigma$ field but -- because we are at nonzero density -- a certain combination with $\omega$.
Therefore the repulsive forces between nucleons should be getting longer range as well. To
tell what happens we need a reliable theory or some dedicated experiments. Fortunately, we can
do it in the coming low energy scan. Isoscalar sigma interacts with scalar -- net baryon -- density $n_s=N_N+N_{\bar{N}}$ ,
while omega interacts with $n_B=N_N-N_{\bar{N}}$. The powers of these differ 
by the non-diagonal terms such as nucleon-antinucleon correlators $C^{m,n}=<N_N^m N_{\bar{N}}^n>$ 
which can and should be measured. Perhaps restricting kinematics of all particles involved -- rapidity and momenta 
differences -- would further enhance the signal. 

 \begin{figure}[b]
\begin{center}
\includegraphics [height=5.cm]{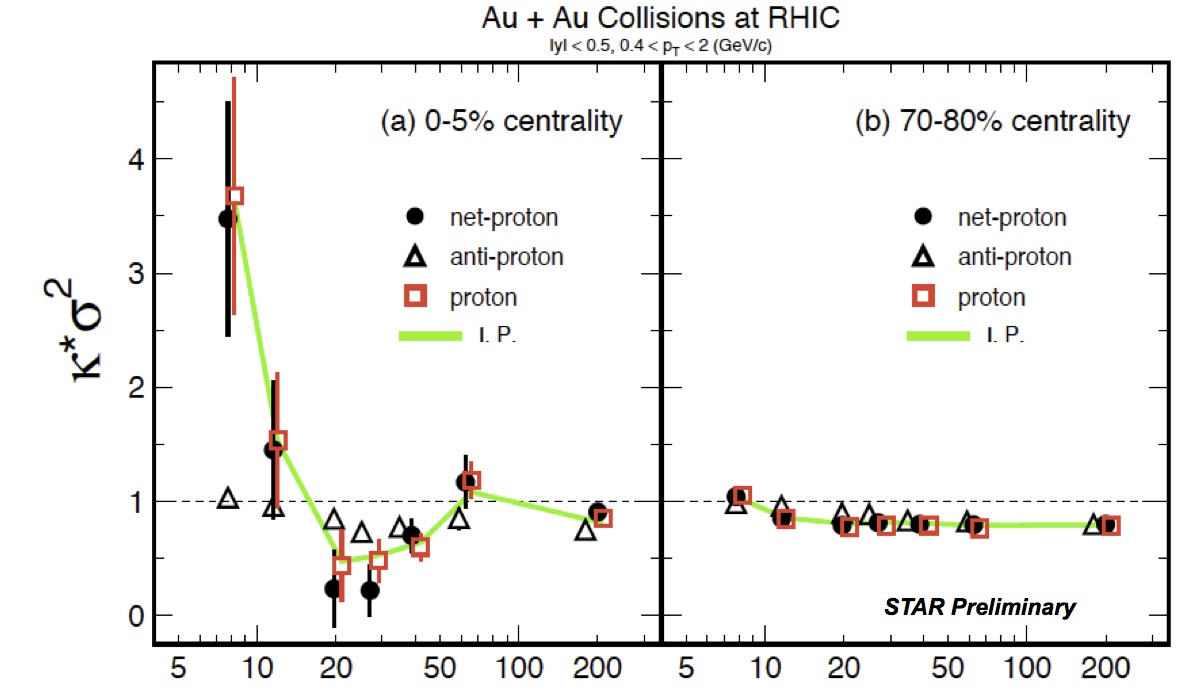}
\caption{
 The kurtosis -- 4 particle correlator -- in units of the width, as a function of the collision energy $\sqrt{s}, GeV$ .}
 \label{fig_star_scan}
\end{center}
\end{figure}

Let me now jump to the latest STAR data, presented at CPOD 2014 by Luo, and shown in Fig.\ref{fig_star_scan}.
The proton 4-point correlator has an interesting structure: a minimum at $\sqrt{s}=20-30 \, GeV$
and perhaps a maximum at low energy (?). Antiprotons have a similar shape but with much smaller amplitude. 
Theory in fact predicts some oscillatory behavior of the kurtosis near the critical point: it can be that.

(However before getting excited by the new large signal -- with large error bar -- let me remind that it appeared as a result of
particle ID improvement, from $0.4< p_t<0.8\, GeV/c$ to now reaching  $p_t=2\, GeV/c$. The newly open 
kinematic window should be sensitive to hydro flow, and potentially to its fluctuations.)

Also near critical point one expect significant modification of attractive (sigma-related) nuclear forces:
can those affect the  4-proton (antiprotons) correlations in question?

Clearly more data at the low energies are needed to understand what is going on, and whether one does
indeed discovered the QCD critical point or not.

\section{Summary and discussion}
\subsection{Progress on the big questions} 
Before we go to summaries of the particular subjects, let me remind my list of ``big questions" mentioned in the Introduction:

I.{\bf Can one locate the ``soft-to-hard" boundary},  where  the transition from strong to  weak coupling (perturbative)
regimes take place?

II.  {\bf  Can one locate the `micro-to- macro" boundary }, where the transition from single-participant to collective 
regime takes place?

III. {\bf Can we experimentally identify signals of the QCD phase transition}, in particularly locate the QCD 
critical point?

Somewhat surprisingly, the sharpest  observed transition which we discussed above
 is in the {\em profile of the $pp$ elastic amplitude}  shown in Fig.\ref{fig_profile2}(a).
Although indirectly, rather sharp transition from black to gray
is claimed to be related to the phase transition from the deconfined (gluonic) to confined (stringy)
regimes of the Pomeron.
At one hand, its sharpness is surprising because it is associated not with macroscopically large with 
quite small system -- the Pomeron or a pair of strings.
At the other, the analogy basically originates from the phase transition in gluodynamics (strings at early stage are considered excitable but not breakable, so no
quarks yet), which is in fact a very strong first order transition. 

   Micro-to-macro transition in $pA$ and $pp$  collisions, as a function of multiplcity, is a debatable case.
 Data on  the mean $p_\perp$ and slopes shown in Fig.\ref{fig_slopesa} does indicate setting of the
 radial flow: but we know the radial flow can be ``faked". At the same time, the $v_2\{2\}$ as a function
 of multiplicity are rather flat. Its more-particle version $v_2\{n\},n>2$ are show rapid changes but
  better understanding of why it happens is still needed. 
 
  The theoretical justification of good agreement with hydro, inside the kinetics and/or hydrodynamics itself,
 is getting under control.   A number of examples show 
rather effective cancellations of all higher-gradient corrections, leading to surprisingly
accurate Navier-Stokes predictions, even in situations in which one hardly expected it to work.

The low energy scan at RHIC  does show experimental evidences for  ``the softest point".
Yet 
attempts to locate the effects of the QCD critical point   are intriguing
but not yet fully successful. 
Dedicated  measurements of event-by-event fluctuations at freezeout
are already accurate enough to put $T,\mu$ points on the phase diagram, but not yet good enough to discover/disprove the critical point.

 \subsection{Sounds} 
The first triumph of hydrodynamics, at the onset of RHIC program,  was  description of the ``Little Bang" in great details: 
the radial and elliptic flows were obtained as a function of $p_\perp$, centrality, rapidity, particle type and collision energy.
The second one, discussed above in detail, is even more spectacular description of higher azimuthal harmonics of the flow,
with $m=3-6$.  As it has been repeatedly emphasized, those are basically sound harmonics, which should become
even more obvious as $m$ grows. The viscous damping of these modes actually agree with acoustic-inspired formulae
very well.

Another phenomenon, well known for  the Big Bang perturbations, is the ``phase factor". Common freezeout time
for all flows does not imply the same phases, as their frequencies are $m$-dependent: and as the phases rotate
one should see maxima/minima. The only experimental indication for that is enhancement of the triangular flow $m=3$
over the damping curve, and even the elliptic one $m=2$ for the ultra-central bin. 

We emphasized that we only observed harmonics with $m<7$ because the higher ones are damped too much
by the freezeout time. Yet we are quite sure that at the initial time the Glauber model produced equally well
harmonics up to $m=20$ or so, and models like IP glasma predict harmonics by the hundreds. 
Perhaps one can invent other observable manifestations of those modes, such as Magneto-Sono-Luminescence 
process in which they are converted into electromagnetic signals. 

The damping of   harmonics with $m>6$ is also an opportunity to observe the sources of sounds other
than the initial state, in particular from inhomogeneities at the phase transition. (Those are important as seeds of the
Big Bang acoustic cascade with possible observable production of the gravity waves).  

Finally, even in equilibrium there must be fluctuations emitting sounds. Those in particular generate 
nontrivial ``loop corrections"
to hydrodynamics. Observation of ``sound background" in hadronic matter is another challenge 
of the field.

\subsection{The conflicting views of the initial state}

Perhaps the most important conceptual controversy in the field remains the conflicting conclusions
coming from weakly coupled and strongly coupled scenarios of the initial state and equilibration.

Significant progress in
 the theory of weakly coupled initial state is the adoption of the concept of
 {\em turbulent cascade}, with stationary and time-dependent self-similar solutions. Both classical glue simulation and gluon cascades
 came up with out-of-equilibrium attractors possessing power spectra with certain indices, 
which are qualitatively different from the equilibrium.   From practical perspective,
these studies suggest that the stress tensor remains anisotropic for a long time.
However more recent works indicate that nontrivial attarctor solution is only approached if the
coupling is unrealistically small.

Strongly coupled approaches, especially based on AdS/CFT and related models, view 
equilibration as  a process dual to the
{\em gravitational collapse} resulting in
 black hole production in the bulk. As soon as some trapped surface (a black hole) is there, the equilibration is very rapid:
 any kind of ``debries" simply falls into it. Mathematically, the non-hydro modes have imaginary parts
 comparable to the real one, which numerically are quite large (\ref{eqn_isotropization}). So, in this scenario,
 equilibration is extremely rapid:  there are no cascades or even gluon quasiparticles
 themselves, the only light  propagating modes
 are sounds. 
 
Whether  the stress tensor remains anisotropic beyond the short initial period or not 
is still an open question. Theoreticall efforts 
 to combine hydrodynamics with
out-of-equilibrium parameterization of the stress tensor were discussed above,
and they will sure allow to model the situation at any realistic anisotropy.
 In order to decide which picture is correct  one needs to think about experimental observables sensitive to
 the early stage. (My specific proposal -- the dilepton polarization -- has been discussed in section \ref{dilepton_polarization}.)

\subsection{The smallest drops of sQGP }
The major experimental discovery which came from the first years of LHC operation has been collective anisotropies
in  high multiplicity $pA$ and $pp$  collisions. 

 One point of view -- admittedly advocated above -- is that in those cases  there are explosive QGP fireballs.
 While small then
 those produced in AA collisions, they are still ``macroscopically large"  and can be described hydrodynamically.
Strong arguments for this are strong radial and elliptic flows in those systems.

The opposite point of view is that from the smallest to the highest multiplicity binds the $pA$ and $pp$  collisions
produce microscopic systems which can be discussed dynamically, by the same models as used
for minimally biased $pp$. The issue is reduced to ``the shape of Pomeron" problem, and models based 
 on pQCD (BFKL or color glass) or confining one (stringy Pomeron) need to be developed
 much further to see if such hopes are correct. High collectivity of angular anisotropies 
may still be a result of  certain shape of the process. Experiments at RHIC with $d$ and $He^3$ 
beams however disfavor such scenario, in my opinion.

Groups working on both scenarios now try to figure out the limits of their approaches, which is always a good thing.
Inside hydrodynamics, for example, one study higher gradients and their effect.
Inside the string-based picture we discussed a string-string interaction -- ignored for  long time by  event generators --
leading to ``spaghetti collapse" at certain density. 

   Meanwhile, phenomenologists describe the data. 
 Hydrodynamical  treatment of high multiplicity $pA,pp$ events  
 seem to be rather successful: but those require really small initial sizes and high temperatures of the fireball
 produced. But we do not really understand how such systems can be produced. In particular,
 the case of central $pA$ collisions is contested between 
the IP-glasma model and a string-based initial state picture.
So far one has very little theoretical control over the initial state of the high multiplicity $pp$ : in anywhere glasma should be there. It is difficult to study it for statistical reasons, but since this is
the highest density system we have by now, it should be pursued. 

\subsection{Heavy quarks  and quarkonia}
LHC data confirmed what has been already hinted by the RHIC data: significant fraction of the observed charmonia  comes from $recombination$ at the 
chemical freezeout of charm quarks. The ``surviving  charmonia"  fraction continue to be reduced.
   Such major change in charm quark behavior, from ``heavy-like" to "light-quark-like" is clearly an important discovery. 

Let me clarify this last statement. It remains true that
$c,b$ quarks are produced differently from the light ones, namely in the initial partonic processes.  
Yet their interaction with the ambient matter is strong.
 At large $p_t$ we observe quenching $R_{AA}^{c,b}$  comparable to that of gluons/light quarks.
 At small $p_t$ we observe an elliptic flow of open charm and changes in  spectra. 
 
Langevin/Fokker-Planck studies however suggest that $c$ quarks are not moving with the flow. At early time
$c,b$  quarks are produced with large $p_t$ and start decelerating, due to drag, while the matter is slowly accelerating due to pressure gradients:
 their velocities move toward each other, yet they do not match even by the end.  As a result, 
charm radial/elliptic flows are $not$ given by the Cooper-Fry expression. 
The recombinant charmonia  are perhaps an exception: whether those actually co-move with the flow
still needs to be established.
 
 On the theory front, Langevin/Fokker-Planck studies has induced new conceptual developments. In particular,  we discussed
 new set of solutions of those for charmonia, the quasi-equilibrium attractors with constant particle flux. Those states, not the original bound states like $J/\psi,\psi'$ etc, provide a convenient basis for evaluation of 
 the speed of relaxation
  and out-of-equilibrium corrections to current charm hadronization models.

 \subsection{Jets}


The theory of hard processes -- jets, charm/bottom production -- were based on factorization
theorems and a concept of structure functions. It is a solid foundation, but a very restrictive one.
When one  asks  questions about say jets in high multiplicity bins of $pp$ collisions,
 one  soon realizes   the ``corresponding structure functions" do not exist:  that concept
has only been defined for the ``untouched proton" in inclusive setting. Universal  
 structure functions,  measured rather than calculated, had served us since 1970's, but now they cannot
 be used anymore.  If certain fluctuation of a nucleon is
selected, new models and measurements are needed. 

Of course, there are practical limits: hard processes costs several orders of magnitude, and central $pA$ costs
also about factor 20-100 down the probability. Yet high LHC luminosity  plus specialized triggers should be enough to get to some of those issues in the near future. 

Jet quenching in central $pA$ remains to be understood. Scaling arguments, like the ones we used for
hydro in smaller-but-hotter systems, can and should be developed and confronted with data.

We argued above that in AA collisions jet quenching parameter  $\hat{q}$
seem to be strongly enhanced at the near-$T_c$ region. Small systems 
evolution and freezeout is very different: this should play a role in jet quenching.  


{\bf Acknowledgements.}
This paper is a summary from multiple conversations with colleagues, at seminars, workshops and conferences.
They are too many to attempt to name them here (see long list of references below), but 
still I need to thank them for patiently teaching me about this or that idea or experimental findings. 
This work was supported in part by the U.S. Department of Energy, Office of Science, under Contract No. DE-FG-88ER40388.

\appendix

 \section{Bjorken flow}
 The idea of rapidity-independent ``scaling" distribution of secondaries
 originates from Feynman's early discussion of the parton model, around 1970.
The existence  of rapidity-independent hydro solution was perhaps first noticed by Landau, who used rapidity
 variable in his classic paper, as a somewhat trivial case.  The
  space-time picture connected with such  scaling  regime  was
 discussed in refs  \cite{preB1,preB2} before Bjorken's famous paper  \cite{Bjorken} 
in which the solution was spelled out explicitly.
 
  It is instructive first to describe it in the original Cartesian coordinates.
 There is no dependence on transverse coordinates $x,y$, only on time $t$ and longitudinal coordinate $z$.
  The 1+1d equations $\partial_\mu T^{\mu\nu}=0$  can be re-written
in the following way 

\be \label{eq_entropy_1d} 
{\partial \over \partial t}(s \, \cosh y)+ 
{\partial \over \partial z}(s \,\sinh y)=0\ee
\be  \label{eq_pressure_1d}
{\partial \over \partial t}(T \,\sinh y)+ 
{\partial \over \partial z}(T \, \cosh y)=0\ee
where $u_\mu=(cosh(y),sinh(y)$,
and $T,s$ are the temperature and the energy density.
The first equation manifests the entropy conservation.

The central point is the  1-d-Hubble  ansatz
for the 4-velocity 
\be  u_\mu=(t,0,0,z)/\tau \ee
where $\tau^2=t^2-z^2$ is the proper time. 
Note that 
 all  volume  elements  are
 expanded linearly with time  and  move  along   straight
 lines from the collision point. The spatial $\eta=\tanh^{-1}(z/t)$
 and the momentum rapidities
$y=\tanh^{-1} v$ are just equal to each other.
 Exactly as in the Big Bang, for each "observer" ( the volume
 element ) the picture is just the same,  with  the  pressure
 from the left  compensated  by  that  from  the  right.  The
 history is also the same for all volume
elements, if  it  is  expressed  in  its  own
 proper time $\tau$. Thus one has $s(\tau),T(\tau)$.
 Using this ansatz,  the
 entropy conservation  
becomes an ordinary differential equation in proper time $\tau$
\be {ds(\tau) \over d\tau}+{s\over \tau}=0 \ee
with an obvious solution 

\be s={const \over \tau}\ee
   So far all dissipative phenomena were ignored. Including
   first dissipative terms into our equations one finds the following source for the entropy current
  \be 
{1 \over\epsilon+p} {d\epsilon \over d\tau} ={1\over s}{ds\over d\tau}=-{1 \over
   \tau}\left(1-{(4/3)\eta+\xi \over (\epsilon+p)  \tau}\right)
      \ee
with shear and bulk viscosities $\eta,\xi$, which tells us that
one has to abandon ideal hydrodynamics at sufficiently early time.

Alternatively, one can start with curved coordinates $\tau,\eta$ from the beginning, and look for $\eta$-independent solution. Those are co-moving coordinates, in those  $u_\mu=(1,0,0,0)$ but the equations obtain
extra term from Christoffel symbols.

 \section{Gubser flow} \label{seq_Gubser} 
The Gubser flow \cite{Gubser:2010ze,Gubser:2010ui} is a solution which keeps the boost-invariance and the
axial symmetry in the transverse plane of the Bjorken flow, but replaces the
translational invariance in the transverse plane
 by symmetry under special conformal transformation.
Therefore, one restriction is that the matter is required to be conformal, with the EOS
$ \epsilon=3p $. Another is that the colliding systems has to be of a particular shape, corresponding to
conformal map of the sphere onto the transverse plane.
 
The ideal hydro solution has three parameters:  One is dimensional  $q$, it
 defines the size of the system (and is roughly corresponding to the radii of
the colliding nuclei). The other two are dimensionless, $f^*$ 
  characterizes the number of degrees of freedom in the matter, and $\hat \epsilon_0$ the amount
of entropy in the system. 

The original setting uses the coordinates we used above, 
  the  proper time -spatial rapidity - transverse radius - azimuthal angle
 $(\bar{\tau},\eta,\bar{r},\phi)$  with the metric
\begin{eqnarray}
ds^2 & = & -d\bar{\tau}^2 + \bar{\tau}^2 d\eta^2 + d\bar{r}^2 +\bar{r}^2d\phi^2,
\end{eqnarray}
The dimensionless coordinates $\bar \tau=q\tau ,\bar r=q r$ are rescaled versions
of the actual coordinates. 
 
 Looking for solutions independent on both ``angles" $\eta,\phi$ and using transverse rapidity 
\begin{eqnarray}
u_{\mu} &  = &
\left(-\cosh{\kappa(\tau,r)},0,\sinh{\kappa(\tau,r)},0\right)
\end{eqnarray}
Gubser obtained the following solution
\begin{eqnarray} \label{G_v}
v_\perp & = & \tanh{\kappa(\tau,r)}  =  \left(\frac{2q^2\tau
r}{1+q^2\tau^2 + q^2r^2}\right)
\end{eqnarray}
\begin{eqnarray}   \label{G_energy}
\epsilon & = & \frac{\hat{\epsilon}_0 (2
q)^{8/3}}{\tau^{4/3}\left(1+2q^2(\tau^2 +
r^2)+q^4(\tau^2-r^2)^2\right)^{4/3}}
\end{eqnarray}
where $\hat{\epsilon}_0 $ is the second parameter.
In \cite{Gubser:2010ui} Gubser and Yarom re-derived the same
solution by going into the co-moving frame. In order to do so they
rescaled the metric 
\begin{eqnarray}
ds^2 & = & \tau^2 d\hat{s}^2
\end{eqnarray}
and performed  a coordinate transformation from the $\tau,r$ to a new set $\rho,\theta$ given by:
\begin{eqnarray}
\sinh{\rho} & = & -\frac{1-q^2\tau^2+q^2r^2}{2q\tau}\label{rho_coord}\\
\tan{\theta} & = &
\frac{2qr}{1+q^2\tau^2-q^2r^2}\label{theta_coord}
\end{eqnarray}

In the new coordinates the rescaled metric reads:
\begin{eqnarray}
d\hat{s}^2 & = &-d\rho^2 + \cosh^2{\rho}\left(d\theta^2 +
\sin^2{\theta}d\phi^2\right)+d\eta^2
\end{eqnarray}
and we will use $\rho$ as the ``new time" coordinate and $\theta$ as a
new ``space" coordinate. In the new coordinates the fluid is at
rest.

The relation between the velocity  in Minkowski space in the
$(\tau, r,\phi,\eta)$ coordinates and the one in the rescaled
metric in $(\rho,\theta,\phi,\eta)$ coordinates corresponds to:
\begin{eqnarray}
u_{\mu} & = & \tau \frac{\partial \hat{x}^{\nu}}{\partial
\hat{x}^{\mu}}\hat{u}_{\nu}\, , \label{u_transf}
\end{eqnarray}
while the  energy density transforms as:
$\epsilon=\tau^{-4}\hat{\epsilon}$.

The  temperature (in the rescaled frame, $\hat{T}=\tau f_*^{1/4}T$, with $f_*=\epsilon/T^4=11$ as in \cite{Gubser:2010ze}) is now dependent only on the
new time $\rho$, in the case with  nonzero viscosity the solution is
\footnotesize
\begin{eqnarray}
\hat{T} & = &\frac{\hat{T}_0}{(\cosh{\rho})^{2/3}} +\frac{H_0
\sinh^3{\rho}}{9 (\cosh{\rho})^{2/3}} \,
_2F_1\left(\frac{3}{2},\frac{7}{6};\frac{5}{2},-\sinh^2{\rho}\right)\nonumber\\
\label{back_T}
\end{eqnarray}\normalsize
where $H_0$ is a dimensionless constant made out of the shear
viscosity and the temperature, $\eta = H_0 T^{3}$ and $_2F_1$ is
the hypergeometric function. In the inviscid case the solution is
just the first term of expression (\ref{back_T}),
and of course it also conserves the entropy in this case.
The picture of the explosion is obtained by transformation from this expression back to $\tau,r$ coordinates.

Small  perturbations to the Gubser flow obey linearized equations which have also been derived in \cite{Gubser:2010ui}. We start with  the zero viscosity case, so that the background temperature (now to be called $T_0$) will be given by just the first term in (\ref{back_T}). The perturbations over the previous solution are defined by
\begin{eqnarray}
\hat{T} & = &  \hat{T}_0(1+\delta)\label{Tpertb}\\
u_{\mu} & = & u_{0 \,\mu} + u_{1\mu}\label{upert}
\end{eqnarray}
with
\begin{eqnarray}
\hat{u}_{0 \,\mu} & = & (-1,0,0,0)\\
\hat{u}_{1\mu} & = & (0,u_{\theta}(\rho,\theta,\phi),u_{\phi}(\rho,\theta,\phi),0)\\
\delta & = & \delta(\rho,\theta,\phi)
\end{eqnarray}

Plugging expressions (\ref{Tpertb}),(\ref{upert}) into the
hydrodynamic equations and only keeping linear terms in the
perturbation, one can get a system of coupled 1-st order
differential equations. Furthermore, if one ignores the viscosity terms, one may exclude velocity and
get the following (second order) closed equation for the
temperature perturbation. 
\begin{eqnarray}
& &\frac{\partial^2 \delta}{\partial \rho^2} -
\frac{1}{3\cosh^2{\rho}} \left( \frac{\partial^2 \delta}{\partial
\theta^2}  +\frac{1}{\tan{\theta}}\frac{\partial \delta}{\partial
\theta}+ \frac{1}{\sin^2{\theta}}\frac{\partial^2 \delta}{\partial
\phi^2}
\right)  \nonumber\\
& & +\frac{4}{3}\tanh{\rho}\frac{\partial \delta}{\partial \rho}=0
\label{T_pert_eqn}
\end{eqnarray}
(Since the initial perturbations are
assumed to be rapidity-independent, we also ignored this coordinate here.)

It has a number of remarkable properties: all 4 coordinates can be separated $\delta(\rho,\theta,\phi)=R(\rho)\Theta(\theta)\Phi(\theta)$ and a general solution is given by
\begin{eqnarray}
R(\rho) & = & {C_1 \over (\cosh{\rho})^{2/3} }   
   P_{-\frac{1}{2} +\frac{1}{6}\sqrt{12\lambda+1}}^{2/3}(\tanh{\rho}) \nonumber \\ 
&&+ {C_2\over (\cosh{\rho})^{2/3} } 
Q_{-\frac{1}{2} +
\frac{1}{6}\sqrt{12\lambda+1}}^{2/3}(\tanh{\rho}  \nonumber\\
\Theta(\theta) & = & C_3P_l^m(\cos{\theta})+C_4Q_l^m(\cos{\theta})\nonumber\\
\Phi(\phi) & = & C_5 e^{im\phi} + C_6 e^{-im\phi}\label{exact}
\end{eqnarray}
\noindent\normalsize where $\lambda=l(l+1)$ and P and Q are
associated Legendre polynomials.  The part of the solution
depending on $\theta$ and $\phi$ can be combined in order to form
spherical harmonics $Y_{lm}(\theta, \phi)$, such that
$\delta(\rho,\theta,\phi)\propto R_l(\rho)Y_{lm}(\theta,\phi)$.
This property should have been anticipated, as one of the main
ideas of Gubser has been to introduce a coordinate which together
with $\phi$ make a map on a 2-d sphere.

Gubser setting was used as a theoretical laboratory ever since.
A complete Green function has been constructed \cite{Staig:2011wj}, leading to 
pictures of sound circles we discussed at the beginning of this review.  Generalization to perturbations by the quenching jets,
with the sounds propagating in the rapidity direction, was done in \cite{Shuryak:2013cja}.
 For the second order (the Israel-Stuart version) of the hydrodynamics 
in it has been done in \cite{Marrochio:2013wla,Pang:2014ipa}. Recently the Boltzmann equation in tau-approximation has also been solved in it, see
\cite{Denicol:2014tha} discussed in section  \ref{sec_anisotropy}.

There are also a number of phenomenological applications. Without going into those, let me just comment that
those are limited by the fact that at large $r$ the power tail of the solution is completely
inadequate for heavy ion collisions. So to say, Gubser solution is like an explosion in atmosphere,
while the real ones are in vacuum. As a result, in applications one basically has to amputate
the unphysical regions.  


\end{document}